\documentclass[5p,twocolumn,authoryear]{elsarticle}

\usepackage{amsmath}            
\usepackage{epstopdf}           
\usepackage{flushend}    

\usepackage{xcolor}		
\usepackage{hyperref}
\hypersetup{
    colorlinks = true,
    linkcolor = magenta,
    urlcolor  = blue,
    citecolor = purple,
    anchorcolor = black,
}   

\usepackage{multirow}
\usepackage[normalem]{ulem}

\usepackage{amssymb}

\usepackage{refcount}
\usepackage{ifthen}
\newcommand{\linerange}[2]{%
\ifthenelse{\equal{\getrefnumber{#1}}{\getrefnumber{#2}}}{%
line \ref{#1}%
}{%
lines \ref{#1}--\ref{#2}%
}%
}

\usepackage[switch]{lineno} 

\biboptions{numbers,square,sort&compress}

\usepackage[figuresright]{rotating}

\usepackage{subfigure}

\usepackage{custom_macros}

\begin{document}

\begin{frontmatter}

\title{
Emergent behavior and neural dynamics \\
in artificial agents tracking turbulent plumes
}

\author[ECE]{Satpreet H. Singh\corref{cor1}}
\ead{satsingh@uw.edu}
\cortext[cor1]{Author for correspondence}
\author[UNR]{Floris van Breugel}
\author[CSE,ECE,CNT]{Rajesh P. N. Rao}
\author[Bio,eScience]{Bingni W. Brunton}

\address[ECE]{Department of Electrical and Computer Engineering, University of Washington, Seattle, USA.}
\address[UNR]{Department of Mechanical Engineering, University of Nevada, Reno, USA.}
\address[CSE]{Paul G. Allen School of Computer Science and Engineering, University of Washington, Seattle, USA.}
\address[CNT]{Center for Neurotechnology, University of Washington, Seattle, USA.}
\address[Bio]{Department of Biology, University of Washington, Seattle, USA.}
\address[eScience]{eScience Institute, University of Washington, Seattle, USA.}

\begin{abstract}
\linenumbers
\begin{linenumbers}
Tracking a turbulent plume to locate its source under variable wind and plume statistics is a complex task; flying insects routinely accomplish such tracking, often over long distances, in pursuit of food or mates.
Several aspects of this remarkable behavior and its underlying neural circuitry have been studied experimentally. 
Here, we take a complementary \emph{in silico} approach to develop an integrated understanding of behavior and neural computations.
Specifically, we train artificial recurrent neural network (RNN) agents using deep reinforcement learning (DRL) to locate the source of simulated turbulent plumes.
Interestingly, the agents' emergent behaviors resemble those of flying insects, and the RNNs learn to compute task-relevant variables with distinct dynamic structures in population activity.
Our analyses put forward a testable behavioral hypothesis for tracking plumes in changing wind direction, and we provide key intuitions for memory requirements and neural dynamics in turbulent plume tracking. \\


\noindent \textit{One sentence summary}: 
Artificial neural network agents provide key intuitions for
behaviors and neural computations that support plume tracking
\end{linenumbers}
\end{abstract}

\end{frontmatter}

\section{Introduction}
Locating the source of an odor in a windy environment is a challenging control problem, where an agent must act to correct course in the face of intermittent odor signals, changing wind directions, and the variability in odor plume shape \citep{reddyannrev,celani2014odor}.
Moreover, an agent tracking an intermittent plume needs memory, where current and past egocentric 
odor, visual, and wind sensory signals must be integrated to determine the next action.
For flying insects, localizing the source of odor plumes emanating from potential food sources or mates is critical for survival and reproduction.
Therefore, many aspects of their plume tracking abilities have been experimentally studied in great detail \citep{baker2018algorithms, park2016neurally, carde2008navigational,currier2020multisensory}.
However, most such studies are limited to one or two levels of analysis such as behavior \citep{van2008insects}, computation \citep{lochmatter2009theoretical,pang2018history} or neural implementation \citep{sun2018analysis}.

Despite the wide adoption of wind tunnel experiments to study odor plume tracking \citep{van2014plume}, generating controlled dynamic turbulent plumes and recording flight trajectories at high resolution is expensive and laborious.
Exciting alternative approaches have been developed using virtual reality \citep{kaushik2020characterizing}
and kilometer-scale outdoor dispersal experiments \citep{leitch2020long}. 
While behavioral experiments are now tractable, collecting significant neural data during free flight in small insects remains technologically infeasible, and larger insects require larger wind tunnels.
Here we are motivated to take a complementary \textit{in silico} approach using artificial recurrent neural network (RNN) agents trained to track simulated turbulent plumes, with the goal of developing an integrated understanding of the behavioral strategies and the associated neural computations that support plume tracking.

\begin{figure*}[h!]
\centering
\includegraphics[width=1.0\linewidth]{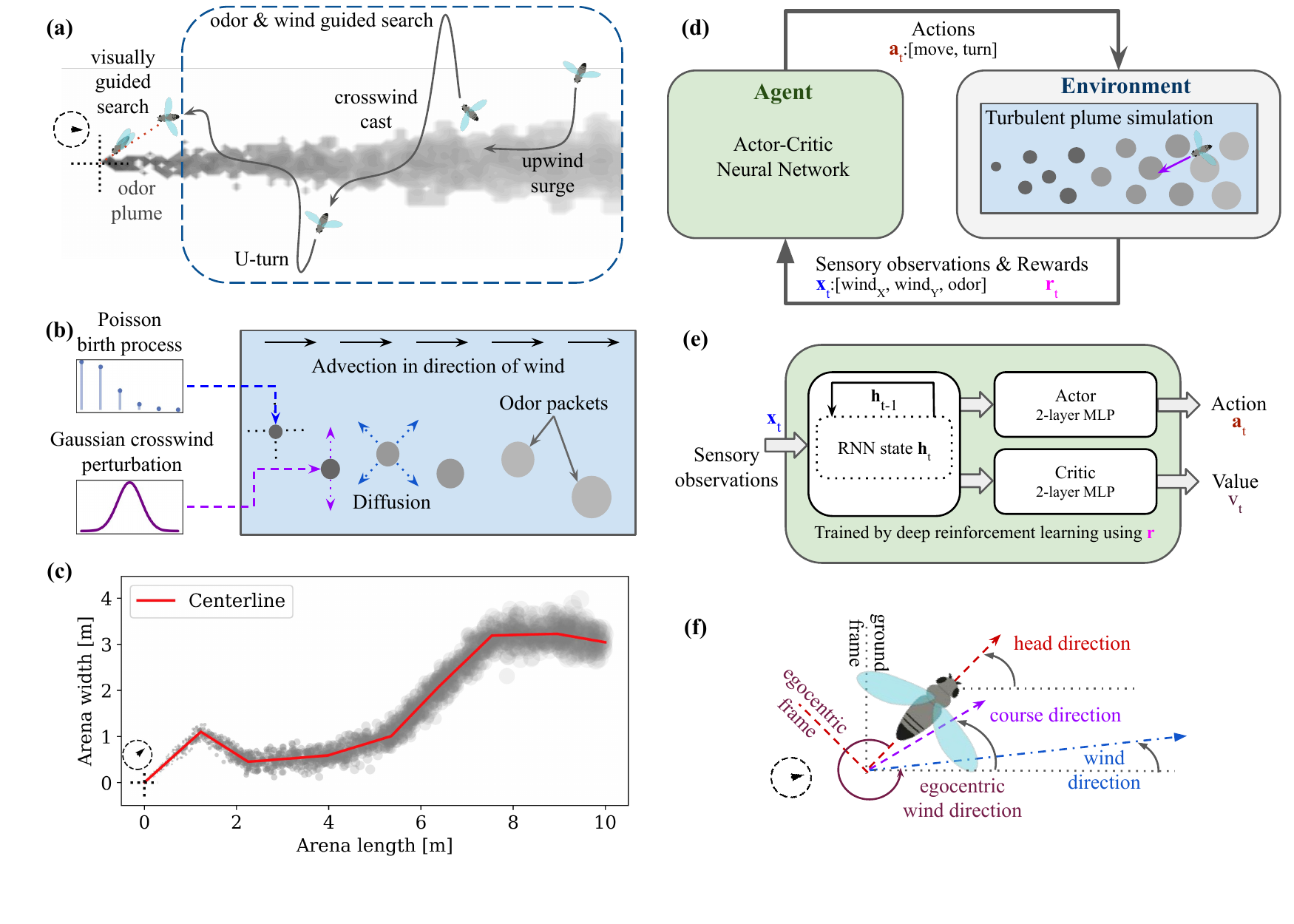}
\caption{
\textbf{Training artificial agents to track turbulent plumes with deep reinforcement learning.}
    \textbf{(a)} 
    A schematic of a flying insect performing a plume tracking task, showing \textit{upwind surges}, \textit{crosswind casts}, and \textit{U-turns} behaviors (inspired by a figure in \cite{baker2018algorithms}).
    In this work, we model the spatial scale (dashed rectangle) where the insect can use only olfactory and mechanosensory (to sense wind velocity) cues for plume tracking.
    \textbf{(b)} 
    The plume simulator models stochastic emission of odor packets from a source carried by wind. 
    Odor packets are subject to advection by wind, random cross-wind perturbation, and radial diffusion.
    \textbf{(c)}
    An example of a plume simulation where the wind direction changed several times.
    \textbf{(d)} 
    A schematic showing how the artificial agent interacts with the environment at each time step. 
    The environment model determines the sensory observations available to the agent $\mathbf{x}$ (egocentric wind direction vector and local odor concentration) and the rewards used in training.
    The agent navigates within the environment with actions $\mathbf{a}$ (turn direction and magnitude of movement).
    \textbf{(e)} Agents are modeled as neural networks and trained by deep reinforcement learning (DRL). 
    A recurrent neural network (RNN) generates an internal state representation from sensory observations, followed by parallel Actor and Critic heads that implement the agent's control policy and predict the state values, respectively.
    The Actor and Critic heads are 2-layer, feedforward multi-layer perceptron (MLP) networks.
    \textbf{(f)} 
    A schematic showing a flying agent's head-direction, course-direction, and the wind direction, all measured with respect to the ground and counter-clockwise from the x-axis.
    Course direction is the direction that the agent actually moves in, accounting for the effect of the wind on the agent's intended direction of movement (head-direction).
    Egocentric wind direction is the direction of the wind as sensed by the agent and is measured counter-clockwise with respect to the agent's current head-direction. 
    }
\label{fig_training}
\end{figure*}

In recent years, artificial neural networks (ANNs) have gained increasing popularity for modeling and understanding aspects of neural function and animal behavior \citep{kietzmann2019deep, cichy2019deep}, including vision \citep{kriegeskorte2015deep}, movement \citep{sussillo2015neural}, navigation   \citep{kanitscheider2017training, cueva2018emergence, cueva2019emergence, haesemeyer2019convergent}, and collective behaviors \citep{verma2018efficient}. 
Whereas many ANNs have been trained using supervised approaches that rely on labeled training data, an alternative emerging algorithmic toolkit known as deep reinforcement learning (DRL) has made it computationally tractable to train ANN agents (Figure \ref{fig_training}d).
In particular, an ANN agent receives sensory observations and task-aligned rewards based on its actions at each step and tries to learn a strategy for its next actions to maximize total expected reward  \citep{arulkumaran2017deep,sutton2018reinforcement}.
Such learning and optimization based models are \textit{normative} in the sense that they can prescribe how a neural system \textit{should} behave, rather than describing how it has been observed to behave.
As neuroscience moves towards studying increasingly naturalistic behaviors \citep{nastase2020keep,sonkusare2019naturalistic,huk_beyond_2018,GOMEZMARIN201925}, such normative approaches are gaining traction as tools to gain insight, rapidly explore hypotheses, and generate ideas for theoretical development  \citep{richards2019deep,le2020towards,merel2019deep,ahrens2019zebrafish,banino2018vector,colabrese2017flow,verma2018efficient}. 

Flying insects search for sources of odor using several strategies, depending on the spatial scale being considered and odor source visibility \citep{baker2018algorithms} (Figure \ref{fig_training}a). 
Close to the odor source, insects can fly to the source guided by vision.
At longer ranges (from several meters up to about $100$ meters \citep{wall1987range}) or when the odor source is not yet visible, their search must be guided by olfaction to detect odors and mechanosensation to estimate wind velocity.
At this larger scale, there are a few stereotyped behavioral sequences that are known to be important for plume tracking: 
\textit{upwind surges} when the insect can sense the odor, and 
\textit{crosswind casts} and \textit{U-turns} to locate the plume body when the insect loses the odor scent \citep{carde2008navigational}.
Here we focus on this larger-scale odor and wind guided regime, where agents have access to only mechanosensory and olfactory cues.

\begin{figure*}[h!]
\begin{center}
\includegraphics[width=1.0\linewidth]{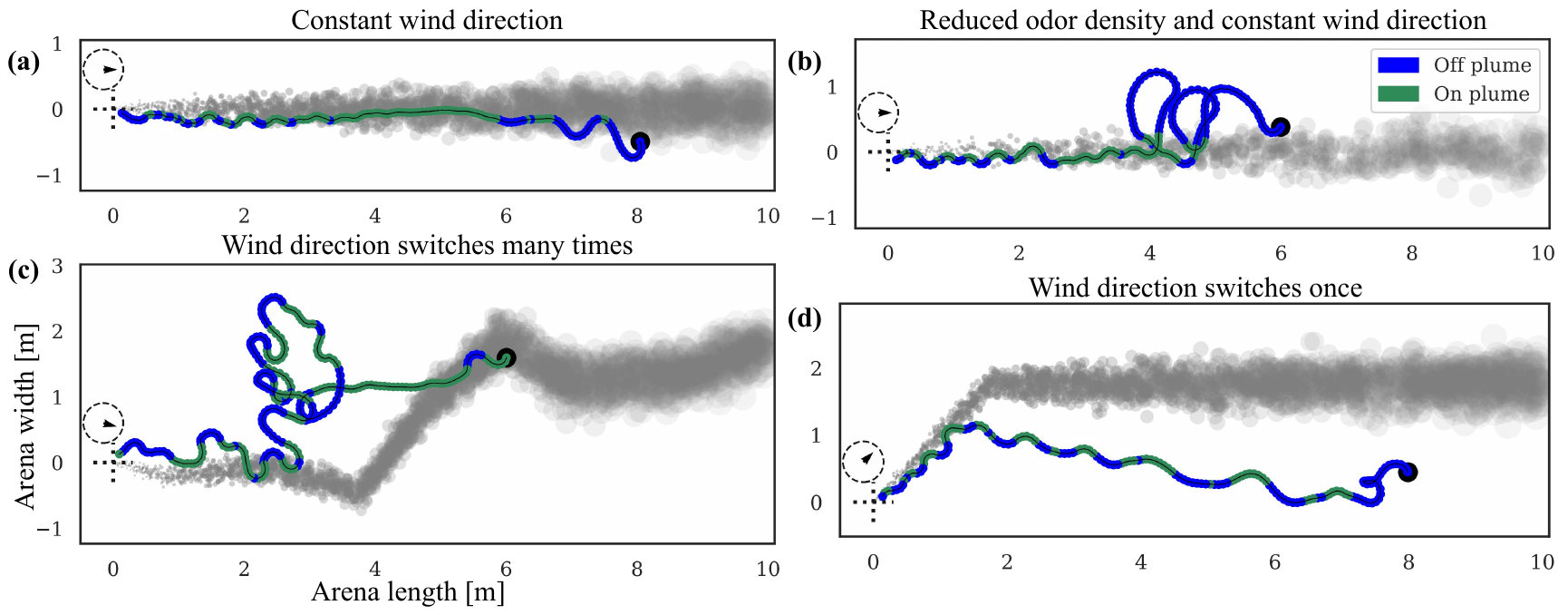}
\caption{
\textbf{Examples of successful plume tracking trajectories under various plume simulator configurations}.  
The odor plumes (grey) are overlaid with RNN agent trajectories, which are colored by whether the agent was able to sense the presence (green) or absence (dark blue) of odor. 
Importantly, because the plume simulation models stochastic odor packets, the agent encounters odors only intermittently.
Trajectories start at filled black circle and end at odor source indicated by dotted cross-hairs in the lower left of each sub-figure.
The plume visualizations are from the end of the tracking episode and thus deviates from the plume as observed by the agent during the episode.
The arrow within dotted circle above the cross-hair shows direction of wind at time of snapshot.
All examples use a 0.5 m/s wind. 
The different plume configurations (Section \ref{sec_methods_plume}) are:
\textbf{(a)} `constant' left-to-right wind direction plume. 
\textbf{(b)} `sparse' plume with same left-to-right constant wind direction but reduced (0.4x) birthrate of odor packets.
\textbf{(c)} `switch-many' plume with wind direction switches occurring every $\sim 3s$ 
\textbf{(d)} `switch-once' plume, which makes one 45$^{\circ}$ counter-clockwise wind direction switch during the tracking episode. 
Supplementary animations provide additional examples of successful and unsuccessful tracking episodes:
\url{https://github.com/BruntonUWBio/plumetracknets}.
}
\label{fig_behavior_qual}
\end{center}
\end{figure*}

In this paper, we describe behaviors that emerge in RNN agents trained to track odors in a flexible plume simulation and analyze the neural dynamics that underlie these behaviors.
At a behavioral level, we find that the agents' actions can be summarized by several  modules that closely resemble 
those observed in flying insects (Section \ref{sec_behavior_qual}).
While odor plumes that do not change in direction can be tracked using a few steps of history, longer timescales of memory are essential for plumes that are non-stationary and change directions unpredictably.
Interestingly, the learned tracking behavior of RNN agents in non-stationary plumes suggests a testable experimental hypothesis:  that tracking is accomplished through local plume shape rather than wind direction (Section \ref{sec_centerline}).
The RNNs learn to represent variables known to be important to flying insect navigation, such as head direction and time between odor encounters (Section \ref{sec_repr}).
Further, the low-dimensional neural activity associated with the emergent behavior modules is structured into two distinct regimes (Section \ref{sec_dynamics}), and transitions between these regimes are asymmetric in duration (Section \ref{sec_ttcs}).

\section{Related work}
In the field of neural computation, an emerging body of work has used DRL to train ANNs that solve tasks closely inspired by tasks from neuroscience.
For instance, agents have been trained to study learning and dynamics in the motor-cortex \citep{weinstein2017structure, song2020deep}, time encoding in the hippocampus \citep{lin2021time}, reward-based learning and meta-learning in the pre-frontal cortex \citep{song2017reward, wang2018prefrontal, botvinick2019reinforcement}, and task-associated representations across multiple brain areas \citep{cross2021using}.
There have been several recent perspectives articulating the relevance of this emerging algorithmic toolkit to neuroscience \citep{botvinick2020deep,gershman2020neurobiology} and ethology \citep{crosby2020building}.

\begin{figure*}[h!]
\begin{center}
\includegraphics[width=1.0\linewidth]{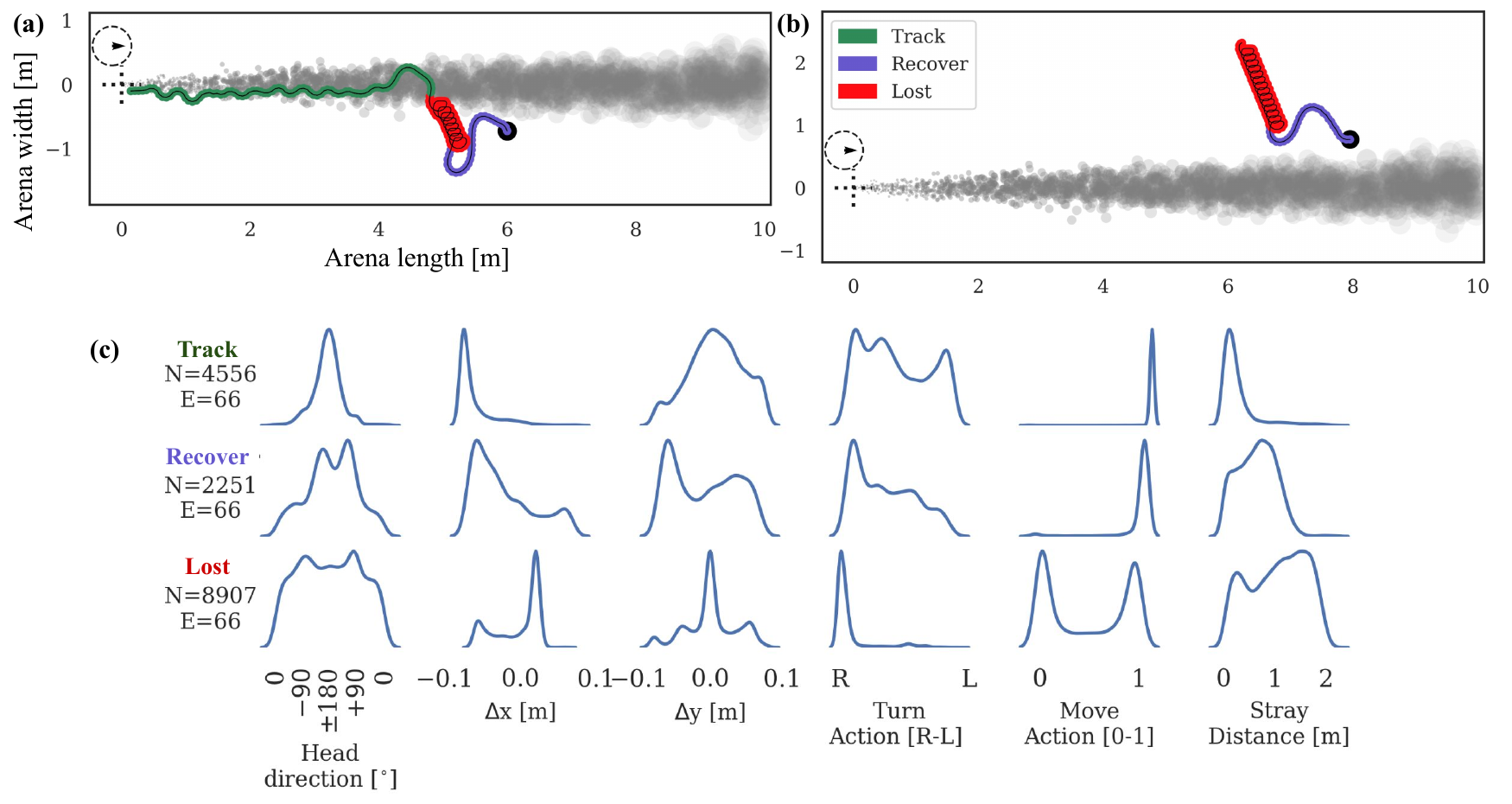} 
\caption{
\textbf{Emergent plume tracking behavior can be decomposed into behavior modules.}
\textbf{(a \& b)} 
Trajectories for a successful \textbf{(a)} and an unsuccessful \textbf{(b)}
plume tracking episodes showing three distinct behavior modules: \textit{tracking} (green), \textit{lost} (red) and \textit{recovering} (purple-blue)
\textbf{(c)}
Kernel density estimates show data aggregated from an equal number of successful and unsuccessful constant wind direction plume tracking episodes (N timesteps, E episodes). 
Plots reveal differences between the three behavior modules across key behavioral measures:
\textbf{Head-direction}: 
Head-direction densities are concentrated around $\pm 180^{\circ}$, a signature of zig-zagging but mostly upwind movement; 
the concentration of density around the upwind direction reduces from tracking to recover to lost, accounting for the more complex trajectories encountered in the latter two behavior modules.
(angle measured counterclockwise with 0$^{\circ}$ indicating directly downwind).
\textbf{$\mathbf{\Delta x}$ and $\mathbf{\Delta y}$}: Density estimates for drift in the x-direction ($\Delta x$) and y-direction ($\Delta y$) per timestep show how tracking is characterized by primarily upwind (negative x-direction) movement in both \textit{tracking} and \textit{recover} modules, but lesser so in the \textit{lost} module.
Y-direction movements are significant in the \textit{tracking} and \textit{recovering} modules, corresponding to more complex turning behaviors, but minimal in the \textit{lost} module.
\textbf{Turn action}: Left/right turning movements are balanced in the \textit{tracking} module as the agent closely tracks the edge of the plume, but is biased towards clockwise movements in the other two modules, especially the \textit{lost} module.
\textbf{Move action}: The agent significantly modulates its forward movement speed in the \textit{lost} module only.
\textbf{Stray distance}: The agent strays from the plume minimally in the \textit{tracking} module, but significantly otherwise.   
See  \ref{sec_supp_module} for equivalent plots for other agents.
}
\label{fig_behavior_modules}
\end{center}
\end{figure*}

Our work is most directly related to three recent research efforts.
\cite{merel2019deep} developed a virtual-reality model of a rodent embodied in a skeleton body and endowed with a deep ANN `brain.' 
They trained this model using DRL to solve four tasks and then analyzed the virtual rodent's emergent behavior and neural activity, finding similarities at an abstract level between their agent and observations from rodent studies. 
\cite{reddy2021sector} studied the trail tracking strategies of terrestrial animals with one (e.g. one antenna) or two (e.g. two nostrils) odor sensors. 
They found that RL agents trained on simulated trails recapitulate the stereotypical zig-zagging tracking behavior seen in such animals. 
Using a static trail model and an explicit (not neural) probabilistic model for sensory integration, they studied the effect of varying agent and task parameters on the emergent stereotypical zig-zagging behavior.     
\cite{rapp2020spiking} used a biologically detailed spiking neural circuit model of a fly mushroom body to study sensory processing, learning, and motor control in flying insects when foraging within turbulent odor plumes. 

We build on the approach of these recent papers that study artificial agents solving neural inspired tasks, and our work is also distinct in several key ways. 
First, we simulate a more computationally challenging task than those tackled in \cite{reddy2021sector} and \cite{rapp2020spiking}, because our odor environment is configurable, dynamic, and stochastic.
Second, we have made several simplifications and abstractions that make analysis more tractable, so that we may focus on the general principles behind plume tracking.
Specifically, we omit biomechanical details, impose no biologically inspired connectivity constraints, and do not use spiking neurons.
Instead, our networks are `Vanilla' RNNs (rather than the gated RNNs used in \cite{merel2019deep} or the spiking neurons in \cite{rapp2020spiking}), which facilitates analyses from the dynamical systems perspective \citep{rajan2006eigenvalue,sussillo2013opening,maheswaranathan2019reverse,maheswaranathan2019universality,vyas2020computation}.
We analyze emergent behaviors and neural dynamics at the network level, which provides us with an abstract understanding of task-relevant neural computations that is robust to small changes in network architecture and training hyperparameters \citep{vyas2020computation, maheswaranathan2019universality, sussillo2013opening}.
Lastly but importantly, since we do not model vision or joint-level motor control as in \cite{merel2019deep}, our neural networks are simpler and can be trained on a computational budget accessible to an academic lab.

\section{Training artificial agents to track turbulent plumes}
\label{sec_methods}

Here we describe how we use DRL to train RNN agents that can track simulated turbulent odor plumes.
Training episodes situate agents at random initial locations within plumes that switch directions multiple times during the course of the episode.  
Agents are actor-critic neural networks that receive sensory observations as inputs---namely, egocentric instantaneous wind velocity and local odor concentration.
Importantly, since the plume simulator models diffusing and advecting odor packets, the agent's encounters with odor packets are intermittent and stochastic.
To train our agents to solve this task where both the observation space and action space are continuous valued, we use the Proximal Policy Gradient (PPO) \cite{schulman2017proximal} algorithm.
We use a reward function that primarily rewards homing in on the odor source and augment that with intermediate rewards for actions that reduce distance to the odor source.
For evaluation, we assess trained agents on additional simulations where the odors are relatively sparse and the wind switches direction at different frequencies. 
More details of our implementation are available in \ref{sec_supp_train_eval} and \ref{sec_supp_hyperparams}.

\subsection{Plume simulation}
\label{sec_methods_plume}
We implement a particle-based two-dimensional plume simulation model (Figure \ref{fig_training}f) that replicates the statistics of real-world turbulent plumes \citep{farrell2002filament}.
This type of simulation has been used in a wide range of domains including olfactory navigation \citep{carde2008navigational}, robotics \citep{kowadlo2008robot} and sensor networks \citep{michaelides2005plume}. 
The simulator (Figure \ref{fig_training}b) comprises a spatially homogeneous wind vector-field (0.5 m/s with configurable direction) and an odor source located at the origin that emits odor puffs as a Poisson process.
Puffs are initialized with a fixed initial radius and undergo radial diffusion.
In addition, each emitted puff is advected downwind at the wind velocity and perturbed randomly by crosswind translation.
In other words, each puff effectively performs a biased random walk downwind over time, while diffusing in concentration spatially.
Our simulated plumes and agents are constrained to two dimensions for simplicity of analysis.
The dimensions of the simulated arena are $[-2m,+10m]$ and $[-5m, +5m]$ in the x and y coordinates respectively, totaling a $120m^2$ arena.
Plumes are simulated at 100 iterations/second. 
The plume's centerline is obtained by simulating puffs that have no random crosswind translation at each iteration (Figure \ref{fig_training}f).  

We simulate the following four wind configurations. 
First, the wind direction is held constant ($0^{\circ}$) throughout the simulation (`constant').
Second, the wind direction makes one 45$^{\circ}$ counter-clockwise switch during a tracking episode (`switch-once'). 
Third, the wind direction switches at multiple random times during a tracking episode (`switch-many').
Each wind direction turn is a random draw from a Gaussian distribution with mean $0$ and 45$^{\circ}$ s.d., truncated at $\pm 60^{\circ}$, and occurs approximately every 3 seconds.
Fourth, the wind direction is held constant, but the puff birth-rate is reduced (0.4x) compared to the 'constant' configuration (`sparse').
See \ref{sec_supp_train_eval} for further details on the plume simulation.

\begin{figure*}[h!]
\centering
\includegraphics[width=1.0\linewidth]{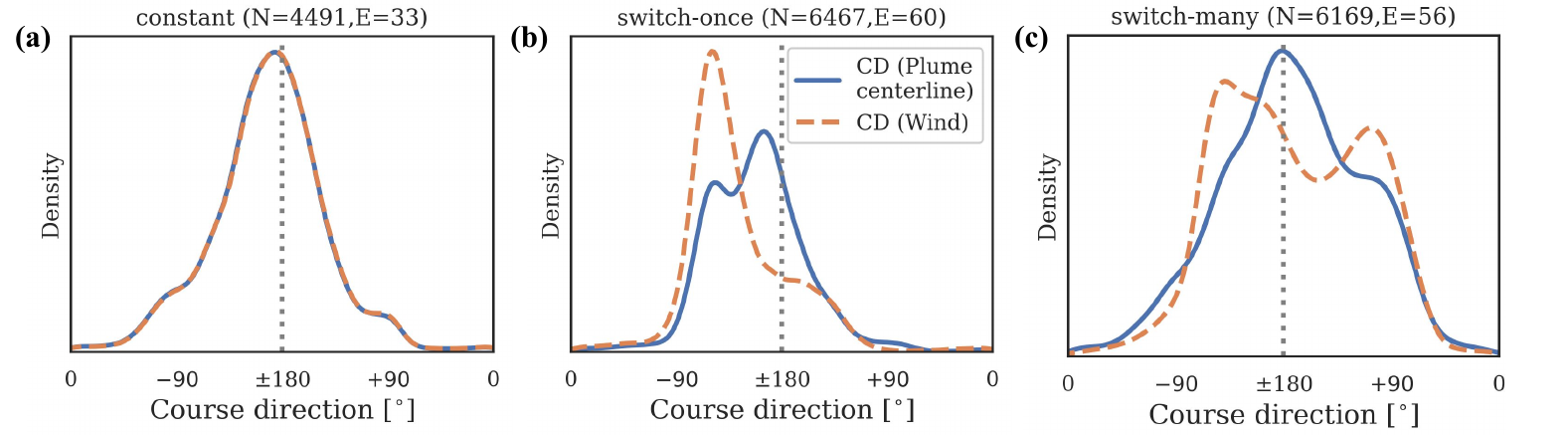} 
\caption{
\textbf{Empirical distributions of course direction suggest that agents track plume with respect to plume centerline rather than current wind direction.}
Kernel density estimates of the agents' course direction (CD) are shown relative to the local plume centerline (solid blue) and to the current wind direction (dashed orange) in three plume configurations.
$\pm180^{\circ}$ implies anti-parallel movement with respect to the plume centerline, or exactly upwind movement with respect to the wind direction.
\textbf{(a)}
For the constant wind direction plume, both the course direction distributions are equivalent.
\textbf{(b)}
In the `switch-once' configuration, a significant proportion of time is spent at a $\approx45^{\circ}$ angle to the wind, but is actually aligned (anti-parallel) with the plume centerline.
\textbf{(c)}
In the `switch-many' configuration, once again,
a significant proportion of time is spent at an angle to the upwind direction, but course direction is actually aligned with the plume centerline.
Plots show data for RNN Agent 3. 
See \ref{sec_supp_centerline} for equivalent plots for all 5 RNN agents.
Subfigure titles indicate how many timesteps (N) and how many successful episodes (E) were summarized in each plot.
}
\label{fig_centerline}
\end{figure*}

\subsection{Agent architecture}
\label{sec_methods_arch}
Our agents are actor-critic networks (Figure \ref{fig_training}e), where a recurrent neural network (RNN) receives sensory observations and passes a transformed representation of them onto parallel Actor and Critic heads
that are both two-layer multi-layer perceptrons (MLPs) \citep{konda2000actor}.
The Actor head implements a control policy to map the RNN's learned state representation to actions, while the Critic head implements a value function that maps the state representation to an estimate of the state's value based on rewards.
This value function is used only during agent training and not thereafter.
In the DRL literature, two-layer deep heads are typically sufficiently expressive for such control problems \citep{hill2018stable}.
At each time step, an agent receives a 3-dimensional real-valued input vector comprising egocentric wind velocities $(x, y)$ and odor concentration at its current location.
In response, the agent produces continuous valued turn (maximum $\pm  6.25 \pi$ radians/s) and forward-movement (maximum 2.5 m/s) actions; these velocities are matched to the capabilities of flying fruit flies \citep{van2014plume,van2008insects}.
In contrast to the orthogonal initialization typically employed in the mainstream machine learning literature \citep{henaff2016recurrent}, we initialize our RNNs with normally distributed weights to facilitate comparisons with the computational neuroscience literature \citep{vogels2005neural,sussillo2014neural,yang2019task}. 

Additionally, to understand the role of memory on tracking performance (Section \ref{sec_behavior_quant}), we compare the RNN-based agents with an alternative feedforward-only network (multi-layer perceptron, MLP) architecture with fixed-length memory, simulated by appending historical sensory observations onto instantaneous network inputs \citep{mnih2013playing}.
Although such MLPs are far from being biologically plausible architectures, they serve as useful tools for abstract comparison since their memory capacities can be controlled precisely. 
Both RNN and MLP layers across all agents are 64 units wide with $tanh$ nonlinearities.

\subsection{Agent training and evaluation}
\label{sec_methods_train_eval}
We train our agents using the Proximal Policy Gradient (PPO) algorithm  \citep{schulman2017proximal}, which is known to robustly solve continuous observation-space continuous action-space control problems without needing significant hyperparameter tuning.  
To guide agent training, we developed a curriculum and a simple reward function that greatly rewards homing in on odor source, mildly rewards actions that reduce the radial distance between agent and odor source, and penalizes longer duration trajectories and straying too far from the plume. 
We train 14 independently randomly initialized networks for each architecture type, i.e. RNNs and MLPs with 2, 4, 6, 8, 10 \& 12 timesteps of observation history.

Next, we evaluate each trained agent's performance with a behavioral assay. 
Each trained agent is evaluated with 240 episodes at different initializations (15 initial locations, 2 initial simulation timestamps, and 8 initial head-directions), and at each of the 'constant', 'switch-once' and `switch-many' plume configurations.
For each architecture type, we proceed to analyze only the top 5 seeds with the best performance, as measured by total number of successful episodes across the four plume configurations.
Agent training/evaluation episodes are run at $25$ frames per second on a sub-sampled plume and limited to 300 frames/timesteps (12 seconds of flight) per episode to accelerate DRL training.
See \ref{sec_supp_train_eval} for additional details on agent training and evaluation, and \ref{sec_supp_hyperparams} for a full list of associated hyperparameters.

\section{Behavior and neural dynamics of trained agents}
\label{sec_analysis}
Our trained agents learn strategies to successfully localize plume sources in turbulent, non-stationary environments.
In this section, we characterize their performance, then highlight their emergent behavioral and neural features.
In addition to comparing artificial agents to biology, we discover behavioral strategies that motivate future experiments and gain intuition about the neural computations underlying these emergent behaviors.

Unless otherwise specified, this section describes results from one agent randomly chosen from the trained RNNs as evaluated on randomly selected episodes in the \textit{evaluation subset}.
This evaluation subset is chosen to include trajectories from test behavioral assays, balanced across successful and unsuccessful tracking episodes, in each of the `constant,' `switch-once,' and `switch-many' configurations
(see \ref{sec_supp_eval} for evaluation details).

\begin{figure*}[h!]
\begin{center}
\includegraphics[width=1.0\linewidth]{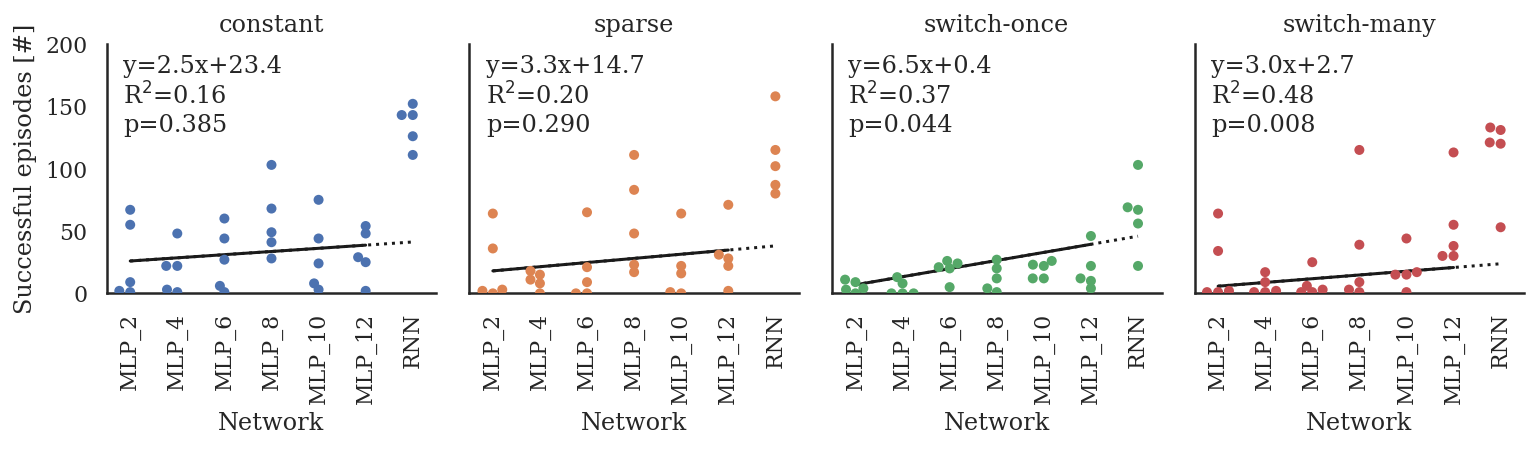}
\caption{
\textbf{Memory capacity improves plume tracking, especially in non-stationary wind direction plumes.}
Number of successful homing episodes for different agent architectures, across different plume configurations for the same set of 240 initial conditions across varying agent starting location and head direction, and plume simulator state (Section \ref{sec_methods_train_eval}).
`MLP\_X' refers to feedforward networks with X timesteps of sensory history (Section \ref{sec_methods_arch}). 
RNNs generally outperform feedforward networks, with more pronounced gains for more complex, switching wind direction (`switch-once', `switch-many') plume tasks.
In feedforward networks, performance on plumes with switching wind direction can improve significantly with increasing memory.
Regression lines (solid black) are fit on only MLP data, but are extended slightly (dotted line) for comparison with RNNs 
(p-values are for a two-sided Wald Test with null hypothesis that the slope is zero).
}
\label{fig_behavior_quant}
\end{center}
\end{figure*}

\subsection{Agents track plumes with varying wind conditions using distinct behavioral modules}
\label{sec_behavior_qual}

Our trained RNN agents are able to complete the plume tracking task with changing wind direction and varying plume sparsity
(Figure \ref{fig_behavior_qual} shows example trajectories). 
The observed trajectories can be summarized as one of three behavior modules, determined approximately by the time elapsed since the agent last sensed odor (Figure \ref{fig_behavior_modules}). 
We refer to these three modules as \textit{tracking}, \textit{lost}, and \textit{recovering}. 
In the \textit{tracking} module, the agent rapidly moves closer to the plume source, using either straight-line trajectories when it is well within the plume, or a quasi-periodic `plume skimming' behavior where it stays close to the edge of the plume while moving in and out of it. 
The interval between the agent's encounters with odor packets in this module is under 0.5 seconds.
\textit{Recovering} corresponds to an irregular behavior where the agent makes large, usually cross-wind, movements after having lost track of the plume for a relatively short period of time (about 0.5 second).
\textit{Lost} corresponds to a periodic behavior that appears variably across trained agents as either a spiraling or slithering/oscillating motion, often with an additional slow drift in an arbitrary direction.
This behavior is seen when the agent has not encountered the plume for a relatively long time, typically over 1 second.
See  \ref{sec_supp_module} for the exact thresholds used to segment each agent's trajectories into behavior modules. 

\begin{figure*}[h!]
\begin{center}
\includegraphics[width=1.0\linewidth]{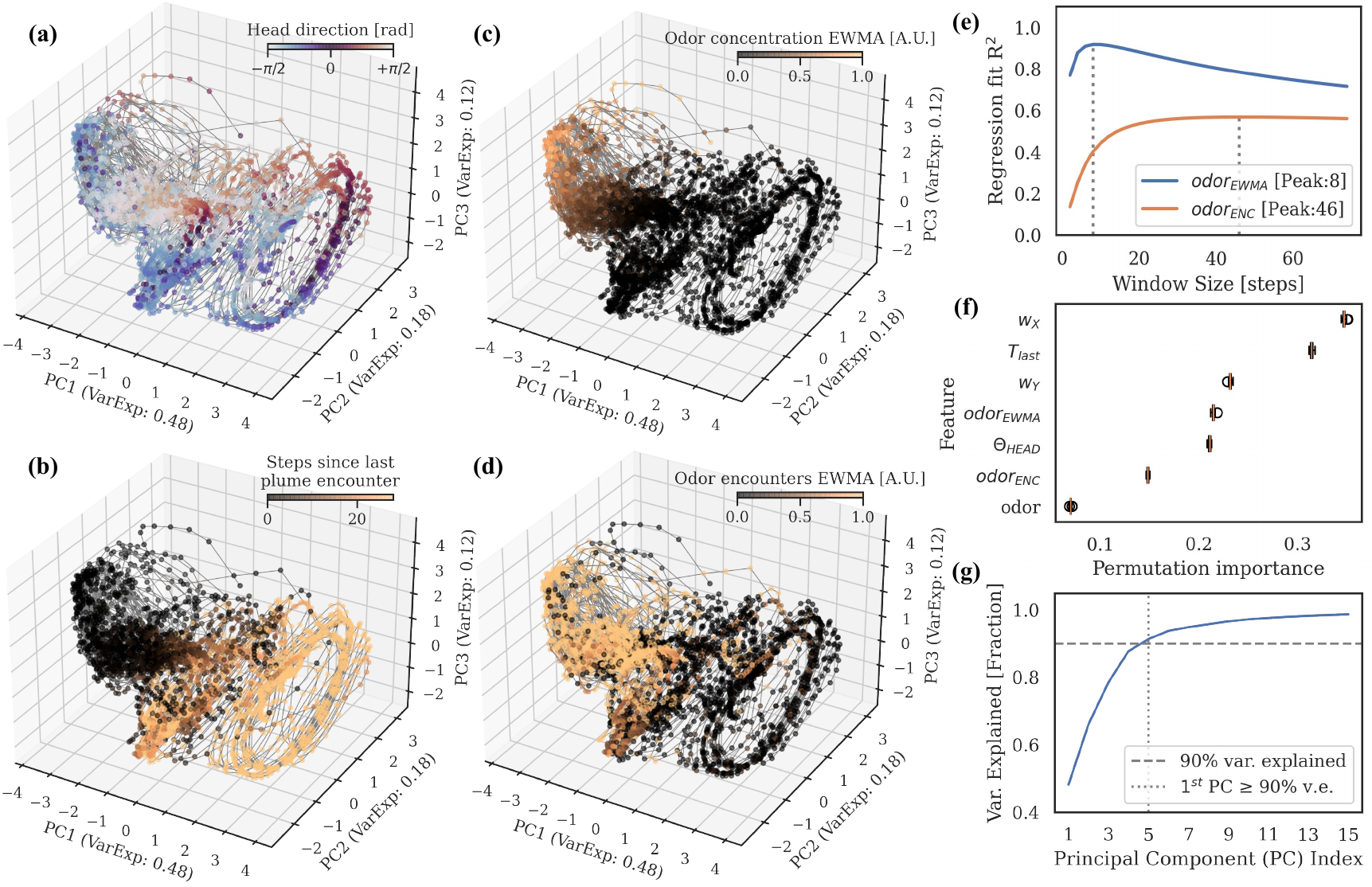} 
\caption{
\textbf{Neural activity is low-dimensional and represents biologically relevant variables.}
(a--d) 
Neural activity trajectories plotted over a diversity of plume conditions and tracking outcomes, 
\textbf{(a)} colored by agent head-direction $\Theta_{HEAD}$, 
\textbf{(b)} steps since last odor encounter $T_{last}$, 
\textbf{(c)} exponentially-weighted moving-average of odor concentration ($odor_{EWMA}$, window-size = 8 steps), and
\textbf{(d)} exponentially-weighted moving-average of recent odor encounters ($odor_{ENC}$, window-size = 46 steps).
\textbf{(e)}
Quality of fit ($R^2$) of a linear model regressing neural activity onto $odor_{EWMA}$ and $odor_{ENC}$ for sliding-windows of varying lengths. 
The sliding-window size for subfigures \textbf{(c)} and \textbf{(d)} are determined by identifying the peaks of these curves. 
\textbf{(e)}
Permutation importance scores of features of a classifier trained to predict agent actions using the aforementioned plotted features 
($T_{last}$, $\Theta_{HEAD}$,  $odor_{EWMA}$, and $odor_{ENC}$), 
and instantaneous sensory observations (wind $w_X, w_Y$ and odor).
\textbf{(f)} 
Plot of cumulative variance explained by top principal components of neural activity aggregated across multiple plume configurations (`constant', 'switch-once' \& `switch-many') suggests a low-dimensional structure. 
\textbf{(g)} 
90$\%$ of the variance of the 64-dimensional neural activity can be explained by the first-5 principal components.
See \ref{sec_supp_module} for equivalent plots for all 5 RNN agents.
}
\label{fig_representations}
\end{center}
\end{figure*}

Agents that are successful in tracking plumes in constant wind direction primarily use the \textit{tracking} and \textit{recovering} modules\footnote{
Animations:~\url{https://github.com/BruntonUWBio/plumetracknets/blob/main/constant.md}}.
Successful trajectories on the `switch-once'\footnote{
\url{https://github.com/BruntonUWBio/plumetracknets/blob/main/switch-once.md}} 
and `switch-many'\footnote{
\url{https://github.com/BruntonUWBio/plumetracknets/blob/main/switch-many.md}} 
plumes reveal that RNN agents use more complex strategies in the face of changing wind directions.
If an agent is in the \textit{tracking} module and well within the plume at the time of wind direction change, it continues along its path until it reaches the edge of the plume without changing its actions.
If it is skimming the edge of the plume when the wind direction switch happens, it tries to compensate for the added movement of the plume by making more pronounced oscillations in and out of the plume.
The shape of these oscillations appears to depend more on the local shape of the plume than on the current direction of the wind (explored further in Section \ref{sec_centerline}). 
Finally, if the agent cannot keep up with the movement of the plume, it typically orchestrates a sequence of large oscillations and spiral-like movements, corresponding to the \textit{recovering} and \textit{lost} modules, to try to find the plume boundary.
On returning to the plume, it resumes the \textit{tracking} module behaviors once again.

\begin{figure*}[h!]
\begin{center}
    \includegraphics[width=1.0\linewidth]{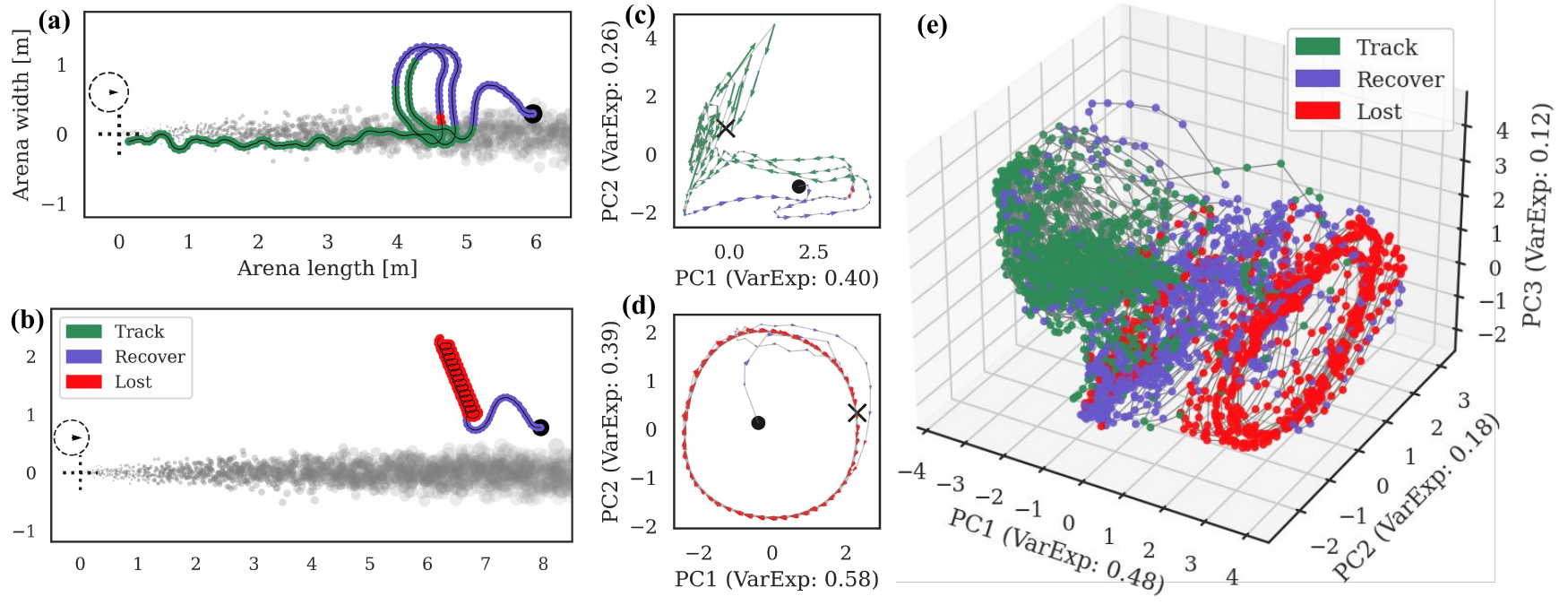}
\caption{
\textbf{Neural dynamics appear to organize themselves into overlapping yet distinct regimes.}
\textbf{(a)} Plume tracking episode that ends in successful homing-in on the odor source, and 
\textbf{(b)} Unsuccessful episode that strays from the plume and ends up exceeding the simulator's bounds. 
\textbf{(c \& d)} Neural activity plots corresponding to each row's trajectory projected on a 2D subspace (state-space) generated from the first 2 principal components of that episode's neural activity.  
Quiver arrows correspond to direction of neural activity gradient, and are colored by the agent's current behavior module.
\textbf{(c)} A `funnel' like structure (in green) emerges in the state-space corresponding to the \textit{tracking} behavior module. 
\textbf{(d)} The agent's periodic \textit{lost} behavior shows up as a limit-cycle in the state-space (red). 
\textbf{(e)} Neural activity plotted over multiple trajectories comprising a diversity of plume conditions and tracking outcomes, projected onto the first 3 principal components of the aggregated neural activity and colored by behavior module (Section \ref{sec_behavior_qual}).
Examples from RNN agent 3.
See  \ref{sec_supp_dynamics} for equivalent plots for all 5 RNN agents.
}
\label{fig_dynamics}
\end{center}
\end{figure*}

\subsection{Agents track plume centerline, not current wind direction}
\label{sec_centerline}

Successful trajectories in plumes that switch direction suggest that agents take the local shape of the plume into account, rather than just the current wind conditions (Figure \ref{fig_behavior_qual}c--d and supplementary animations).
To quantify this, we look at the empirical distributions of the agent's course direction computed with respect to the current wind direction, and with respect to the centerline of the nearby plume (Figure \ref{fig_training}f).
The agent's course direction (Figure \ref{fig_training}c) is defined as the direction of its instantaneous movement with respect to the ground.
Subtracting the current wind direction angle from the course direction  provides the course direction with respect to the wind.
To find the course direction with respect to the centerline, we first find the median local centerline angle using centerline puffs (Section \ref{sec_methods_plume}) within a $\pm 2$ c.m. band of the x-coordinate of the agent's location, then subtract this from the course-direction with respect to the ground.
The empirical distributions include aggregate data from when agents are in the \textit{tracking} behavior module from up to 60 random successful trajectories from the `constant', `switch-once' and `switch-many' plume configurations.
Additionally, for the 'switch-once' configuration, we trim trajectories to consider only the timesteps after the wind direction switch has occurred. 

Figure \ref{fig_centerline} shows the empirical course direction distributions are much better aligned with the plume centerline than to the wind for one example agent.
For `switch-once' plumes, 
the peak of the course direction distribution is much closer to $\pm180^{\circ}$ when considered relative to the centerline than relative to the wind direction.
This observation indicates that the agent's flight is on average aligned (anti-parallel) with the plume centerline, but at an $\approx 45^{\circ}$ angle with respect to the current wind direction.
Similarly, the same trend holds in the `switch-many' configuration, where the course direction distribution is aligned with the plume centerline, but diverges from the wind direction.
In \ref{sec_supp_centerline}, we see that this trend holds across all 5 RNN agents.

\subsection{Recurrence and memory enable plume tracking}
\label{sec_behavior_quant}

To understand the role of memory capacity in plume tracking, we compare the performance of our trained RNNs to trained feedforward networks (multi-layer perceptrons, MLPs) that receive varying timescales of sensory history (see Section \ref{sec_methods_arch} for more information on the MLP architecture). 
As seen in Figure \ref{fig_behavior_quant}, RNNs outperform MLPs for every plume tracking task, with the performance gains being largest in the most challenging tasks.
For MLPs, longer duration sensory memories support much better performance on tougher tracking tasks, where the plumes switch more often or odor packets are sparser.

\subsection{Neural activity is low-dimensional and represents task-relevant variables}
\label{sec_repr}

We now turn our attention to the neural dynamics of the RNNs as agents perform plume tracking.  
Rather than characterizing the activity of individual units, we consider the population activity of the network \citep{ebitz2021population,saxena2019towards}.

First, we reduce and visualize the population activity of our RNN across the `constant', 'switch-once' \& `switch-many' plume configurations and find that the neural activity is low-dimensional (Figure \ref{fig_representations}g), with the first 5--8 principal components explaining 90$\%$ of the variance in the 64-dimensional population activity. 
In \ref{sec_supp_repr}, we see that this trend holds across all 5 agents.

To gain insights into the computations supporting the plume tracking behavior, we look for variables represented in this low-dimensional population activity that are relevant for solving the task. 
We find that the RNNs have learned to represent task-relevant quantities beyond the instantaneous sensory observations that are provided to it by the simulator (Figure \ref{fig_representations}a--d).

Interestingly, these quantities reflect information necessary for solving these challenging plume tracking tasks and require memories of past sensory cues encountered by the agent.
First, the agent's \textit{head-direction}, or its the orientation with respect to the ground, is evident in Figure~\ref{fig_representations}a.
The \textit{time since plume was last encountered} is encoded as in Figure~\ref{fig_representations}b and may be involved in determining transitions between behavior modules.
Whereas the agent only receives local odor concentrations as a sensory input, we find that an exponentially-weighted moving-average of sensed \textit{odor concentrations} is present in Figure~\ref{fig_representations}c.
We conjecture this quantity may be useful as a memory in the face of an intermittent odor signal arising from a turbulent plume.
Similarly, an exponentially-weighted moving-average of a discretized \textit{odor encounters} signal is evident in Figure~\ref{fig_representations}d.

We determine the window-sizes \citep{ewma_pandas} for odor concentrations and encounters by linearly regressing neural activity onto them for sliding-windows of varying lengths, and we chose the window-size that produces the best fit as measured by the coefficient of determination $R^2$ (Figure \ref{fig_representations}e).
The best moving-average window length for time-averaged odor concentrations (7 timesteps or 0.3s on average across all 5 agents) is significantly shorter than that for time-averaged odor encounters (47 timesteps or 1.9s on average across all 5 agents). 
Time-averaged odor concentrations are also better encoded ($R^2$=0.91 on average across 5 agents) than time-averaged odor encounters ($R^2$=0.59 on average across 5 agents).
See Appendix Table~\ref{table_supp_ewma} for data on each individual agent.

\begin{figure*}[t!]
\begin{center}
\includegraphics[width=1.0\linewidth]{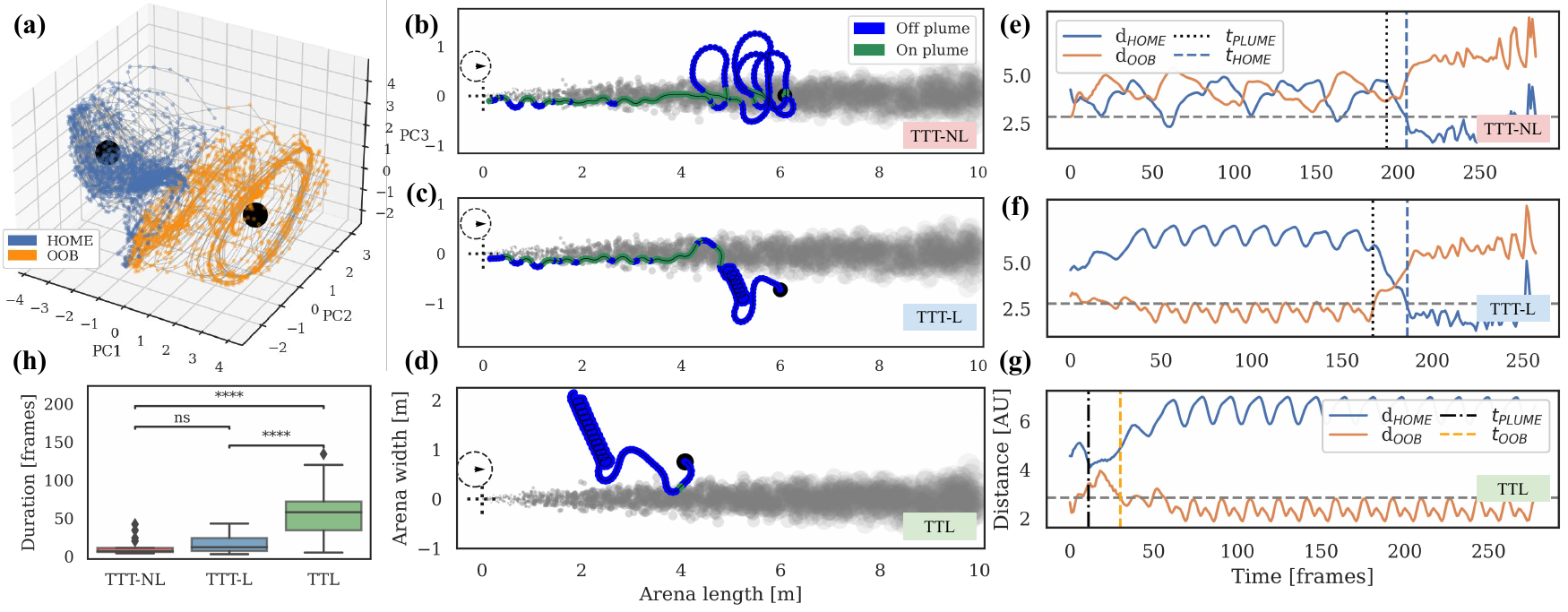}
\caption{\textbf{Transitions between neural activity regimes are asymmetric in duration.}
\textbf{(a)} 
Neural activity from multiple tracking episodes plotted on same top-3 principal component subspace as in Figure \ref{fig_dynamics}.
Points are colored by the neural regime centroid they are closest to (see \ref{sec_supp_ttcs} for centroid definitions; orange color for points nearer to Out-Of-Bounds/OOB centroid and blue for those closer to HOME centroid; black circles denote centroid locations). 
(b--d) 
Example trajectories where the agent 
\textbf{(b)} enters the \textit{tracking} behavior module after entering the plume from the \textit{recovering} behavior module,
\textbf{(c)} enters the \textit{tracking} behavior module after entering the plume from the \textit{lost} behavior module, and
\textbf{(d)} enters the \textit{lost} behavior module after a brief encounter with the plume.
(e--g)
Time courses of neural activity distances to the HOME and OOB centroid (in blue and orange respectively), associated with the respective trajectories in the center column.
Dotted vertical lines show time of entering or leaving plume, while dashed vertical lines show time when the agent has entered the target neural activity regime, i.e.
neural activity is less than D/2 units away from target centroid, where D is the distance between centroids.   
\textbf{(h)} Box plots compare transition times to target regimes over a large set of trajectories across varying plume conditions (`constant', `switch-once', `switch-many'). 
Transitions into the \textit{lost} neural activity regime (TTL) tend to take longer than transitions into the \textit{tracking} neural activity regime (TTT-NL or TTT-L)
(two-sided Mann-Whitney-Wilcoxon test with Bonferroni correction, ****: p $\leq$ 1.00e-04, ns: not significant).
See  \ref{sec_supp_ttcs} for plots of all 5 RNNs. 
}
\label{fig_ttcs}
\end{center}
\end{figure*}

To quantify how important these represented variables are to actual task performance, we train a Random Forest (RF) \citep{breiman2001random} classifier to predict the (discretized) actions taken by the agent over successful trajectories (see \ref{sec_supp_repr} for details).
We also estimate the relative importance of each input feature by calculating its permutation importance score \citep{strobl2008conditional, breiman2001random}, which is an estimate of the reduction in the classifier’s accuracy across several (N=30) randomized permutations of that feature.
Classifier accuracies using all aforementioned represented features (Figure \ref{fig_representations}) along with instantaneous sensory features is 10--18\% higher across all agents than that using classifiers receiving just instantaneous sensory observations, and 26--51$\%$ higher across all agents than that produced by a majority-class classifier 
(See \ref{table_supp_repr} for each agent's classifier accuracies). 
Represented variables have permutation importance scores within the range covered by the importance scores of the instantaneous sensory inputs.
\textit{Time since plume was last encountered} is always one of the top two most important features, close to the wind velocity (x-axis component).
The two time-averaged odor features always easily dwarf the importance of the instantaneous odor feature.
Furthermore, time-averaged \textit{odor concentrations} are more important than time-averaged \textit{odor encounters} in 4 out of 5 agents.
\textit{Head-direction} has an importance intermediate to the two   time-averaged odor features in 4 out of 5 agents.
Note that the estimates provided by this analysis are approximate due to the discretization of the action data and correlations between features.
See \ref{sec_supp_repr} for results at the individual agent level for all 5 agents.

\begin{figure*}[htbp!]
\begin{center}
\includegraphics[width=0.9\linewidth]{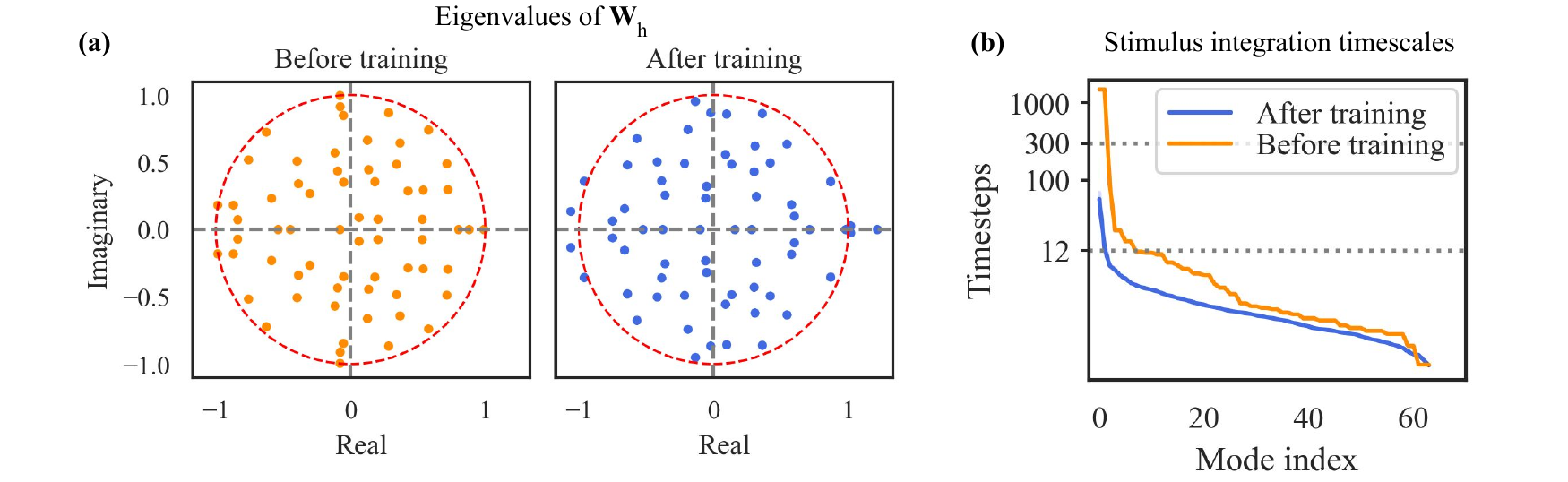}
\caption{
 \textbf{Eigenvalues and stimulus integration timescales.} 
\textbf{(a)} Eigenvalue spectra of the RNN recurrence matrix $\mathbf{W_{h}}$ (for Agent 3) before and after training show how training results in the generation of unstable modes.
\textbf{(b)} Time-averaged (over 6 episodes and 1738 timesteps) stimulus integration timescales associated with stable eigenmodes of recurrence Jacobian $\mathbf{J}^{\mathrm{rec}}$ show a bulk of relatively short timescales (within $12$ timesteps).  
Top 5 integration timescales for the agent shown are 56.5, 13.0, 7.7, 6.8 \& 5.8 timesteps.
Before training, timescales associated with $\mathbf{W_{h}}$'s eigenmodes can be large, even exceeding the length of the training/evaluation episodes (300 steps or 12 seconds).
99\% confidence interval bands have been plotted for the post-training timescales curve, but these bands are of negligible magnitude and therefore invisible. 
See  \ref{sec_supp_eigen} for equivalent plots and data for all RNN agents.
}
\label{fig_eigen}
\end{center}
\end{figure*}

\subsection{Neural dynamics organized into two structured regimes and a transition region}
\label{sec_dynamics}

We now examine the dynamics of the RNN's hidden state $\mathbf{h}$ and how it evolves over the course of tracking episodes.
This analysis is inspired by previous work characterizing the nonlinear dynamics of RNN agents by their fixed-points and transitions among them \citep{vyas2020computation, sussillo2013opening, maheswaranathan2019universality}.
However, in a noteworthy deviation from these structures, we did not find any fixed points in our RNNs. 
Instead, our RNNs adopt neural dynamics that are better described by dynamical regimes.
Specifically, the dynamics appear to organize themselves into overlapping but distinctly structures associated with the \textit{tracking} and \textit{lost} behavioral modules (Figure \ref{fig_dynamics}). 
Interestingly, the periodic spiral or oscillatory movements seen in the \textit{lost} behavioral module appear to also have a quasi-periodic limit-cycle structure in the neural state space (Figure \ref{fig_dynamics}d), while 
the neural dynamics associated with the \textit{tracking} behavior are represented as quasi-periodic `funnel' like structures (Figure \ref{fig_dynamics}c).
We also see an amorphous transition region associated with the \textit{recovering} behavioral module.
We see the same approximate structures (limit-cycles and funnel) emerge in the neural dynamics for 4 of the 5 RNN agents. 
See \ref{sec_supp_dynamics} and associated animations\footnote{ \url{https://github.com/BruntonUWBio/plumetracknets/}} for data on all 5 agents. 

\subsection{Macroscopic transitions between neural activity regimes are asymmetric in duration}
\label{sec_ttcs}

After having found distinct neural activity regimes for the \textit{tracking} and \textit{lost} behaviors in the previous section, we now explore transitions between these two regimes.
Specifically, we look at differences in the duration between 
(1) when an agent enters the plume and when it `enters' the \textit{tracking} neural activity regime, and 
(2) when an agent leaves the plume and when it `enters' the \textit{lost} neural activity regime.
Entry into a neural activity regime is determined by when the neural activity is within a pre-specified distance from a `centroid' corresponding to that regime (see details in \ref{sec_supp_ttcs}).    
As shown in Figure \ref{fig_ttcs}, the time taken to enter the \textit{lost} neural activity regime after the agent leaves the plume is significantly longer than the time taken to enter the \textit{tracking} neural activity regime after the agent enters the plume.
In \ref{sec_supp_ttcs}, we see that this trend holds across 4 out of 5 agents.

\subsection{Connectivity of trained RNNs reveal signatures of instability and memory }
\label{sec_eigen}

The weight matrices and recurrence Jacobians of our RNNs after training offer some theoretical insights into how the neural dynamics of the artificial agents are shaped to track turbulent plumes.

The update rule for a Vanilla RNN with hidden state vector $\bh_t$ is given by
\begin{align*}
    \bh_t = F (\bh_{t-1}, \bx_t) = \tanh \left(\bW_h \bh_{t-1} + \bW_x \bx_t + b \right),
\end{align*}
where $\bW_h$ is recurrence (connectivity) matrix of the hidden layer,
$\bx_t$ are the network's inputs,
$\bW_x$ is the input-to-hidden layer matrix,
and $b$ is a bias term \citep{sussillo2013opening}.

We find that the training process reorganizes the eigenvalue spectrum of the RNN recurrence matrix $\bW_h$ (Figure~\ref{fig_eigen}a).
Before training, weights are initialized as normally distributed random variables.
After training, there are multiple eigenvalues outside the unit circle in the complex plane.
Interestingly, for all 5 agents, there is at least one strictly real-valued eigenvalue larger than unity.
Along with external stimuli, these unstable eigenvalues drive the network's hidden dynamics\footnote{Animations showing how recurrence Jacobian eigenspectra change over the course of  tracking episodes: \url{https://github.com/BruntonUWBio/plumetracknets/blob/main/VRNN3-eigen.md}}.

Next, we consider a linearization of this nonlinear system around arbitrary expansion points. 
The RNN update equation can be linearized around an arbitrary expansion point $\left(\mathbf{h}^{\mathrm{e}}, \mathbf{x}^{\mathrm{e}}\right)$ to get a linear dynamical system approximated by: 
\begin{align*}
\mathbf{h}_{t} \approx & F\left(\mathbf{h}^{\mathrm{e}}, \mathbf{x}^{\mathrm{e}}\right) \\
&+ \left.\mathbf{J}^{\mathrm{rec}}\right|_{\left(\mathbf{h}^{\mathrm{e}}, \mathbf{x}^{\mathrm{e}}\right)} \Delta \mathbf{h}_{t-1}+\left.\mathbf{J}^{\mathrm{inp}}\right|_{\left(\mathbf{h}^{\mathrm{e}}, \mathbf{x}^{\mathrm{e}}\right)} \Delta \mathbf{x}_{t},
\end{align*}
where $\Delta \mathbf{h}_{t-1}=\mathbf{h}_{t-1}-\mathbf{h}^{\mathrm{e}}$ is the state of the linearized system, 
\mbox{$\Delta \mathbf{x}_{t}=\mathbf{x}_{t}-\mathbf{x}^{\mathrm{e}}$} is the linearized system's input,
$\mathbf{J}^{\mathrm{rec}}$ is the recurrence Jacobian, and $\mathbf{J}^{\mathrm{rec}}$ is the input Jacobian \citep{maheswaranathan2019reverse}.
To be explicit,
\begin{align*}
\left.J_{i j}^{\mathrm{rec}}\right|_{\left(\mathbf{h}^{\mathrm{}{e}}, \mathbf{x}^{\mathrm{e}}\right)} =& \frac{\partial F(\mathbf{h}, \mathbf{x})_{i}}{\partial h_{j}}, \\ 
\left.J_{i j}^{\mathrm{inp}}\right|_{\left(\mathbf{h}^{e}, \mathbf{x}^{e}\right)}=&\frac{\partial F(\mathbf{h}, \mathbf{x})_{i}}{\partial x_{j}}.
\end{align*}
Note that 
\mbox{$\left.\mathbf{J}^{\mathrm{rec}}\right|_{\left(0, 0\right)} = \bW_h$} and
\mbox{$\left.\mathbf{J}^{\mathrm{inp}}\right|_{\left(0, 0\right)} = \bW_x$}.

Following the approach of \citep{maheswaranathan2019reverse}, we consider the eigenvalues of the recurrence Jacobian and associated stimulus integration timescales along the trajectories of several episodes.
This timescale governs the integration of stimuli in the direction of the corresponding eigenvectors.
We chose at random one successful and one unsuccessful episode from each of three plume configurations (`constant,' `switch-once,' and `switch-many').
At each time step of the trajectory, we computed the recurrence Jacobian assuming zero input $\left.\mathbf{J}^{\mathrm{rec}}\right|_{(\bh,0)}$.
The stimulus integration timescale $\tau_i$ associated with a stable eigenvalue $\lambda_i$ (i.e. $|\lambda_i| \leq 1$) can be interpreted as a timescale with the conversion \mbox{$\tau_i = \left| ( 1/ \ln|\lambda_i| ) \right|$}.

Comparing the time-averaged stimulus integration timescales with those from the untrained RNN reveals that training adjusts these timescales to lie well within the maximum episode length of 300 timesteps (Figure~\ref{fig_eigen}b).
Furthermore, we see that the bulk of these timescales are within about $12$ timesteps ($\approx 0.5$s), suggesting that the plume tracking task predominantly needs short timescale memories.
In \ref{sec_supp_eigen}, we see that this trend holds across all 5 RNNs.

\section{Connections to tracking turbulent plumes in biology} 
Our artificial RNN agents exhibit similarities to biology at the levels of behavior, computation, and neural dynamics. 
In this section, we draw these comparisons, discuss the significance of these connections, and suggest theoretical insights that may be relevant for researchers interested in biological plume tracking.

\subsection{Behavioral features}
The complex behavior exhibited by our agents can be decomposed into simpler modules, 
sequenced by the time elapsed since the agent last encountered the plume (Section \ref{sec_behavior_qual}).
These modules show features similar to \textit{upwind surging}, \textit{crosswind casting} and \textit{U-turn} behaviors previously reported in many studies on moths, fruit flies, and other flying insects \citep{baker2018algorithms,van2014plume,carde2008navigational, budick2006free}.
The spiraling behavior seen in the agent's \textit{lost} behavior module has been previously proposed as a plume reacquisition strategy \citep{lochmatter2009theoretical}.
Furthermore, the variable sequencing behavior modules resembles the odor loss activated clock mechanism that has been previously proposed to drive changes in flight behavior in moths \citep{kennedy1974pheromone,kennedy1983zigzagging,baker1990upwind}.

Our observations make a behavioral hypothesis that agent track plumes with respect to the centerline 
rather than with respect to the current wind direction 
(Section \ref{sec_centerline}).
In a previous study on tracking in constant wind direction plumes, \cite{grunbaum2015spatial} proposed a model where insects explicitly performed upwind surges when close to the plume centerline.
However, a later study by \cite{pang2018history} failed to find support for this model.
Our analysis provides intuition for the role of centerline tracking in non-stationary plumes and suggests a testable hypothesis: we predict that centerline tracking behaviors will be more apparent in flying insects when they track plumes in wind that switches direction.

\subsection{Algorithms for odor localization}
How biological organisms search and localize odor sources has a long and rich literature, and a variety of algorithms has been developed to explain this capability of single-celled organisms, cells in an organ, and animals in complex environments.
Where gradients exist, these smoothly varying rates of changes in concentration may be exploited to localize odor sources by chemotaxis and related algorithms \citep{adler1966chemotaxis,friedrich2007chemotaxis,cremer2019chemotaxis}.
However, in intermittent odor landscapes, gradient-based algorithms cannot be successful, and the \emph{Infotaxis} algorithm was developed as an alternative \citep{balkovsky2002olfactory,vergassola2007infotaxis,masson2009chasing,barbieri2011trajectories,calhoun2014maximally}. 

Both Infotaxis \citep{vergassola2007infotaxis} and our approach are formulated as solutions to plume tracking as  a Partially Observable Markov Decision Process (POMDP) \citep{sutton2018reinforcement}. 
Infotaxis crafts a control policy that chooses actions (movements) to maximally reduce the expected entropy of the odor source location probability on the next time step.
This policy makes two computational requirements of the agent.
First, Infotaxis agents must store a probability distribution for the source location spanning the size of the arena being navigated.
Second, agents are able to perform Bayesian inference \citep{reddyannrev}.
In contrast, here our approach is to learn this control policy from only locally available measurements, and actions are chosen to maximize the expected discounted reward over a trajectory.
Compared to Infotaxis, our approach produces trajectories with a stronger semblance to biology and a control policy that reacts to changing wind conditions.
It also uses a neural implementation that does not make any (potentially biologically implausible) assumptions about which variables are implemented or how inference is performed.

\subsection{Neural representations}
Our RNN agents learn to represent variables that have been previously reported to be crucial to odor navigation (Section \ref{sec_repr}).
First, \textit{agent head-direction} has been found to be implemented as a ring attractor circuit in the central complex of many flying insects and is implicated in navigation \citep{pfeiffer2014organization,seelig2015neural,green2017neural,kim2019generation,okubo2020neural}.
Second, \textit{time since plume was last encountered}
is analogous to the hypothesized internal-clock that determines behavior switching in moths \citep{kennedy1974pheromone,kennedy1983zigzagging,baker1990upwind}.
Additionally, \cite{park2016neurally} showed how this variable is encoded by the bursting olfactory receptor neurons (bORNs) in many animals, and that it contains information relevant to navigating in turbulent odors.

Third, \textit{exponential moving-average of odor encounters} was found by \cite{demir2020walking} to determine the probability of turn and stop behaviors in walking flies navigating in turbulent plumes.
Specifically, higher odor encounter rates were associated with more frequent saccadic upwind turns
\citep{celani2020olfactory}.
Fourth, \textit{exponentially moving-average of sensed odor concentration}  
is motivated by theoretical work by \cite{maheswaranathan2019reverse} that posits exponentially-weighted moving averages to be good canonical models for stimulus integration in RNNs.
Between these two time-averaged odor variables, the best represented window length for time-averaged concentration is significantly shorter ($\approx0.3$s) than that for time-averaged encounters ($\approx 1.9$s). 
Furthermore, we find that time-averaged odor concentration is relatively better represented and more important 
in predicting agent behavior, 
corroborating the intuition that turn decisions during flight would require quick decision making on sub-second timescales.
We note that alternative variables beyond these four may exist that better explain agent navigation decisions.

\subsection{Neural dynamics}
The agents' neural activity is low dimensional and structured, with an interesting  asymmetry in macroscopic transitions between these structures.
Like often seen in neurobiological recordings \citep{pang2016dimensionality,cunningham2014dimensionality}, the population activity of our RNNs is low-dimensional, with the top 5--8 principal components explaining an overwhelming majority of the 64-dimensional population's total variance (Section \ref{sec_repr}).

The neural dynamics associated with behavior modules further exhibit interesting structure.
\textit{Lost} behaviors are represented as quasi-limit-cycles, while \textit{tracking} behaviors show a `funnel' like structure (Section \ref{sec_dynamics}).
Similar 1-D circular manifolds and 2-D funnels \citep{vyas2020computation,kriegeskorte2021neural} have been previously reported on the representational geometry of sensory populations, but not, to the best of our knowledge, in the closed-loop agent setting.

Finally, we find that the interval between entering the neural activity cluster associated with the \textit{lost} behavior and leaving the plume, is significantly longer than the interval between entering neural activity cluster associated with the \textit{tracking} behavior and entering the plume.
This asymmetry in macro-scale transitions in the neural state resembles an asymmetry in behavior transitions reported in \cite{van2014plume}, where the authors experimentally observe that flies take about twice as long to cast crosswind after plume loss, than to surge upwind on encountering attractive odors.

\subsection{The role of memory}
Two independent analyses give us insight into the memory requirements of the plume tracking task.
First, our comparison of RNN agents to MLPs with fixed amounts of sensory history input (Section \ref{sec_behavior_quant}) suggests that longer sensory histories and recurrence lead to better performance on tougher tracking tasks, such as those with plumes that switch one or more times.
Second, analyzing stable eigenmodes of the RNN recurrence Jacobians suggests that only a couple of long stimulus integration timescales 
are involved in the neural computation. The bulk of stimulus integration timescales are within $\sim 12$ steps or 0.5s (Section \ref{sec_eigen}).
Together, we believe that memory is crucial for tracking non-stationary wind direction plumes, but short timescale (under $\sim 0.5s$) and reflexive mechanisms may be sufficient for tracking constant wind direction plumes.
This corroborates results by \cite{pang2018history} and \cite{grunbaum2015spatial} and extends them by highlighting the importance of longer term memory in cases where wind changes direction.

\section{Limitations and future work} 
Our results motivate further development in using DRL to model and understand complex behaviors in several ways.
First, here we used vanilla recurrent units with no biomechanical body model, and models that incorporate known complexity from biology as constraints may give rise to further insights.
For instance, DRL agents may be trained using spiking neural networks \citep{jia2021neuronal,yuan2019reinforcement,recanatesi2019dimensionality}.
Further, the wealth of architectural insights emerging from the fly connectome may be used to constrain wiring motifs in artificial networks \citep{hulse2020connectome,scheffer2020connectome}.
Modeling multiple antennae \cite{Kadakia2021}, or more generally a biomechanical body, would enrich the interactions between the agent and the simulation environment \citep{rios2021neuromechfly,plum2021scant,merel2019deep}.

Second, multi-task training should produce agents with richer behaviors and more complex neural activity structures with shared and
task-specific adaptations \citep{crawshaw2020multi,yang2019task,duncker2020organizing,mlynarski2018adaptive,weber2019role}.
Adding other sensory modalities like vision and training the agents in a 3D virtual reality environment could produce more realistic perceptual representations in the agent \citep{crosby2019animal, crosby2020building}.

Finally, future work could explore learning algorithms that respect biological constraints like excitation-inhibition balance and Dale's law \citep{GOULAS2021,ehrlich2021psychrnn,delahunt2018biological}.
More complex training curricula \citep{bengio2009curriculum} or
alternative training algorithms using evolutionary techniques \citep{de2013evolutionary,stanley2019designing,gupta2021embodied} might be able to mitigate the significant performance variability we observed in our agents (Section \ref{sec_behavior_quant}).

Our analyses also motivate further methodological development in theoretical tools to understand actor-critic RNNs.
Currently available reverse-engineering methods that characterize RNNs using discrete dynamical features such as fixed-points  \citep{sussillo2013opening,maheswaranathan2019universality,maheswaranathan2019reverse} are not applicable to the continuous and amorphous dynamical structures that we encountered in our analyses (Section \ref{sec_dynamics}).
New methods are also needed for comparing multiple agents at the behavioral level, specifically taking into account the compounding differences that arise from small differences in action-stimulus loops.
Finally, further theoretical work is required to understand the role of training-induced unstable RNN connectivity eigenmodes, such as those observed in Section~\ref{sec_eigen}, including extensions of analytic techniques developed to understand RNNs trained by supervised-learning \citep{sussillo2009generating,rajan2006eigenvalue,maheswaranathan2019reverse,mikulik2020meta,schaeffer2020reverse}.

\section{Conclusion} 
In this paper, we used deep reinforcement learning to train recurrent neural network agents to solve a stochastic plume tracking task.
We find several behavioral and neural features that emerge in these trained agents and connect these features with how flying insects track turbulent plumes.
Our findings motivate future experiments and theoretical developments, and provide a foundation for more nuanced future work.
We hope our approach will contribute to the growing convergence in the understanding of artificial and biological networks \citep{hasson2020direct,hassabis2017neuroscience}.
Efforts to reverse engineer such neural network agents will help accelerate the development of similar methods for biological agents \citep{kwon2020inverse,ashwood2020inferring}.
Moreover, our RNN agents may serve as generative models of complex naturalistic behaviors, which may facilitate the development of behavior analysis tools for biology \citep{berman2016predictability,singh2021mining,nassar2018tree}.
Insights from these studies may also inspire the development of robotic agents with artificial \citep{vouloutsi2013synthetic} or hybrid \citep{anderson2019smellicopter} olfactory sensing.

\section*{Online supplement}
Animations accompanying this manuscript can be found at: \url{https://github.com/BruntonUWBio/plumetracknets}.
Code will be released at this location on manuscript publication.

\section*{Author Contributions}
SHS, FvB, RPNR and BWB conceived of the study/analysis. 
SHS engineered the agents. 
SHS performed the data analysis. 
SHS, FvB, RPNR, and BWB interpreted the results. 
SHS and BWB wrote the manuscript. 
All authors reviewed and edited the manuscript.

\section*{Acknowledgements}
We thank Aravind Rajeswaran, Stefano Recanatesi, Scott Sterrett, Niru Maheswaranathan, and Steve Brunton for helpful comments and discussions.
The plume tracking task graphic in Figure \ref{fig_training} is inspired by a similar figure in \cite{baker2018algorithms}, and uses parts of an open source fruitfly graphic from scidraw.io \citep{costa_gil_2020_3926137}.
We are grateful for the well documented open source implementation of PPO by Ilya Kostrikov \citep{kostrikov2018pytorch}, which we heavily adapted and built upon for our work.

This work has been funded by 
the Air Force Research Lab award FA8651-20-1-0002 and FA9550-21-0122 (FvB); 
the National Institutes of Health award NIH P20GM103650 (FvB);
the Defense Advanced Research Projects Agency HR001120C0021 (RPNR); 
the Templeton World Charity Foundation (RPNR); 
the National Science Foundation award EEC-1028725 (RPNR); 
the Air Force Office of Scientific Research awards FA9550-19-1-0386 and FA9550-18-1-0114 (BWB); 
and the Washington Research Foundation (BWB);

\small
\bibliographystyle{elsarticle-harv}
\bibliography{plume}

\begin{thebibliography}{128}
\expandafter\ifx\csname natexlab\endcsname\relax\def\natexlab#1{#1}\fi
\providecommand{\url}[1]{\texttt{#1}}
\providecommand{\href}[2]{#2}
\providecommand{\path}[1]{#1}
\providecommand{\DOIprefix}{doi:}
\providecommand{\ArXivprefix}{arXiv:}
\providecommand{\URLprefix}{URL: }
\providecommand{\Pubmedprefix}{pmid:}
\providecommand{\doi}[1]{\href{http://dx.doi.org/#1}{\path{#1}}}
\providecommand{\Pubmed}[1]{\href{pmid:#1}{\path{#1}}}
\providecommand{\bibinfo}[2]{#2}
\ifx\xfnm\relax \def\xfnm[#1]{\unskip,\space#1}\fi
\bibitem[{Adler(1966)}]{adler1966chemotaxis}
\bibinfo{author}{Adler, J.}, \bibinfo{year}{1966}.
\newblock \bibinfo{title}{Chemotaxis in bacteria}.
\newblock \bibinfo{journal}{Science} \bibinfo{volume}{153},
  \bibinfo{pages}{708--716}.
\bibitem[{Ahrens(2019)}]{ahrens2019zebrafish}
\bibinfo{author}{Ahrens, M.B.}, \bibinfo{year}{2019}.
\newblock \bibinfo{title}{Zebrafish neuroscience: Using artificial neural
  networks to help understand brains}.
\newblock \bibinfo{journal}{Current Biology} \bibinfo{volume}{29},
  \bibinfo{pages}{R1138--R1140}.
\bibitem[{Anderson et~al.(2019)Anderson, Sullivan, Talley, Brink, Fuller and
  Daniel}]{anderson2019smellicopter}
\bibinfo{author}{Anderson, M.J.}, \bibinfo{author}{Sullivan, J.G.},
  \bibinfo{author}{Talley, J.L.}, \bibinfo{author}{Brink, K.M.},
  \bibinfo{author}{Fuller, S.B.}, \bibinfo{author}{Daniel, T.L.},
  \bibinfo{year}{2019}.
\newblock \bibinfo{title}{The “smellicopter,” a bio-hybrid odor localizing
  nano air vehicle}, in: \bibinfo{booktitle}{2019 IEEE/RSJ International
  Conference on Intelligent Robots and Systems (IROS)},
  \bibinfo{organization}{IEEE}. pp. \bibinfo{pages}{6077--6082}.
\bibitem[{Arulkumaran et~al.(2017)Arulkumaran, Deisenroth, Brundage and
  Bharath}]{arulkumaran2017deep}
\bibinfo{author}{Arulkumaran, K.}, \bibinfo{author}{Deisenroth, M.P.},
  \bibinfo{author}{Brundage, M.}, \bibinfo{author}{Bharath, A.A.},
  \bibinfo{year}{2017}.
\newblock \bibinfo{title}{Deep reinforcement learning: A brief survey}.
\newblock \bibinfo{journal}{IEEE Signal Processing Magazine}
  \bibinfo{volume}{34}, \bibinfo{pages}{26--38}.
\bibitem[{Ashwood et~al.(2020)Ashwood, Roy, Bak and
  Pillow}]{ashwood2020inferring}
\bibinfo{author}{Ashwood, Z.}, \bibinfo{author}{Roy, N.A.},
  \bibinfo{author}{Bak, J.H.}, \bibinfo{author}{Pillow, J.W.},
  \bibinfo{year}{2020}.
\newblock \bibinfo{title}{Inferring learning rules from animal
  decision-making}.
\newblock \bibinfo{journal}{Advances in Neural Information Processing Systems}
  \bibinfo{volume}{33}.
\bibitem[{Baker et~al.(2018)Baker, Dickinson, Findley, Gire, Louis, Suver,
  Verhagen, Nagel and Smear}]{baker2018algorithms}
\bibinfo{author}{Baker, K.L.}, \bibinfo{author}{Dickinson, M.},
  \bibinfo{author}{Findley, T.M.}, \bibinfo{author}{Gire, D.H.},
  \bibinfo{author}{Louis, M.}, \bibinfo{author}{Suver, M.P.},
  \bibinfo{author}{Verhagen, J.V.}, \bibinfo{author}{Nagel, K.I.},
  \bibinfo{author}{Smear, M.C.}, \bibinfo{year}{2018}.
\newblock \bibinfo{title}{Algorithms for olfactory search across species}.
\newblock \bibinfo{journal}{Journal of Neuroscience} \bibinfo{volume}{38},
  \bibinfo{pages}{9383--9389}.
\bibitem[{Baker(1990)}]{baker1990upwind}
\bibinfo{author}{Baker, T.}, \bibinfo{year}{1990}.
\newblock \bibinfo{title}{Upwind flight and casting flight: complementary
  phasic and tonic systems used for location of sex pheromone sources by male
  moth}, in: \bibinfo{booktitle}{Proc. 10th Int. Symp. Olfaction and Taste,
  Oslo, 1990}, pp. \bibinfo{pages}{18--25}.
\bibitem[{Balkovsky and Shraiman(2002)}]{balkovsky2002olfactory}
\bibinfo{author}{Balkovsky, E.}, \bibinfo{author}{Shraiman, B.I.},
  \bibinfo{year}{2002}.
\newblock \bibinfo{title}{Olfactory search at high reynolds number}.
\newblock \bibinfo{journal}{Proceedings of the national academy of sciences}
  \bibinfo{volume}{99}, \bibinfo{pages}{12589--12593}.
\bibitem[{Banino et~al.(2018)Banino, Barry, Uria, Blundell, Lillicrap,
  Mirowski, Pritzel, Chadwick, Degris, Modayil et~al.}]{banino2018vector}
\bibinfo{author}{Banino, A.}, \bibinfo{author}{Barry, C.},
  \bibinfo{author}{Uria, B.}, \bibinfo{author}{Blundell, C.},
  \bibinfo{author}{Lillicrap, T.}, \bibinfo{author}{Mirowski, P.},
  \bibinfo{author}{Pritzel, A.}, \bibinfo{author}{Chadwick, M.J.},
  \bibinfo{author}{Degris, T.}, \bibinfo{author}{Modayil, J.}, et~al.,
  \bibinfo{year}{2018}.
\newblock \bibinfo{title}{Vector-based navigation using grid-like
  representations in artificial agents}.
\newblock \bibinfo{journal}{Nature} \bibinfo{volume}{557},
  \bibinfo{pages}{429--433}.
\bibitem[{Barbieri et~al.(2011)Barbieri, Cocco and
  Monasson}]{barbieri2011trajectories}
\bibinfo{author}{Barbieri, C.}, \bibinfo{author}{Cocco, S.},
  \bibinfo{author}{Monasson, R.}, \bibinfo{year}{2011}.
\newblock \bibinfo{title}{On the trajectories and performance of infotaxis, an
  information-based greedy search algorithm}.
\newblock \bibinfo{journal}{EPL (Europhysics Letters)} \bibinfo{volume}{94},
  \bibinfo{pages}{20005}.
\bibitem[{Bengio et~al.(2009)Bengio, Louradour, Collobert and
  Weston}]{bengio2009curriculum}
\bibinfo{author}{Bengio, Y.}, \bibinfo{author}{Louradour, J.},
  \bibinfo{author}{Collobert, R.}, \bibinfo{author}{Weston, J.},
  \bibinfo{year}{2009}.
\newblock \bibinfo{title}{Curriculum learning}, in:
  \bibinfo{booktitle}{Proceedings of the 26th annual international conference
  on machine learning}, pp. \bibinfo{pages}{41--48}.
\bibitem[{Berman et~al.(2016)Berman, Bialek and
  Shaevitz}]{berman2016predictability}
\bibinfo{author}{Berman, G.J.}, \bibinfo{author}{Bialek, W.},
  \bibinfo{author}{Shaevitz, J.W.}, \bibinfo{year}{2016}.
\newblock \bibinfo{title}{Predictability and hierarchy in drosophila behavior}.
\newblock \bibinfo{journal}{Proceedings of the National Academy of Sciences}
  \bibinfo{volume}{113}, \bibinfo{pages}{11943--11948}.
\bibitem[{Botvinick et~al.(2019)Botvinick, Ritter, Wang, Kurth-Nelson, Blundell
  and Hassabis}]{botvinick2019reinforcement}
\bibinfo{author}{Botvinick, M.}, \bibinfo{author}{Ritter, S.},
  \bibinfo{author}{Wang, J.X.}, \bibinfo{author}{Kurth-Nelson, Z.},
  \bibinfo{author}{Blundell, C.}, \bibinfo{author}{Hassabis, D.},
  \bibinfo{year}{2019}.
\newblock \bibinfo{title}{Reinforcement learning, fast and slow}.
\newblock \bibinfo{journal}{Trends in Cognitive Sciences} .
\bibitem[{Botvinick et~al.(2020)Botvinick, Wang, Dabney, Miller and
  Kurth-Nelson}]{botvinick2020deep}
\bibinfo{author}{Botvinick, M.}, \bibinfo{author}{Wang, J.X.},
  \bibinfo{author}{Dabney, W.}, \bibinfo{author}{Miller, K.J.},
  \bibinfo{author}{Kurth-Nelson, Z.}, \bibinfo{year}{2020}.
\newblock \bibinfo{title}{Deep reinforcement learning and its neuroscientific
  implications}.
\newblock \bibinfo{journal}{Neuron} .
\bibitem[{Breiman(2001)}]{breiman2001random}
\bibinfo{author}{Breiman, L.}, \bibinfo{year}{2001}.
\newblock \bibinfo{title}{Random forests}.
\newblock \bibinfo{journal}{Machine learning} \bibinfo{volume}{45},
  \bibinfo{pages}{5--32}.
\bibitem[{van Breugel and Dickinson(2014)}]{van2014plume}
\bibinfo{author}{van Breugel, F.}, \bibinfo{author}{Dickinson, M.H.},
  \bibinfo{year}{2014}.
\newblock \bibinfo{title}{Plume-tracking behavior of flying drosophila emerges
  from a set of distinct sensory-motor reflexes}.
\newblock \bibinfo{journal}{Current Biology} \bibinfo{volume}{24},
  \bibinfo{pages}{274--286}.
\bibitem[{van Breugel et~al.(2008)van Breugel, Regan and
  Lipson}]{van2008insects}
\bibinfo{author}{van Breugel, F.}, \bibinfo{author}{Regan, W.},
  \bibinfo{author}{Lipson, H.}, \bibinfo{year}{2008}.
\newblock \bibinfo{title}{From insects to machines}.
\newblock \bibinfo{journal}{IEEE Robotics \& Automation Magazine}
  \bibinfo{volume}{15}, \bibinfo{pages}{68--74}.
\bibitem[{Brockman et~al.(2016)Brockman, Cheung, Pettersson, Schneider,
  Schulman, Tang and Zaremba}]{brockman2016openai}
\bibinfo{author}{Brockman, G.}, \bibinfo{author}{Cheung, V.},
  \bibinfo{author}{Pettersson, L.}, \bibinfo{author}{Schneider, J.},
  \bibinfo{author}{Schulman, J.}, \bibinfo{author}{Tang, J.},
  \bibinfo{author}{Zaremba, W.}, \bibinfo{year}{2016}.
\newblock \bibinfo{title}{Open{AI} gym}.
\newblock \bibinfo{journal}{arXiv preprint arXiv:1606.01540} .
\bibitem[{Budick and Dickinson(2006)}]{budick2006free}
\bibinfo{author}{Budick, S.A.}, \bibinfo{author}{Dickinson, M.H.},
  \bibinfo{year}{2006}.
\newblock \bibinfo{title}{Free-flight responses of drosophila melanogaster to
  attractive odors}.
\newblock \bibinfo{journal}{Journal of Experimental Biology}
  \bibinfo{volume}{209}, \bibinfo{pages}{3001--3017}.
\bibitem[{Calhoun et~al.(2014)Calhoun, Chalasani and
  Sharpee}]{calhoun2014maximally}
\bibinfo{author}{Calhoun, A.J.}, \bibinfo{author}{Chalasani, S.H.},
  \bibinfo{author}{Sharpee, T.O.}, \bibinfo{year}{2014}.
\newblock \bibinfo{title}{Maximally informative foraging by caenorhabditis
  elegans}.
\newblock \bibinfo{journal}{Elife} \bibinfo{volume}{3},
  \bibinfo{pages}{e04220}.
\bibitem[{Card{\'e} and Willis(2008)}]{carde2008navigational}
\bibinfo{author}{Card{\'e}, R.T.}, \bibinfo{author}{Willis, M.A.},
  \bibinfo{year}{2008}.
\newblock \bibinfo{title}{Navigational strategies used by insects to find
  distant, wind-borne sources of odor}.
\newblock \bibinfo{journal}{Journal of Chemical Ecology} \bibinfo{volume}{34},
  \bibinfo{pages}{854--866}.
\bibitem[{Celani(2020)}]{celani2020olfactory}
\bibinfo{author}{Celani, A.}, \bibinfo{year}{2020}.
\newblock \bibinfo{title}{Olfactory navigation: Tempo is the key}.
\newblock \bibinfo{journal}{Elife} \bibinfo{volume}{9},
  \bibinfo{pages}{e63385}.
\bibitem[{Celani et~al.(2014)Celani, Villermaux and
  Vergassola}]{celani2014odor}
\bibinfo{author}{Celani, A.}, \bibinfo{author}{Villermaux, E.},
  \bibinfo{author}{Vergassola, M.}, \bibinfo{year}{2014}.
\newblock \bibinfo{title}{Odor landscapes in turbulent environments}.
\newblock \bibinfo{journal}{Physical Review X} \bibinfo{volume}{4},
  \bibinfo{pages}{041015}.
\bibitem[{Cichy and Kaiser(2019)}]{cichy2019deep}
\bibinfo{author}{Cichy, R.M.}, \bibinfo{author}{Kaiser, D.},
  \bibinfo{year}{2019}.
\newblock \bibinfo{title}{Deep neural networks as scientific models}.
\newblock \bibinfo{journal}{Trends in Cognitive Sciences} .
\bibitem[{Colabrese et~al.(2017)Colabrese, Gustavsson, Celani and
  Biferale}]{colabrese2017flow}
\bibinfo{author}{Colabrese, S.}, \bibinfo{author}{Gustavsson, K.},
  \bibinfo{author}{Celani, A.}, \bibinfo{author}{Biferale, L.},
  \bibinfo{year}{2017}.
\newblock \bibinfo{title}{Flow navigation by smart microswimmers via
  reinforcement learning}.
\newblock \bibinfo{journal}{Physical Review Letters} \bibinfo{volume}{118},
  \bibinfo{pages}{158004}.
\bibitem[{Costa(2020)}]{costa_gil_2020_3926137}
\bibinfo{author}{Costa, G.}, \bibinfo{year}{2020}.
\newblock \bibinfo{title}{Flies mating dance}.
\newblock \URLprefix \url{https://doi.org/10.5281/zenodo.3926137},
  \DOIprefix\doi{10.5281/zenodo.3926137}.
\bibitem[{Crawshaw(2020)}]{crawshaw2020multi}
\bibinfo{author}{Crawshaw, M.}, \bibinfo{year}{2020}.
\newblock \bibinfo{title}{Multi-task learning with deep neural networks: A
  survey}.
\newblock \bibinfo{journal}{arXiv preprint arXiv:2009.09796} .
\bibitem[{Cremer et~al.(2019)Cremer, Honda, Tang, Wong-Ng, Vergassola and
  Hwa}]{cremer2019chemotaxis}
\bibinfo{author}{Cremer, J.}, \bibinfo{author}{Honda, T.},
  \bibinfo{author}{Tang, Y.}, \bibinfo{author}{Wong-Ng, J.},
  \bibinfo{author}{Vergassola, M.}, \bibinfo{author}{Hwa, T.},
  \bibinfo{year}{2019}.
\newblock \bibinfo{title}{Chemotaxis as a navigation strategy to boost range
  expansion}.
\newblock \bibinfo{journal}{Nature} \bibinfo{volume}{575},
  \bibinfo{pages}{658--663}.
\bibitem[{de~Croon et~al.(2013)de~Croon, O'connor, Nicol and
  Izzo}]{de2013evolutionary}
\bibinfo{author}{de~Croon, G.C.}, \bibinfo{author}{O'connor, L.},
  \bibinfo{author}{Nicol, C.}, \bibinfo{author}{Izzo, D.},
  \bibinfo{year}{2013}.
\newblock \bibinfo{title}{Evolutionary robotics approach to odor source
  localization}.
\newblock \bibinfo{journal}{Neurocomputing} \bibinfo{volume}{121},
  \bibinfo{pages}{481--497}.
\bibitem[{Crosby(2020)}]{crosby2020building}
\bibinfo{author}{Crosby, M.}, \bibinfo{year}{2020}.
\newblock \bibinfo{title}{Building thinking machines by solving animal
  cognition tasks}.
\newblock \bibinfo{journal}{Minds and Machines} , \bibinfo{pages}{1--27}.
\bibitem[{Crosby et~al.(2019)Crosby, Beyret and Halina}]{crosby2019animal}
\bibinfo{author}{Crosby, M.}, \bibinfo{author}{Beyret, B.},
  \bibinfo{author}{Halina, M.}, \bibinfo{year}{2019}.
\newblock \bibinfo{title}{The {A}nimal-{AI} olympics}.
\newblock \bibinfo{journal}{Nature Machine Intelligence} \bibinfo{volume}{1},
  \bibinfo{pages}{257--257}.
\bibitem[{Cross et~al.(2021)Cross, Cockburn, Yue and
  O’Doherty}]{cross2021using}
\bibinfo{author}{Cross, L.}, \bibinfo{author}{Cockburn, J.},
  \bibinfo{author}{Yue, Y.}, \bibinfo{author}{O’Doherty, J.P.},
  \bibinfo{year}{2021}.
\newblock \bibinfo{title}{Using deep reinforcement learning to reveal how the
  brain encodes abstract state-space representations in high-dimensional
  environments}.
\newblock \bibinfo{journal}{Neuron} \bibinfo{volume}{109},
  \bibinfo{pages}{724--738}.
\bibitem[{Cueva et~al.(2019)Cueva, Wang, Chin and Wei}]{cueva2019emergence}
\bibinfo{author}{Cueva, C.J.}, \bibinfo{author}{Wang, P.Y.},
  \bibinfo{author}{Chin, M.}, \bibinfo{author}{Wei, X.X.},
  \bibinfo{year}{2019}.
\newblock \bibinfo{title}{Emergence of functional and structural properties of
  the head direction system by optimization of recurrent neural networks}, in:
  \bibinfo{booktitle}{International Conference on Learning Representations}.
\bibitem[{Cueva and Wei(2018)}]{cueva2018emergence}
\bibinfo{author}{Cueva, C.J.}, \bibinfo{author}{Wei, X.X.},
  \bibinfo{year}{2018}.
\newblock \bibinfo{title}{Emergence of grid-like representations by training
  recurrent neural networks to perform spatial localization}, in:
  \bibinfo{booktitle}{International Conference on Learning Representations}.
\bibitem[{Cunningham and Byron(2014)}]{cunningham2014dimensionality}
\bibinfo{author}{Cunningham, J.P.}, \bibinfo{author}{Byron, M.Y.},
  \bibinfo{year}{2014}.
\newblock \bibinfo{title}{Dimensionality reduction for large-scale neural
  recordings}.
\newblock \bibinfo{journal}{Nature Neuroscience} \bibinfo{volume}{17},
  \bibinfo{pages}{1500--1509}.
\bibitem[{Currier and Nagel(2020)}]{currier2020multisensory}
\bibinfo{author}{Currier, T.A.}, \bibinfo{author}{Nagel, K.I.},
  \bibinfo{year}{2020}.
\newblock \bibinfo{title}{Multisensory control of navigation in the fruit fly}.
\newblock \bibinfo{journal}{Current Opinion in Neurobiology}
  \bibinfo{volume}{64}, \bibinfo{pages}{10--16}.
\bibitem[{Delahunt et~al.(2018)Delahunt, Riffell and
  Kutz}]{delahunt2018biological}
\bibinfo{author}{Delahunt, C.B.}, \bibinfo{author}{Riffell, J.A.},
  \bibinfo{author}{Kutz, J.N.}, \bibinfo{year}{2018}.
\newblock \bibinfo{title}{Biological mechanisms for learning: a computational
  model of olfactory learning in the manduca sexta moth, with applications to
  neural nets}.
\newblock \bibinfo{journal}{Frontiers in computational neuroscience}
  \bibinfo{volume}{12}, \bibinfo{pages}{102}.
\bibitem[{Demir et~al.(2020)Demir, Kadakia, Anderson, Clark and
  Emonet}]{demir2020walking}
\bibinfo{author}{Demir, M.}, \bibinfo{author}{Kadakia, N.},
  \bibinfo{author}{Anderson, H.D.}, \bibinfo{author}{Clark, D.A.},
  \bibinfo{author}{Emonet, T.}, \bibinfo{year}{2020}.
\newblock \bibinfo{title}{Walking drosophila navigate complex plumes using
  stochastic decisions biased by the timing of odor encounters}.
\newblock \bibinfo{journal}{bioRxiv} .
\bibitem[{Duncker et~al.(2020)Duncker, Driscoll, Shenoy, Sahani and
  Sussillo}]{duncker2020organizing}
\bibinfo{author}{Duncker, L.}, \bibinfo{author}{Driscoll, L.},
  \bibinfo{author}{Shenoy, K.V.}, \bibinfo{author}{Sahani, M.},
  \bibinfo{author}{Sussillo, D.}, \bibinfo{year}{2020}.
\newblock \bibinfo{title}{Organizing recurrent network dynamics by
  task-computation to enable continual learning}.
\newblock \bibinfo{journal}{Advances in Neural Information Processing Systems}
  \bibinfo{volume}{33}.
\bibitem[{Ebitz and Hayden(2021)}]{ebitz2021population}
\bibinfo{author}{Ebitz, R.B.}, \bibinfo{author}{Hayden, B.Y.},
  \bibinfo{year}{2021}.
\newblock \bibinfo{title}{The population doctrine in cognitive neuroscience}.
\newblock \bibinfo{journal}{Neuron} .
\bibitem[{Ehrlich et~al.(2021)Ehrlich, Stone, Brandfonbrener, Atanasov and
  Murray}]{ehrlich2021psychrnn}
\bibinfo{author}{Ehrlich, D.B.}, \bibinfo{author}{Stone, J.T.},
  \bibinfo{author}{Brandfonbrener, D.}, \bibinfo{author}{Atanasov, A.},
  \bibinfo{author}{Murray, J.D.}, \bibinfo{year}{2021}.
\newblock \bibinfo{title}{Psychrnn: An accessible and flexible python package
  for training recurrent neural network models on cognitive tasks}.
\newblock \bibinfo{journal}{Eneuro} \bibinfo{volume}{8}.
\bibitem[{Farrell et~al.(2002)Farrell, Murlis, Long, Li and
  Card{\'e}}]{farrell2002filament}
\bibinfo{author}{Farrell, J.A.}, \bibinfo{author}{Murlis, J.},
  \bibinfo{author}{Long, X.}, \bibinfo{author}{Li, W.},
  \bibinfo{author}{Card{\'e}, R.T.}, \bibinfo{year}{2002}.
\newblock \bibinfo{title}{Filament-based atmospheric dispersion model to
  achieve short time-scale structure of odor plumes}.
\newblock \bibinfo{journal}{Environmental Fluid Mechanics} \bibinfo{volume}{2},
  \bibinfo{pages}{143--169}.
\bibitem[{Friedrich and J{\"u}licher(2007)}]{friedrich2007chemotaxis}
\bibinfo{author}{Friedrich, B.M.}, \bibinfo{author}{J{\"u}licher, F.},
  \bibinfo{year}{2007}.
\newblock \bibinfo{title}{Chemotaxis of sperm cells}.
\newblock \bibinfo{journal}{Proceedings of the National Academy of Sciences}
  \bibinfo{volume}{104}, \bibinfo{pages}{13256--13261}.
\bibitem[{Gershman and {\"O}lveczky(2020)}]{gershman2020neurobiology}
\bibinfo{author}{Gershman, S.J.}, \bibinfo{author}{{\"O}lveczky, B.P.},
  \bibinfo{year}{2020}.
\newblock \bibinfo{title}{The neurobiology of deep reinforcement learning}.
\newblock \bibinfo{journal}{Current Biology} \bibinfo{volume}{30},
  \bibinfo{pages}{R629--R632}.
\bibitem[{Gomez-Marin and Ghazanfar(2019)}]{GOMEZMARIN201925}
\bibinfo{author}{Gomez-Marin, A.}, \bibinfo{author}{Ghazanfar, A.A.},
  \bibinfo{year}{2019}.
\newblock \bibinfo{title}{The life of behavior}.
\newblock \bibinfo{journal}{Neuron} \bibinfo{volume}{104},
  \bibinfo{pages}{25--36}.
\bibitem[{Goulas et~al.(2021)Goulas, Damicelli and Hilgetag}]{GOULAS2021}
\bibinfo{author}{Goulas, A.}, \bibinfo{author}{Damicelli, F.},
  \bibinfo{author}{Hilgetag, C.C.}, \bibinfo{year}{2021}.
\newblock \bibinfo{title}{Bio-instantiated recurrent neural networks:
  Integrating neurobiology-based network topology in artificial networks}.
\newblock \bibinfo{journal}{Neural Networks} .
\bibitem[{Green et~al.(2017)Green, Adachi, Shah, Hirokawa, Magani and
  Maimon}]{green2017neural}
\bibinfo{author}{Green, J.}, \bibinfo{author}{Adachi, A.},
  \bibinfo{author}{Shah, K.K.}, \bibinfo{author}{Hirokawa, J.D.},
  \bibinfo{author}{Magani, P.S.}, \bibinfo{author}{Maimon, G.},
  \bibinfo{year}{2017}.
\newblock \bibinfo{title}{A neural circuit architecture for angular integration
  in drosophila}.
\newblock \bibinfo{journal}{Nature} \bibinfo{volume}{546},
  \bibinfo{pages}{101--106}.
\bibitem[{Gr{\"u}nbaum and Willis(2015)}]{grunbaum2015spatial}
\bibinfo{author}{Gr{\"u}nbaum, D.}, \bibinfo{author}{Willis, M.A.},
  \bibinfo{year}{2015}.
\newblock \bibinfo{title}{Spatial memory-based behaviors for locating sources
  of odor plumes}.
\newblock \bibinfo{journal}{Movement Ecology} \bibinfo{volume}{3},
  \bibinfo{pages}{1--21}.
\bibitem[{Gupta et~al.(2021)Gupta, Savarese, Ganguli and
  Fei-Fei}]{gupta2021embodied}
\bibinfo{author}{Gupta, A.}, \bibinfo{author}{Savarese, S.},
  \bibinfo{author}{Ganguli, S.}, \bibinfo{author}{Fei-Fei, L.},
  \bibinfo{year}{2021}.
\newblock \bibinfo{title}{Embodied intelligence via learning and evolution}.
\newblock \bibinfo{journal}{arXiv preprint arXiv:2102.02202} .
\bibitem[{Haesemeyer et~al.(2019)Haesemeyer, Schier and
  Engert}]{haesemeyer2019convergent}
\bibinfo{author}{Haesemeyer, M.}, \bibinfo{author}{Schier, A.F.},
  \bibinfo{author}{Engert, F.}, \bibinfo{year}{2019}.
\newblock \bibinfo{title}{Convergent temperature representations in artificial
  and biological neural networks}.
\newblock \bibinfo{journal}{Neuron} \bibinfo{volume}{103},
  \bibinfo{pages}{1123--1134}.
\bibitem[{Hassabis et~al.(2017)Hassabis, Kumaran, Summerfield and
  Botvinick}]{hassabis2017neuroscience}
\bibinfo{author}{Hassabis, D.}, \bibinfo{author}{Kumaran, D.},
  \bibinfo{author}{Summerfield, C.}, \bibinfo{author}{Botvinick, M.},
  \bibinfo{year}{2017}.
\newblock \bibinfo{title}{Neuroscience-inspired artificial intelligence}.
\newblock \bibinfo{journal}{Neuron} \bibinfo{volume}{95},
  \bibinfo{pages}{245--258}.
\bibitem[{Hasson et~al.(2020)Hasson, Nastase and Goldstein}]{hasson2020direct}
\bibinfo{author}{Hasson, U.}, \bibinfo{author}{Nastase, S.A.},
  \bibinfo{author}{Goldstein, A.}, \bibinfo{year}{2020}.
\newblock \bibinfo{title}{Direct fit to nature: An evolutionary perspective on
  biological and artificial neural networks}.
\newblock \bibinfo{journal}{Neuron} \bibinfo{volume}{105},
  \bibinfo{pages}{416--434}.
\bibitem[{Henaff et~al.(2016)Henaff, Szlam and LeCun}]{henaff2016recurrent}
\bibinfo{author}{Henaff, M.}, \bibinfo{author}{Szlam, A.},
  \bibinfo{author}{LeCun, Y.}, \bibinfo{year}{2016}.
\newblock \bibinfo{title}{Recurrent orthogonal networks and long-memory tasks},
  in: \bibinfo{booktitle}{International Conference on Machine Learning},
  \bibinfo{organization}{PMLR}. pp. \bibinfo{pages}{2034--2042}.
\bibitem[{Hill et~al.(2018)Hill, Raffin, Ernestus, Gleave, Kanervisto, Traore,
  Dhariwal, Hesse, Klimov, Nichol, Plappert, Radford, Schulman, Sidor and
  Wu}]{hill2018stable}
\bibinfo{author}{Hill, A.}, \bibinfo{author}{Raffin, A.},
  \bibinfo{author}{Ernestus, M.}, \bibinfo{author}{Gleave, A.},
  \bibinfo{author}{Kanervisto, A.}, \bibinfo{author}{Traore, R.},
  \bibinfo{author}{Dhariwal, P.}, \bibinfo{author}{Hesse, C.},
  \bibinfo{author}{Klimov, O.}, \bibinfo{author}{Nichol, A.},
  \bibinfo{author}{Plappert, M.}, \bibinfo{author}{Radford, A.},
  \bibinfo{author}{Schulman, J.}, \bibinfo{author}{Sidor, S.},
  \bibinfo{author}{Wu, Y.}, \bibinfo{year}{2018}.
\newblock \bibinfo{title}{Stable baselines}.
\newblock
  \bibinfo{howpublished}{\url{https://github.com/hill-a/stable-baselines}}.
\bibitem[{{H}uk et~al.(2018){H}uk, {B}onnen and {H}e}]{huk_beyond_2018}
\bibinfo{author}{{H}uk, A.}, \bibinfo{author}{{B}onnen, K.},
  \bibinfo{author}{{H}e, B.}, \bibinfo{year}{2018}.
\newblock \bibinfo{title}{{B}eyond {T}rial-{B}ased {P}aradigms:
  {{{C}ontinuous}} {B}ehavior, {O}ngoing {N}eural {A}ctivity, and {N}atural
  {S}timuli}.
\newblock \bibinfo{journal}{{T}he {J}ournal of {N}euroscience} ,
  \bibinfo{pages}{1920--17}.
\bibitem[{Hulse et~al.(2020)Hulse, Haberkern, Franconville, Turner-Evans,
  Takemura, Wolff, Noorman, Dreher, Dan, Parekh et~al.}]{hulse2020connectome}
\bibinfo{author}{Hulse, B.K.}, \bibinfo{author}{Haberkern, H.},
  \bibinfo{author}{Franconville, R.}, \bibinfo{author}{Turner-Evans, D.B.},
  \bibinfo{author}{Takemura, S.}, \bibinfo{author}{Wolff, T.},
  \bibinfo{author}{Noorman, M.}, \bibinfo{author}{Dreher, M.},
  \bibinfo{author}{Dan, C.}, \bibinfo{author}{Parekh, R.}, et~al.,
  \bibinfo{year}{2020}.
\newblock \bibinfo{title}{A connectome of the drosophila central complex
  reveals network motifs suitable for flexible navigation and context-dependent
  action selection}.
\newblock \bibinfo{journal}{bioRxiv} .
\bibitem[{Jia et~al.(2021)Jia, Zhang, Cheng, Liu and Xu}]{jia2021neuronal}
\bibinfo{author}{Jia, S.}, \bibinfo{author}{Zhang, T.}, \bibinfo{author}{Cheng,
  X.}, \bibinfo{author}{Liu, H.}, \bibinfo{author}{Xu, B.},
  \bibinfo{year}{2021}.
\newblock \bibinfo{title}{Neuronal-plasticity and reward-propagation improved
  recurrent spiking neural networks}.
\newblock \bibinfo{journal}{Frontiers in Neuroscience} \bibinfo{volume}{15},
  \bibinfo{pages}{205}.
\bibitem[{Kadakia et~al.(2021)Kadakia, Demir, Michaelis, Reidenbach, Clark and
  Emonet}]{Kadakia2021}
\bibinfo{author}{Kadakia, N.}, \bibinfo{author}{Demir, M.},
  \bibinfo{author}{Michaelis, B.T.}, \bibinfo{author}{Reidenbach, M.A.},
  \bibinfo{author}{Clark, D.A.}, \bibinfo{author}{Emonet, T.},
  \bibinfo{year}{2021}.
\newblock \bibinfo{title}{Odor motion sensing enables complex plume
  navigation}.
\newblock \bibinfo{journal}{bioRxiv} \URLprefix
  \url{https://www.biorxiv.org/content/early/2021/12/11/2021.09.29.462473},
  \DOIprefix\doi{10.1101/2021.09.29.462473},
  \href{http://arxiv.org/abs/https://www.biorxiv.org/content/early/2021/12/11/2021.09.29.462473.full.pdf}{{\tt
  arXiv:https://www.biorxiv.org/content/early/2021/12/11/2021.09.29.462473.full.pdf}}.
\bibitem[{Kanitscheider and Fiete(2017)}]{kanitscheider2017training}
\bibinfo{author}{Kanitscheider, I.}, \bibinfo{author}{Fiete, I.},
  \bibinfo{year}{2017}.
\newblock \bibinfo{title}{Training recurrent networks to generate hypotheses
  about how the brain solves hard navigation problems}, in:
  \bibinfo{booktitle}{Proceedings of the 31st International Conference on
  Neural Information Processing Systems}, pp. \bibinfo{pages}{4532--4541}.
\bibitem[{Kaushik et~al.(2020)Kaushik, Renz and
  Olsson}]{kaushik2020characterizing}
\bibinfo{author}{Kaushik, P.K.}, \bibinfo{author}{Renz, M.},
  \bibinfo{author}{Olsson, S.B.}, \bibinfo{year}{2020}.
\newblock \bibinfo{title}{Characterizing long-range search behavior in diptera
  using complex 3d virtual environments}.
\newblock \bibinfo{journal}{Proceedings of the National Academy of Sciences}
  \bibinfo{volume}{117}, \bibinfo{pages}{12201--12207}.
\bibitem[{Kennedy(1983)}]{kennedy1983zigzagging}
\bibinfo{author}{Kennedy, J.}, \bibinfo{year}{1983}.
\newblock \bibinfo{title}{Zigzagging and casting as a programmed response to
  wind-borne odour: a review}.
\newblock \bibinfo{journal}{Physiological Entomology} \bibinfo{volume}{8},
  \bibinfo{pages}{109--120}.
\bibitem[{Kennedy and Marsh(1974)}]{kennedy1974pheromone}
\bibinfo{author}{Kennedy, J.S.}, \bibinfo{author}{Marsh, D.},
  \bibinfo{year}{1974}.
\newblock \bibinfo{title}{Pheromone-regulated anemotaxis in flying moths}.
\newblock \bibinfo{journal}{Science} \bibinfo{volume}{184},
  \bibinfo{pages}{999--1001}.
\bibitem[{Kietzmann et~al.(2019)Kietzmann, McClure and
  Kriegeskorte}]{kietzmann2019deep}
\bibinfo{author}{Kietzmann, T.C.}, \bibinfo{author}{McClure, P.},
  \bibinfo{author}{Kriegeskorte, N.}, \bibinfo{year}{2019}.
\newblock \bibinfo{title}{Deep neural networks in computational neuroscience},
  in: \bibinfo{booktitle}{Oxford research encyclopedia of neuroscience}.
  \bibinfo{publisher}{Oxford University Press}.
\bibitem[{Kim et~al.(2019)Kim, Hermundstad, Romani, Abbott and
  Jayaraman}]{kim2019generation}
\bibinfo{author}{Kim, S.S.}, \bibinfo{author}{Hermundstad, A.M.},
  \bibinfo{author}{Romani, S.}, \bibinfo{author}{Abbott, L.},
  \bibinfo{author}{Jayaraman, V.}, \bibinfo{year}{2019}.
\newblock \bibinfo{title}{Generation of stable heading representations in
  diverse visual scenes}.
\newblock \bibinfo{journal}{Nature} \bibinfo{volume}{576},
  \bibinfo{pages}{126--131}.
\bibitem[{Konda and Tsitsiklis(2000)}]{konda2000actor}
\bibinfo{author}{Konda, V.R.}, \bibinfo{author}{Tsitsiklis, J.N.},
  \bibinfo{year}{2000}.
\newblock \bibinfo{title}{Actor-critic algorithms}, in:
  \bibinfo{booktitle}{Advances in Neural Information Processing Systems}, pp.
  \bibinfo{pages}{1008--1014}.
\bibitem[{Kostrikov(2021)}]{kostrikov2018pytorch}
\bibinfo{author}{Kostrikov, I.}, \bibinfo{year}{2021}.
\newblock \bibinfo{title}{Pytorch implementations of reinforcement learning
  algorithms}.
\newblock \URLprefix
  \url{https://github.com/ikostrikov/pytorch-a2c-ppo-acktr-gail}.
\bibitem[{Kowadlo and Russell(2008)}]{kowadlo2008robot}
\bibinfo{author}{Kowadlo, G.}, \bibinfo{author}{Russell, R.A.},
  \bibinfo{year}{2008}.
\newblock \bibinfo{title}{Robot odor localization: a taxonomy and survey}.
\newblock \bibinfo{journal}{The International Journal of Robotics Research}
  \bibinfo{volume}{27}, \bibinfo{pages}{869--894}.
\bibitem[{Kriegeskorte(2015)}]{kriegeskorte2015deep}
\bibinfo{author}{Kriegeskorte, N.}, \bibinfo{year}{2015}.
\newblock \bibinfo{title}{{Deep Neural Networks: A new framework for modeling
  biological vision and brain information processing}}.
\newblock \bibinfo{journal}{Annual Review of Vision Science}
  \bibinfo{volume}{1}, \bibinfo{pages}{417--446}.
\bibitem[{Kriegeskorte and Wei(2021)}]{kriegeskorte2021neural}
\bibinfo{author}{Kriegeskorte, N.}, \bibinfo{author}{Wei, X.X.},
  \bibinfo{year}{2021}.
\newblock \bibinfo{title}{Neural tuning and representational geometry}.
\newblock \bibinfo{journal}{Nature Neuroscience} , \bibinfo{pages}{1--16}.
\bibitem[{Kwon et~al.(2020)Kwon, Daptardar, Schrater and
  Pitkow}]{kwon2020inverse}
\bibinfo{author}{Kwon, M.}, \bibinfo{author}{Daptardar, S.},
  \bibinfo{author}{Schrater, P.R.}, \bibinfo{author}{Pitkow, Z.},
  \bibinfo{year}{2020}.
\newblock \bibinfo{title}{Inverse rational control with partially observable
  continuous nonlinear dynamics}.
\newblock \bibinfo{journal}{Advances in Neural Information Processing Systems}
  \bibinfo{volume}{33}.
\bibitem[{Le~Mo{\"e}l and Wystrach(2020)}]{le2020towards}
\bibinfo{author}{Le~Mo{\"e}l, F.}, \bibinfo{author}{Wystrach, A.},
  \bibinfo{year}{2020}.
\newblock \bibinfo{title}{Towards a multi-level understanding in insect
  navigation}.
\newblock \bibinfo{journal}{Current Opinion in Insect Science} .
\bibitem[{Leitch et~al.(2020)Leitch, Ponce, van Breugel and
  Dickinson}]{leitch2020long}
\bibinfo{author}{Leitch, K.}, \bibinfo{author}{Ponce, F.}, \bibinfo{author}{van
  Breugel, F.}, \bibinfo{author}{Dickinson, M.H.}, \bibinfo{year}{2020}.
\newblock \bibinfo{title}{The long-distance flight behavior of drosophila
  suggests a general model for wind-assisted dispersal in insects}.
\newblock \bibinfo{journal}{bioRxiv} .
\bibitem[{Lin and Richards(2021)}]{lin2021time}
\bibinfo{author}{Lin, D.}, \bibinfo{author}{Richards, B.A.},
  \bibinfo{year}{2021}.
\newblock \bibinfo{title}{Time cell encoding in deep reinforcement learning
  agents depends on mnemonic demands}.
\newblock \bibinfo{journal}{bioRxiv} .
\bibitem[{Lochmatter and Martinoli(2009)}]{lochmatter2009theoretical}
\bibinfo{author}{Lochmatter, T.}, \bibinfo{author}{Martinoli, A.},
  \bibinfo{year}{2009}.
\newblock \bibinfo{title}{Theoretical analysis of three bio-inspired plume
  tracking algorithms}, in: \bibinfo{booktitle}{2009 IEEE International
  Conference on Robotics and Automation}, \bibinfo{organization}{IEEE}. pp.
  \bibinfo{pages}{2661--2668}.
\bibitem[{Maheswaranathan et~al.(2019a)Maheswaranathan, Williams, Golub,
  Ganguli and Sussillo}]{maheswaranathan2019reverse}
\bibinfo{author}{Maheswaranathan, N.}, \bibinfo{author}{Williams, A.},
  \bibinfo{author}{Golub, M.}, \bibinfo{author}{Ganguli, S.},
  \bibinfo{author}{Sussillo, D.}, \bibinfo{year}{2019}a.
\newblock \bibinfo{title}{Reverse engineering recurrent networks for sentiment
  classification reveals line attractor dynamics}, in:
  \bibinfo{booktitle}{Advances in Neural Information Processing Systems}, pp.
  \bibinfo{pages}{15696--15705}.
\bibitem[{Maheswaranathan et~al.(2019b)Maheswaranathan, Williams, Golub,
  Ganguli and Sussillo}]{maheswaranathan2019universality}
\bibinfo{author}{Maheswaranathan, N.}, \bibinfo{author}{Williams, A.},
  \bibinfo{author}{Golub, M.}, \bibinfo{author}{Ganguli, S.},
  \bibinfo{author}{Sussillo, D.}, \bibinfo{year}{2019}b.
\newblock \bibinfo{title}{Universality and individuality in neural dynamics
  across large populations of recurrent networks}, in:
  \bibinfo{booktitle}{Advances in Neural Information Processing Systems}, pp.
  \bibinfo{pages}{15629--15641}.
\bibitem[{Masson et~al.(2009)Masson, Bechet and Vergassola}]{masson2009chasing}
\bibinfo{author}{Masson, J.}, \bibinfo{author}{Bechet, M.B.},
  \bibinfo{author}{Vergassola, M.}, \bibinfo{year}{2009}.
\newblock \bibinfo{title}{Chasing information to search in random
  environments}.
\newblock \bibinfo{journal}{Journal of Physics A: Mathematical and Theoretical}
  \bibinfo{volume}{42}, \bibinfo{pages}{434009}.
\bibitem[{Merel et~al.(2019)Merel, Aldarondo, Marshall, Tassa, Wayne and
  Olveczky}]{merel2019deep}
\bibinfo{author}{Merel, J.}, \bibinfo{author}{Aldarondo, D.},
  \bibinfo{author}{Marshall, J.}, \bibinfo{author}{Tassa, Y.},
  \bibinfo{author}{Wayne, G.}, \bibinfo{author}{Olveczky, B.},
  \bibinfo{year}{2019}.
\newblock \bibinfo{title}{Deep neuroethology of a virtual rodent}, in:
  \bibinfo{booktitle}{International Conference on Learning Representations}.
\bibitem[{Michaelides and Panayiotou(2005)}]{michaelides2005plume}
\bibinfo{author}{Michaelides, M.P.}, \bibinfo{author}{Panayiotou, C.G.},
  \bibinfo{year}{2005}.
\newblock \bibinfo{title}{Plume source position estimation using sensor
  networks}, in: \bibinfo{booktitle}{Proceedings of the 2005 IEEE International
  Symposium on, Mediterrean Conference on Control and Automation Intelligent
  Control, 2005.}, \bibinfo{organization}{IEEE}. pp. \bibinfo{pages}{731--736}.
\bibitem[{Mikulik et~al.(2020)Mikulik, Del{\'e}tang, McGrath, Genewein, Martic,
  Legg and Ortega}]{mikulik2020meta}
\bibinfo{author}{Mikulik, V.}, \bibinfo{author}{Del{\'e}tang, G.},
  \bibinfo{author}{McGrath, T.}, \bibinfo{author}{Genewein, T.},
  \bibinfo{author}{Martic, M.}, \bibinfo{author}{Legg, S.},
  \bibinfo{author}{Ortega, P.A.}, \bibinfo{year}{2020}.
\newblock \bibinfo{title}{Meta-trained agents implement bayes-optimal agents}.
\newblock \bibinfo{journal}{arXiv preprint arXiv:2010.11223} .
\bibitem[{M{\l}ynarski and Hermundstad(2018)}]{mlynarski2018adaptive}
\bibinfo{author}{M{\l}ynarski, W.F.}, \bibinfo{author}{Hermundstad, A.M.},
  \bibinfo{year}{2018}.
\newblock \bibinfo{title}{Adaptive coding for dynamic sensory inference}.
\newblock \bibinfo{journal}{Elife} \bibinfo{volume}{7},
  \bibinfo{pages}{e32055}.
\bibitem[{Mnih et~al.(2013)Mnih, Kavukcuoglu, Silver, Graves, Antonoglou,
  Wierstra and Riedmiller}]{mnih2013playing}
\bibinfo{author}{Mnih, V.}, \bibinfo{author}{Kavukcuoglu, K.},
  \bibinfo{author}{Silver, D.}, \bibinfo{author}{Graves, A.},
  \bibinfo{author}{Antonoglou, I.}, \bibinfo{author}{Wierstra, D.},
  \bibinfo{author}{Riedmiller, M.}, \bibinfo{year}{2013}.
\newblock \bibinfo{title}{Playing atari with deep reinforcement learning}.
\newblock \bibinfo{journal}{arXiv preprint arXiv:1312.5602} .
\bibitem[{Nassar et~al.(2018)Nassar, Linderman, Bugallo and
  Park}]{nassar2018tree}
\bibinfo{author}{Nassar, J.}, \bibinfo{author}{Linderman, S.},
  \bibinfo{author}{Bugallo, M.}, \bibinfo{author}{Park, I.M.},
  \bibinfo{year}{2018}.
\newblock \bibinfo{title}{Tree-structured recurrent switching linear dynamical
  systems for multi-scale modeling}, in: \bibinfo{booktitle}{International
  Conference on Learning Representations}.
\bibitem[{Nastase et~al.(2020)Nastase, Goldstein and Hasson}]{nastase2020keep}
\bibinfo{author}{Nastase, S.A.}, \bibinfo{author}{Goldstein, A.},
  \bibinfo{author}{Hasson, U.}, \bibinfo{year}{2020}.
\newblock \bibinfo{title}{Keep it real: rethinking the primacy of experimental
  control in cognitive neuroscience}.
\newblock \bibinfo{journal}{NeuroImage} \bibinfo{volume}{222},
  \bibinfo{pages}{117254}.
\bibitem[{Ni et~al.(2021)Ni, Eysenbach and Salakhutdinov}]{ni2021recurrent}
\bibinfo{author}{Ni, T.}, \bibinfo{author}{Eysenbach, B.},
  \bibinfo{author}{Salakhutdinov, R.}, \bibinfo{year}{2021}.
\newblock \bibinfo{title}{Recurrent model-free rl is a strong baseline for many
  pomdps}.
\newblock \bibinfo{journal}{arXiv preprint arXiv:2110.05038} .
\bibitem[{Okubo et~al.(2020)Okubo, Patella, D’Alessandro and
  Wilson}]{okubo2020neural}
\bibinfo{author}{Okubo, T.S.}, \bibinfo{author}{Patella, P.},
  \bibinfo{author}{D’Alessandro, I.}, \bibinfo{author}{Wilson, R.I.},
  \bibinfo{year}{2020}.
\newblock \bibinfo{title}{A neural network for wind-guided compass navigation}.
\newblock \bibinfo{journal}{Neuron} .
\bibitem[{Pandas(2021)}]{ewma_pandas}
\bibinfo{author}{Pandas}, \bibinfo{year}{2021}.
\newblock \bibinfo{title}{Pandas documentation on exponentially weighted
  windows}.
\newblock \URLprefix
  \url{https://pandas.pydata.org/docs/user_guide/window.html#window-exponentially-weighted}.
\bibitem[{Pang et~al.(2018)Pang, van Breugel, Dickinson, Riffell and
  Fairhall}]{pang2018history}
\bibinfo{author}{Pang, R.}, \bibinfo{author}{van Breugel, F.},
  \bibinfo{author}{Dickinson, M.}, \bibinfo{author}{Riffell, J.A.},
  \bibinfo{author}{Fairhall, A.}, \bibinfo{year}{2018}.
\newblock \bibinfo{title}{History dependence in insect flight decisions during
  odor tracking}.
\newblock \bibinfo{journal}{PLoS Computational Biology} \bibinfo{volume}{14},
  \bibinfo{pages}{e1005969}.
\bibitem[{Pang et~al.(2016)Pang, Lansdell and
  Fairhall}]{pang2016dimensionality}
\bibinfo{author}{Pang, R.}, \bibinfo{author}{Lansdell, B.J.},
  \bibinfo{author}{Fairhall, A.L.}, \bibinfo{year}{2016}.
\newblock \bibinfo{title}{Dimensionality reduction in neuroscience}.
\newblock \bibinfo{journal}{Current Biology} \bibinfo{volume}{26},
  \bibinfo{pages}{R656--R660}.
\bibitem[{Park et~al.(2016)Park, Hein, Bobkov, Reidenbach, Ache and
  Principe}]{park2016neurally}
\bibinfo{author}{Park, I.J.}, \bibinfo{author}{Hein, A.M.},
  \bibinfo{author}{Bobkov, Y.V.}, \bibinfo{author}{Reidenbach, M.A.},
  \bibinfo{author}{Ache, B.W.}, \bibinfo{author}{Principe, J.C.},
  \bibinfo{year}{2016}.
\newblock \bibinfo{title}{Neurally encoding time for olfactory navigation}.
\newblock \bibinfo{journal}{PLoS Computational Biology} \bibinfo{volume}{12}.
\bibitem[{Pfeiffer and Homberg(2014)}]{pfeiffer2014organization}
\bibinfo{author}{Pfeiffer, K.}, \bibinfo{author}{Homberg, U.},
  \bibinfo{year}{2014}.
\newblock \bibinfo{title}{Organization and functional roles of the central
  complex in the insect brain}.
\newblock \bibinfo{journal}{Annual Review of Entomology} \bibinfo{volume}{59},
  \bibinfo{pages}{165--184}.
\bibitem[{Plum and Labonte(2021)}]{plum2021scant}
\bibinfo{author}{Plum, F.}, \bibinfo{author}{Labonte, D.},
  \bibinfo{year}{2021}.
\newblock \bibinfo{title}{scant—an open-source platform for the creation of
  3d models of arthropods (and other small objects)}.
\newblock \bibinfo{journal}{PeerJ} \bibinfo{volume}{9},
  \bibinfo{pages}{e11155}.
\bibitem[{Rajan and Abbott(2006)}]{rajan2006eigenvalue}
\bibinfo{author}{Rajan, K.}, \bibinfo{author}{Abbott, L.F.},
  \bibinfo{year}{2006}.
\newblock \bibinfo{title}{Eigenvalue spectra of random matrices for neural
  networks}.
\newblock \bibinfo{journal}{Physical Review Letters} \bibinfo{volume}{97},
  \bibinfo{pages}{188104}.
\bibitem[{Rapp and Nawrot(2020)}]{rapp2020spiking}
\bibinfo{author}{Rapp, H.}, \bibinfo{author}{Nawrot, M.P.},
  \bibinfo{year}{2020}.
\newblock \bibinfo{title}{A spiking neural program for sensorimotor control
  during foraging in flying insects}.
\newblock \bibinfo{journal}{Proceedings of the National Academy of Sciences}
  \bibinfo{volume}{117}, \bibinfo{pages}{28412--28421}.
\bibitem[{Recanatesi et~al.(2019)Recanatesi, Ocker, Buice and
  Shea-Brown}]{recanatesi2019dimensionality}
\bibinfo{author}{Recanatesi, S.}, \bibinfo{author}{Ocker, G.K.},
  \bibinfo{author}{Buice, M.A.}, \bibinfo{author}{Shea-Brown, E.},
  \bibinfo{year}{2019}.
\newblock \bibinfo{title}{Dimensionality in recurrent spiking networks: global
  trends in activity and local origins in connectivity}.
\newblock \bibinfo{journal}{PLoS Computational Biology} \bibinfo{volume}{15},
  \bibinfo{pages}{e1006446}.
\bibitem[{Reddy et~al.(2022)Reddy, Murthy and Vergassola}]{reddyannrev}
\bibinfo{author}{Reddy, G.}, \bibinfo{author}{Murthy, V.N.},
  \bibinfo{author}{Vergassola, M.}, \bibinfo{year}{2022}.
\newblock \bibinfo{title}{Olfactory sensing and navigation in turbulent
  environments}.
\newblock \bibinfo{journal}{Annual Review of Condensed Matter Physics}
  \bibinfo{volume}{13}.
\newblock \DOIprefix\doi{10.1146/annurev-conmatphys-031720-032754}.
\bibitem[{Reddy et~al.(2021)Reddy, Shraiman and Vergassola}]{reddy2021sector}
\bibinfo{author}{Reddy, G.}, \bibinfo{author}{Shraiman, B.I.},
  \bibinfo{author}{Vergassola, M.}, \bibinfo{year}{2021}.
\newblock \bibinfo{title}{Sector search strategies for odor trail tracking}.
\newblock \bibinfo{journal}{bioRxiv} .
\bibitem[{Richards et~al.(2019)Richards, Lillicrap, Beaudoin, Bengio, Bogacz,
  Christensen, Clopath, Costa, de~Berker, Ganguli et~al.}]{richards2019deep}
\bibinfo{author}{Richards, B.A.}, \bibinfo{author}{Lillicrap, T.P.},
  \bibinfo{author}{Beaudoin, P.}, \bibinfo{author}{Bengio, Y.},
  \bibinfo{author}{Bogacz, R.}, \bibinfo{author}{Christensen, A.},
  \bibinfo{author}{Clopath, C.}, \bibinfo{author}{Costa, R.P.},
  \bibinfo{author}{de~Berker, A.}, \bibinfo{author}{Ganguli, S.}, et~al.,
  \bibinfo{year}{2019}.
\newblock \bibinfo{title}{A deep learning framework for neuroscience}.
\newblock \bibinfo{journal}{Nature Neuroscience} \bibinfo{volume}{22},
  \bibinfo{pages}{1761--1770}.
\bibitem[{R{\'\i}os et~al.(2021)R{\'\i}os, {\"O}zdil, Ramalingasetty, Arreguit,
  Rosset, Knott, Ijspeert and Ramdya}]{rios2021neuromechfly}
\bibinfo{author}{R{\'\i}os, V.L.}, \bibinfo{author}{{\"O}zdil, P.G.},
  \bibinfo{author}{Ramalingasetty, S.T.}, \bibinfo{author}{Arreguit, J.},
  \bibinfo{author}{Rosset, S.C.}, \bibinfo{author}{Knott, G.},
  \bibinfo{author}{Ijspeert, A.J.}, \bibinfo{author}{Ramdya, P.},
  \bibinfo{year}{2021}.
\newblock \bibinfo{title}{Neuromechfly, a neuromechanical model of adult
  drosophila melanogaster}.
\newblock \bibinfo{journal}{bioRxiv} .
\bibitem[{Saxena and Cunningham(2019)}]{saxena2019towards}
\bibinfo{author}{Saxena, S.}, \bibinfo{author}{Cunningham, J.P.},
  \bibinfo{year}{2019}.
\newblock \bibinfo{title}{Towards the neural population doctrine}.
\newblock \bibinfo{journal}{Current Opinion in Neurobiology}
  \bibinfo{volume}{55}, \bibinfo{pages}{103--111}.
\bibitem[{Schaeffer et~al.(2020)Schaeffer, Khona, Meshulam, Fiete
  et~al.}]{schaeffer2020reverse}
\bibinfo{author}{Schaeffer, R.}, \bibinfo{author}{Khona, M.},
  \bibinfo{author}{Meshulam, L.}, \bibinfo{author}{Fiete, I.}, et~al.,
  \bibinfo{year}{2020}.
\newblock \bibinfo{title}{Reverse-engineering recurrent neural network
  solutions to a hierarchical inference task for mice}.
\newblock \bibinfo{journal}{Advances in Neural Information Processing Systems}
  \bibinfo{volume}{33}.
\bibitem[{Scheffer et~al.(2020)Scheffer, Xu, Januszewski, Lu, Takemura,
  Hayworth, Huang, Shinomiya, Maitlin-Shepard, Berg
  et~al.}]{scheffer2020connectome}
\bibinfo{author}{Scheffer, L.K.}, \bibinfo{author}{Xu, C.S.},
  \bibinfo{author}{Januszewski, M.}, \bibinfo{author}{Lu, Z.},
  \bibinfo{author}{Takemura, S.y.}, \bibinfo{author}{Hayworth, K.J.},
  \bibinfo{author}{Huang, G.B.}, \bibinfo{author}{Shinomiya, K.},
  \bibinfo{author}{Maitlin-Shepard, J.}, \bibinfo{author}{Berg, S.}, et~al.,
  \bibinfo{year}{2020}.
\newblock \bibinfo{title}{A connectome and analysis of the adult drosophila
  central brain}.
\newblock \bibinfo{journal}{Elife} \bibinfo{volume}{9},
  \bibinfo{pages}{e57443}.
\bibitem[{Schulman et~al.(2015)Schulman, Moritz, Levine, Jordan and
  Abbeel}]{schulman2015high}
\bibinfo{author}{Schulman, J.}, \bibinfo{author}{Moritz, P.},
  \bibinfo{author}{Levine, S.}, \bibinfo{author}{Jordan, M.},
  \bibinfo{author}{Abbeel, P.}, \bibinfo{year}{2015}.
\newblock \bibinfo{title}{High-dimensional continuous control using generalized
  advantage estimation}.
\newblock \bibinfo{journal}{arXiv preprint arXiv:1506.02438} .
\bibitem[{Schulman et~al.(2017)Schulman, Wolski, Dhariwal, Radford and
  Klimov}]{schulman2017proximal}
\bibinfo{author}{Schulman, J.}, \bibinfo{author}{Wolski, F.},
  \bibinfo{author}{Dhariwal, P.}, \bibinfo{author}{Radford, A.},
  \bibinfo{author}{Klimov, O.}, \bibinfo{year}{2017}.
\newblock \bibinfo{title}{Proximal policy optimization algorithms}.
\newblock \bibinfo{journal}{arXiv preprint arXiv:1707.06347} .
\bibitem[{Seelig and Jayaraman(2015)}]{seelig2015neural}
\bibinfo{author}{Seelig, J.D.}, \bibinfo{author}{Jayaraman, V.},
  \bibinfo{year}{2015}.
\newblock \bibinfo{title}{Neural dynamics for landmark orientation and angular
  path integration}.
\newblock \bibinfo{journal}{Nature} \bibinfo{volume}{521},
  \bibinfo{pages}{186--191}.
\bibitem[{Singh et~al.(2021)Singh, Peterson, Rao and Brunton}]{singh2021mining}
\bibinfo{author}{Singh, S.H.}, \bibinfo{author}{Peterson, S.M.},
  \bibinfo{author}{Rao, R.P.}, \bibinfo{author}{Brunton, B.W.},
  \bibinfo{year}{2021}.
\newblock \bibinfo{title}{Mining naturalistic human behaviors in long-term
  video and neural recordings}.
\newblock \bibinfo{journal}{Journal of Neuroscience Methods}
  \bibinfo{volume}{358}, \bibinfo{pages}{109199}.
\bibitem[{Song et~al.(2017)Song, Yang and Wang}]{song2017reward}
\bibinfo{author}{Song, H.F.}, \bibinfo{author}{Yang, G.R.},
  \bibinfo{author}{Wang, X.J.}, \bibinfo{year}{2017}.
\newblock \bibinfo{title}{Reward-based training of recurrent neural networks
  for cognitive and value-based tasks}.
\newblock \bibinfo{journal}{Elife} \bibinfo{volume}{6},
  \bibinfo{pages}{e21492}.
\bibitem[{Song et~al.(2020)Song, Kidzi{\'n}ski, Peng, Ong, Hicks, Levine,
  Atkeson and Delp}]{song2020deep}
\bibinfo{author}{Song, S.}, \bibinfo{author}{Kidzi{\'n}ski, {\L}.},
  \bibinfo{author}{Peng, X.B.}, \bibinfo{author}{Ong, C.},
  \bibinfo{author}{Hicks, J.L.}, \bibinfo{author}{Levine, S.},
  \bibinfo{author}{Atkeson, C.}, \bibinfo{author}{Delp, S.},
  \bibinfo{year}{2020}.
\newblock \bibinfo{title}{Deep reinforcement learning for modeling human
  locomotion control in neuromechanical simulation}.
\newblock \bibinfo{journal}{bioRxiv} .
\bibitem[{Sonkusare et~al.(2019)Sonkusare, Breakspear and
  Guo}]{sonkusare2019naturalistic}
\bibinfo{author}{Sonkusare, S.}, \bibinfo{author}{Breakspear, M.},
  \bibinfo{author}{Guo, C.}, \bibinfo{year}{2019}.
\newblock \bibinfo{title}{Naturalistic stimuli in neuroscience: critically
  acclaimed}.
\newblock \bibinfo{journal}{Trends in Cognitive Sciences} \bibinfo{volume}{23},
  \bibinfo{pages}{699--714}.
\bibitem[{Stanley et~al.(2019)Stanley, Clune, Lehman and
  Miikkulainen}]{stanley2019designing}
\bibinfo{author}{Stanley, K.O.}, \bibinfo{author}{Clune, J.},
  \bibinfo{author}{Lehman, J.}, \bibinfo{author}{Miikkulainen, R.},
  \bibinfo{year}{2019}.
\newblock \bibinfo{title}{Designing neural networks through neuroevolution}.
\newblock \bibinfo{journal}{Nature Machine Intelligence} \bibinfo{volume}{1},
  \bibinfo{pages}{24--35}.
\bibitem[{Strobl et~al.(2008)Strobl, Boulesteix, Kneib, Augustin and
  Zeileis}]{strobl2008conditional}
\bibinfo{author}{Strobl, C.}, \bibinfo{author}{Boulesteix, A.L.},
  \bibinfo{author}{Kneib, T.}, \bibinfo{author}{Augustin, T.},
  \bibinfo{author}{Zeileis, A.}, \bibinfo{year}{2008}.
\newblock \bibinfo{title}{Conditional variable importance for random forests}.
\newblock \bibinfo{journal}{BMC Bioinformatics} \bibinfo{volume}{9},
  \bibinfo{pages}{1--11}.
\bibitem[{Sun et~al.(2018)Sun, Mangan and Yue}]{sun2018analysis}
\bibinfo{author}{Sun, X.}, \bibinfo{author}{Mangan, M.}, \bibinfo{author}{Yue,
  S.}, \bibinfo{year}{2018}.
\newblock \bibinfo{title}{An analysis of a ring attractor model for cue
  integration}, in: \bibinfo{booktitle}{Conference on Biomimetic and Biohybrid
  Systems}, \bibinfo{organization}{Springer}. pp. \bibinfo{pages}{459--470}.
\bibitem[{Sussillo(2014)}]{sussillo2014neural}
\bibinfo{author}{Sussillo, D.}, \bibinfo{year}{2014}.
\newblock \bibinfo{title}{Neural circuits as computational dynamical systems}.
\newblock \bibinfo{journal}{Current Opinion in Neurobiology}
  \bibinfo{volume}{25}, \bibinfo{pages}{156--163}.
\bibitem[{Sussillo and Abbott(2009)}]{sussillo2009generating}
\bibinfo{author}{Sussillo, D.}, \bibinfo{author}{Abbott, L.F.},
  \bibinfo{year}{2009}.
\newblock \bibinfo{title}{Generating coherent patterns of activity from chaotic
  neural networks}.
\newblock \bibinfo{journal}{Neuron} \bibinfo{volume}{63},
  \bibinfo{pages}{544--557}.
\bibitem[{Sussillo and Barak(2013)}]{sussillo2013opening}
\bibinfo{author}{Sussillo, D.}, \bibinfo{author}{Barak, O.},
  \bibinfo{year}{2013}.
\newblock \bibinfo{title}{Opening the black box: low-dimensional dynamics in
  high-dimensional recurrent neural networks}.
\newblock \bibinfo{journal}{Neural Computation} \bibinfo{volume}{25},
  \bibinfo{pages}{626--649}.
\bibitem[{Sussillo et~al.(2015)Sussillo, Churchland, Kaufman and
  Shenoy}]{sussillo2015neural}
\bibinfo{author}{Sussillo, D.}, \bibinfo{author}{Churchland, M.M.},
  \bibinfo{author}{Kaufman, M.T.}, \bibinfo{author}{Shenoy, K.V.},
  \bibinfo{year}{2015}.
\newblock \bibinfo{title}{A neural network that finds a naturalistic solution
  for the production of muscle activity}.
\newblock \bibinfo{journal}{Nature Neuroscience} \bibinfo{volume}{18},
  \bibinfo{pages}{1025--1033}.
\bibitem[{Sutton and Barto(2018)}]{sutton2018reinforcement}
\bibinfo{author}{Sutton, R.S.}, \bibinfo{author}{Barto, A.G.},
  \bibinfo{year}{2018}.
\newblock \bibinfo{title}{Reinforcement Learning: An Introduction}.
\newblock \bibinfo{publisher}{MIT press}.
\bibitem[{Vergassola et~al.(2007)Vergassola, Villermaux and
  Shraiman}]{vergassola2007infotaxis}
\bibinfo{author}{Vergassola, M.}, \bibinfo{author}{Villermaux, E.},
  \bibinfo{author}{Shraiman, B.I.}, \bibinfo{year}{2007}.
\newblock \bibinfo{title}{‘infotaxis’ as a strategy for searching without
  gradients}.
\newblock \bibinfo{journal}{Nature} \bibinfo{volume}{445},
  \bibinfo{pages}{406--409}.
\bibitem[{Verma et~al.(2018)Verma, Novati and
  Koumoutsakos}]{verma2018efficient}
\bibinfo{author}{Verma, S.}, \bibinfo{author}{Novati, G.},
  \bibinfo{author}{Koumoutsakos, P.}, \bibinfo{year}{2018}.
\newblock \bibinfo{title}{Efficient collective swimming by harnessing vortices
  through deep reinforcement learning}.
\newblock \bibinfo{journal}{Proceedings of the National Academy of Sciences}
  \bibinfo{volume}{115}, \bibinfo{pages}{5849--5854}.
\bibitem[{Vogels et~al.(2005)Vogels, Rajan and Abbott}]{vogels2005neural}
\bibinfo{author}{Vogels, T.P.}, \bibinfo{author}{Rajan, K.},
  \bibinfo{author}{Abbott, L.F.}, \bibinfo{year}{2005}.
\newblock \bibinfo{title}{Neural network dynamics}.
\newblock \bibinfo{journal}{Annual Review of Neuroscience}
  \bibinfo{volume}{28}, \bibinfo{pages}{357--376}.
\bibitem[{Vouloutsi et~al.(2013)Vouloutsi, LopezSerrano, Mathews, Chimeno,
  Ziyatdinov, i~Lluna, i~Badia and Verschure}]{vouloutsi2013synthetic}
\bibinfo{author}{Vouloutsi, V.}, \bibinfo{author}{LopezSerrano, L.L.},
  \bibinfo{author}{Mathews, Z.}, \bibinfo{author}{Chimeno, A.E.},
  \bibinfo{author}{Ziyatdinov, A.}, \bibinfo{author}{i~Lluna, A.P.},
  \bibinfo{author}{i~Badia, S.B.}, \bibinfo{author}{Verschure, P.F.J.},
  \bibinfo{year}{2013}.
\newblock \bibinfo{title}{The synthetic moth: a neuromorphic approach toward
  artificial olfaction in robots}.
\newblock \bibinfo{journal}{Neuromorphic Olfaction} ,
  \bibinfo{pages}{117--152}.
\bibitem[{Vyas et~al.(2020)Vyas, Golub, Sussillo and
  Shenoy}]{vyas2020computation}
\bibinfo{author}{Vyas, S.}, \bibinfo{author}{Golub, M.D.},
  \bibinfo{author}{Sussillo, D.}, \bibinfo{author}{Shenoy, K.V.},
  \bibinfo{year}{2020}.
\newblock \bibinfo{title}{Computation through neural population dynamics}.
\newblock \bibinfo{journal}{Annual Review of Neuroscience}
  \bibinfo{volume}{43}, \bibinfo{pages}{249--275}.
\bibitem[{Wall and Perry(1987)}]{wall1987range}
\bibinfo{author}{Wall, C.}, \bibinfo{author}{Perry, J.}, \bibinfo{year}{1987}.
\newblock \bibinfo{title}{Range of action of moth sex-attractant sources}.
\newblock \bibinfo{journal}{Entomologia experimentalis et applicata}
  \bibinfo{volume}{44}, \bibinfo{pages}{5--14}.
\bibitem[{Wang et~al.(2018)Wang, Kurth-Nelson, Kumaran, Tirumala, Soyer, Leibo,
  Hassabis and Botvinick}]{wang2018prefrontal}
\bibinfo{author}{Wang, J.X.}, \bibinfo{author}{Kurth-Nelson, Z.},
  \bibinfo{author}{Kumaran, D.}, \bibinfo{author}{Tirumala, D.},
  \bibinfo{author}{Soyer, H.}, \bibinfo{author}{Leibo, J.Z.},
  \bibinfo{author}{Hassabis, D.}, \bibinfo{author}{Botvinick, M.},
  \bibinfo{year}{2018}.
\newblock \bibinfo{title}{Prefrontal cortex as a meta-reinforcement learning
  system}.
\newblock \bibinfo{journal}{Nature Neuroscience} \bibinfo{volume}{21},
  \bibinfo{pages}{860--868}.
\bibitem[{Weber and Fairhall(2019)}]{weber2019role}
\bibinfo{author}{Weber, A.I.}, \bibinfo{author}{Fairhall, A.L.},
  \bibinfo{year}{2019}.
\newblock \bibinfo{title}{The role of adaptation in neural coding}.
\newblock \bibinfo{journal}{Current Opinion in Neurobiology}
  \bibinfo{volume}{58}, \bibinfo{pages}{135--140}.
\bibitem[{Weinstein and Botvinick(2017)}]{weinstein2017structure}
\bibinfo{author}{Weinstein, A.}, \bibinfo{author}{Botvinick, M.M.},
  \bibinfo{year}{2017}.
\newblock \bibinfo{title}{Structure learning in motor control: A deep
  reinforcement learning model}.
\newblock \bibinfo{journal}{arXiv preprint arXiv:1706.06827} .
\bibitem[{Yang et~al.(2019)Yang, Joglekar, Song, Newsome and
  Wang}]{yang2019task}
\bibinfo{author}{Yang, G.R.}, \bibinfo{author}{Joglekar, M.R.},
  \bibinfo{author}{Song, H.F.}, \bibinfo{author}{Newsome, W.T.},
  \bibinfo{author}{Wang, X.J.}, \bibinfo{year}{2019}.
\newblock \bibinfo{title}{Task representations in neural networks trained to
  perform many cognitive tasks}.
\newblock \bibinfo{journal}{Nature Neuroscience} \bibinfo{volume}{22},
  \bibinfo{pages}{297--306}.
\bibitem[{Yuan et~al.(2019)Yuan, Wu, Yan and Tang}]{yuan2019reinforcement}
\bibinfo{author}{Yuan, M.}, \bibinfo{author}{Wu, X.}, \bibinfo{author}{Yan,
  R.}, \bibinfo{author}{Tang, H.}, \bibinfo{year}{2019}.
\newblock \bibinfo{title}{Reinforcement learning in spiking neural networks
  with stochastic and deterministic synapses}.
\newblock \bibinfo{journal}{Neural Computation} \bibinfo{volume}{31},
  \bibinfo{pages}{2368--2389}.

\end{thebibliography}

\appendix
\onecolumn

\clearpage
\section{Supplementary details on agent training and evaluation}
\label{sec_supp_train_eval}

\begin{figure*}[h!]
\centering
\includegraphics[width=0.85\linewidth]{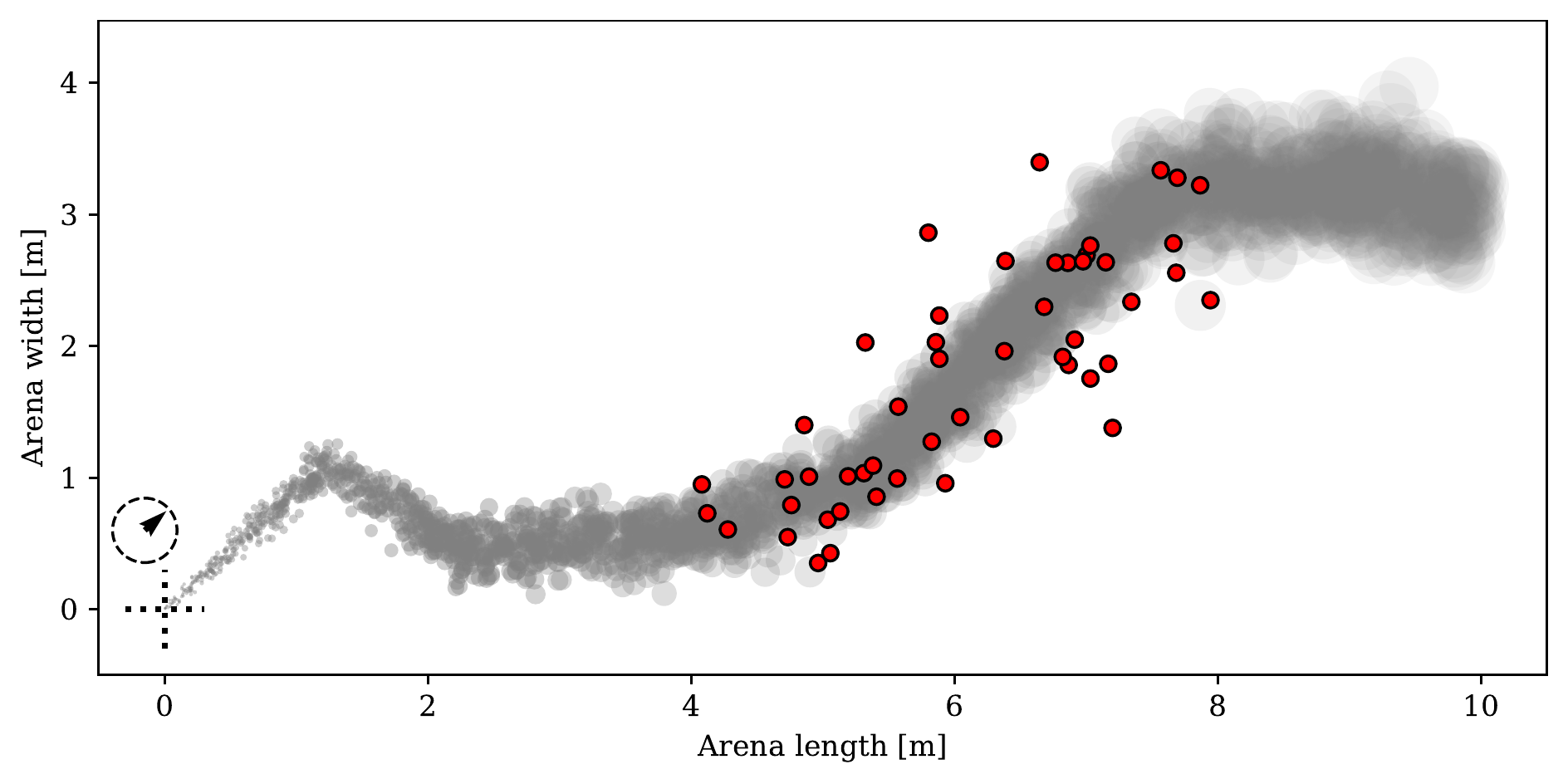}
\caption{\textbf{Snapshot of training plume:} 
Plume (dull grey) originating at crosshairs (dotted black lines, bottom left).
Current wind direction shown by arrow in dashed circle (bottom left).
50 randomly chosen initialization points (red) overlayed on plume.
Agent is initialized with a uniformly randomly chosen head-direction at a random location near or on the plume. 
Wind direction switches by a random amount at random times as described in Section \ref{supp_plumegen}
}
\label{training_plume}
\end{figure*}

[We repeat some details from the main text for the sake of readability.]

\noindent \subsection{Plume pre-computation:}
\label{supp_plumegen}
Puffs are generated at the source located at $(0,0)$, at the rate of $r_t \sim Poisson(R)$ puffs/step, where $R=1.0$.
Each puff's location $p_t = (x_t,y_t)$ is henceforth governed by the stochastic differential equation, $p_t = p_{t-1} + w_t \delta + \xi$, where $w_t$ is the wind-velocity at time $t$,   $\xi \sim \mathcal{N}(0, \sigma)$ is cross wind i.i.d. random Gaussian noise added per the turbulent plume model of \citep{farrell2002filament}. 
Each puff trajectory is integrated using a simple forward Euler integrator at 100 frames/sec.
Furthermore, each puff starts with a radius $r_0 = 0.01$ m and undergoes a diffusion process that increases it's radius at the rate of $0.01$ m/s.

We compute 120-second long (clock time) plumes ahead of training/evaluation time for the `constant',  `switch-once' and `switch-many' configurations.
A 40s window (60s - 100s) of the `switch-many' and `constant' plumes are used for training agents (Figure \ref{training_plume}).
`Sparse' plumes are simulated by downsampling the number of puffs simulated in a `constant' plume simulation.

Wind velocity $w_t$ is held constant at $(0.5, 0.0)$ m/s for all $t$ for the `constant' plume configuration.
For the `switch-once' plume configuration, wind velocity $w_t$ starts at $(0.5, 0.0)$ m/s till $t=60.00$ s, when it makes a single $45^{\circ}$ counter-clockwise turn and stays there for the rest of the simulation.
For the `switch-many' plume configuration, wind velocity $w_t$ changes once every $3.0 + \tau$ seconds, where $\tau \sim Uniform(-0.3, +0.3)$ is a random shift added i.i.d. at each change.
wind direction turns are sampled i.i.d. from a $\mathcal{N}(0^{\circ}, 30^{\circ})$ Gaussian distribution truncated at $\pm60^{\circ}$.

\noindent \subsection{Training}

\textbf{Partially Observable Markov Decision Process (POMDP):} 
To train agents using DRL, we define a Partially Observable Markov Decision Process (POMDP) \citep{sutton2018reinforcement} as follows (also see Figure~\ref{fig_training}): 

\begin{itemize}
    \item 
    \textit{Action space:} Agents provide a two dimensional output $a_t$ at each timestep corresponding to how much they want to turn and how much they want to move forward.
    $$ \mathbf{a_t} = [a_{\theta}, a_m], \text{where } a_{\theta} \in [-\theta_{max}, +\theta_{max}], a_m \in [0, \Delta_{max}]$$ 
    The maximum turn capacity of an agent ($\theta_{max}$) is $6.25\pi$ radians/s (1125 $^\circ$/s), and the maximum forward movement capacity ($\Delta_{max}$) of an agent is 2.5 m/s.
    
    \item 
    \textit{Observation space:} Agents receive a 3-dimensional egocentric sensory observation vector $o_t$ at timestep $t$, comprising odor-concentration and (x, y) coordinates of relative wind-velocity at the agent's current location and orientation in the plume.
    Note that the agent's current location and orientation in the plume are tracked and updated by the training environment code. 
    $$ \mathbf{o_t} = [o_c, o_x, o_y], \text{where } o_c \in [c_{min}, c_{max}], o_x,o_y \in [-(\Delta_{max} + |v_{wind}|), (\Delta_{max} + |v_{wind}|)]$$ 
    Here $c_{min}$ and $c_{max}$ are the minimum and maximum perceivable odor concentrations, that have been manually set to be 0.0001 and 1.0 arbitrary units respectively. 

    \item 
    \textit{Reward function:}:
    Rewards are given to encourage task completion, i.e. home in on the plume source.
    The agent receives:
    $+100$ when it reaches within a small fixed radius $r_{homed} = 0.2m$ of the source, 
    $-\epsilon$ per timestep to simulate a `metabolic cost' to flying and therefore encourage faster homing.
    We also provide the agent two shaping rewards, without which the training process is infeasibly slow:
    First, a reward proportional to the decrease in radial distance to the source per timestep $(r_{t-1} - r_{t})$ as a form of shaping reward.
    Here, $r_{t} = \sqrt{x_t^2 + y_t^2}$ is the euclidean distance of the agent to the source at timestep $t$.
    Second, a fixed negative reward of $-10$ if the agent strays more than $r_{stray} = 2$m away from the plume (i.e. the center of the nearest puff is greater than $r_{stray}$). 
    
    \item 
    \textit{Transition function:} 
    The agent's location and orientation within the plume is randomly initialized at the beginning of each training episode (see Figure \ref{training_plume} for example locations).
    The environment then deterministically updates the agent's location and orientation at each timestep taking into account its actions and the wind velocity. 
    Episodes end if the agent reaches within a radial distance $r_{homed}$ of odor source, or if the agent strays more than $r_{stray}$ from the plume, or if the episode exceeds 300 timesteps (12 seconds of clock time).
    
    \item 
    \textit{Augmented observation space for MLPs:}
    To understand the role of memory on tracking performance, in Section \ref{sec_behavior_quant}, we use feedforward-only networks (MLPs) with fixed-length memory. 
    Memory is simulated by appending historical sensory observations into the MLPs' inputs (known as `frame stacking' in the DRL literature \citep{mnih2013playing}).
    Therefore $\mathbf{o_t}$ for an MLP with $L$ timesteps history is now $[o^{(0)}_{c}, o^{(0)}_{x}, o^{(0)}_{y}, \dots o^{(L)}_{c}, o^{(L)}_{x}, o^{(L)}_{y} ] $.

\end{itemize}

We implement the POMDP environment using the OpenAI Gym \citep{brockman2016openai} and stable-baselines \citep{hill2018stable} libraries. \\

\textbf{Training curricula:}
We adapt an open source implementation \citep{kostrikov2018pytorch} of the Proximal Policy Gradient algorithm with Generalized Advantage Estimation (PPO-GAE) \citep{schulman2015high,schulman2017proximal} to train our agents.

To train our agents to perform across dynamically varying plumes, we randomize the agent's location, agent's  orientation, plume state and plume sparsity at the start of each training episode.
Agents are initialized at random starting locations $(x,y)$, where x is chosen uniformly randomly in the range $[30, 80]$ percentile of puff locations;
y is chosen by sampling from a normal distribution with mean given by the median y-coordinate of odor puffs in the range $[x-1, x+1]$, and variance given by the $5^{th} - 50^{th}$ percentile y-coordinate difference of the aforementioned odor puffs.
Initial agent orientation is selected at random from [$-\pi, \pi$] radians.
The `switch-many' plume, which changes direction every $\approx 3$ seconds, is used for training.
Initial plume state is randomized by choosing a random time between 60s - 90s, at which to initialize the precomputed plume.
The simulation is sparsified by downsampling the number of puffs to a fraction randomly uniformly chosen in the range $[0.3, 1.0]$. 
The plume is randomly flipped about the x-axis to mitigate any y-directional biases that might have crept into the finite plume simulation. 

Curriculum based training methods are known to improve training performance by gradually increasing the difficulty of the training task over the course of the training process \citep{bengio2009curriculum}.
We train our RNNs using a two stage curriculum, where we first train the RNN for 1 million timesteps on the constant wind direction plume, and then train it for another 4 million timesteps on the 'switch-many' plume.
This two stage process improved the stability and performance of the training process for RNNs, but not for MLPs.
MLPs are directly trained for 2 million timesteps on the 'switch-many' plume. 
Training durations have been chosen such that training updates reliably converge within these times. \\

\begin{figure}[h!]
\centering
\includegraphics[width=0.45\linewidth]{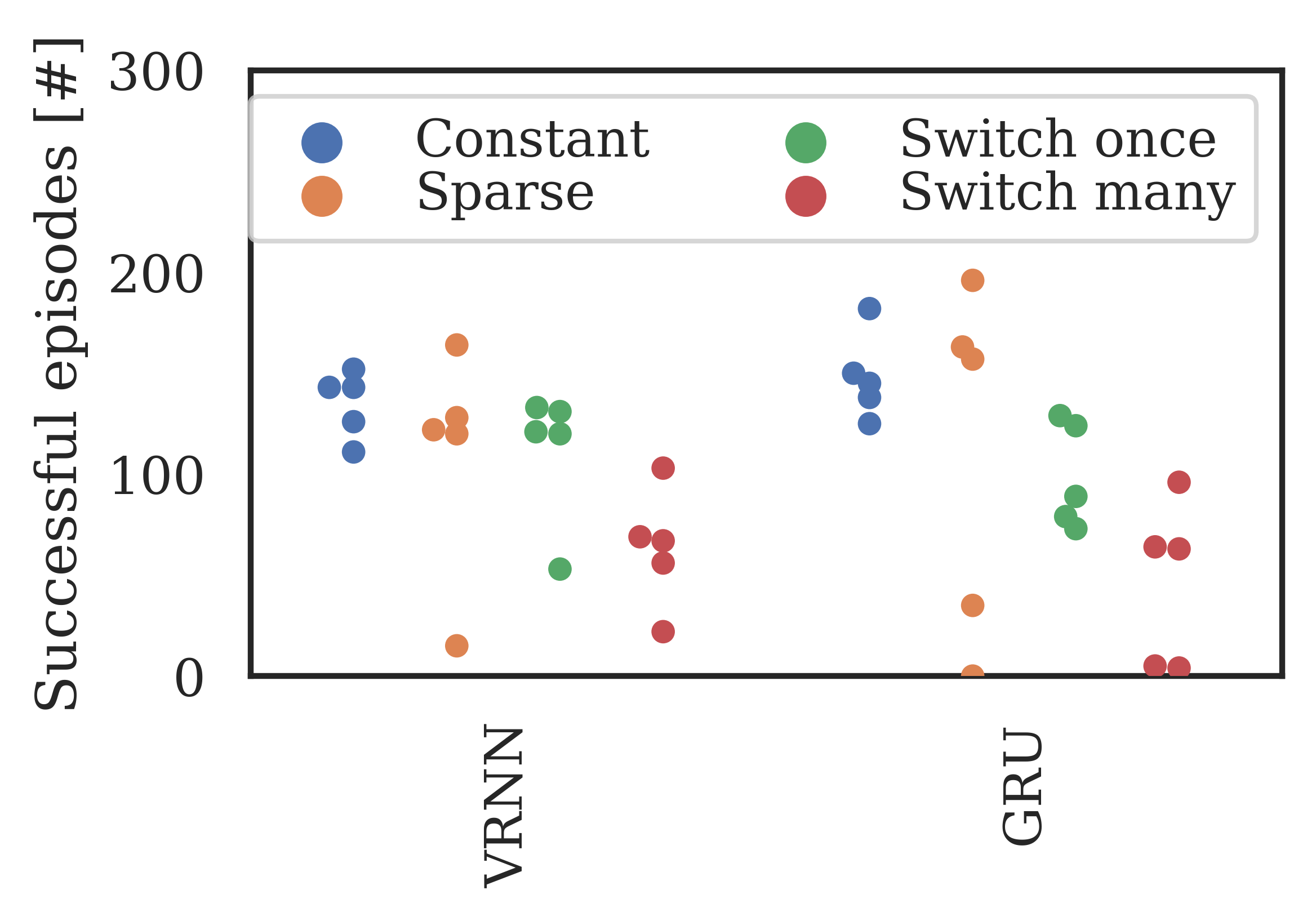}
\caption{Comparison of Vanilla RNNs and GRUs across 4 plume configurations. Vanilla RNN data is same as that in Figure \ref{fig_behavior_quant}}.
\label{fig_supp_gru}
\end{figure}

\textbf{Hyperparameter selection}:
Our training process has hyperparameters relating to (1) training algorithm hyperparameters, (2) training plume parameters, and (3) neural network architecture. 
(See \ref{sec_supp_hyperparams} for a list of all [hyper]parameters).
While PPO is not a very sample efficient algorithm, it is known to work robustly across a wide range of continuous control (continuous observation and action space) problems without needing extensive hyperparameter tuning \citep{schulman2017proximal,ni2021recurrent}.
Furthermore, exhaustive hyperparameter tuning is computationally unfeasible on our budget.
However, we do try to tweak hyperparameters one-by-one starting off from the parameters suggested in the PPO manuscript for continuous control problems.
We also trained Gated Recurrent Units (GRUs) in the same manner as we did our Vanilla RNNs (RNNs), and found that the performance of the GRUs did not significantly exceed that of the RNNs (see Figure \ref{fig_supp_gru}). \\

\textbf{Other shaping rewards explored}: Flying insects are known to exhibit a significant range of speeds \citep{van2014plume}. 
However, our trained agents mostly fly at either their maximum speed or very slowly (see Figure \ref{fig_behavior_modules}).
As additional reward shaping, we did try to add movement-related penalties to the reward function to induce some speed modulation, however, did not use these agents because of drastically worse performance compared to unpenalized agents. 
Future work could explore ways of skewing DRL reward functions towards such auxiliary goals that are not aligned with the primary plume tracking task. \\

\textbf{Computational resources}: All models are trained and evaluated on an Ubuntu Linux v20.04 workstation with Intel Core i9-9940X CPU and a TITAN RTX GPU.
Each seed takes $\approx 16$ hours to train and evaluate, with MLP and RNN models using 1 and 4 cores in parallel respectively. \\

\subsection{Evaluation}
\label{sec_supp_eval}
We evaluate trained agents over a behavioral assay comprising fixed set of initial locations, initial simulation timestamps and initial agent directions across the aforementioned plume wind direction and birth-rate configurations, each comprising 240 episodes.

The same set of $240$ initial conditions for each episode are used to initialize the agent and simulator, for each agent and dataset evaluated: 
\begin{itemize}
    \item Initial agent head angle (with respect to ground): $0, \frac{1}{4}\pi, \frac{1}{2}\pi, \frac{3}{4}\pi, \pi, \frac{5}{4}\pi, \frac{3}{2}\pi, \frac{7}{4}\pi$ radians
    \item Initial x-coordinate: 4, 6, and 8 meters
    \item Initial y-coordinate: 0$^{th}$, 25$^{th}$, 50$^{th}$, 75$^{th}$, and 100$^{th}$ percentile of the minimum and maximum y-coordinate of the puffs located in a 1-meter band around the initial x-coordinate. 
    For 'constant'  wind direction plumes (including sparse plumes), the task is made harder by selecting only 0$^{th}$, 50$^{th}$, and 100$^{th}$ percentiles as described before (i.e. $y_{min}, y_{median}, y_{max}$) and then adding two other locations that are $\pm 0.5$ m outside the plume (i.e. $y_{min} - 0.5$ m and $y_{max} + 0.5$ m)]
    \item Initial timestamp: $60.00$s and $61.00$s ($58.00$s and $59.00$s for the 'switch-once' plume as it switches at exactly $60.00$s)
\end{itemize}

\textbf{Agent selection}:  We train 14 seeds per model type (RNNs, and MLPs with 2, 4, 6 ..., 12 timesteps of history) and select the top-5 best performing seeds for analysis.
Performance here is measured by counting the number of successful episodes across `constant', 'switch-once' and 'switch-many' plumes. \\ 

\textbf{Evaluation subset}: For our analyses in Sections \ref{sec_repr} -- \ref{sec_ttcs}, 
we use a randomly selected 120 episode \textit{evaluation subset} of the 240 evaluation episodes for each of the constant, switch-once and switch-many plume configurations.
The selected episodes are balanced to include an equal number (60 episodes each) of successful and unsuccessful plume tracking episode outcomes.
Whenever there are fewer than 60 episodes of either outcome type (successful or unsuccessful) for any plume configuration, then the selection is trimmed to use an equal number of episodes of the smaller outcome type.
PCAs tend to be sensitive to imbalances in the data, and this balancing process enables visualizations to be consistently compared across agents.  
In the analysis described in Section \ref{sec_behavior_quant}, we use all 240 evaluation episodes per agent. \\

\clearpage
\section{Key parameters for simulation, agent, model, and training and evaluation}
\label{sec_supp_hyperparams}

\begin{table}[h!]
    \centering
    \begin{tabular}{ccccc}
     \hline\hline
     \textbf{Parameter description} & \textbf{Value/Range} \\
     \hline   
       Simulation integration time-step & 0.01s  \\ \hline
       Wind speed & 0.5 m/s  \\  \hline
       Wind speed crosswind noise & $\mathcal{N}(0, 0.005)$ m/s (per timestep)  \\ \hline
       Puff birth rate (Poisson mean) & 1.0 puffs/timestep (at 100 FPS)  \\ \hline
       Puff initial radius & 0.01m  \\ \hline
       Puff radius growth rate (diffusion) & 0.01m/s  \\ \hline
       Maximum plume extent simulated (x, y) & (-2/+10m, $\pm$ 5m)  \\ \hline
     \hline
    \end{tabular}
    \caption{Plume parameters}
\end{table}

\begin{table}[h!]
    \centering
    \begin{tabular}{ccccc}
     \hline\hline
     \textbf{Parameter description} & \textbf{Value/Range} \\
     \hline   
       Environment frame rate  & 25 FPS  \\ \hline
       Sensor sampling rate & 25 Hz  \\ \hline
       Forward movement capacity ($\Delta_{max}$) & 2.5 m/s  \\ \hline
       Turn capacity ($\theta_{max}$) & $\pm$ 6.25 $\pi$ radians/s ($\pm$ 1125 $^\circ$/sec)  \\ \hline
       Homing radius & 0.2 m  \\ \hline
       Max. stray from plume allowed & 2 m  \\ \hline
       Odor sensing thresholds (minimum, maximum) & (0.0001, 1.0) (A.U).  \\ \hline
     \hline
    \end{tabular}
    \caption{Agent and environment parameters}
\end{table}

\begin{table}[h!]
    \centering
    \begin{tabular}{ccccc}
     \hline\hline
     \textbf{Parameter description} & \textbf{Value/Range} \\
     \hline   
       RNN hidden layer width & 64 units  \\ \hline
       Feedforward hidden layer width(s) & 64 units  \\ \hline
       Neural network nonlinearity & tanh  \\ \hline
       Layer initialization (Recurrent, Feedforward) & (Normal, Orthogonal)  \\ \hline
     \hline
    \end{tabular}
    \caption{Model (neural network) parameters}
\end{table}

\begin{table}[h!]
    \centering
    \begin{tabular}{ccccc}
     \hline\hline
     \textbf{Parameter description} & \textbf{Value/Range} \\
     \hline   
        RNN training steps & 5M  \\ \hline
        MLP training steps & 2M  \\ \hline
        Learning Rate & 0.0003 (with linear decay) \\ \hline
        PPO Entropy Coefficient & 0.05 \\ \hline
        PPO Value Loss Coefficient & 0.5 \\ \hline
        PPO Epochs & 10 \\ \hline
        PPO Gamma & 0.99 \\ \hline
        PPO max. gradient norm & 0.5 \\ \hline
        GAE Lambda & 0.95 \\ \hline
        GAE steps & 2048 \\ \hline
     \hline
    \end{tabular}
    \caption{Training algorithm, training curriculum and evaluation parameters}
\end{table}

\clearpage
\section{Behavior module metadata distributions}
\label{sec_supp_module}

\begin{table}[h!]
    \centering
    \begin{tabular}{ccccc}
     \hline\hline
     \textbf{Agent} & \textbf{Agent ID} & \textbf{Lost module threshold}   \\
     \hline   
        RNN 1 & 2760377 & 30 steps (1.2 s) \\ \hline
        RNN 2 & 3199993 & 25 steps (1.0 s) \\ \hline  
        RNN 3 & 3307e9 & 35 steps (1.4 s) \\ \hline
        RNN 4 & 541058 & 38 steps (1.52 s) \\ \hline
        RNN 5 & 9781ba & 25 steps (1.0 s) \\ \hline
     \hline
    \end{tabular}
    \caption[Thresholds defining behavioral module changes]{Thresholds for defining when the \textit{lost} behavior module kicks in i.e. duration (in timesteps or seconds) since the plume was last encountered.}
\label{table_supp_module_threshold}
\end{table}

\begin{figure*}[h!]
\centering
\includegraphics[width=0.85\linewidth]{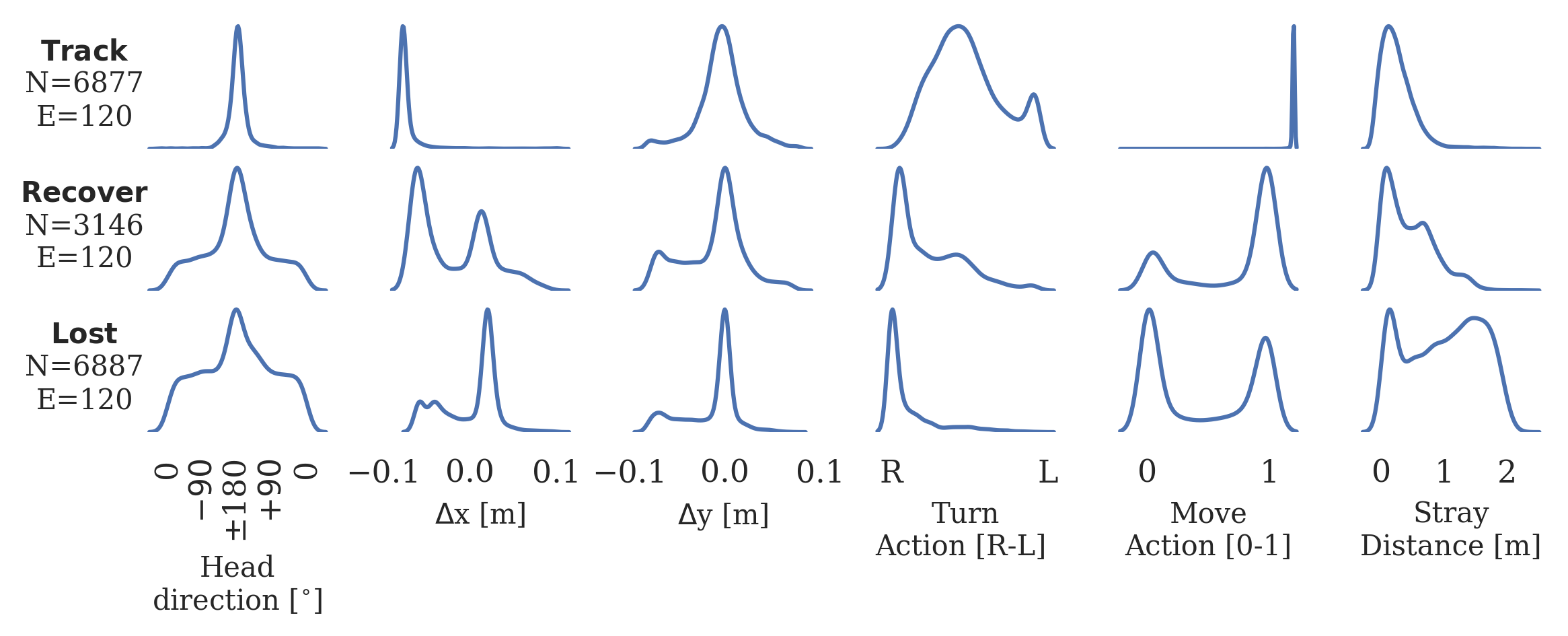}
\caption[Behavior modules - Agent 1]{Behavior modules - Agent 1 (See Figure \ref{fig_behavior_modules} for equivalent data on Agent 3 and figure details)}
\end{figure*}

\begin{figure*}[h!]
\centering
\includegraphics[width=0.85\linewidth]{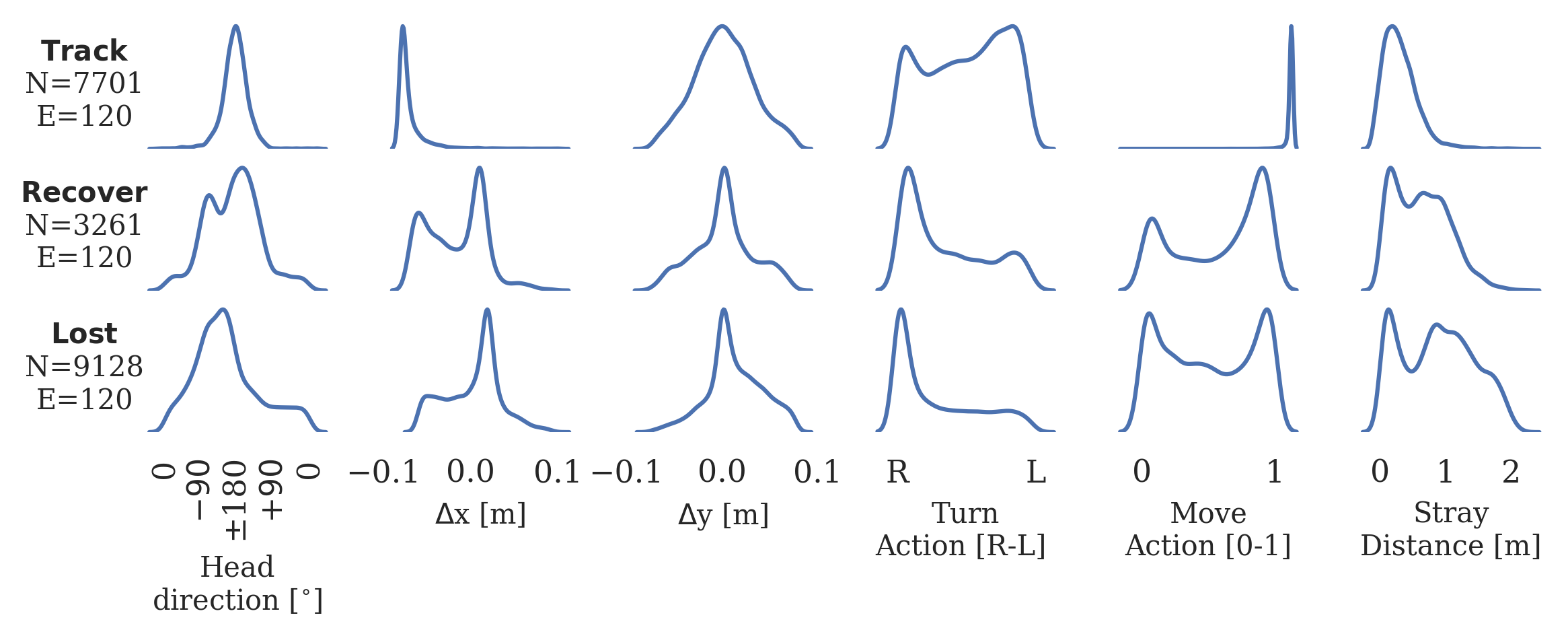}
\caption[Behavior modules - Agent 2]{Behavior modules - Agent 2}
\end{figure*}

\begin{figure*}[h!]
\centering
\includegraphics[width=0.85\linewidth]{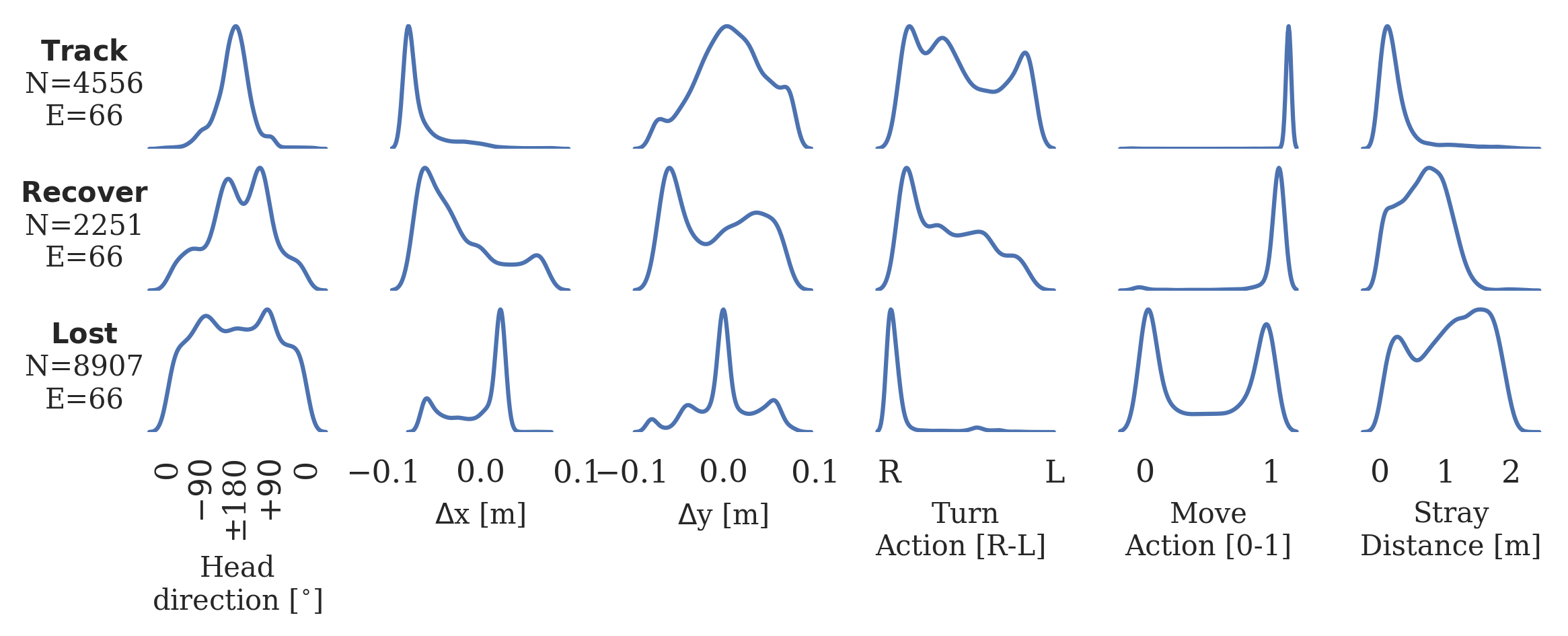}
\caption[Behavior modules - Agent 3]{Behavior modules - Agent 3 (same as Figure \ref{fig_behavior_modules})}
\end{figure*}

\begin{figure*}[h!]
\centering
\includegraphics[width=0.85\linewidth]{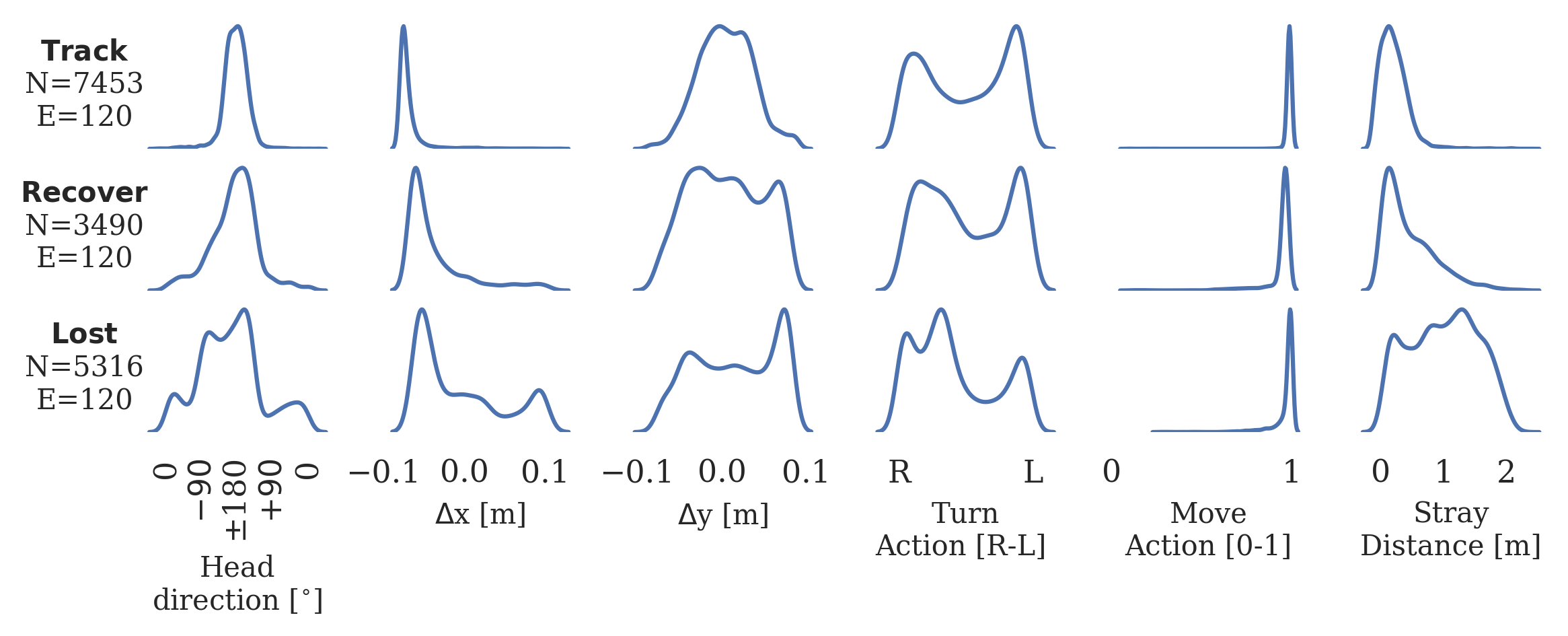}
\caption[Behavior modules - Agent 4]{Behavior modules - Agent 4}
\end{figure*}

\begin{figure*}[h!]
\centering
\includegraphics[width=0.85\linewidth]{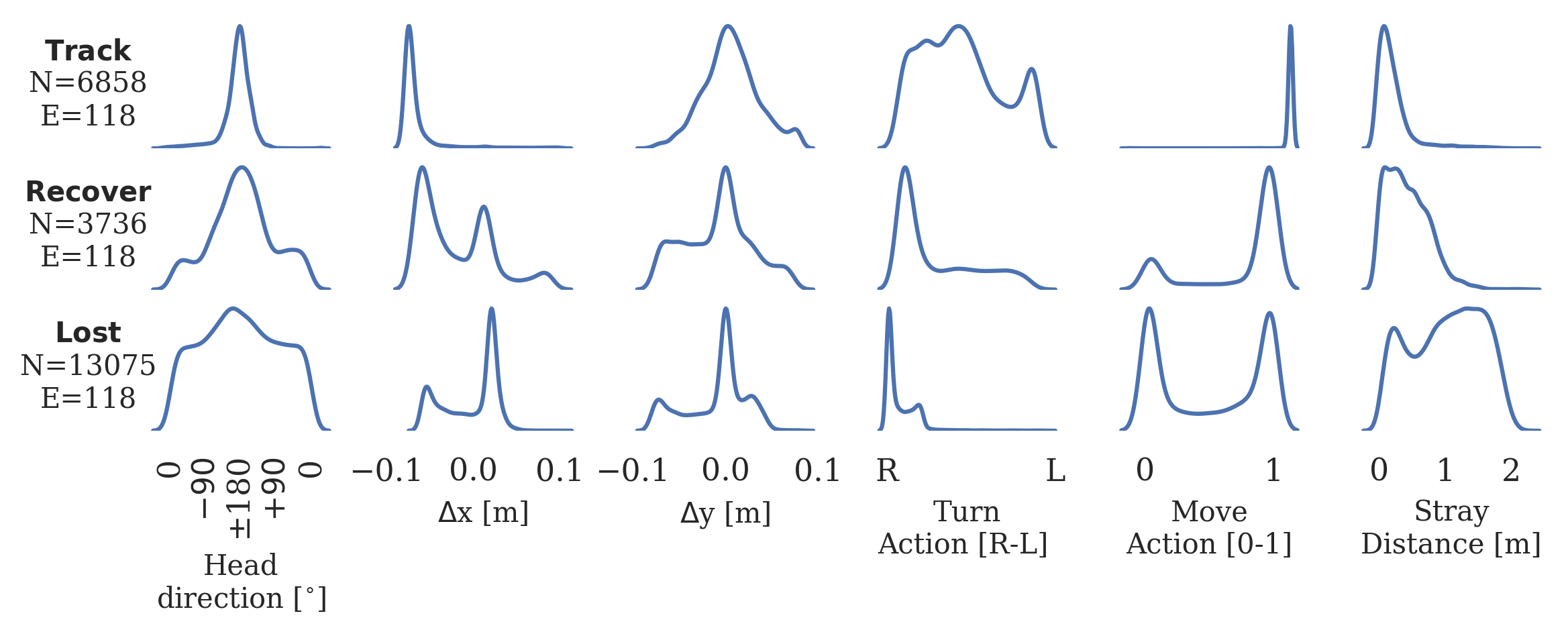}
\caption[Behavior modules - Agent 5]{Behavior modules - Agent 5}
\end{figure*}

\clearpage
\section{Comparing reference frames for plume tracking}
\label{sec_supp_centerline}

\begin{figure*}[h!]
\centering
\includegraphics[width=0.28\linewidth]{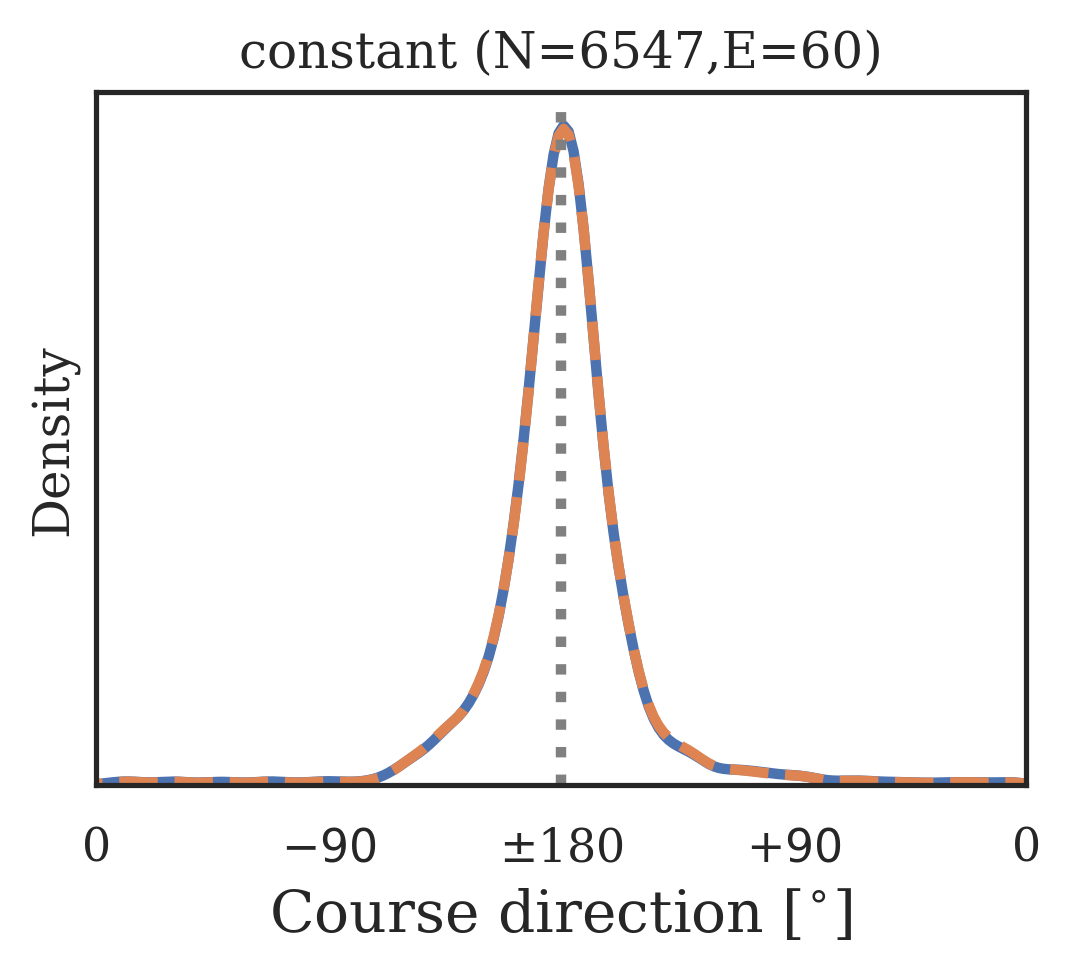}
\includegraphics[width=0.28\linewidth]{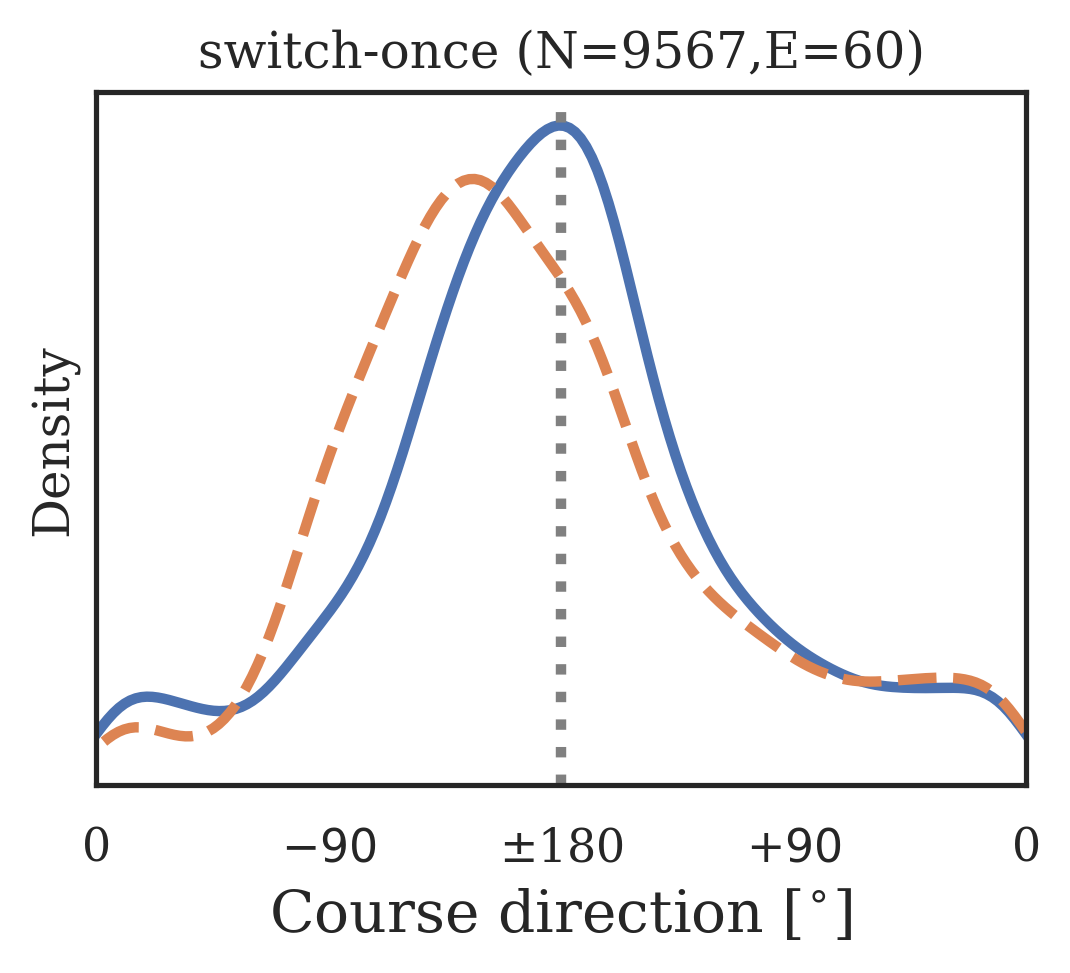}
\includegraphics[width=0.405\linewidth]{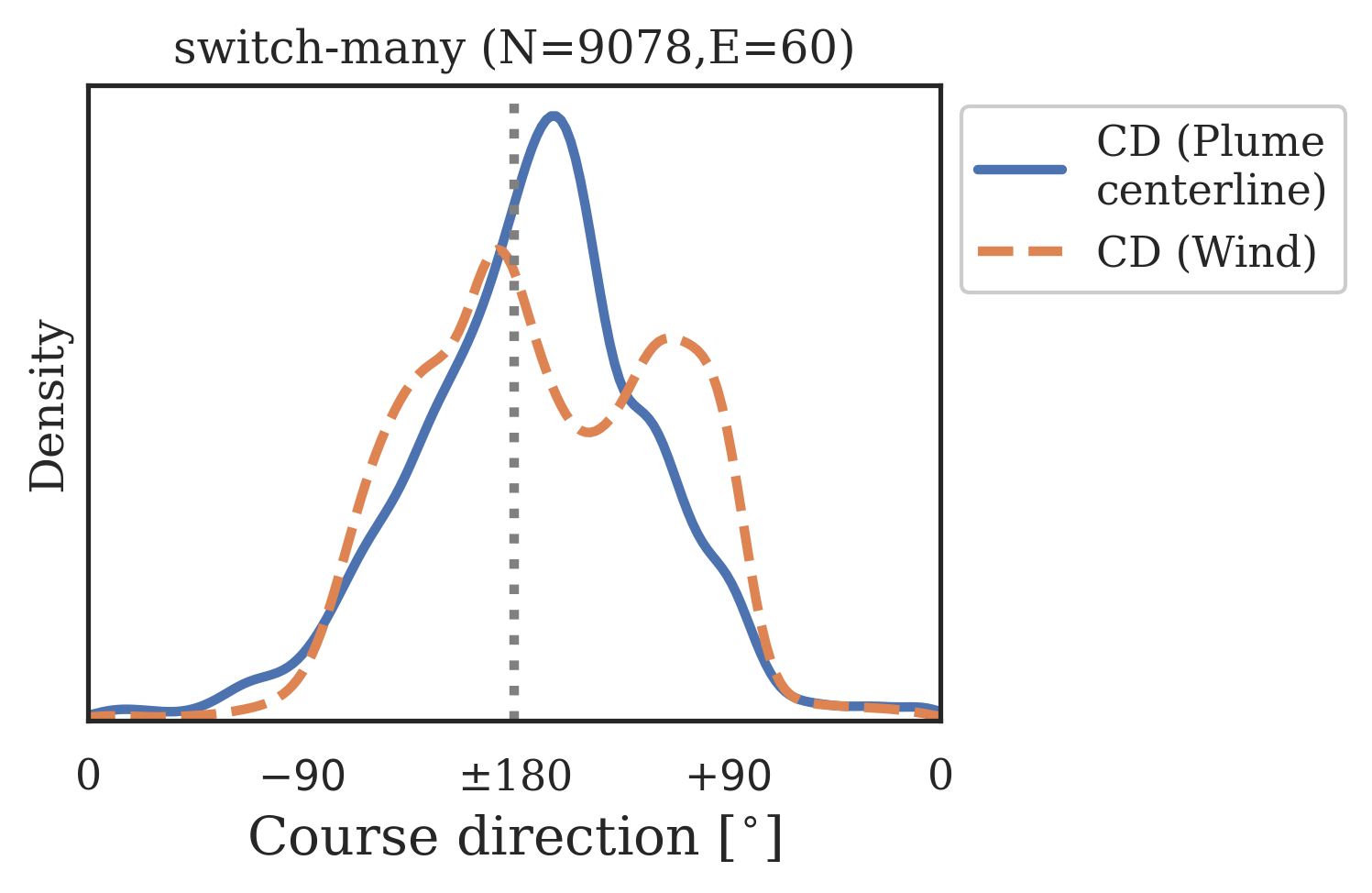}
\caption{Empirical course-direction (CD) distribution - Agent 1 (See Figure \ref{fig_centerline} for equivalent data on Agent 3 and figure details)} 
\end{figure*}

\begin{figure*}[h!]
\centering
\includegraphics[width=0.28\linewidth]{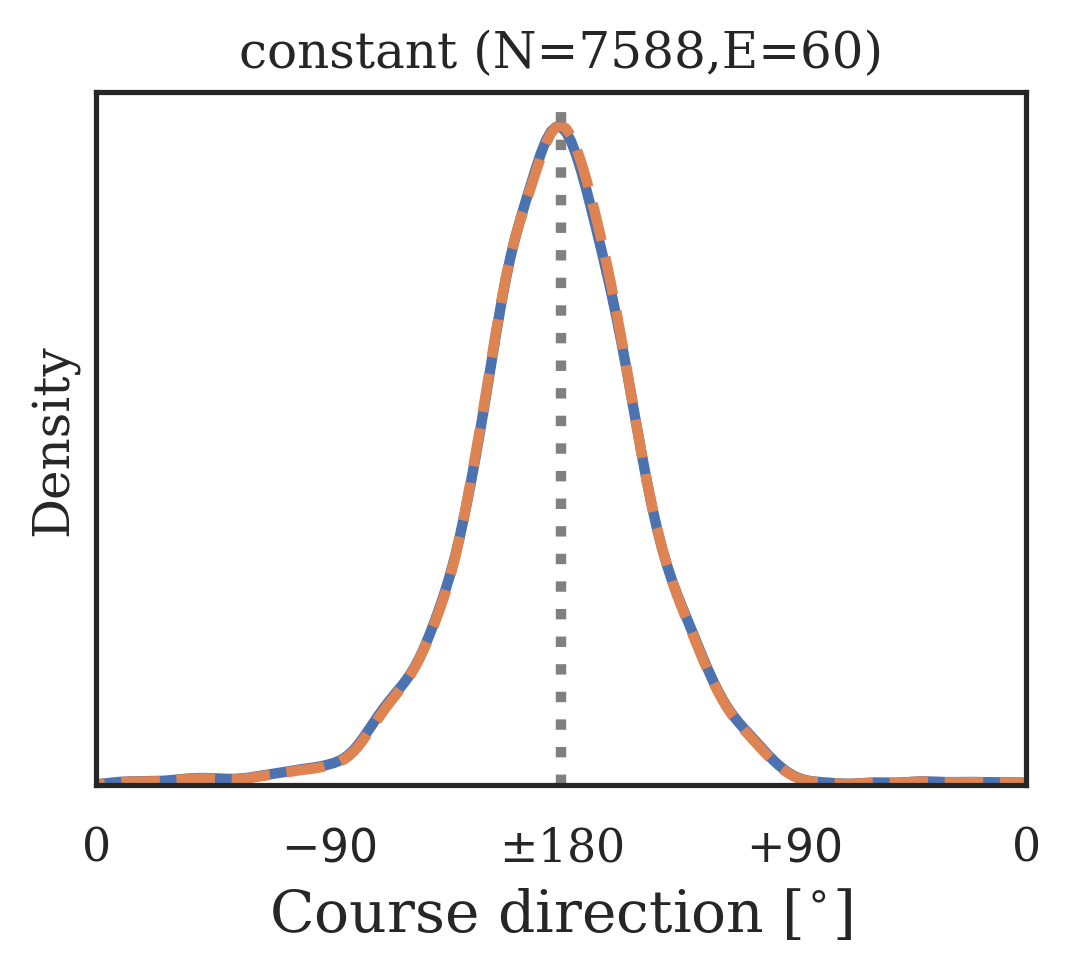}
\includegraphics[width=0.28\linewidth]{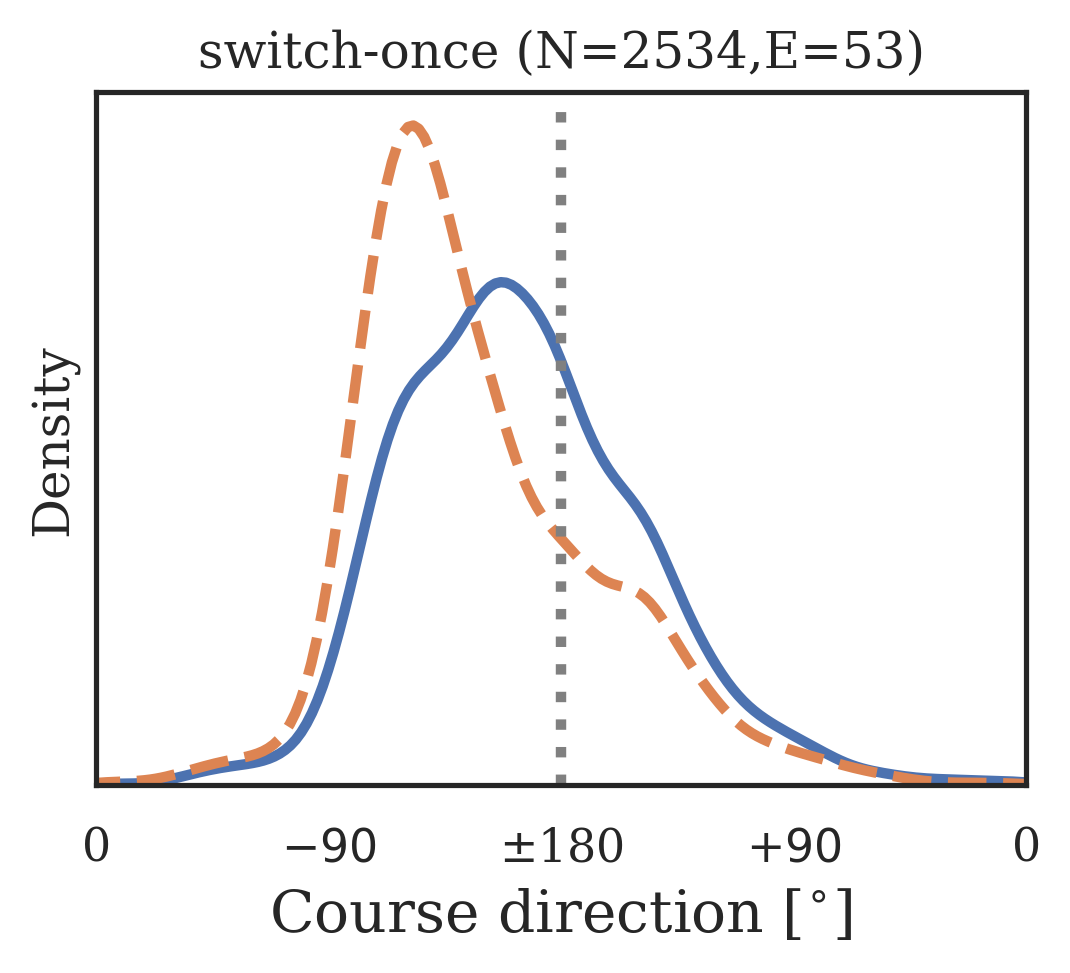}
\includegraphics[width=0.405\linewidth]{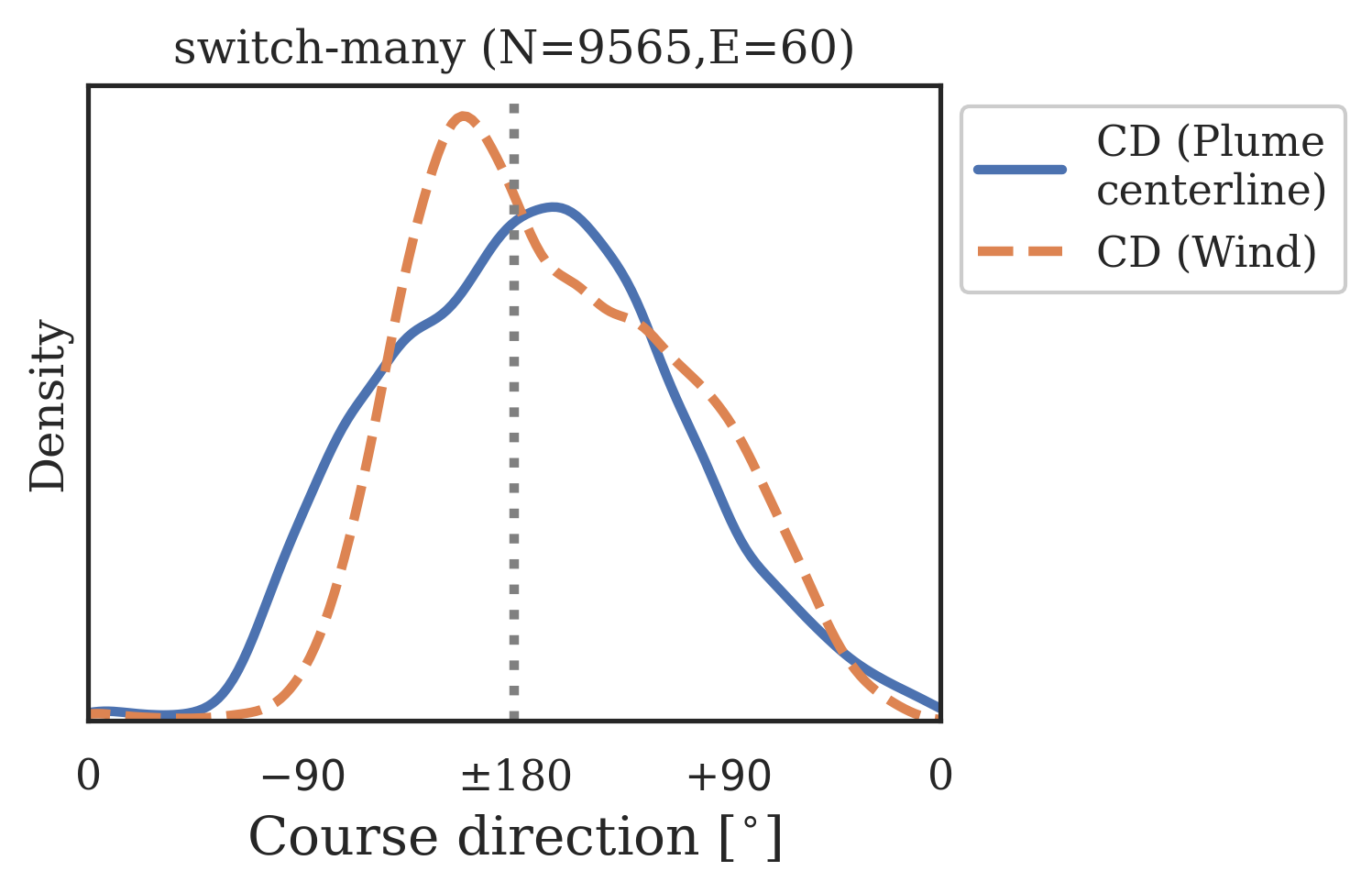}
\caption{Empirical course-direction (CD) distribution - Agent 2} 
\end{figure*}

\begin{figure*}[h!]
\centering
\includegraphics[width=0.28\linewidth]{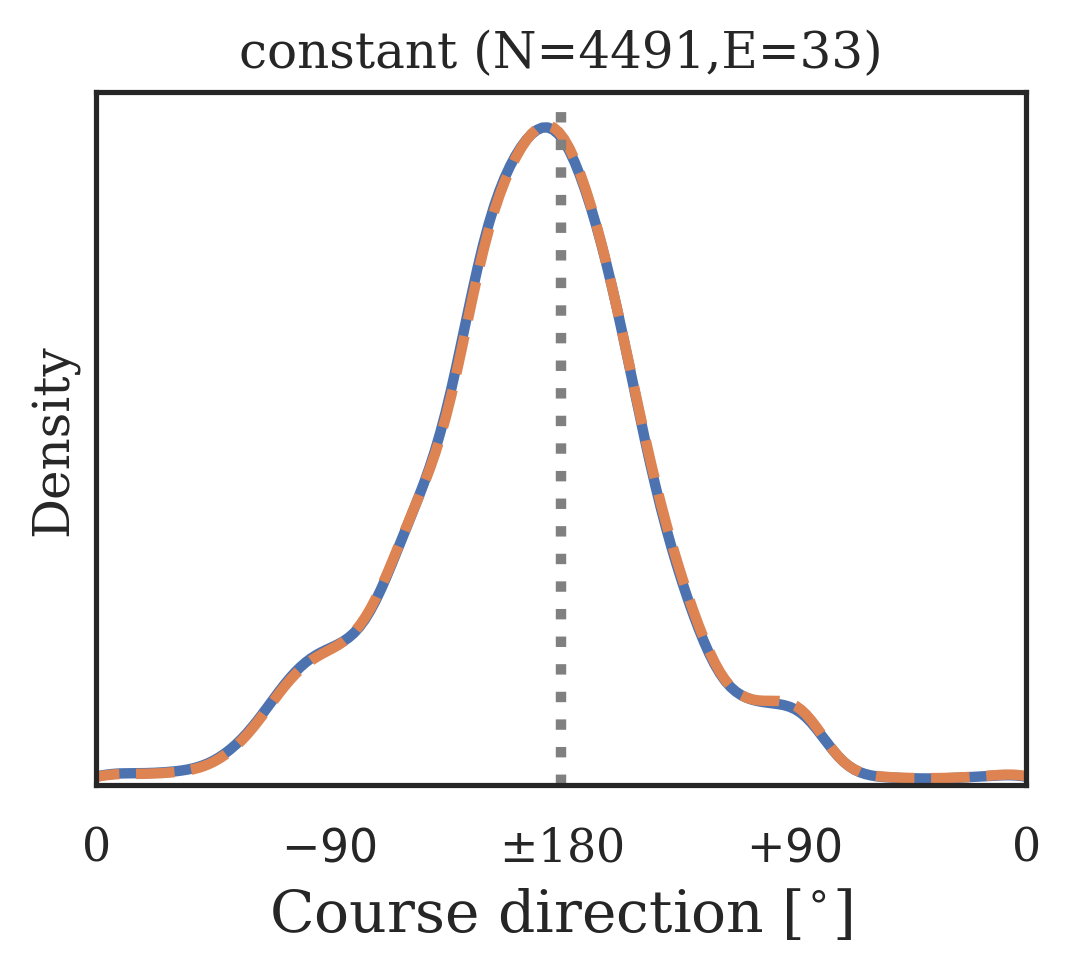}
\includegraphics[width=0.28\linewidth]{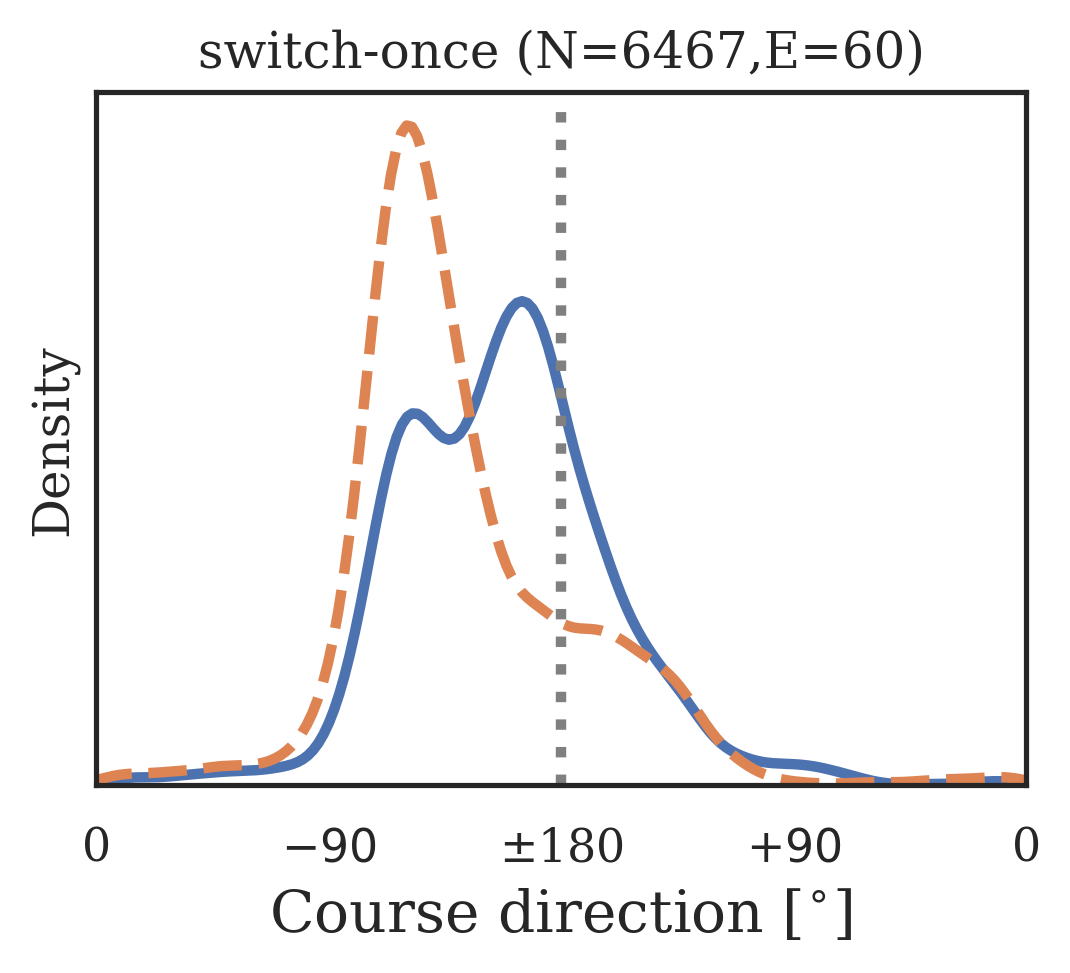}
\includegraphics[width=0.405\linewidth]{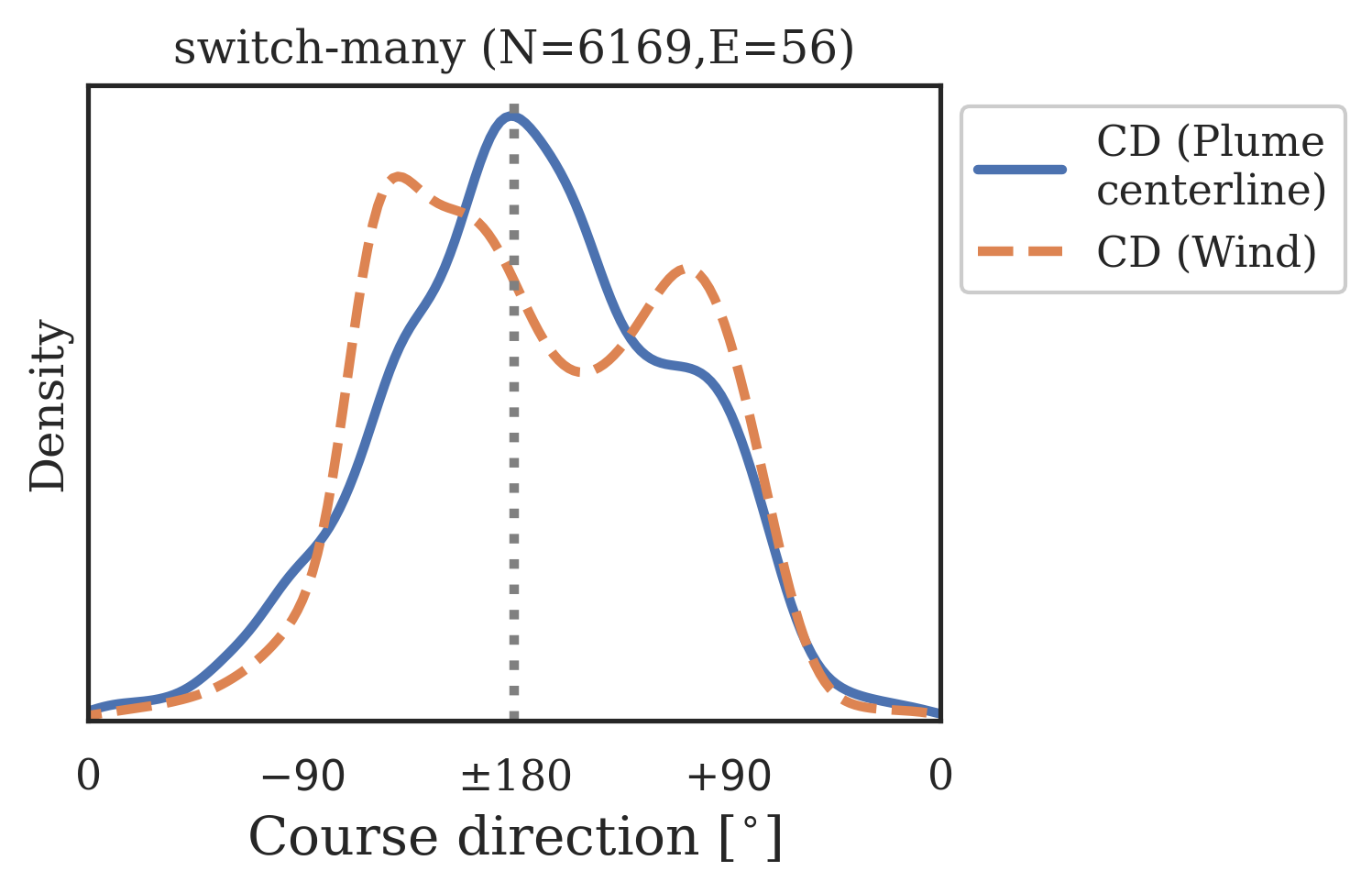}
\caption[Agent 3: Empirical course-direction (CD) distribution]{ 
Empirical course-direction (CD) distribution - Agent 3 (Same as Figure \ref{fig_centerline}) 
}
\end{figure*}

\begin{figure*}[h!]
\centering
\includegraphics[width=0.28\linewidth]{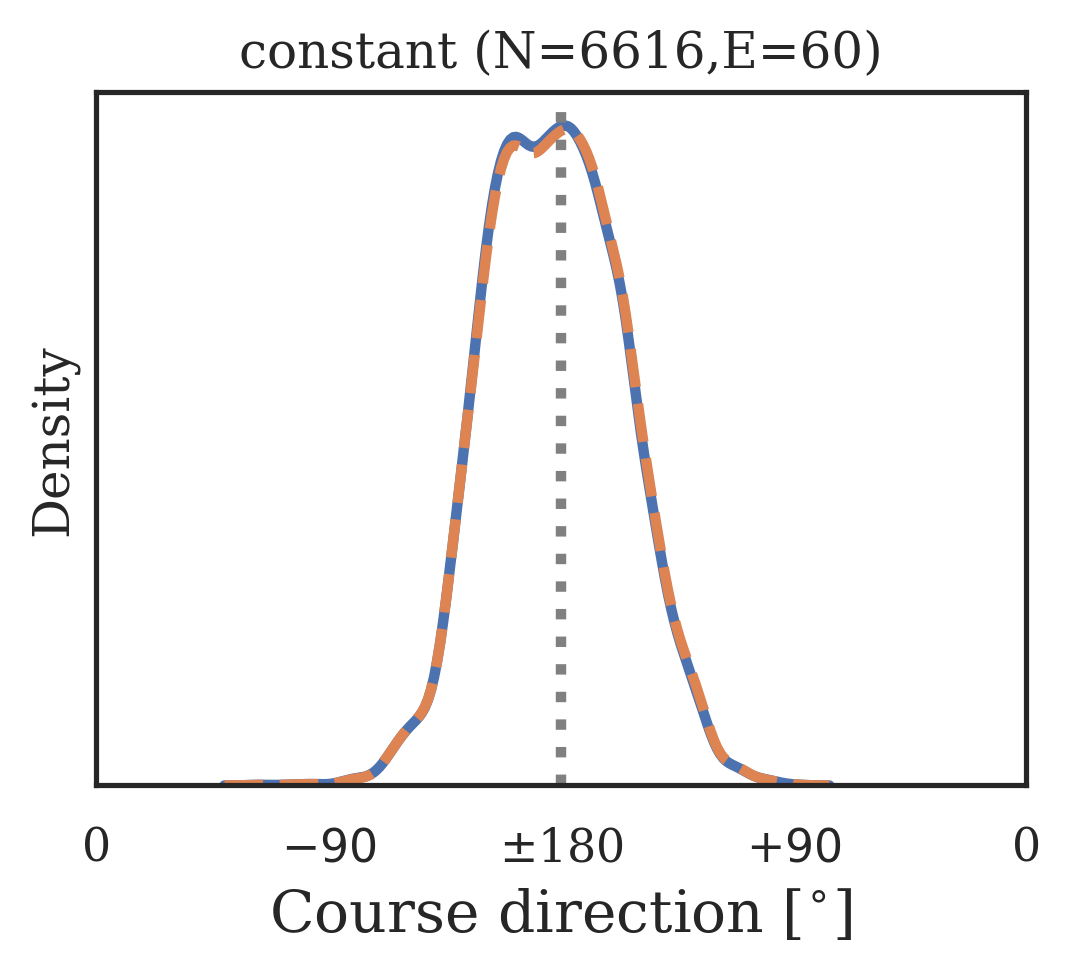}
\includegraphics[width=0.28\linewidth]{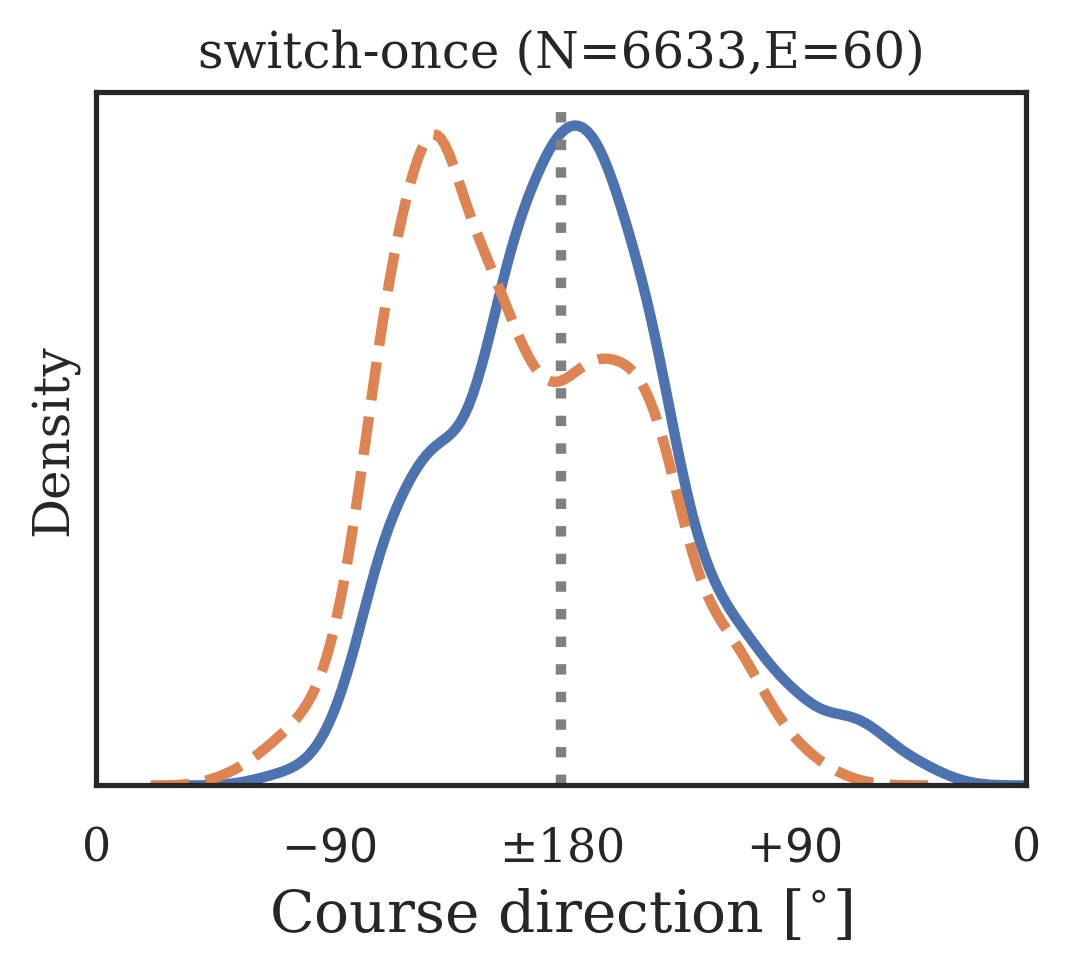}
\includegraphics[width=0.405\linewidth]{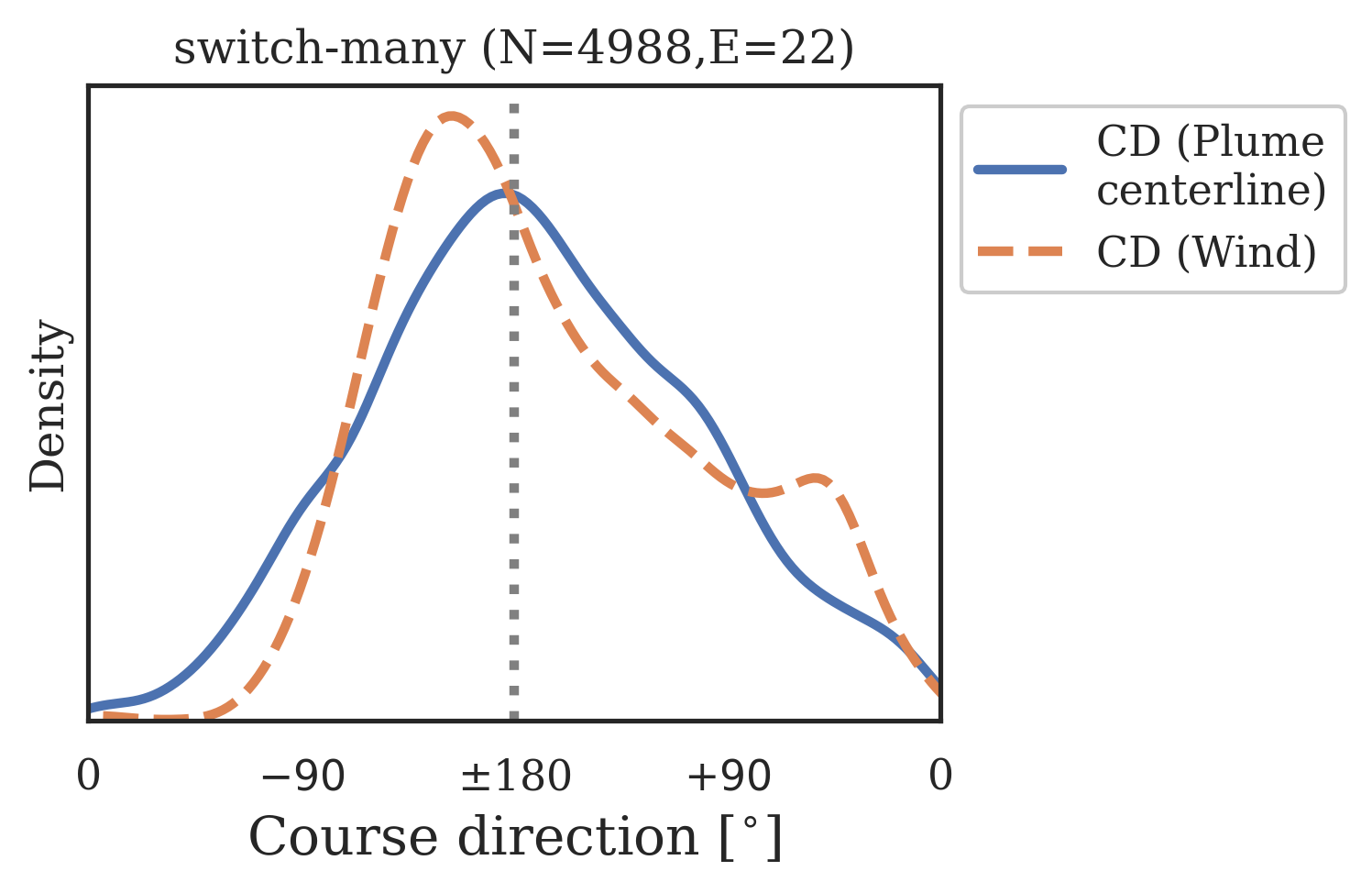}
\caption{Empirical course-direction (CD) distribution - Agent 4} 
\end{figure*}

\begin{figure*}[h!]
\centering
\includegraphics[width=0.28\linewidth]{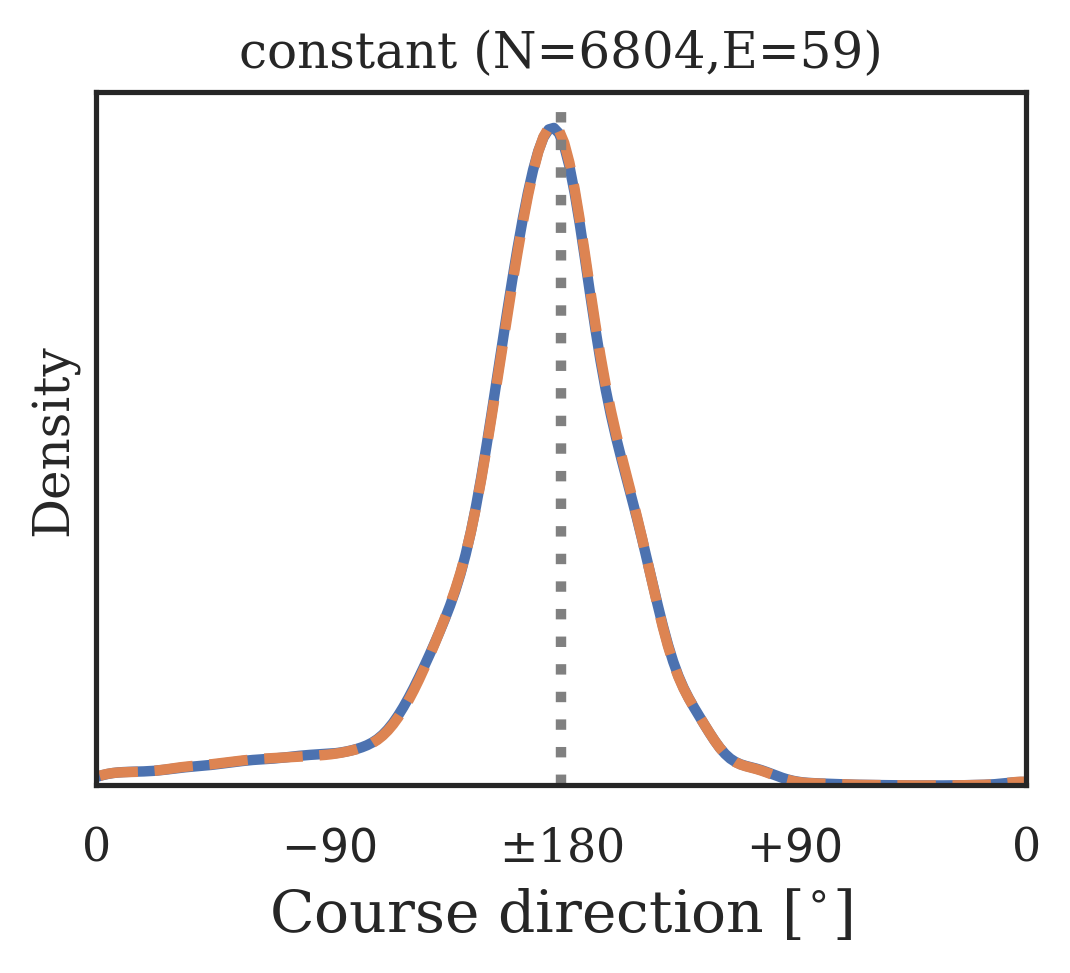}
\includegraphics[width=0.28\linewidth]{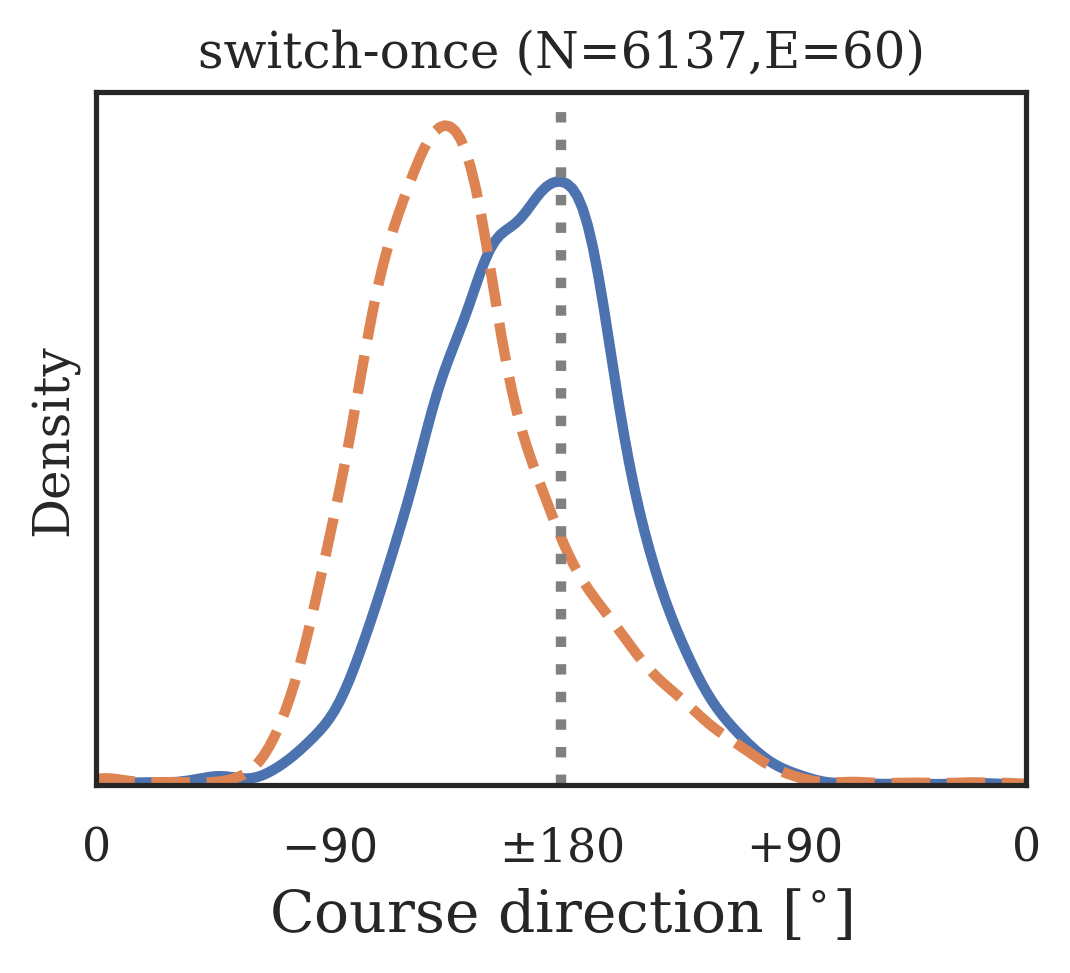}
\includegraphics[width=0.405\linewidth]{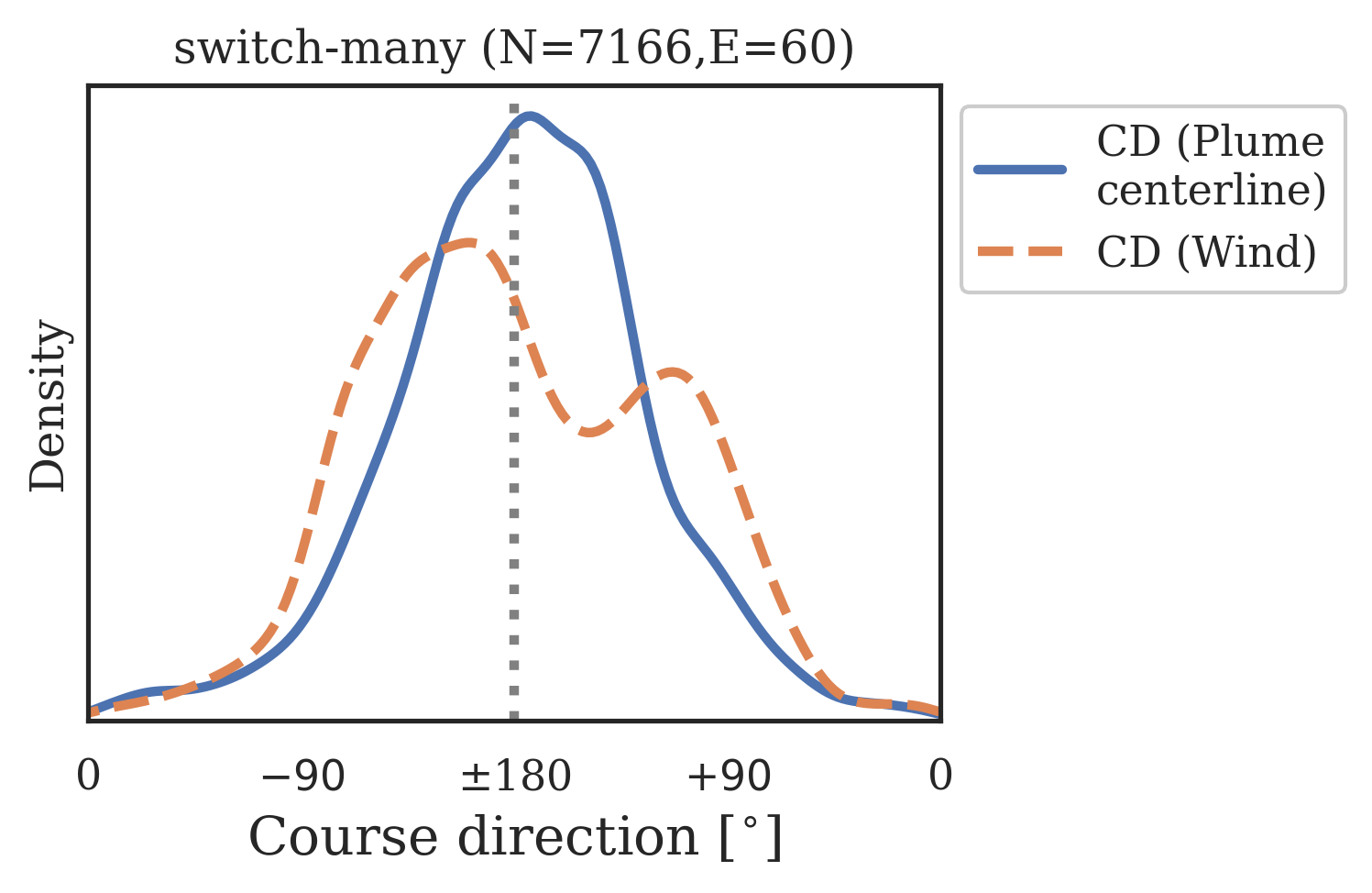}
\caption{Empirical course-direction (CD) distribution - Agent 5} 
\end{figure*}

\clearpage
\section{Neural activity dimensionality and neural representations}
\label{sec_supp_repr}

\textbf{Odor encounters}: Our definition of odor encounters is identical to that used in \cite{demir2020walking}. 
The stream of odor inputs is discretized to be 1 at the first timestep of the stream where the odor is perceptible and 0 for the remaining contiguous steps where it is still perceptible. 
Other processing is as described in Section \ref{sec_repr}.

\begin{table}[h!]
    \centering
    \begin{tabular}{cccccc}
     \hline\hline
     \textbf{Agent} & 
     \textbf{Agent ID} & 
     $odor_{EWMA}$ & 
     $odor_{EWMA}$ & 
     $odor_{ENC}$ & 
     $odor_{ENC}$  \\
     & 
     &
     window [steps] & 
     $R^2$ &
     window [steps] &
     $R^2$ \\
     \hline   
        RNN 1 & 2760377 & 8  & 0.91 & 62 & 0.57 \\ \hline
        RNN 2 & 3199993 & 10 & 0.86 & 44 & 0.71 \\ \hline  
        RNN 3 & 3307e9  & 8  & 0.92 & 46 & 0.57 \\ \hline
        RNN 4 & 541058  & 6  & 0.88 & 40 & 0.51 \\ \hline
        RNN 5 & 9781ba  & 12 & 0.91 & 44 & 0.59 \\ \hline
     \hline
    \end{tabular}
    \caption{Moving window lengths and linear regression fit $R^2$ for two represented variables: $odor_{EWMA}$ and $odor_{ENC}$. 
    See Section \ref{sec_repr} for further details. 
    (Recall that 25 timesteps = 1.0 second).
    }
\label{table_supp_ewma}
\end{table}

\textbf{Agent action classifier}: To quantify how important these represented variables are to actual task performance, we train a Random Forest (RF) \citep{breiman2001random} classifier to predict actions taken by the agent over successful trajectories.
We uniformly partition the Turn and Move action variable, which are continuous valued, into domains of 3 and 2 discrete classes respectively.
These classes correspond roughly to `left', `center' and `right' turns, and to `fast' and `slow' forward movements.
These are concatenated to form a 6-class independent variable.
The classifier receives instantaneous sensory observations (egocentric wind speed x and y coordinates $w_X, w_Y$ and odor concentration) and the four aforementioned encoded features as inputs. 
Training and test sets are a randomized non-overlapping 80$\%$--20$\%$ split of evaluation episodes, balanced across plume configuration and episode outcomes.
We do a 20-trial 3-fold cross-validated randomized search over the number-of-estimators (range: [10,50]) hyperparameter, and then train a classifier using the best hyperparameter on the whole training set. 
We next estimate the relative importance of each input feature by calculating its permutation importance score \citep{strobl2008conditional, breiman2001random}, which is an estimate of the reduction in the classifier’s accuracy across several (N=30) randomized permutations of that feature.
Note again that the estimates provided by this analysis are approximate due to the discretization of the action data and correlations between features.

\begin{table}[h!]
    \centering
    \begin{tabular}{ccccc}
     \hline\hline
     \textbf{Agent} & \textbf{Agent ID} & \textbf{Test set accuracy} & \textbf{Test set accuracy} & \textbf{Test set accuracy}   \\
     & & \textbf{(All features)} & \textbf{(Instantaneous only)} & \textbf{(Most freq. class)}   \\
     \hline   
        RNN 1 & 2760377 & 0.84 & 0.74 (0.10) & 0.33 (0.51) \\ \hline
        RNN 2 & 3199993 & 0.67 & 0.49 (0.18) & 0.28 (0.39) \\ \hline  
        RNN 3 & 3307e9 & 0.82 & 0.69 (0.13) & 0.39 (0.43) \\ \hline
        RNN 4 & 541058 & 0.70 & 0.53 (0.17) & 0.44 (0.26) \\ \hline
        RNN 5 & 9781ba & 0.84 & 0.74 (0.10) & 0.40 (0.44) \\ \hline
     \hline
    \end{tabular}
    \caption{Classifier based quantification of contribution of represented features: 
    In last two columns, quantity in parentheses is the difference in accuracy with respect to classifier that has all features (4 represented features and instantaneous features). 
    Represented features contribute to higher test accuracy. 
    See Section \ref{sec_repr} for further details.
    }
\label{table_supp_repr}
\end{table}

\begin{figure*}[h!]
\begin{center}
\includegraphics[width=0.40\linewidth]{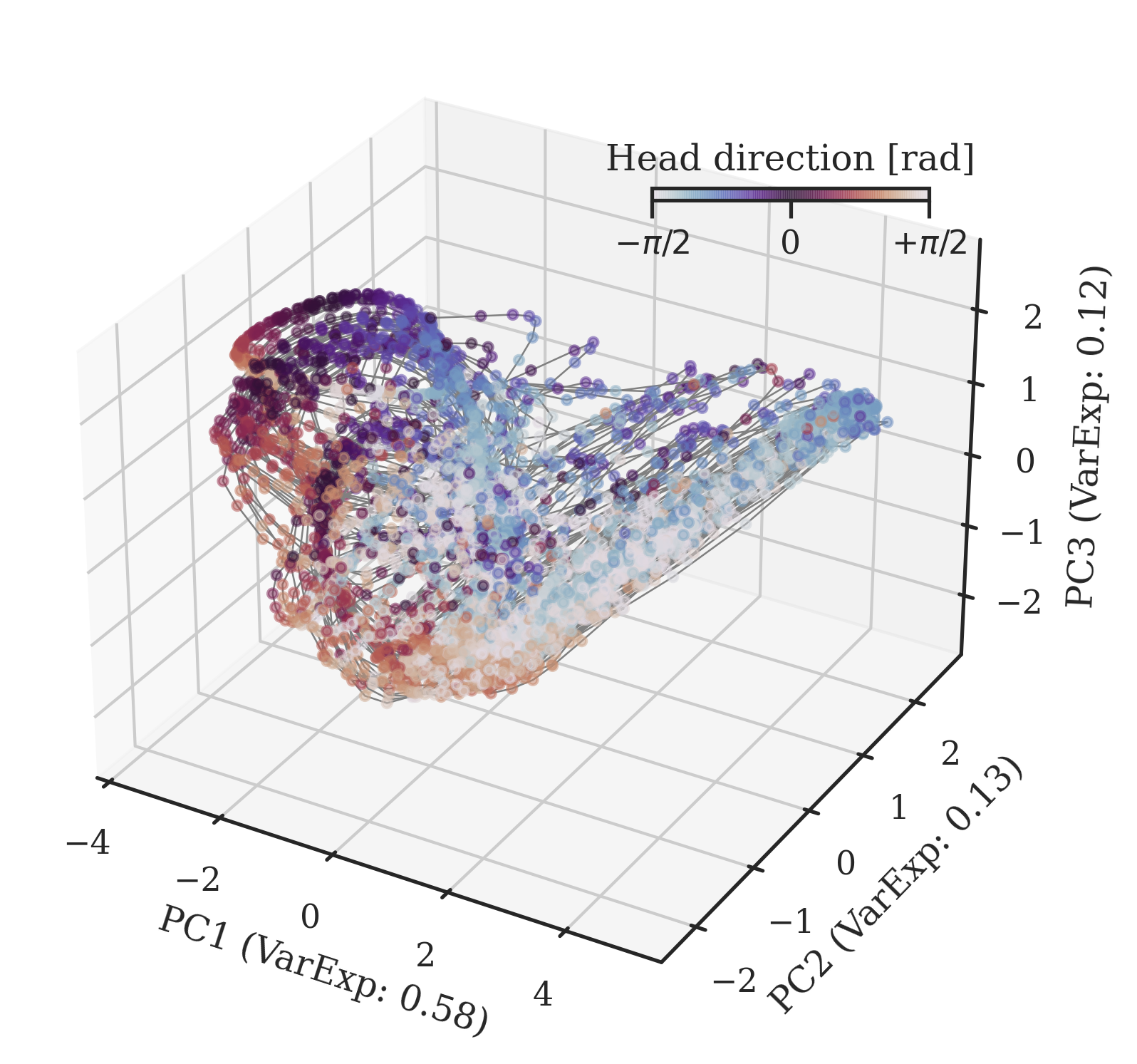}
\includegraphics[width=0.40\linewidth]{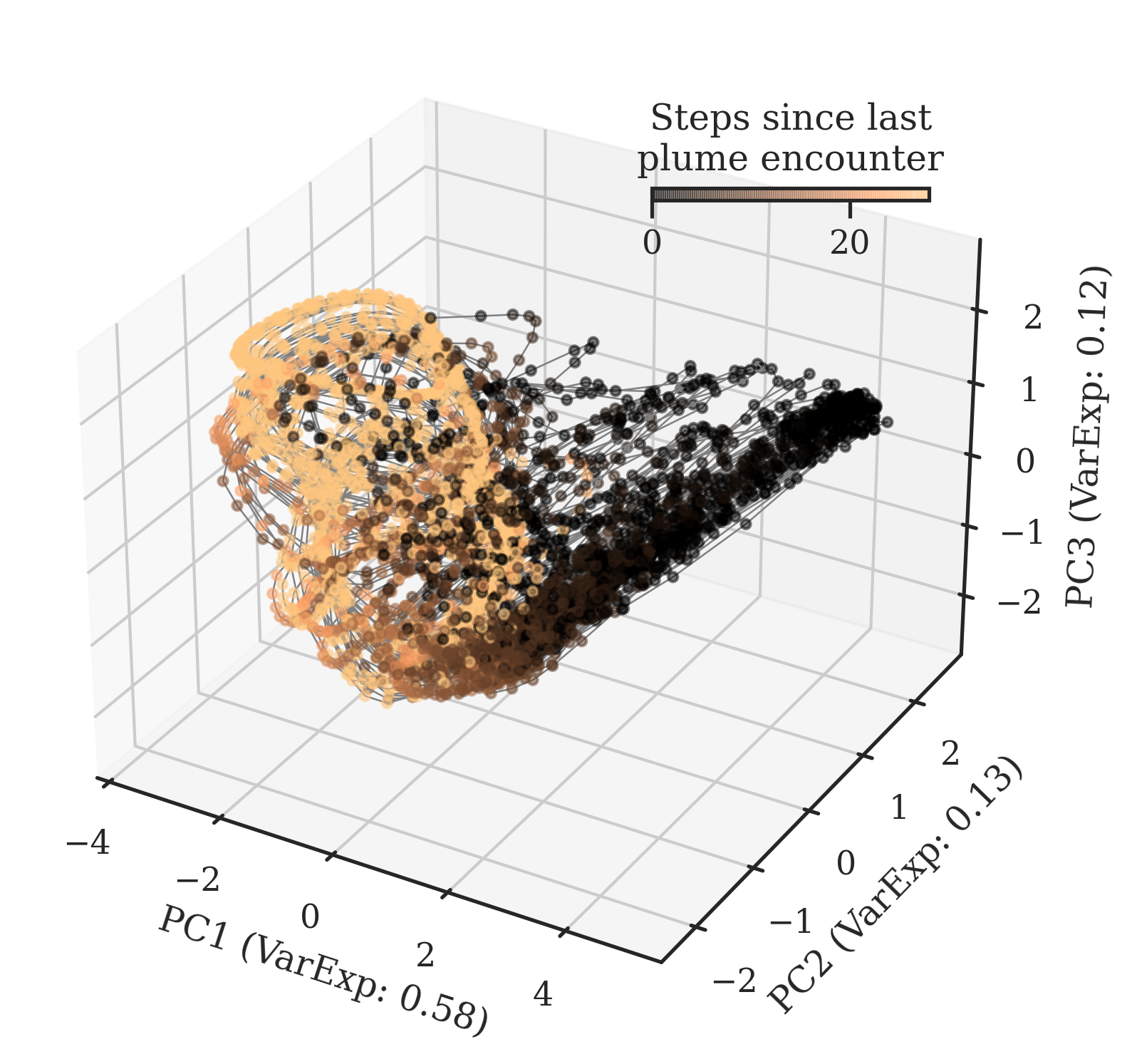} \\
\includegraphics[width=0.40\linewidth]{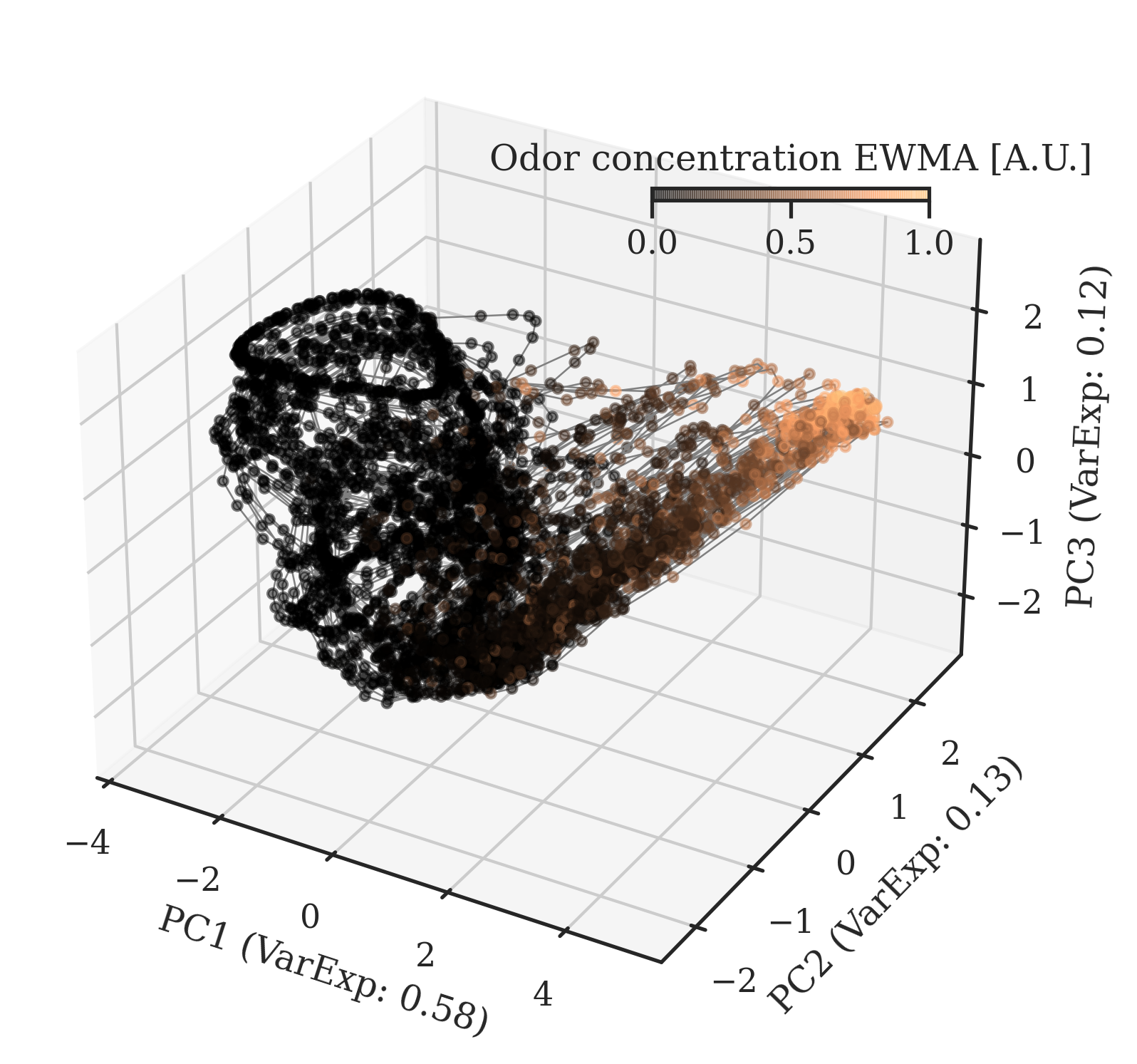}
\includegraphics[width=0.40\linewidth]{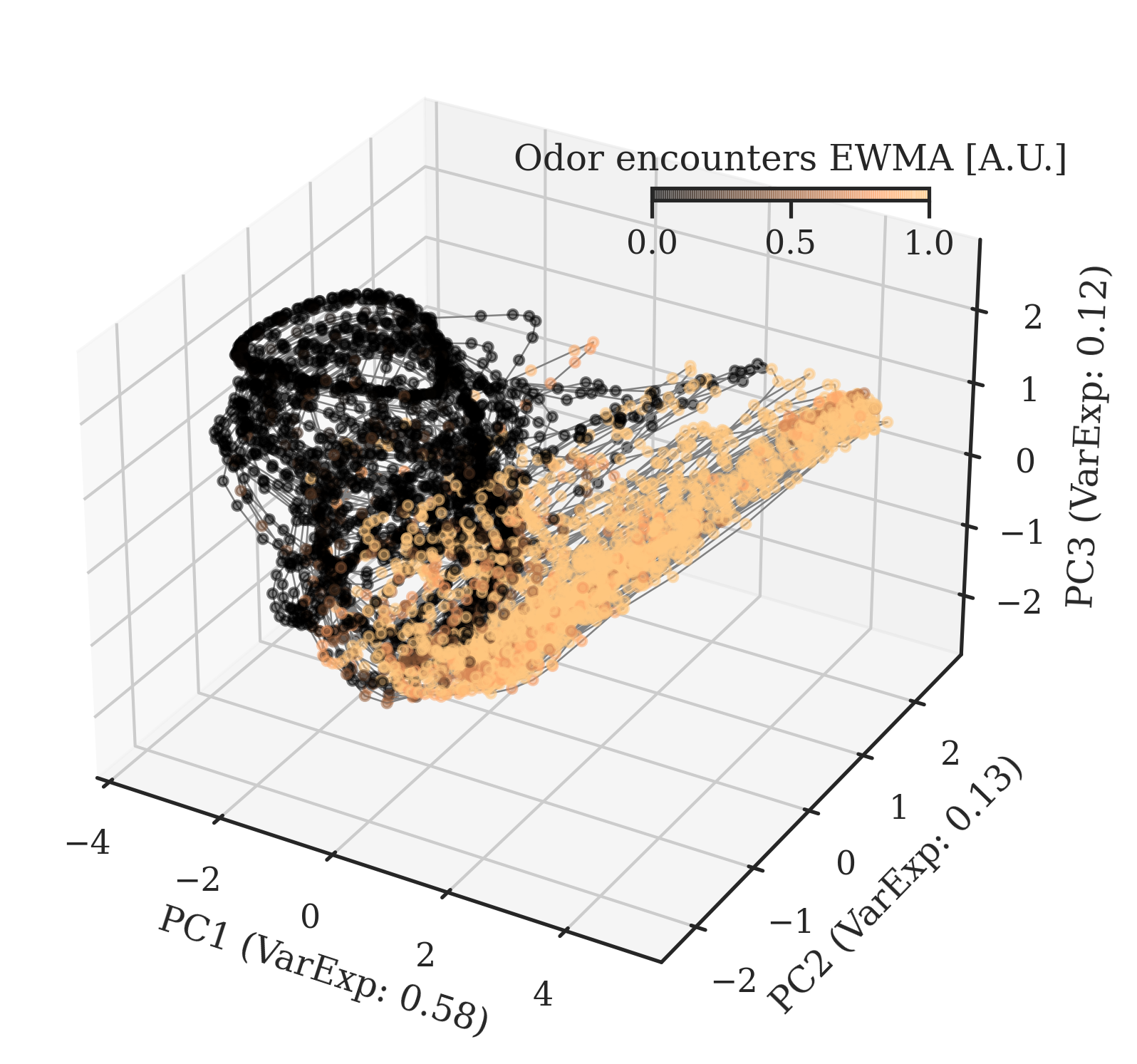} \\
\includegraphics[width=0.30\linewidth]{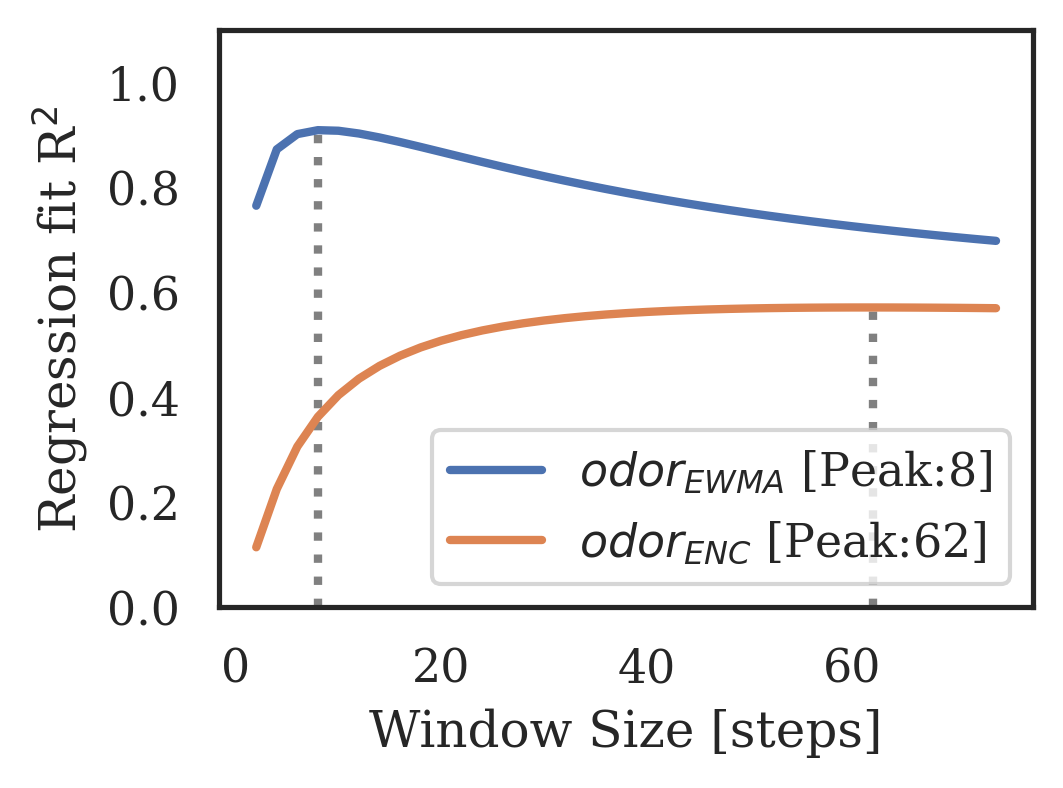}
\includegraphics[width=0.34\linewidth]{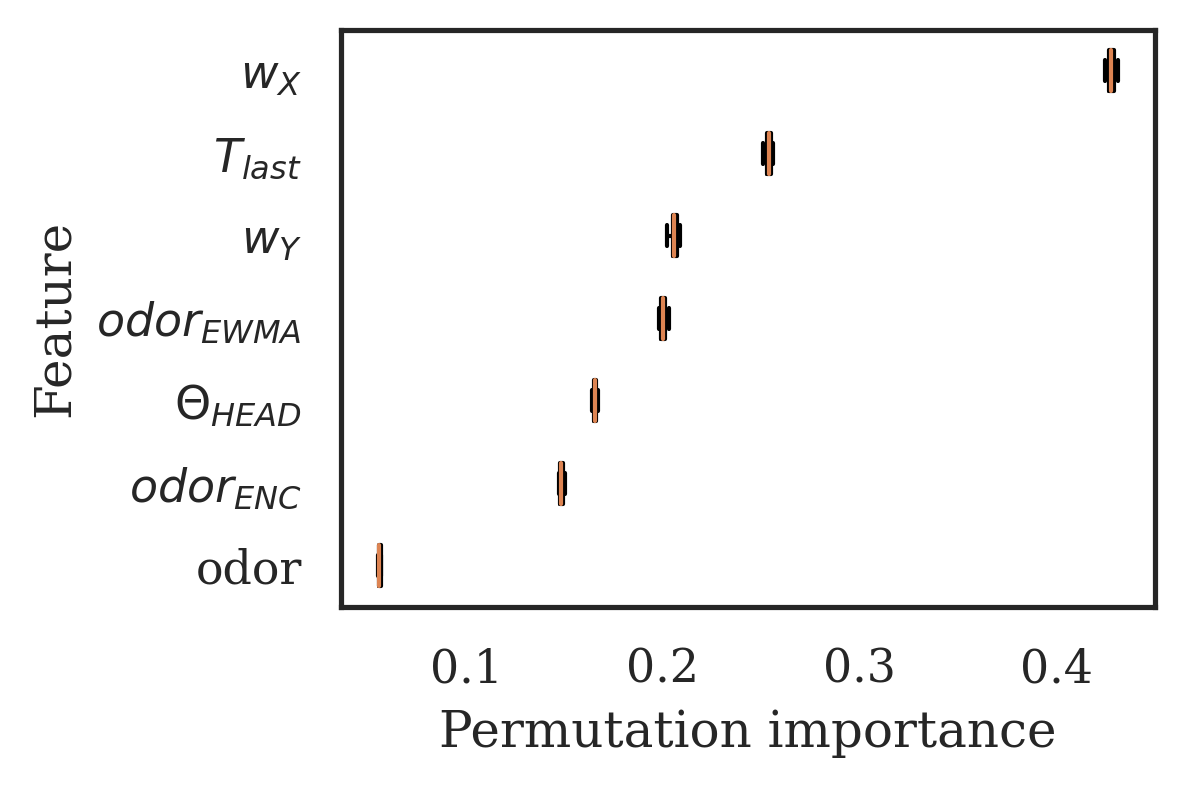}
\includegraphics[width=0.30\linewidth]{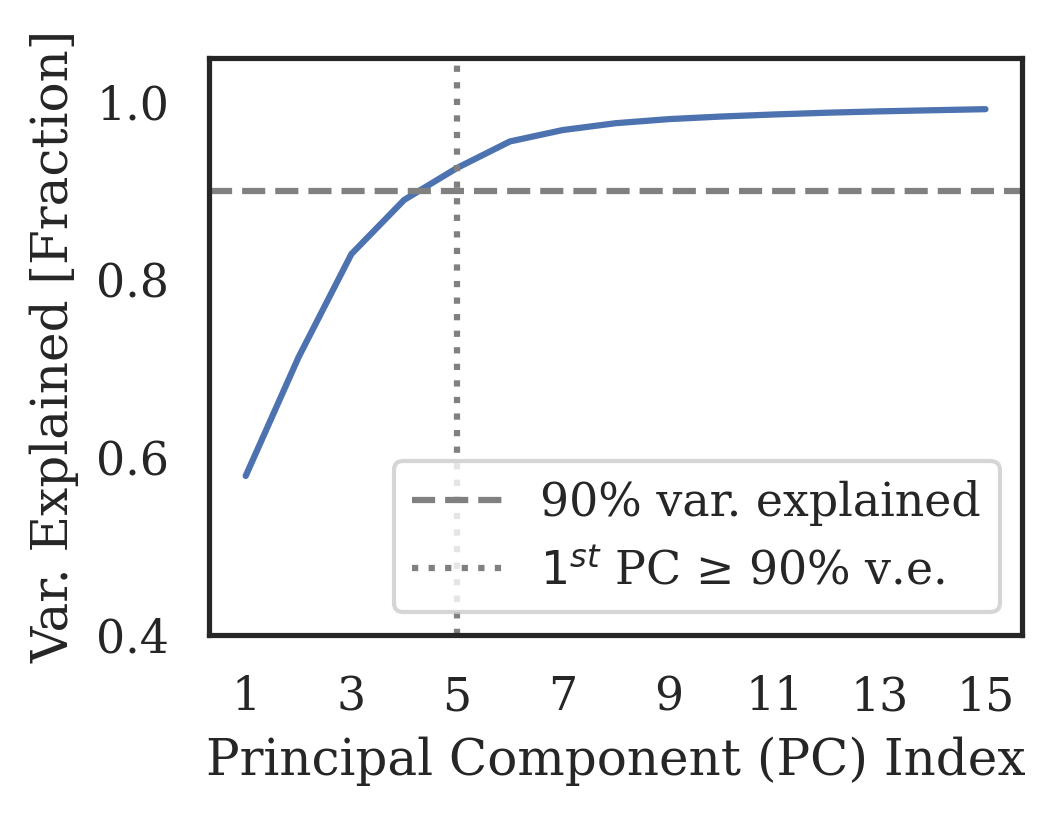}
\caption[Neural representations -- Agent 1]{Neural representations -- Agent 1 (See Figure \ref{fig_representations} for equivalent data on Agent 3 and figure details)}
\end{center}
\end{figure*}

\begin{figure*}[h!]
\begin{center}
\includegraphics[width=0.40\linewidth]{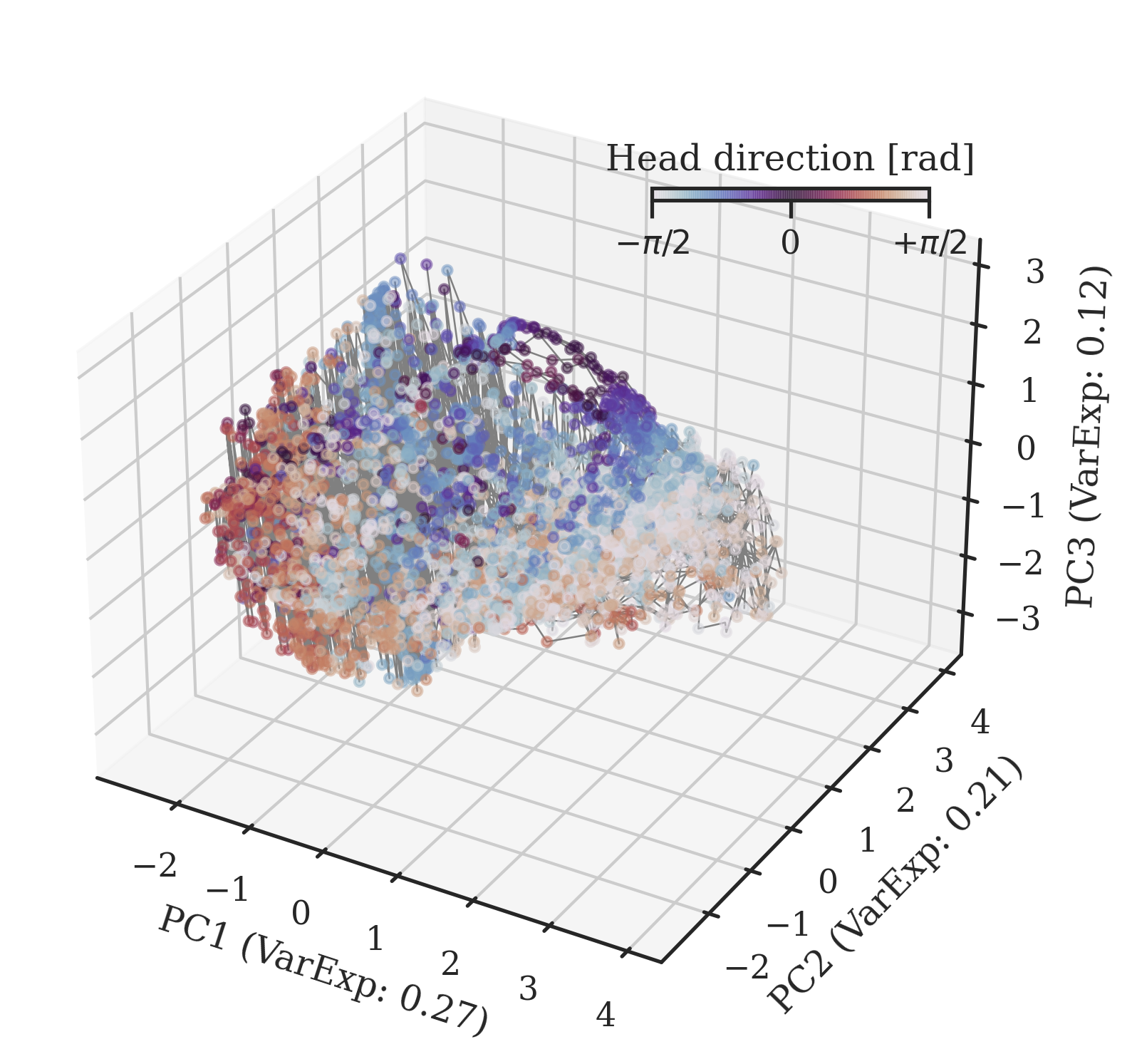}
\includegraphics[width=0.40\linewidth]{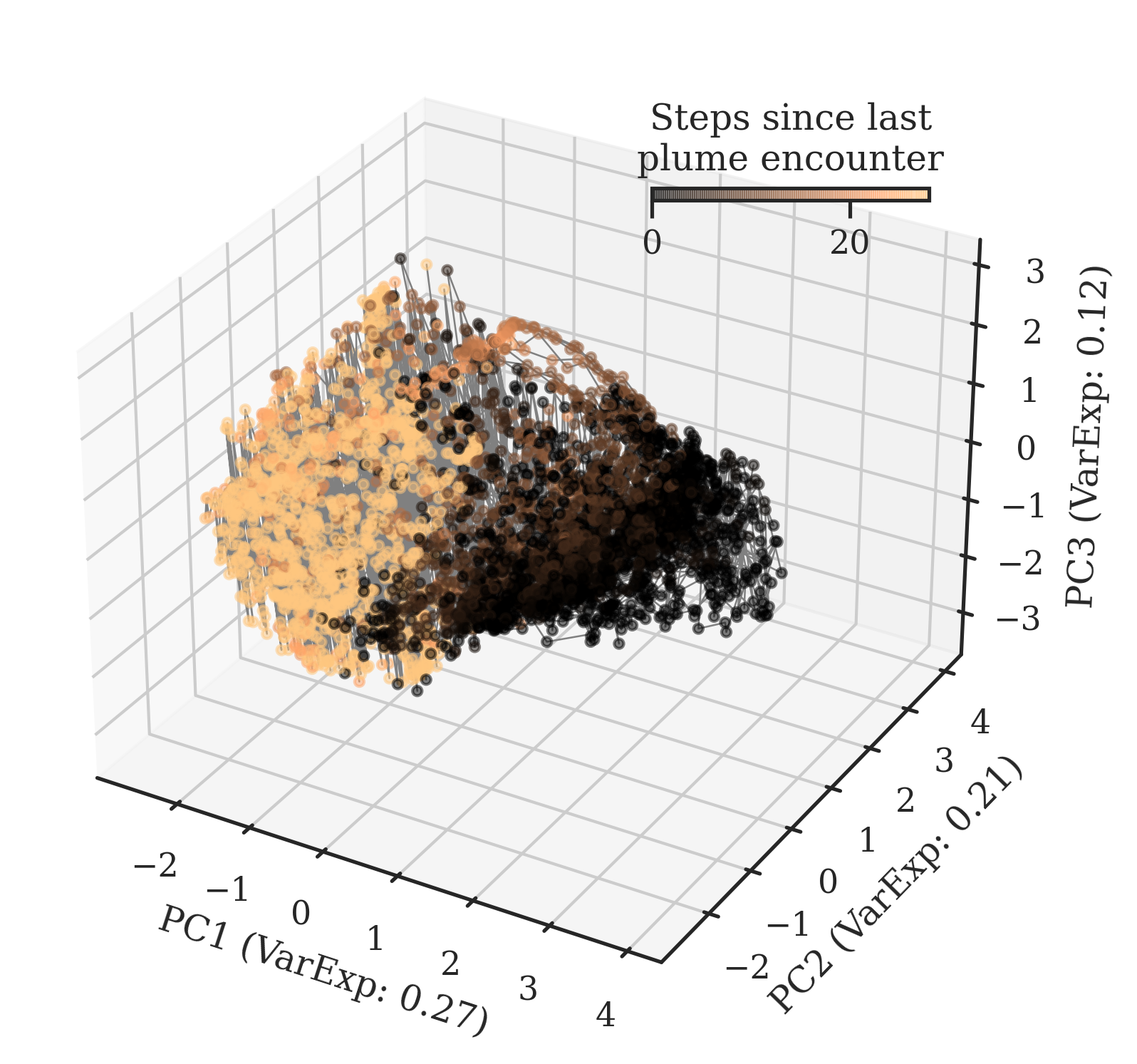} \\
\includegraphics[width=0.40\linewidth]{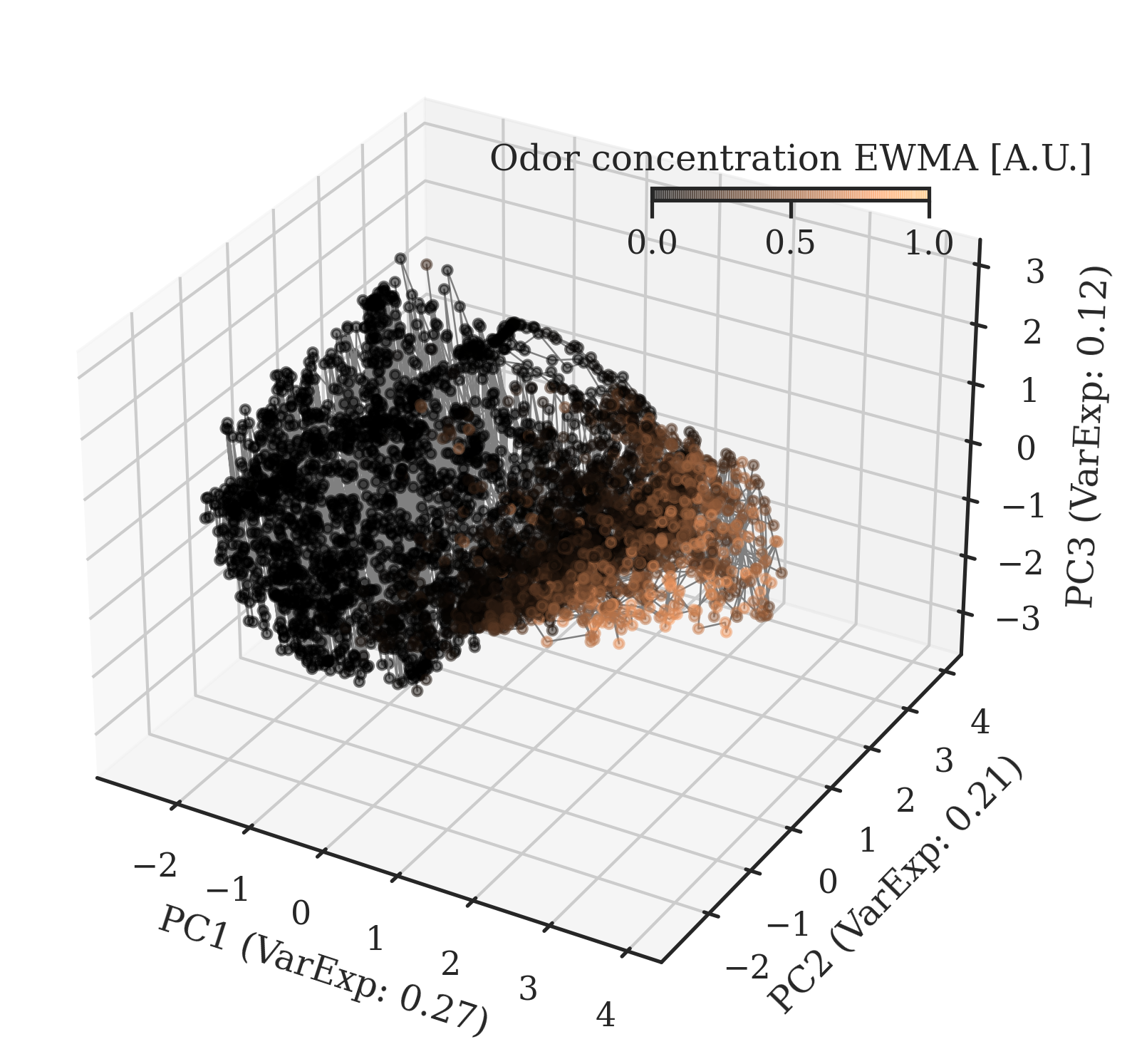}
\includegraphics[width=0.40\linewidth]{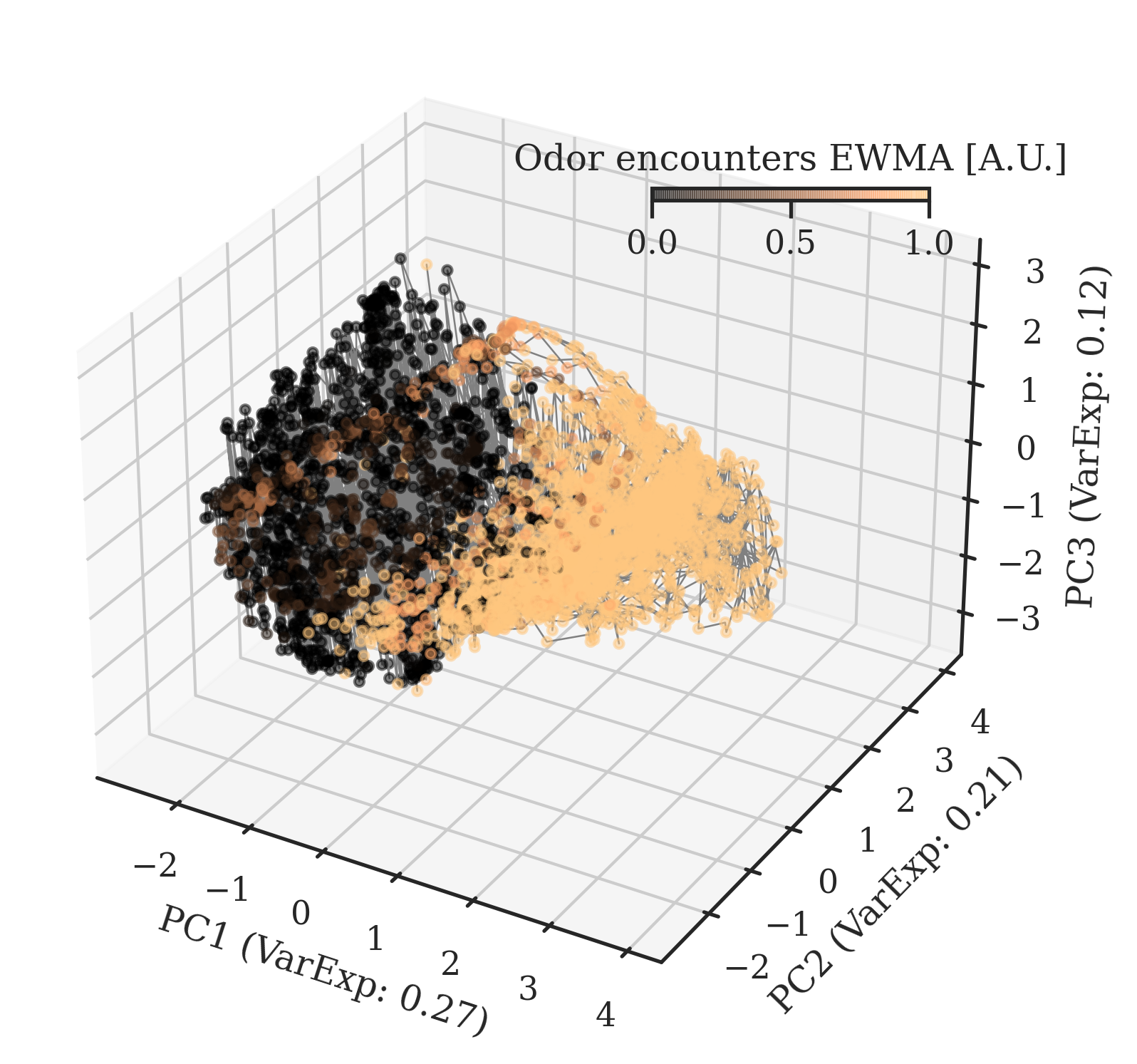} \\
\includegraphics[width=0.30\linewidth]{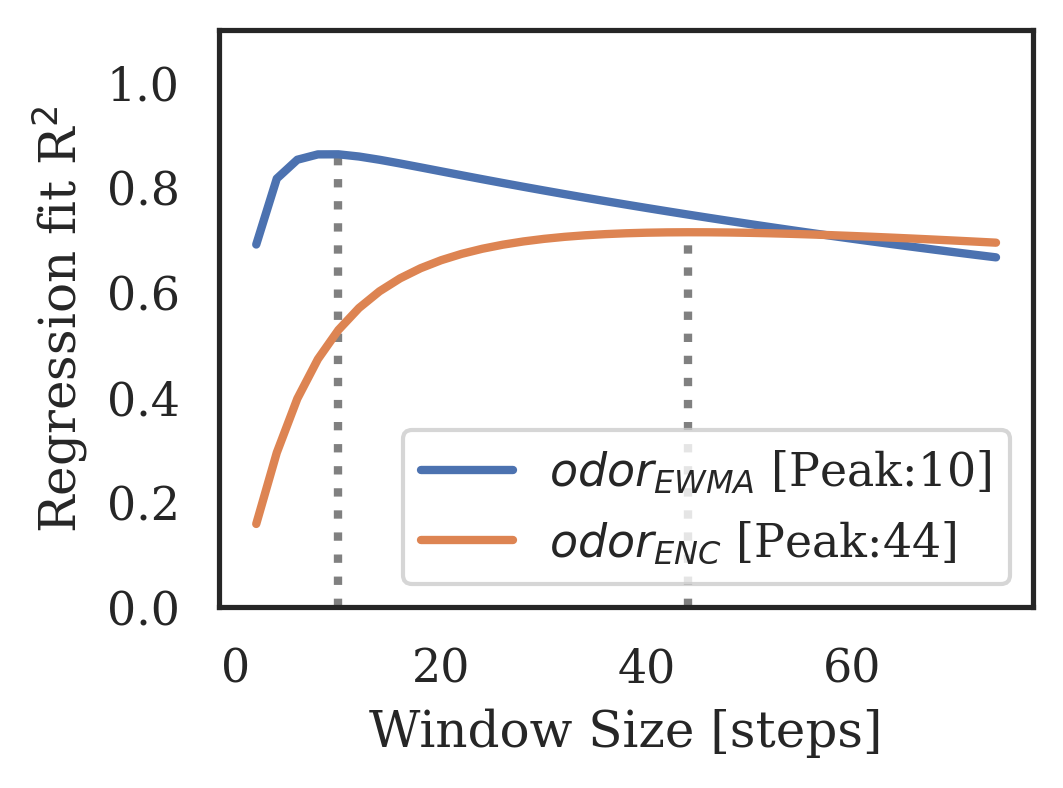}
\includegraphics[width=0.34\linewidth]{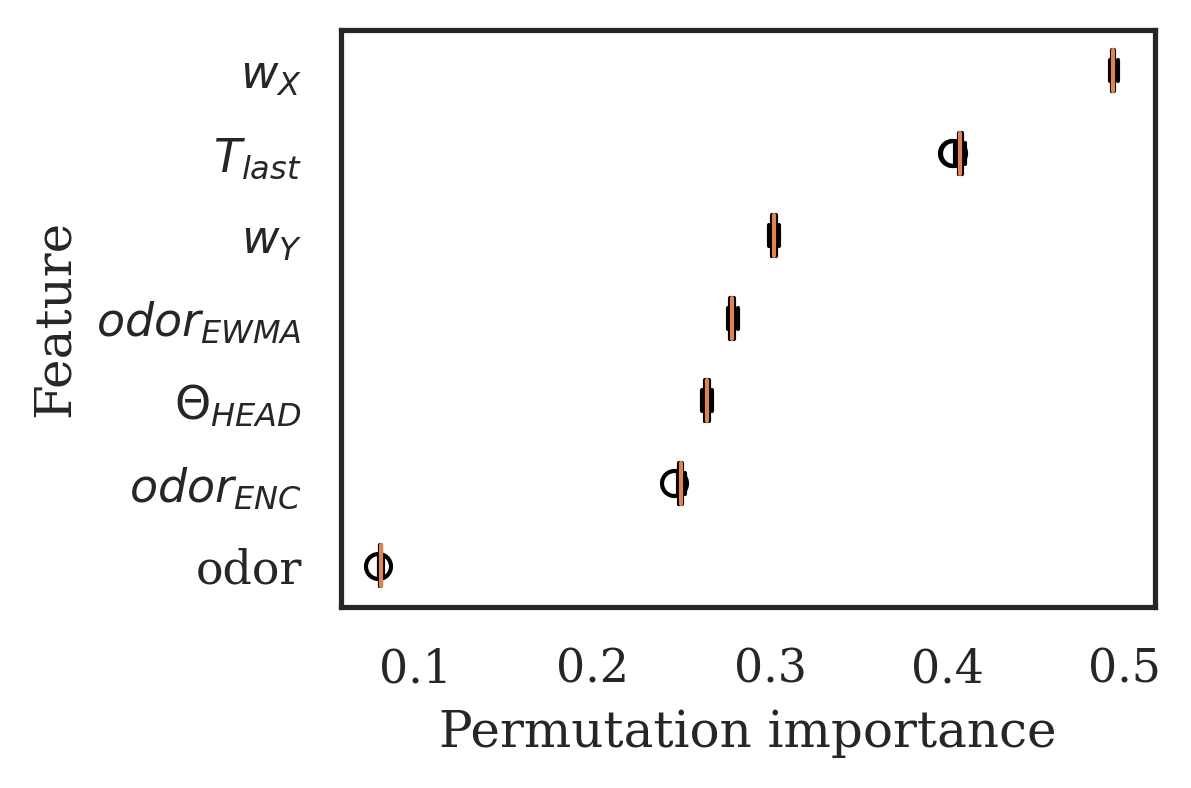}
\includegraphics[width=0.30\linewidth]{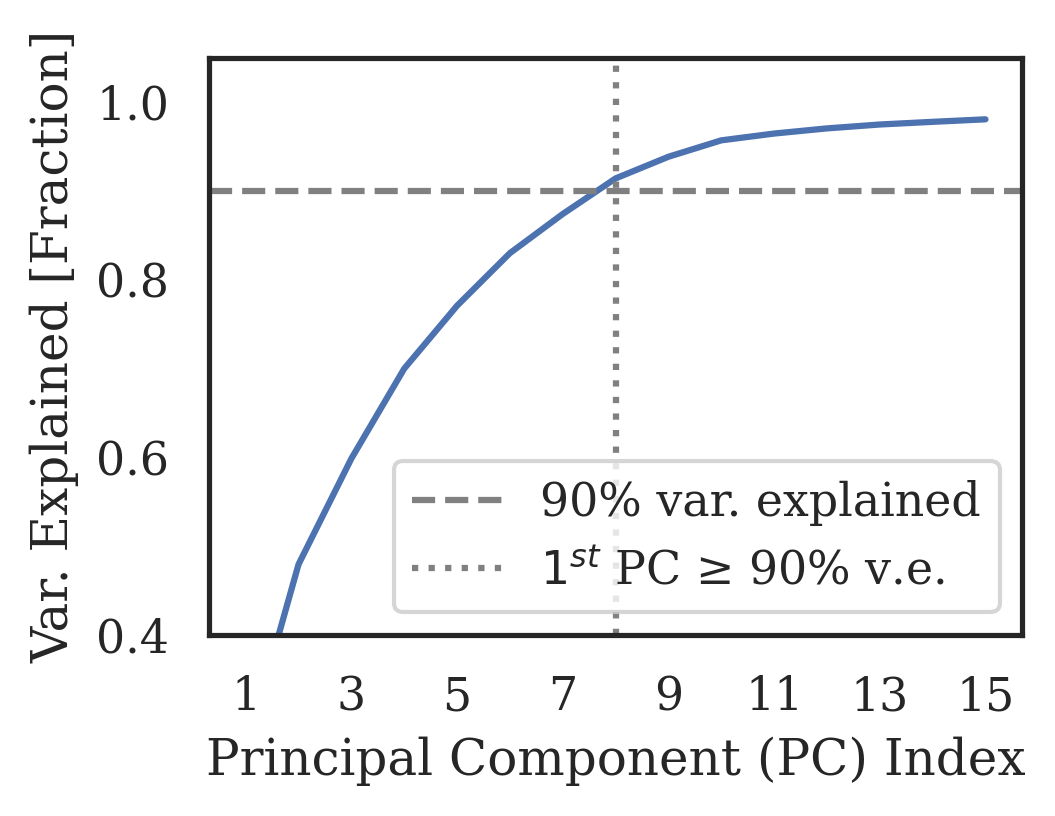}
\caption{Neural representations -- Agent 2}
\end{center}
\end{figure*}

\begin{figure*}[h!]
\begin{center}
\includegraphics[width=0.40\linewidth]{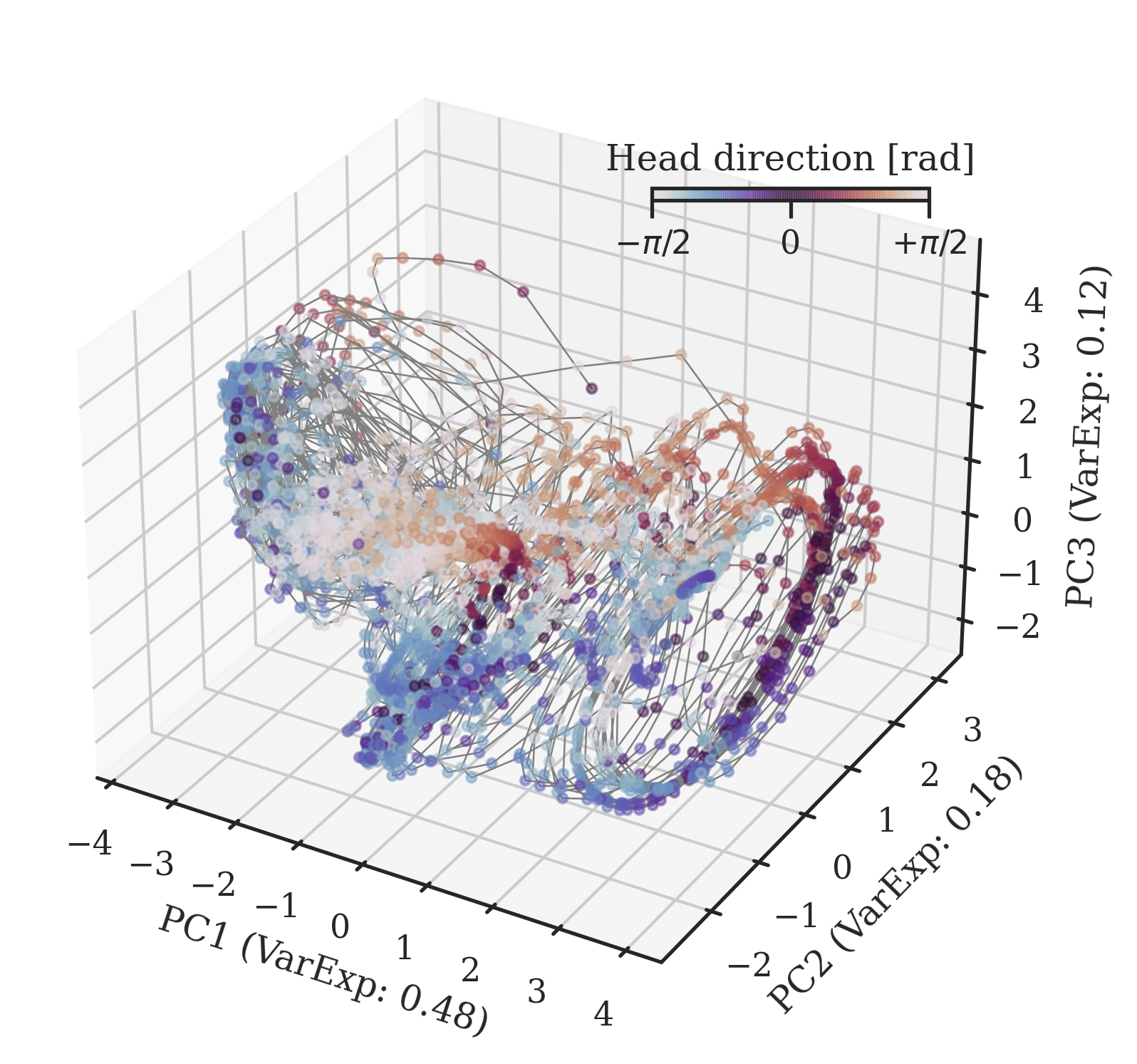}
\includegraphics[width=0.40\linewidth]{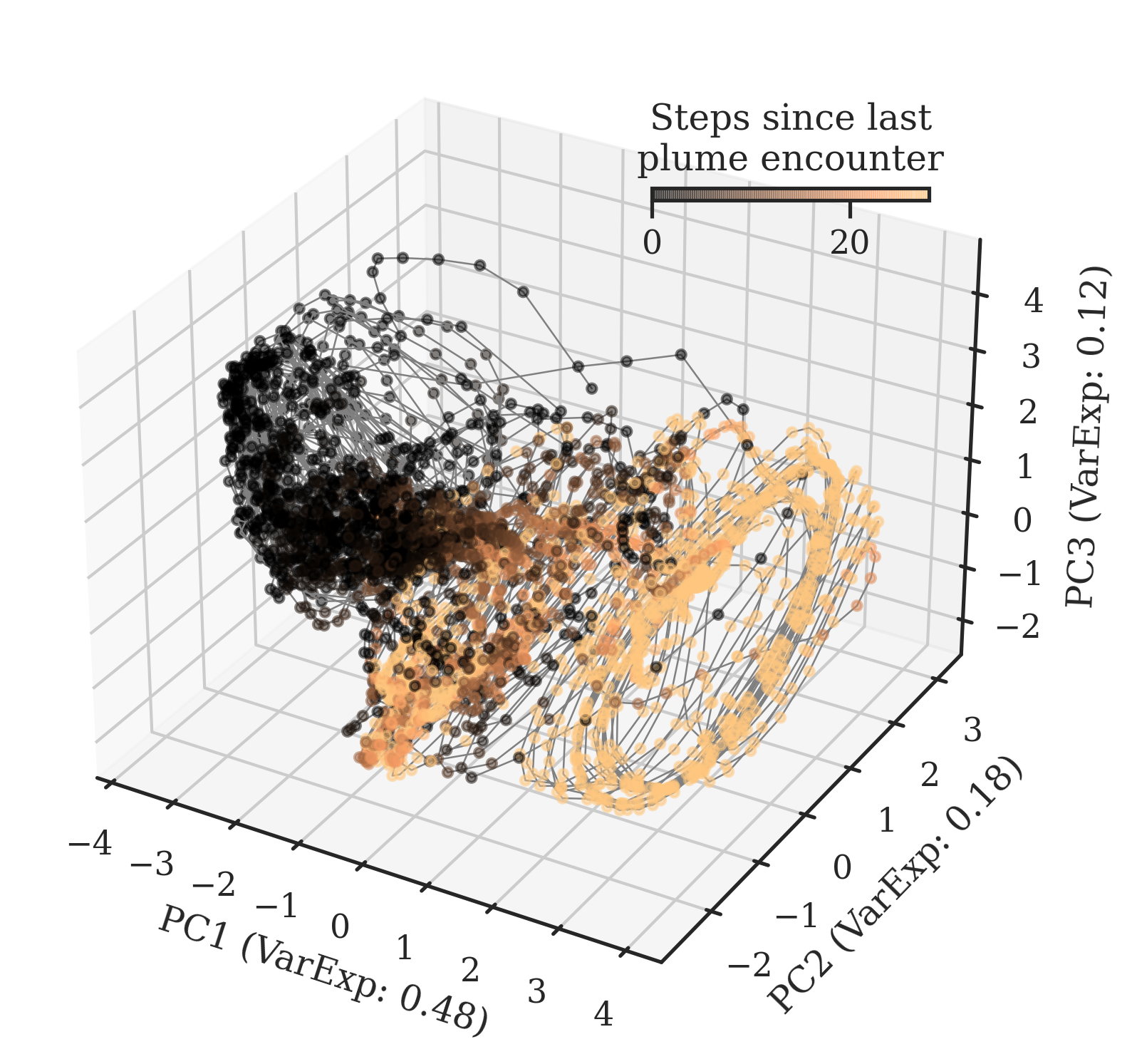} \\
\includegraphics[width=0.40\linewidth]{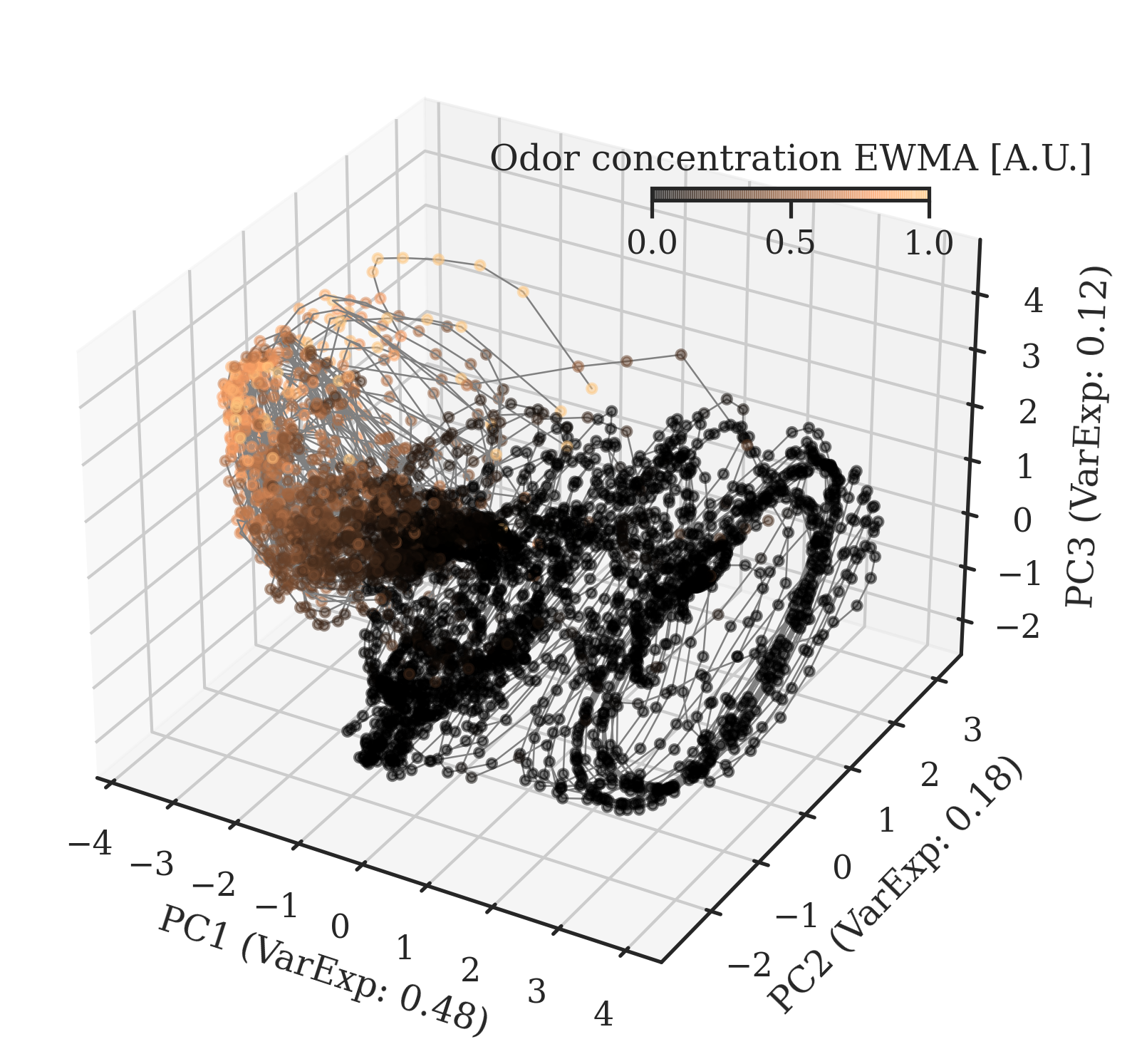}
\includegraphics[width=0.40\linewidth]{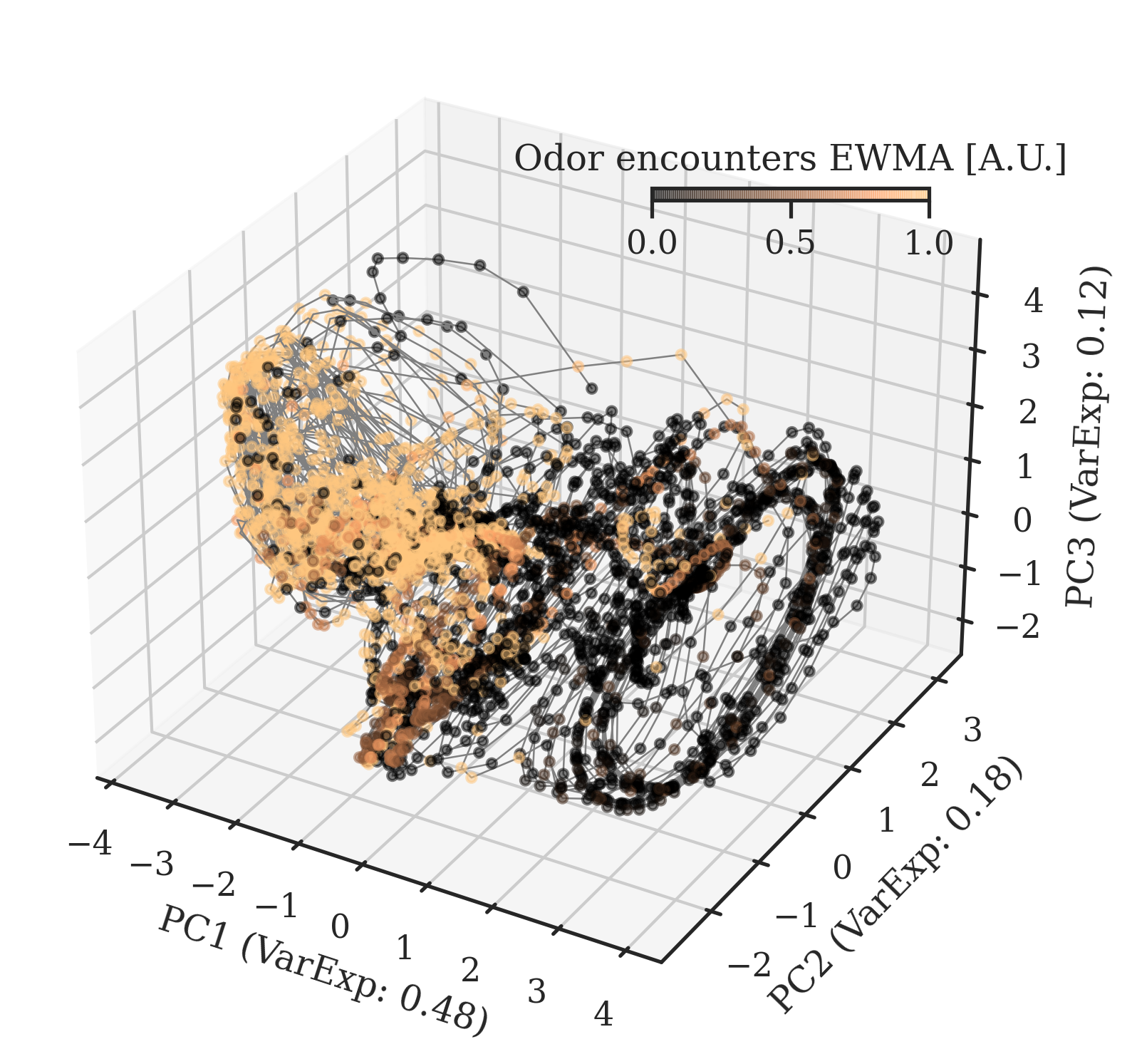} \\
\includegraphics[width=0.30\linewidth]{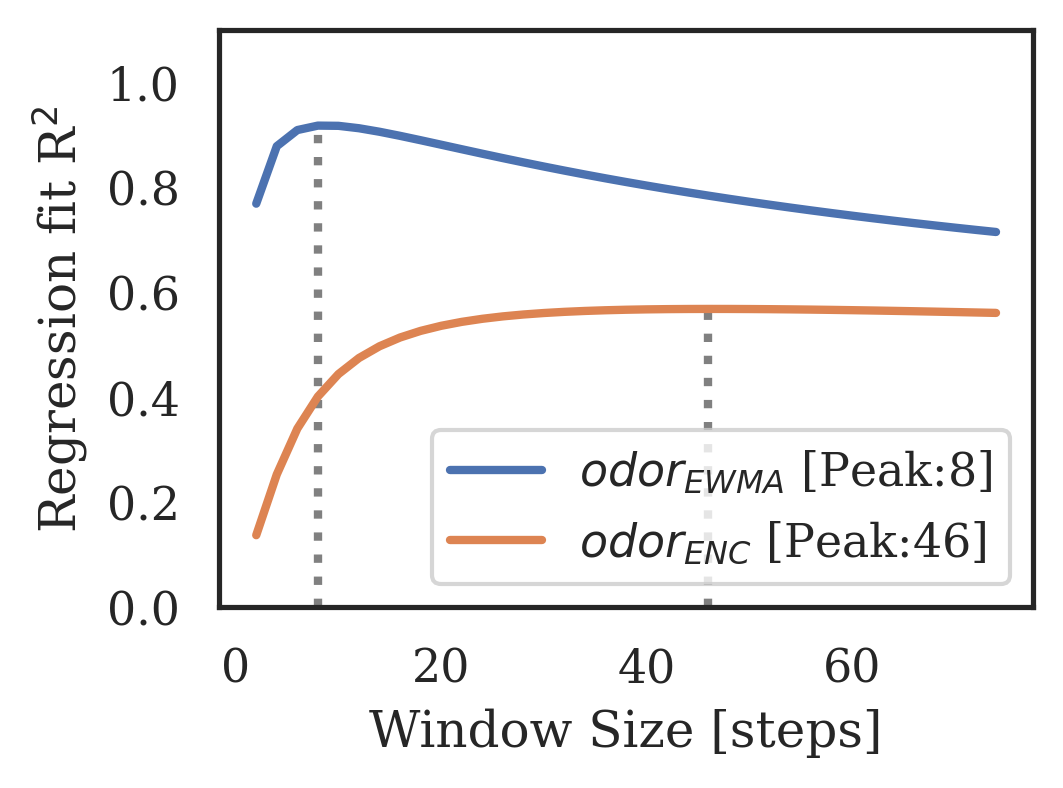}
\includegraphics[width=0.34\linewidth]{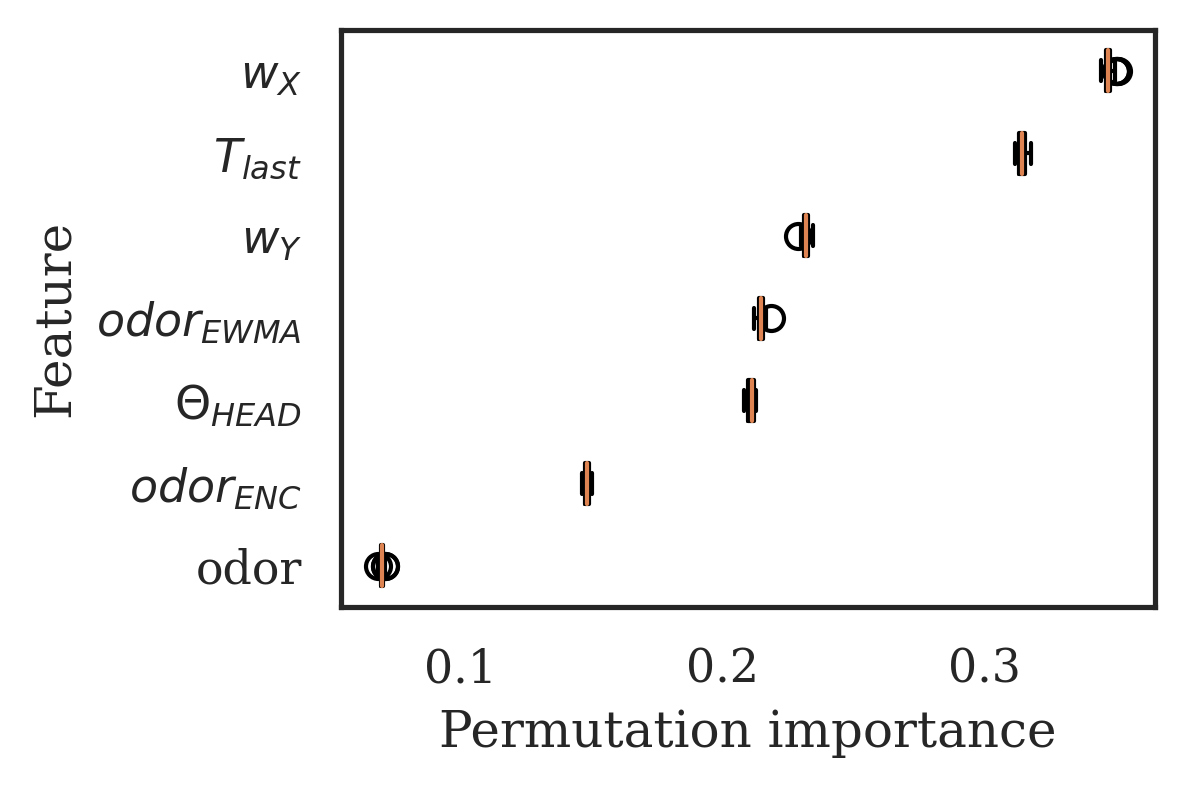}
\includegraphics[width=0.30\linewidth]{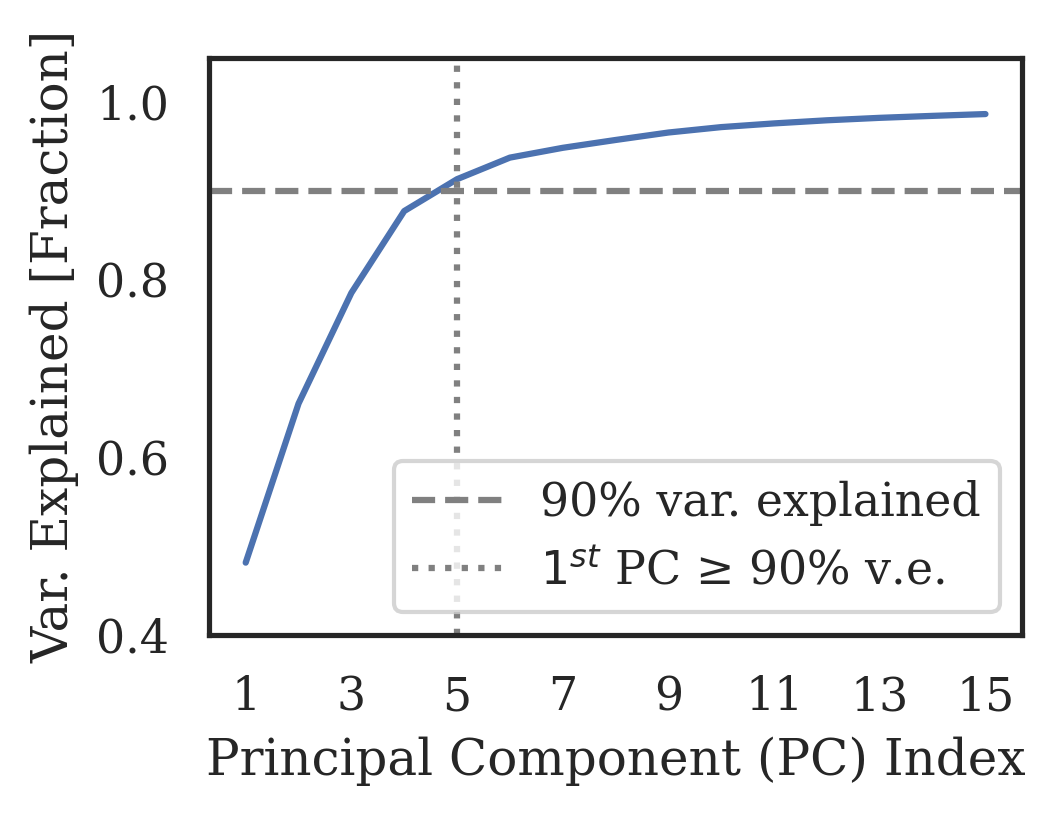}
\caption[Neural representations -- Agent 3]{Neural representations -- Agent 3 (Same agent as in Figure \ref{fig_representations})}
\end{center}
\end{figure*}

\begin{figure*}[h!]
\begin{center}
\includegraphics[width=0.40\linewidth]{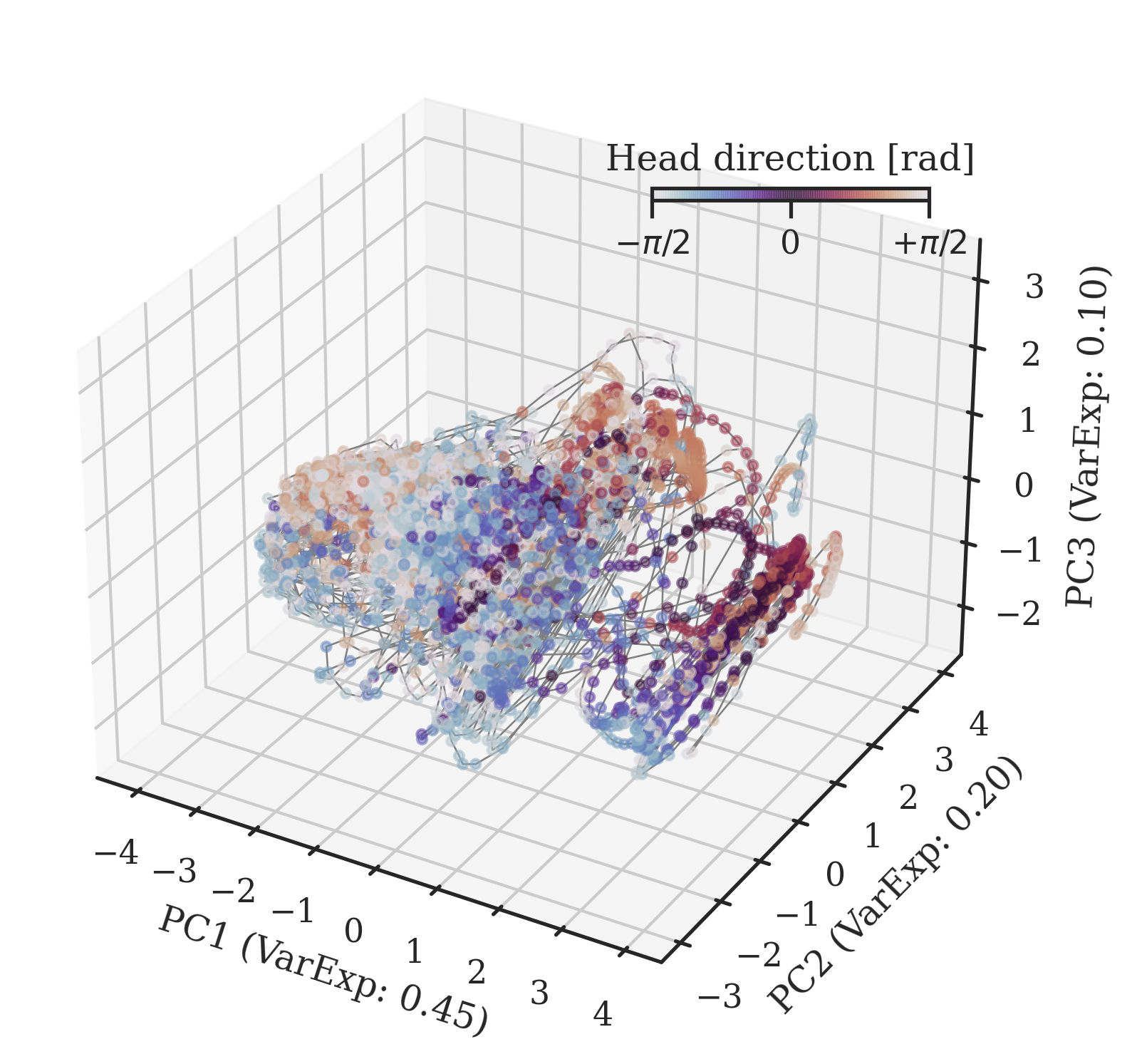}
\includegraphics[width=0.40\linewidth]{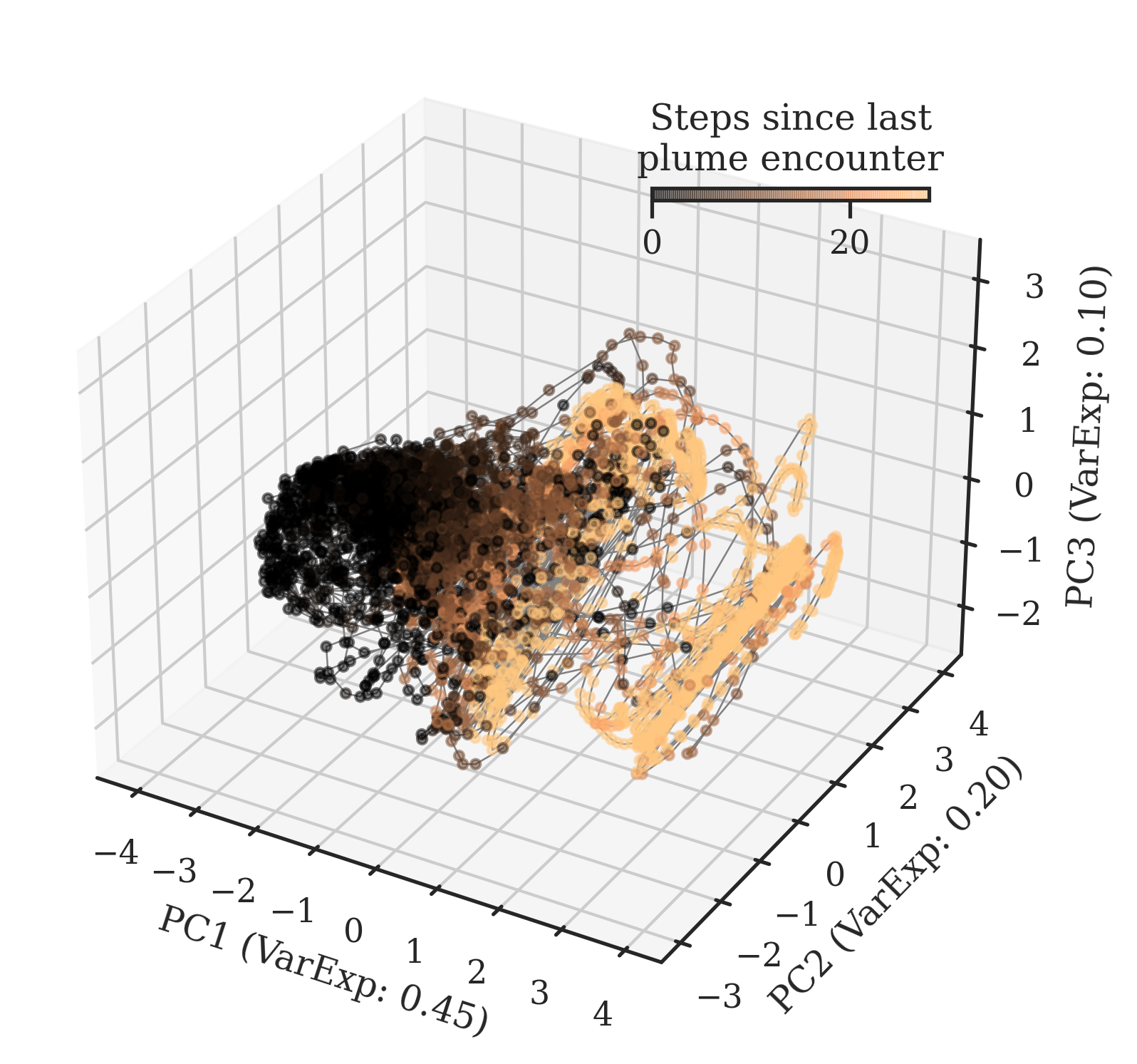} \\
\includegraphics[width=0.40\linewidth]{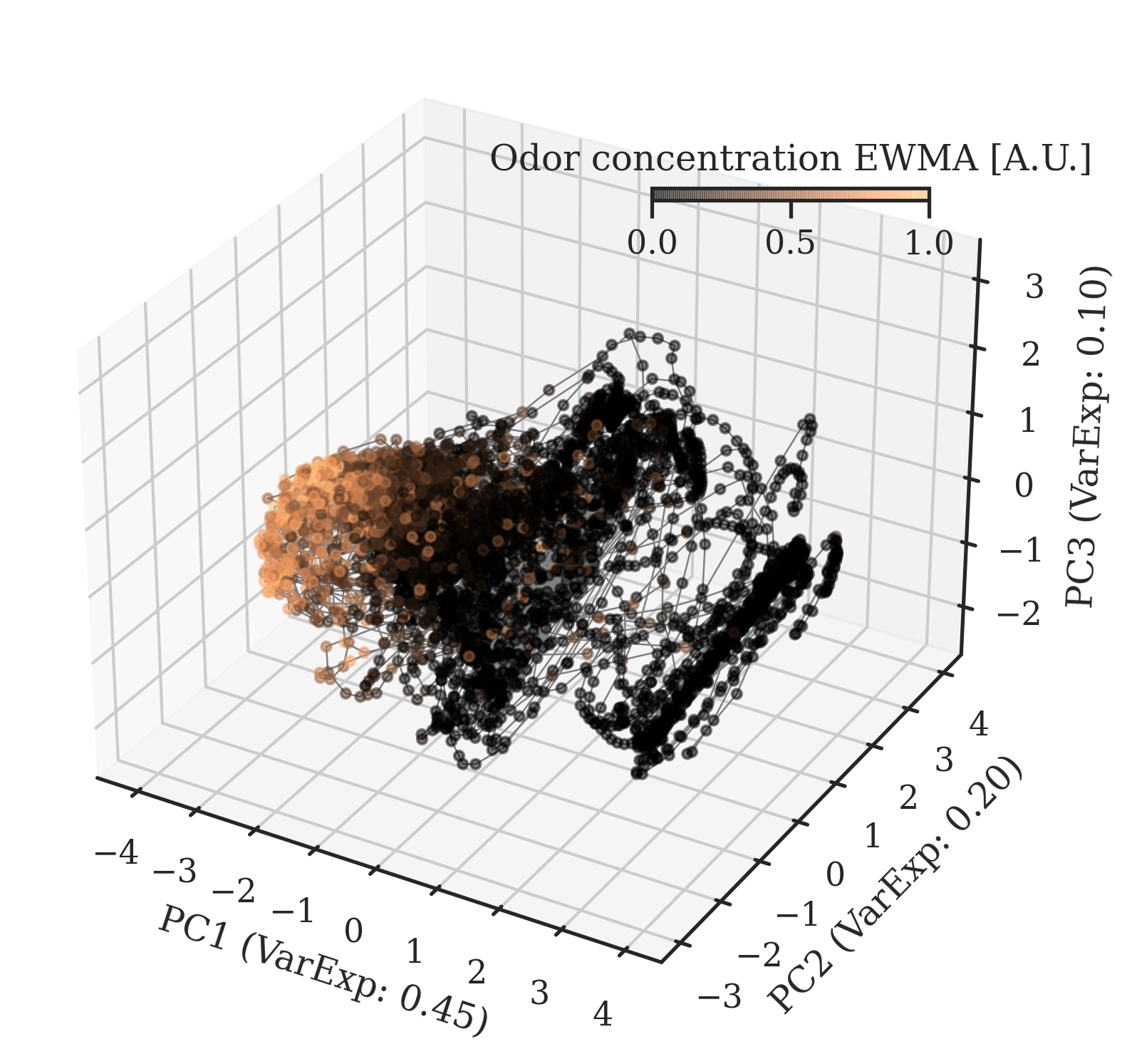}
\includegraphics[width=0.40\linewidth]{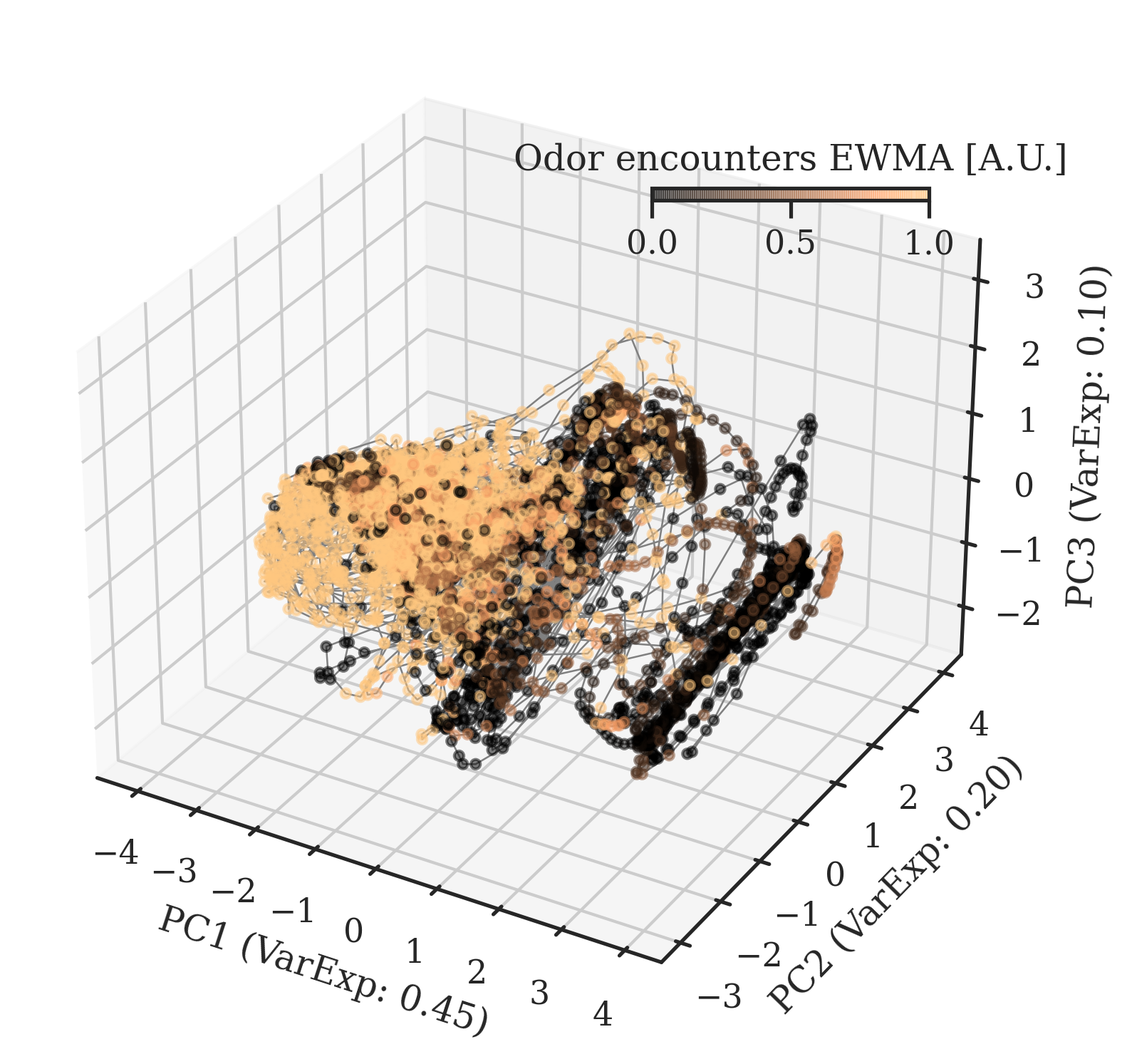} \\
\includegraphics[width=0.30\linewidth]{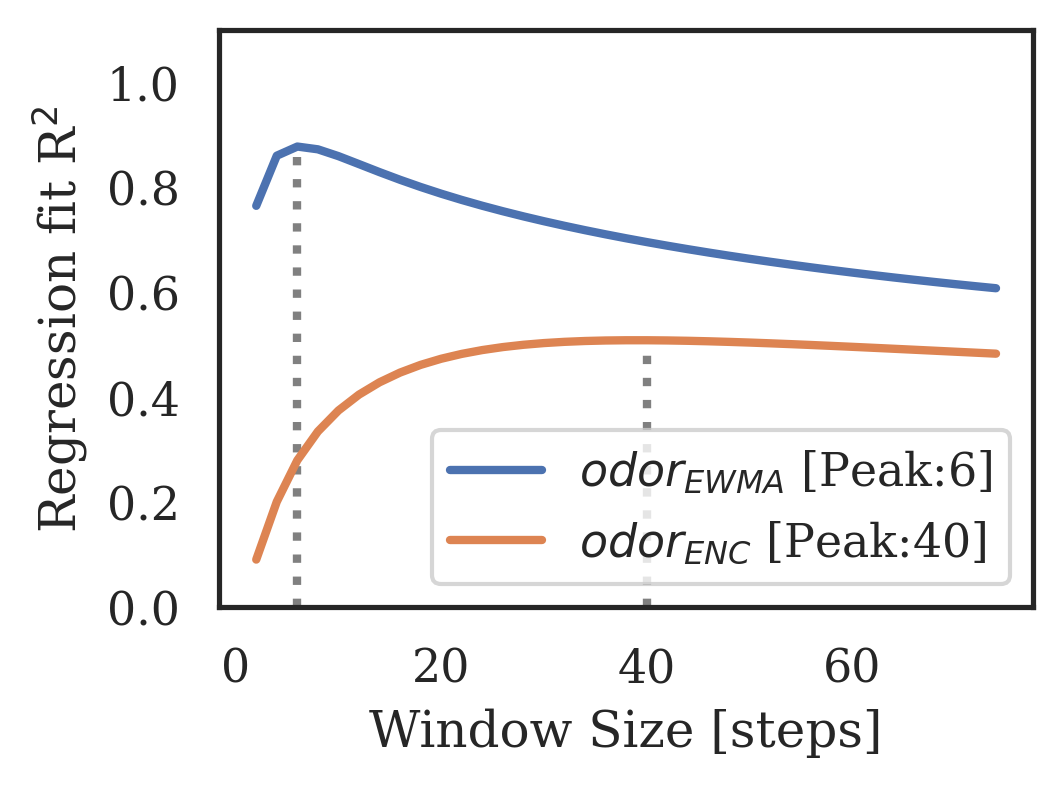}
\includegraphics[width=0.34\linewidth]{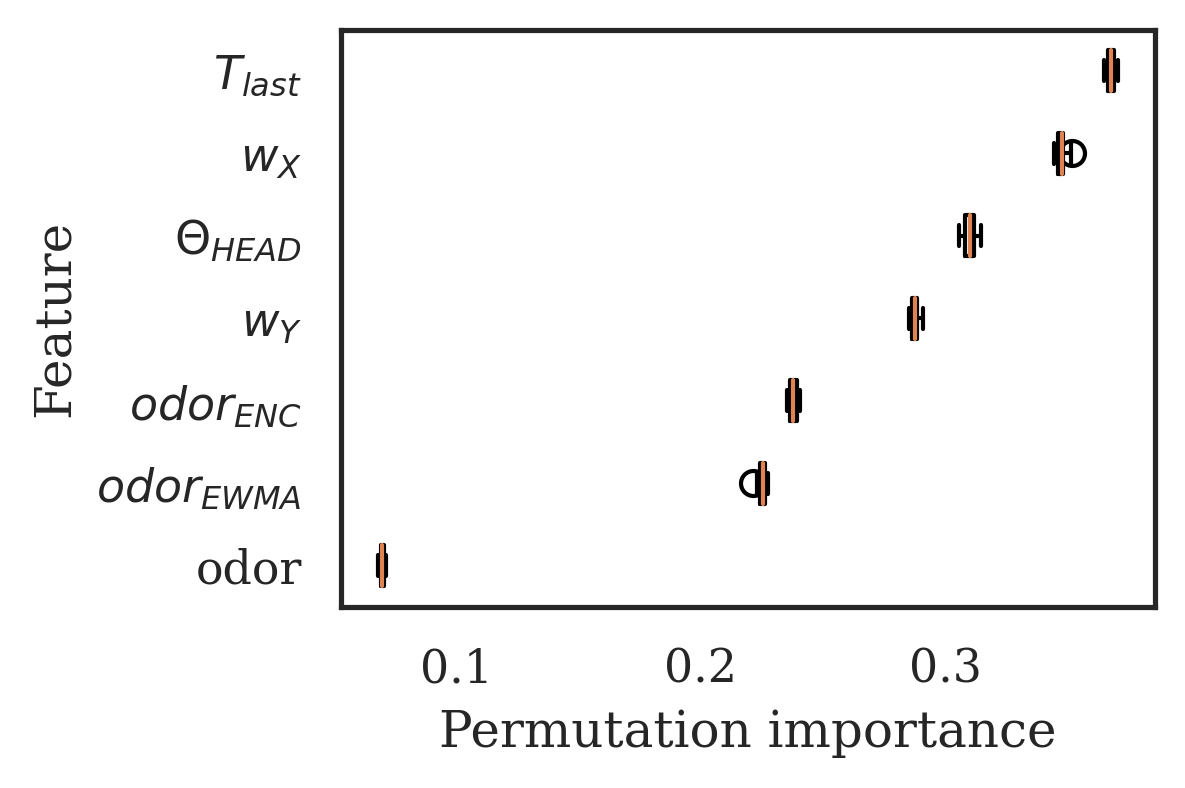}
\includegraphics[width=0.30\linewidth]{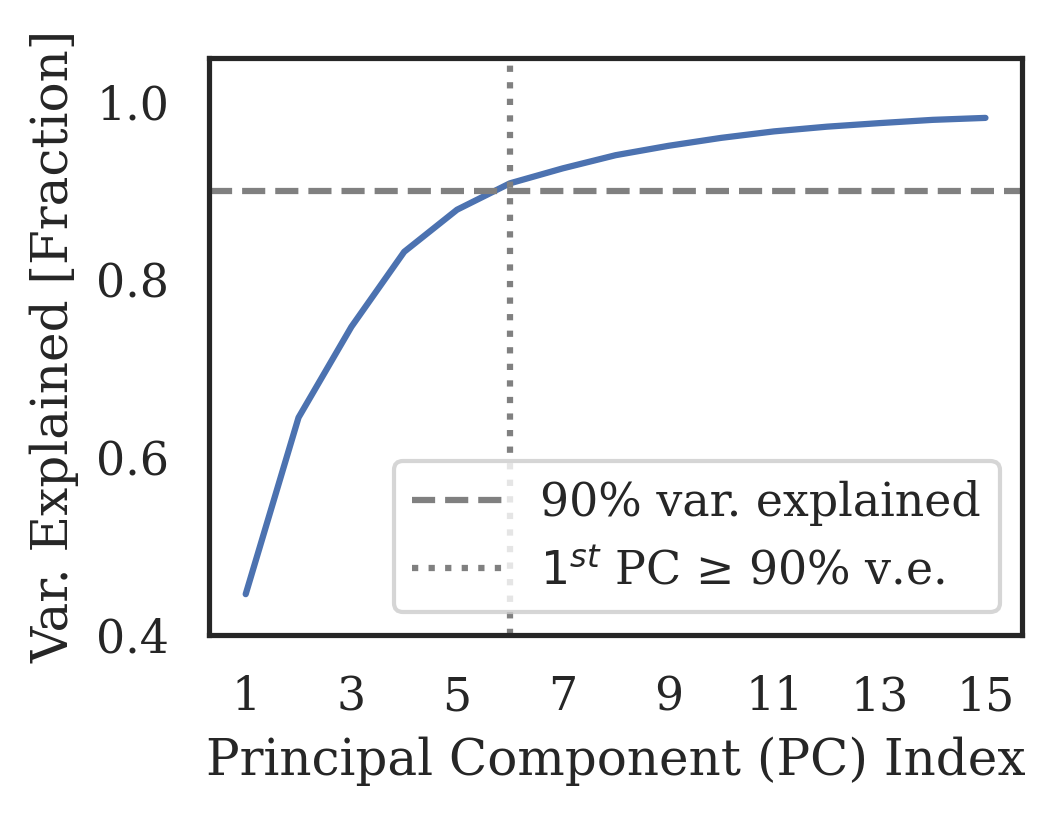}
\caption{Neural representations -- Agent 4}
\end{center}
\end{figure*}

\begin{figure*}[h!]
\begin{center}
\includegraphics[width=0.40\linewidth]{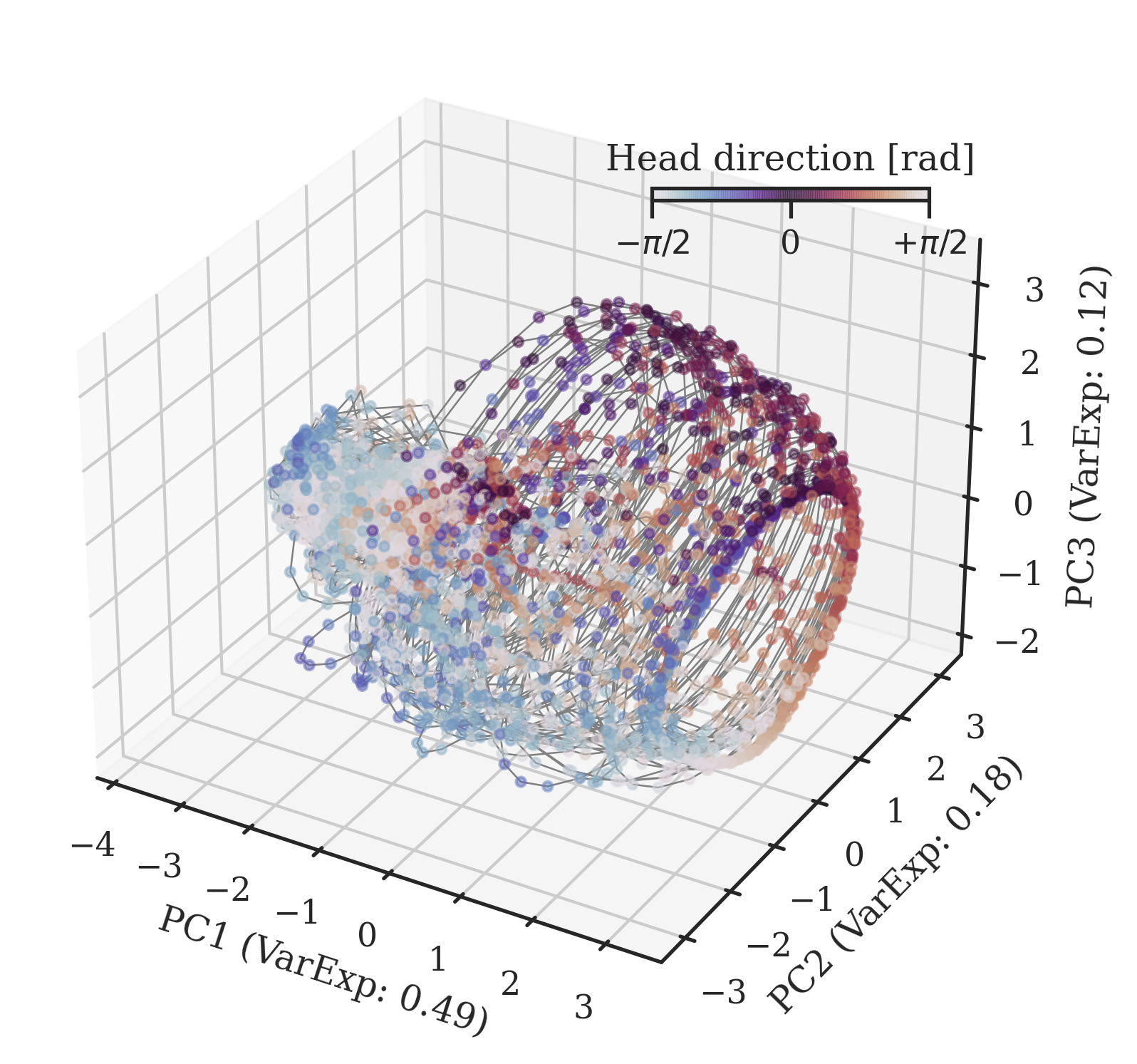}
\includegraphics[width=0.40\linewidth]{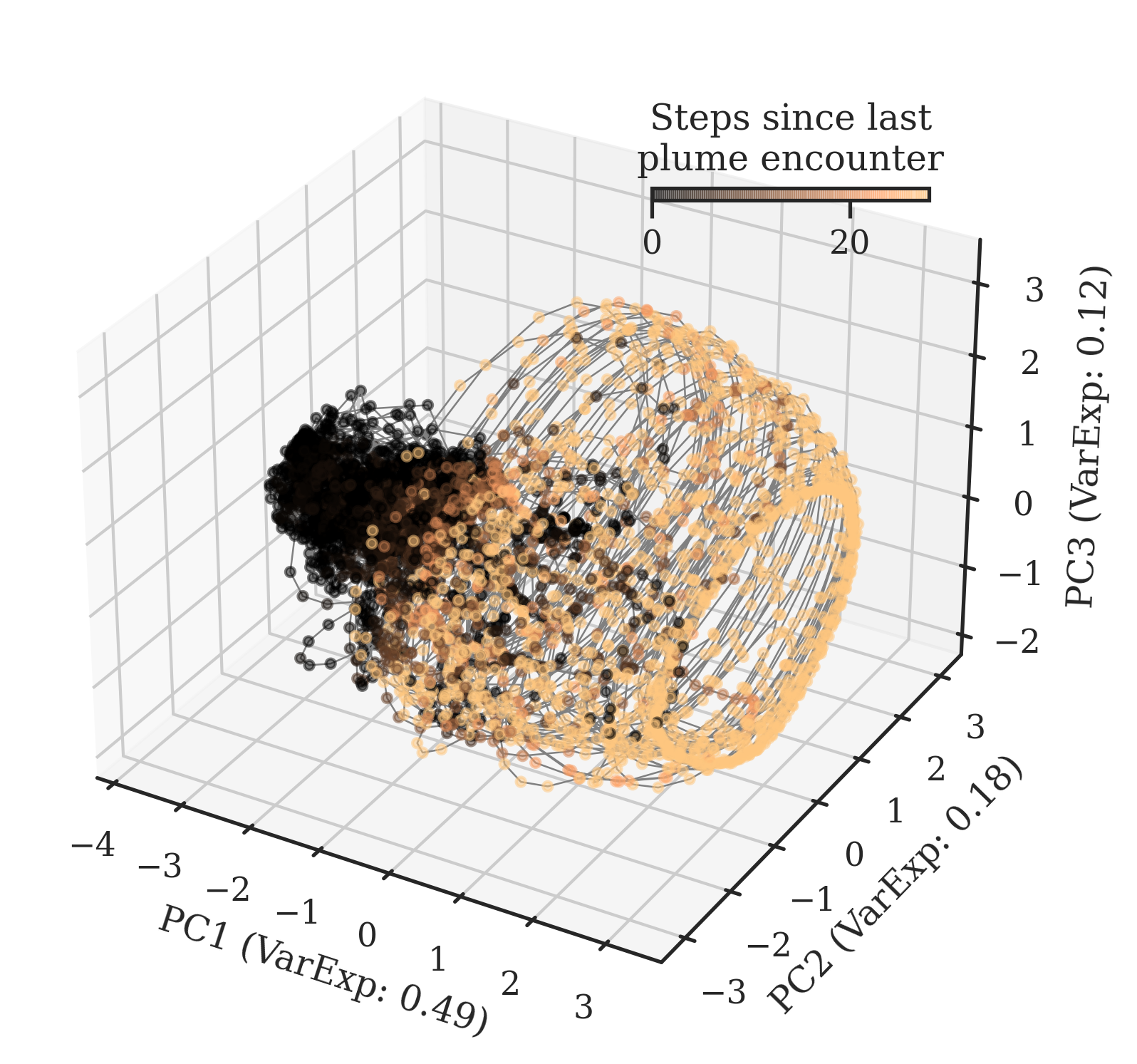} \\
\includegraphics[width=0.40\linewidth]{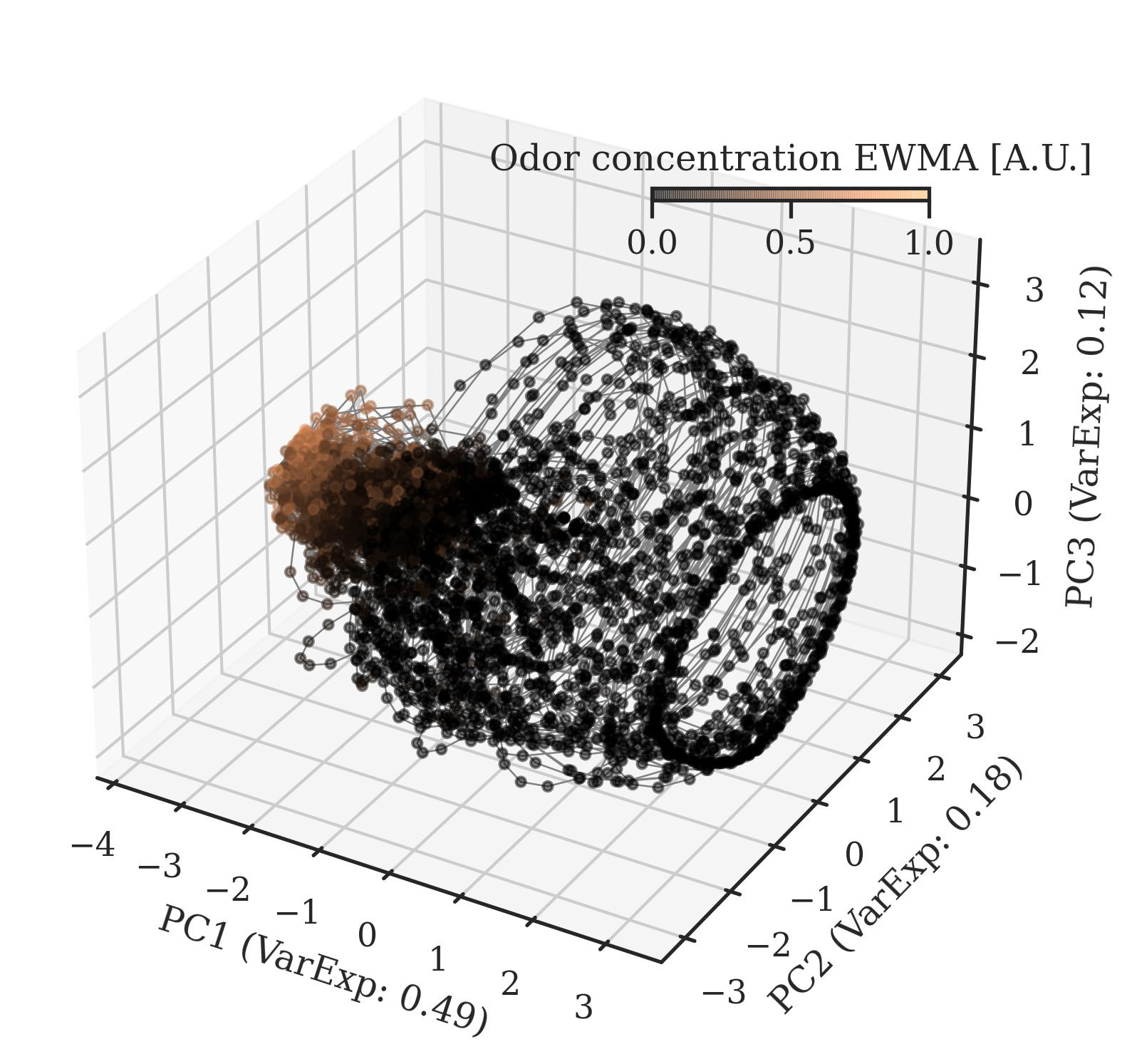}
\includegraphics[width=0.40\linewidth]{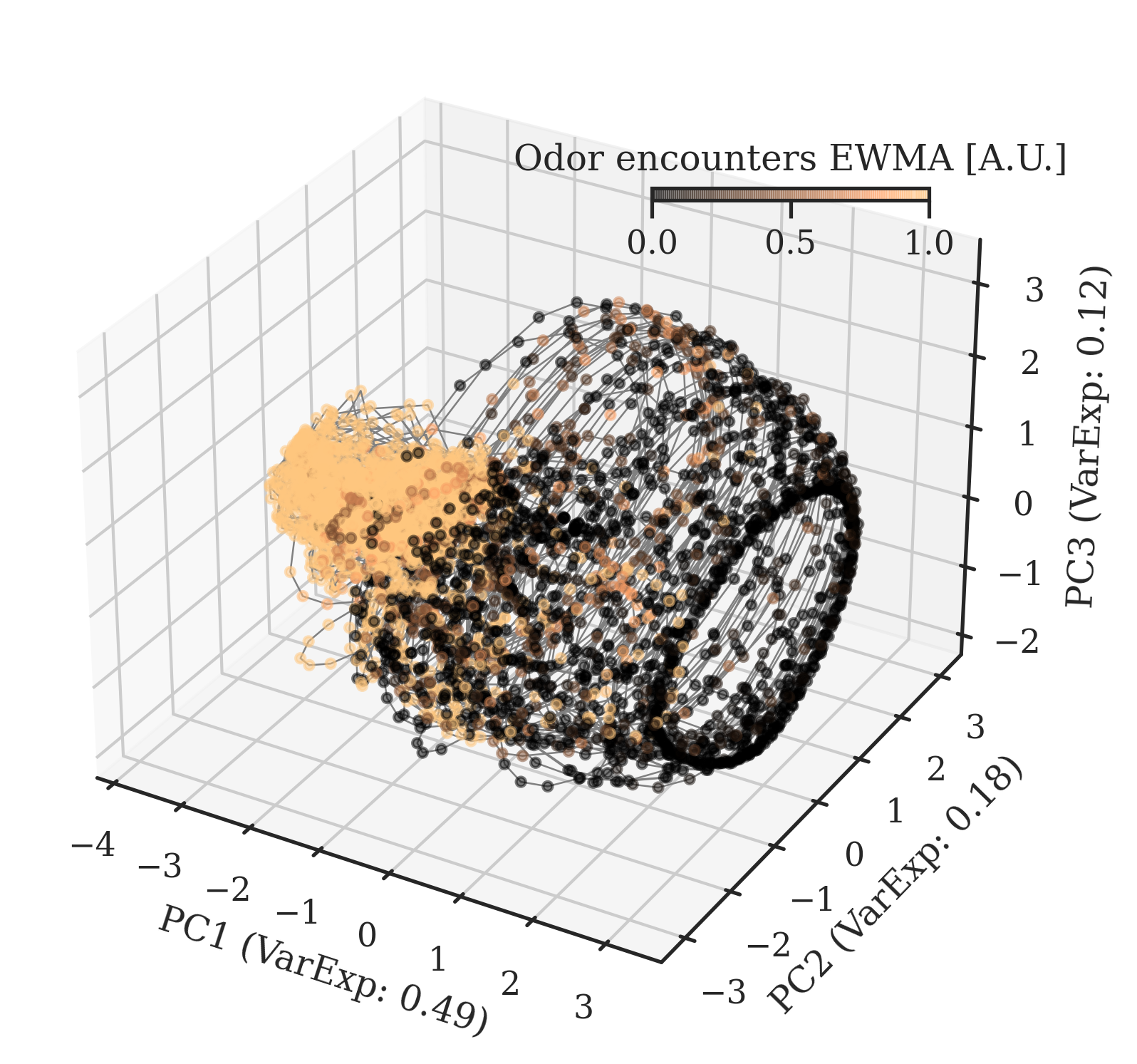} \\
\includegraphics[width=0.30\linewidth]{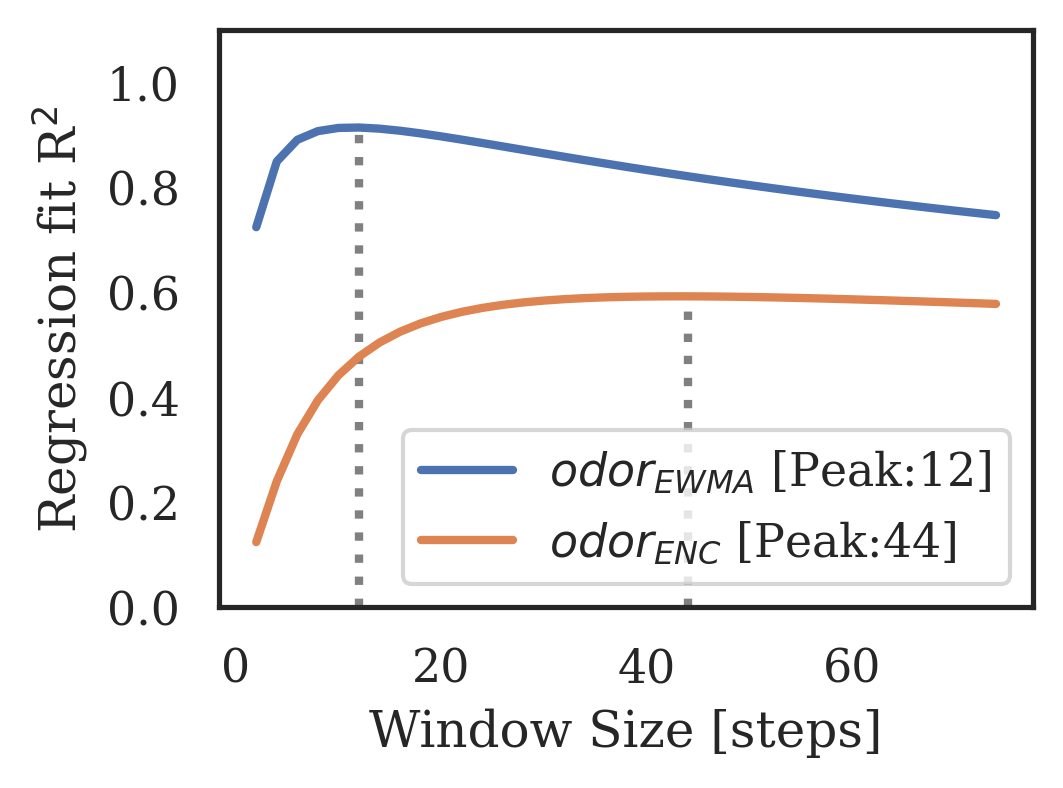}
\includegraphics[width=0.34\linewidth]{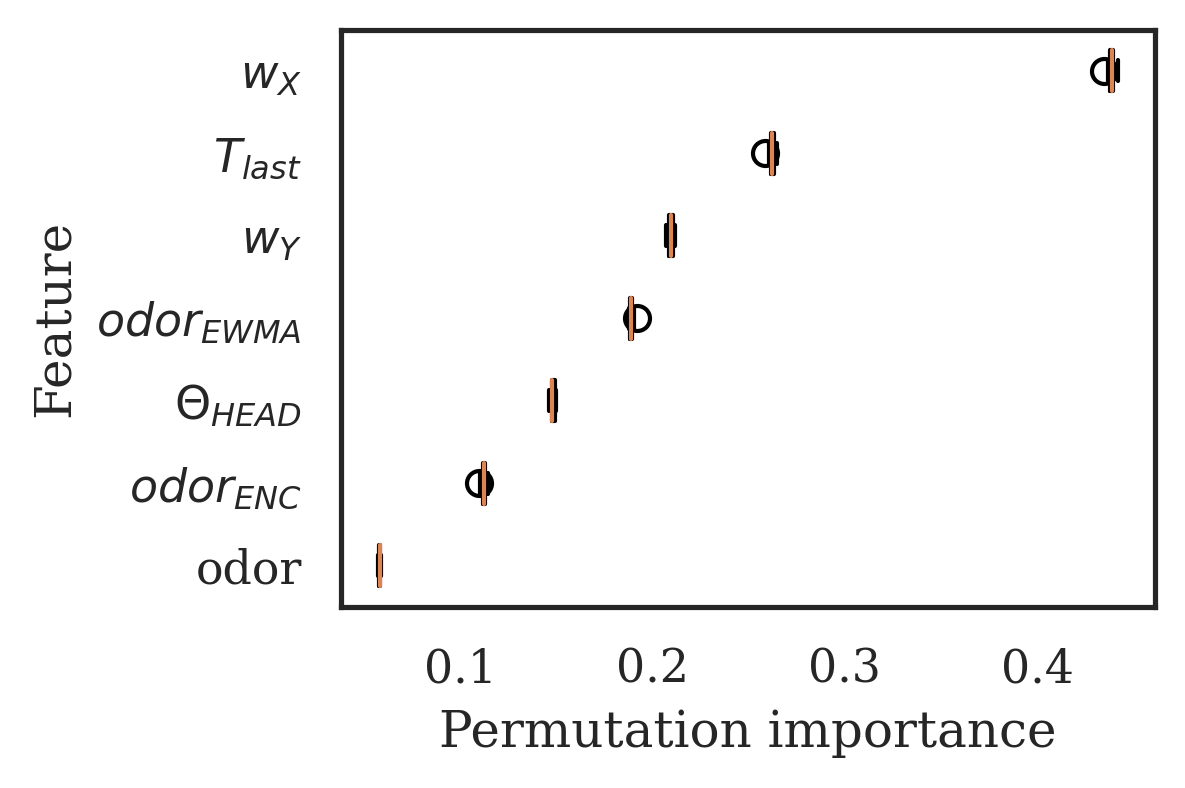}
\includegraphics[width=0.30\linewidth]{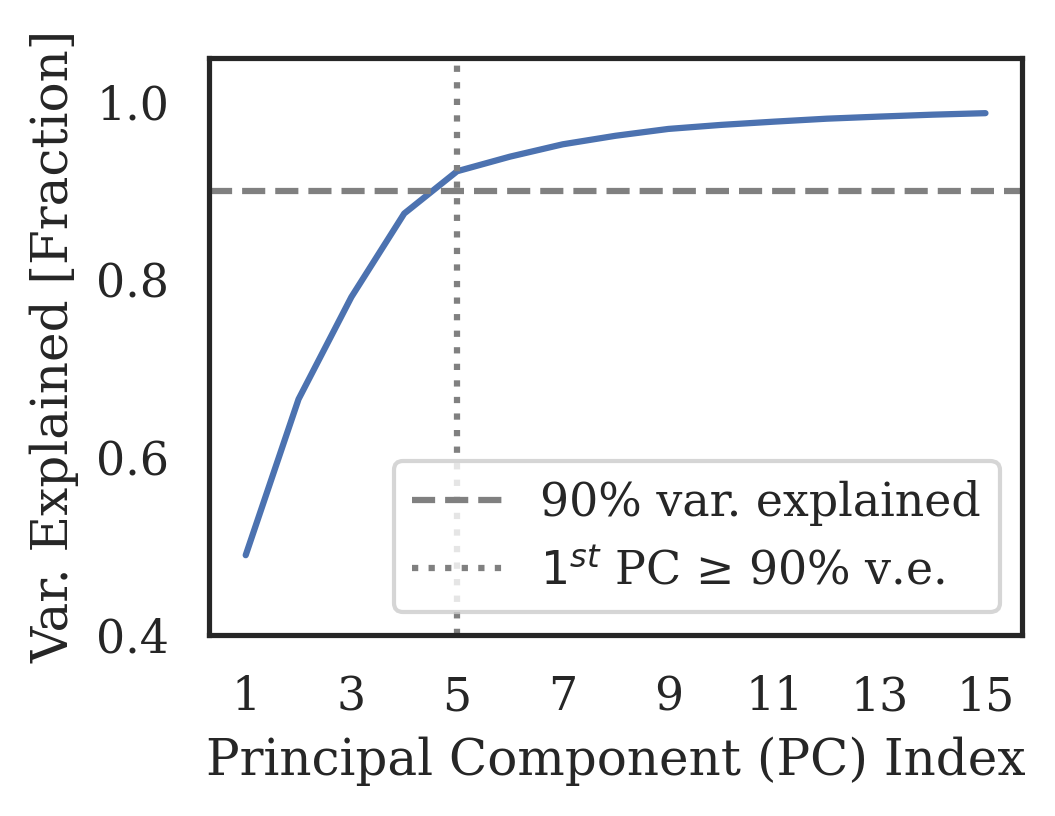}
\caption{Neural representations -- Agent 5}
\end{center}
\end{figure*}
\clearpage
\section{Structured neural dynamics}
\label{sec_supp_dynamics}

\begin{table}[h!]
    \centering
    \begin{tabular}{ccccc}
     \hline\hline
     \textbf{Agent} & \textbf{Agent ID} & \textbf{Limit-cycle period}  \\
     \hline   
        RNN 1 & 2760377 & 19 steps (0.76 s) \\ \hline
        RNN 2 & 3199993 & NA (clear periodic structure not observed)  \\ \hline
        RNN 3 & 3307e9 & 17 steps (0.68 s) \\ \hline
        RNN 4 & 541058 & 28 steps (1.12 s) \\ \hline
        RNN 5 & 9781ba & 18 steps (0.72 s) \\ \hline
     \hline
    \end{tabular}
    \caption{Limit cycle periods for each RNN agent}
\label{table_supp_LC}
\end{table}

\begin{figure*}[h!]
\centering
\includegraphics[width=0.40\linewidth]{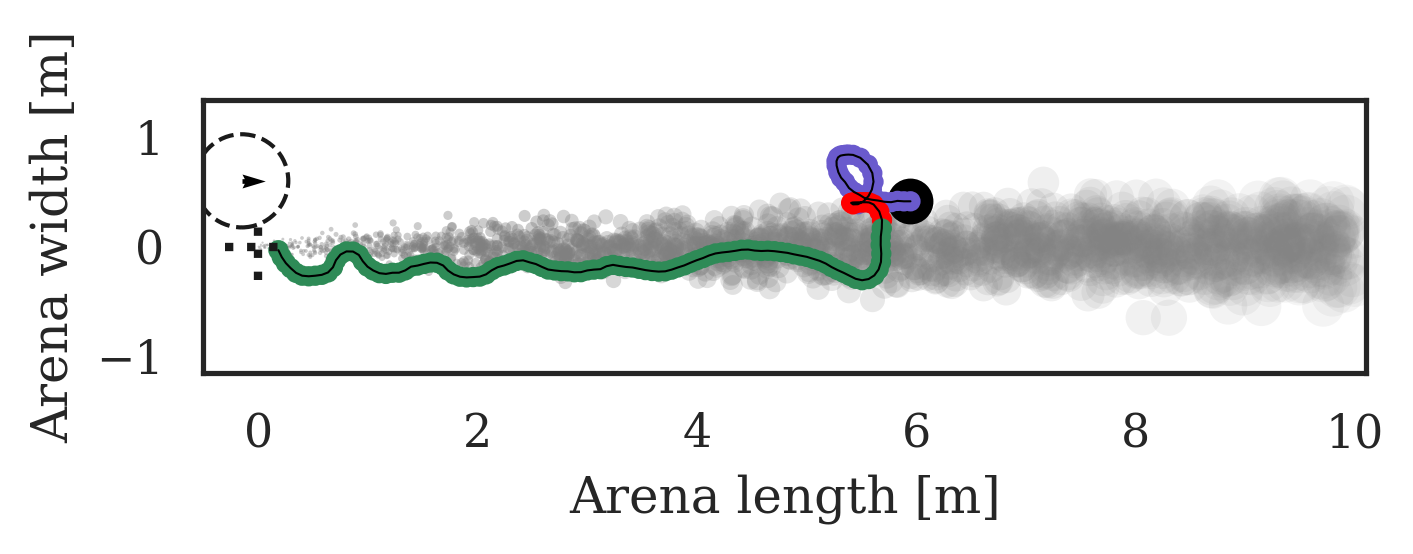}
\includegraphics[width=0.20\linewidth]{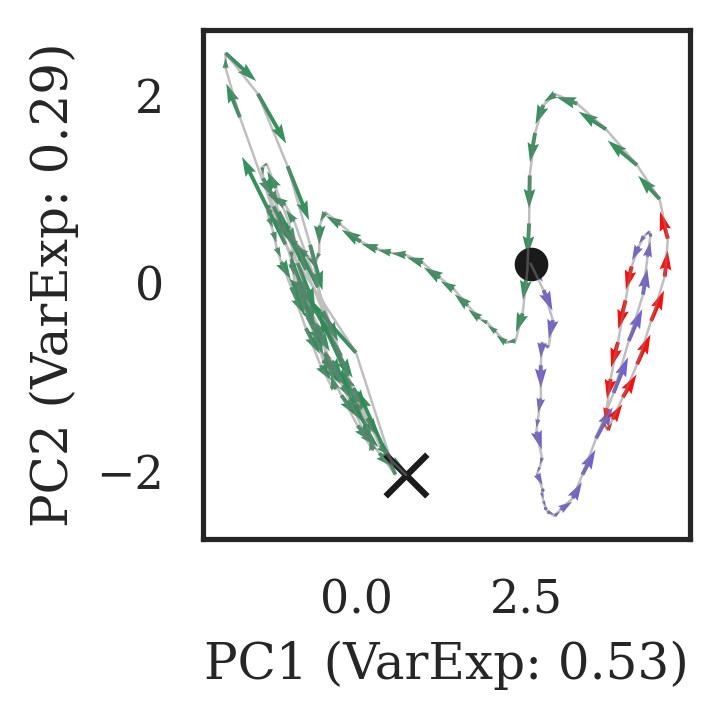} \\
\includegraphics[width=0.40\linewidth]{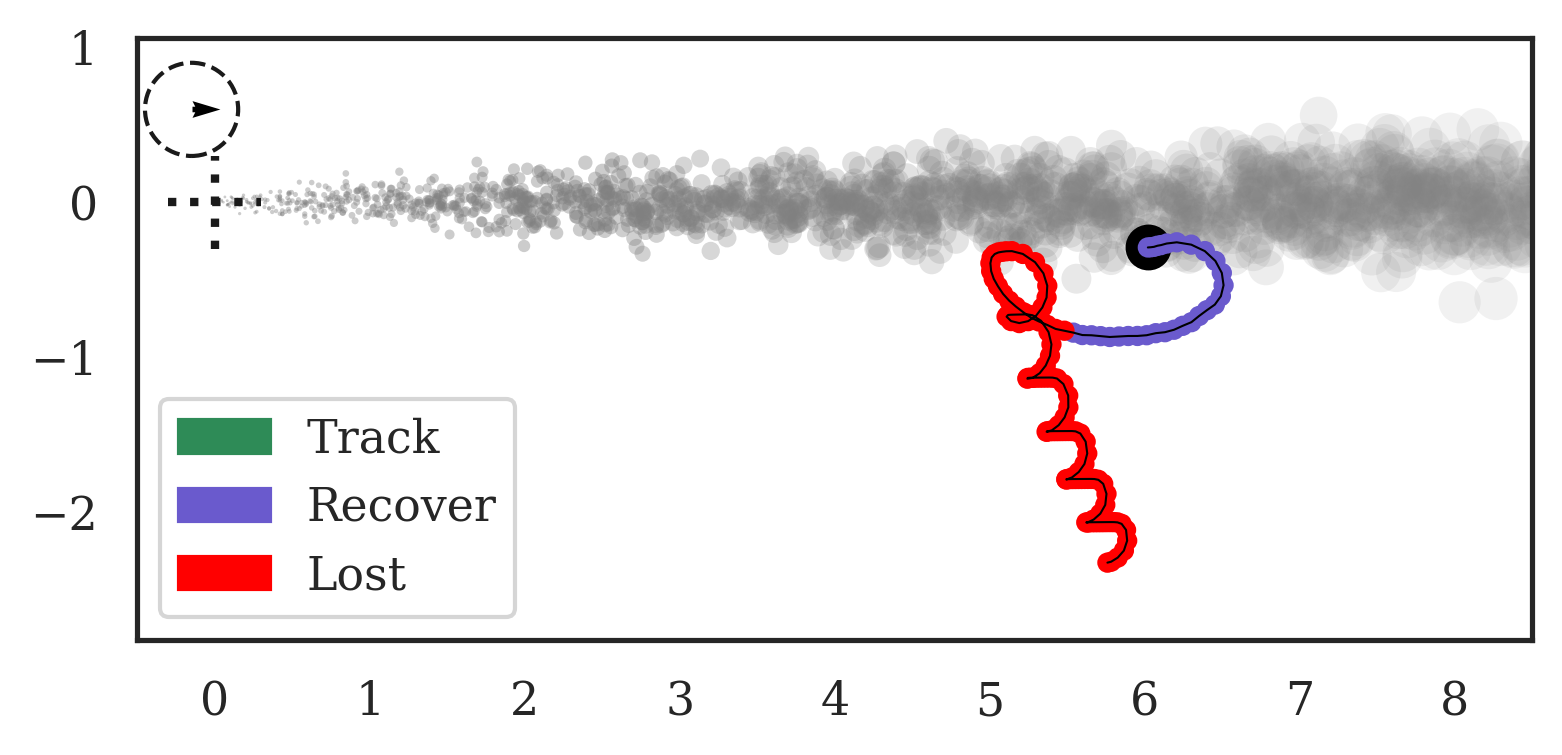}
\includegraphics[width=0.20\linewidth]{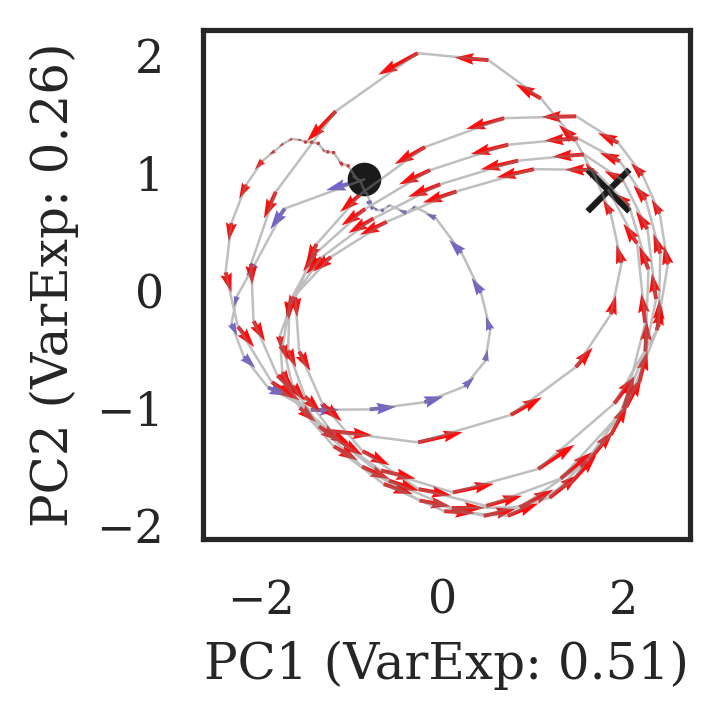} \\
\includegraphics[width=0.40\linewidth]{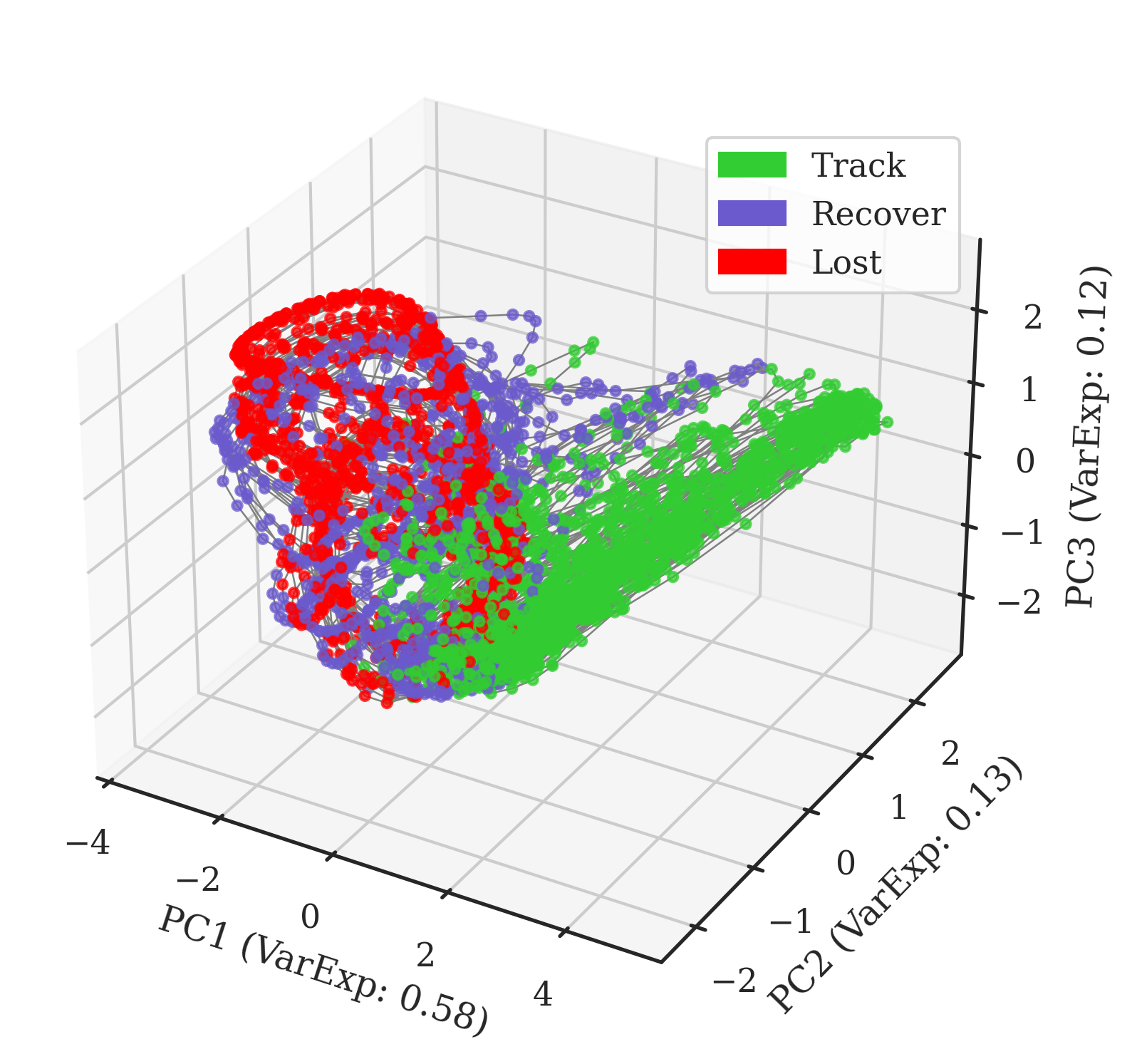}
\caption[Neural dynamics -- Agent 1]{Neural dynamics -- Agent 1 (See Figure \ref{fig_dynamics} for equivalent data on Agent 3 and figure details)}
\end{figure*}

\begin{figure*}[h!]
\centering
\includegraphics[width=0.40\linewidth]{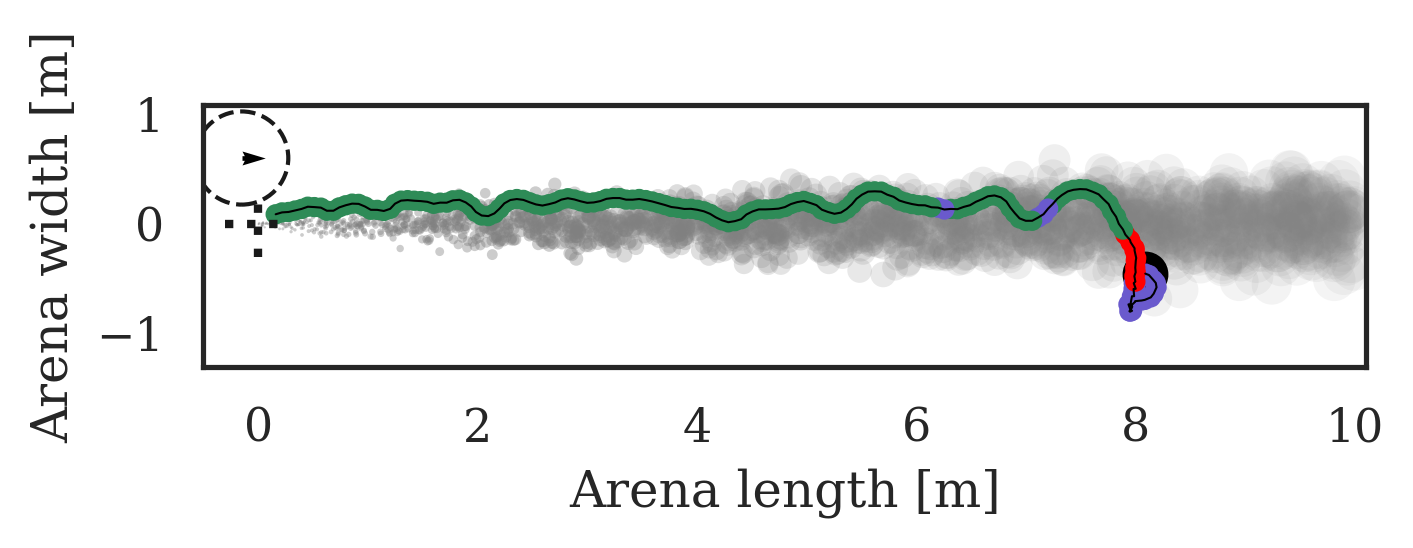}
\includegraphics[width=0.20\linewidth]{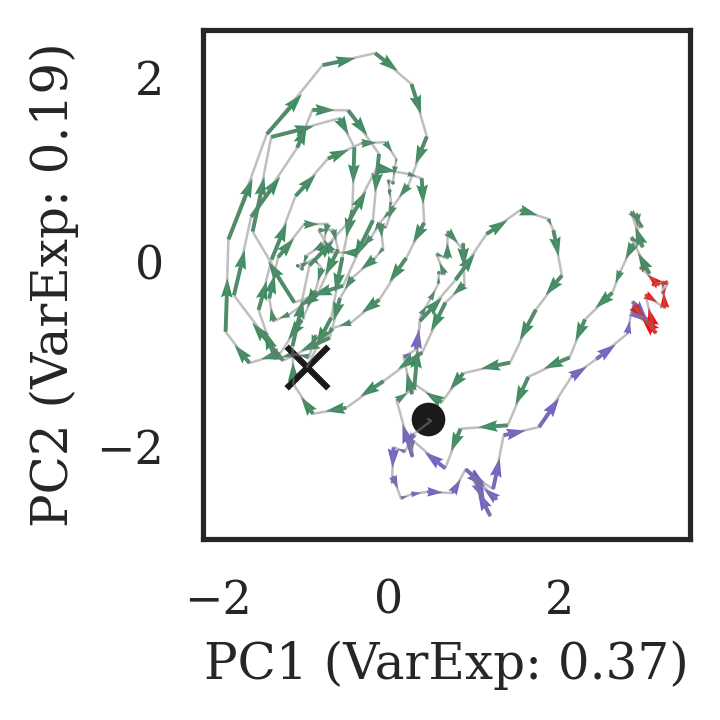} \\
\includegraphics[width=0.40\linewidth]{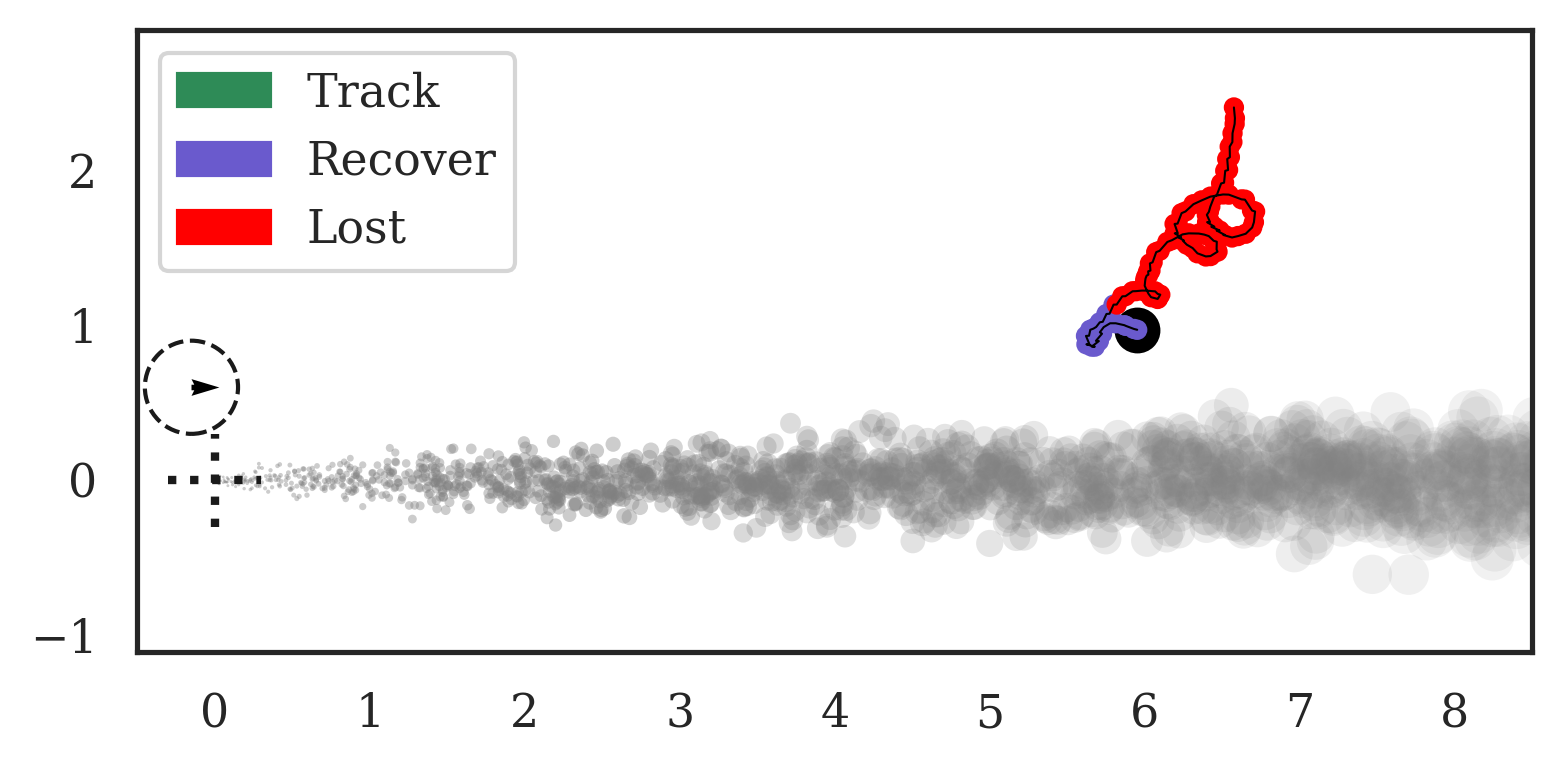}
\includegraphics[width=0.20\linewidth]{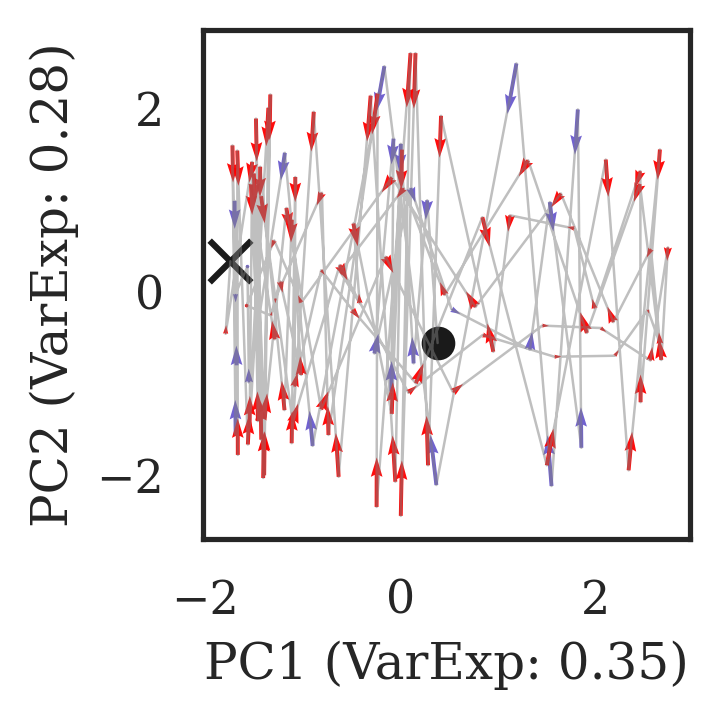} \\
\includegraphics[width=0.40\linewidth]{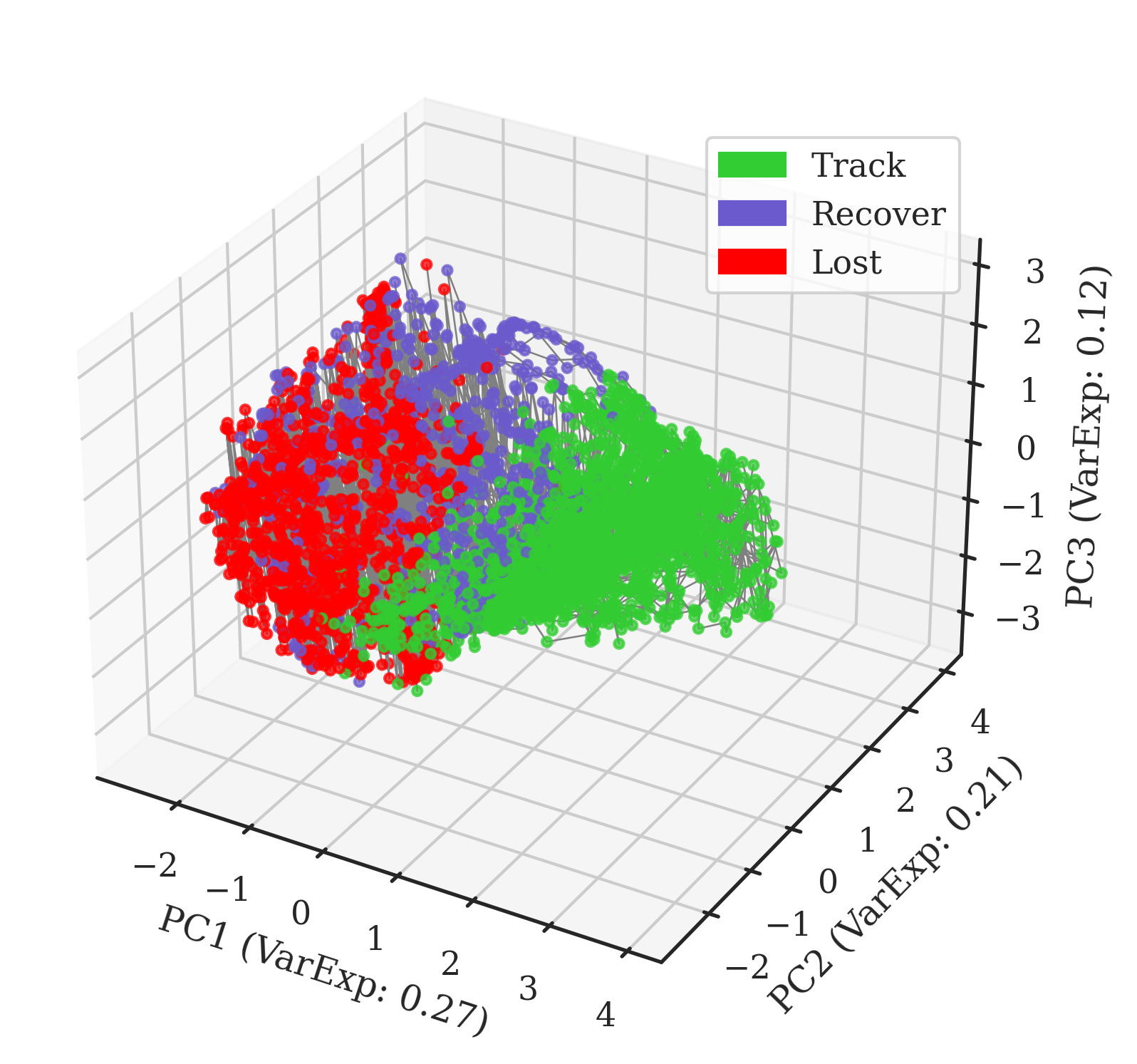}
\caption{Neural dynamics -- Agent 2}
\end{figure*}

\begin{figure*}[h!]
\centering
\includegraphics[width=0.40\linewidth]{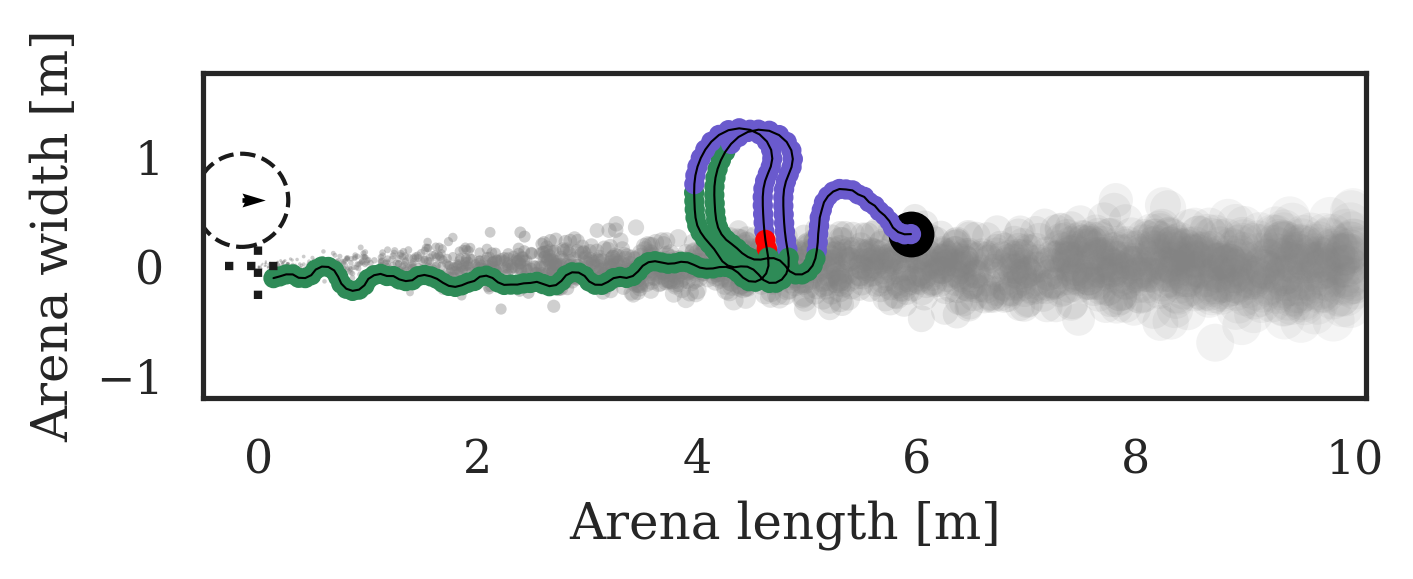}
\includegraphics[width=0.20\linewidth]{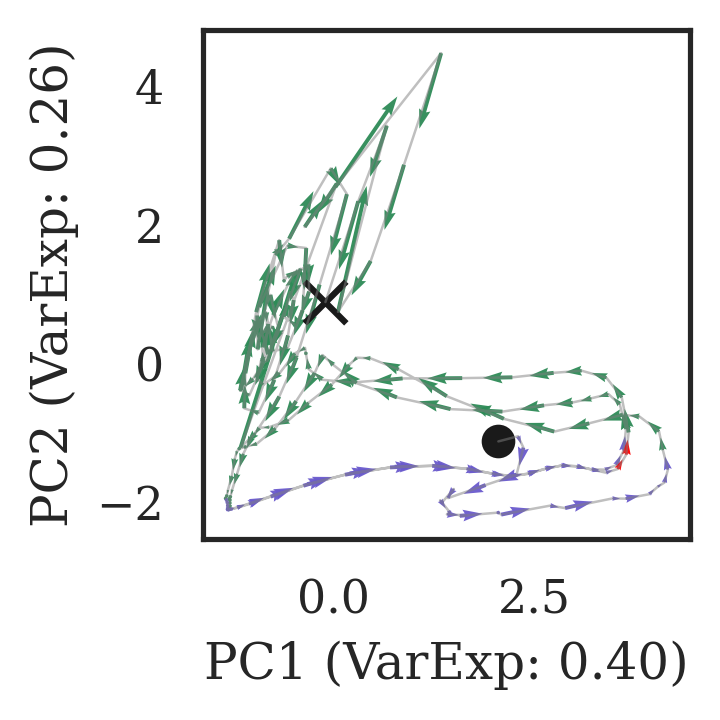} \\
\includegraphics[width=0.40\linewidth]{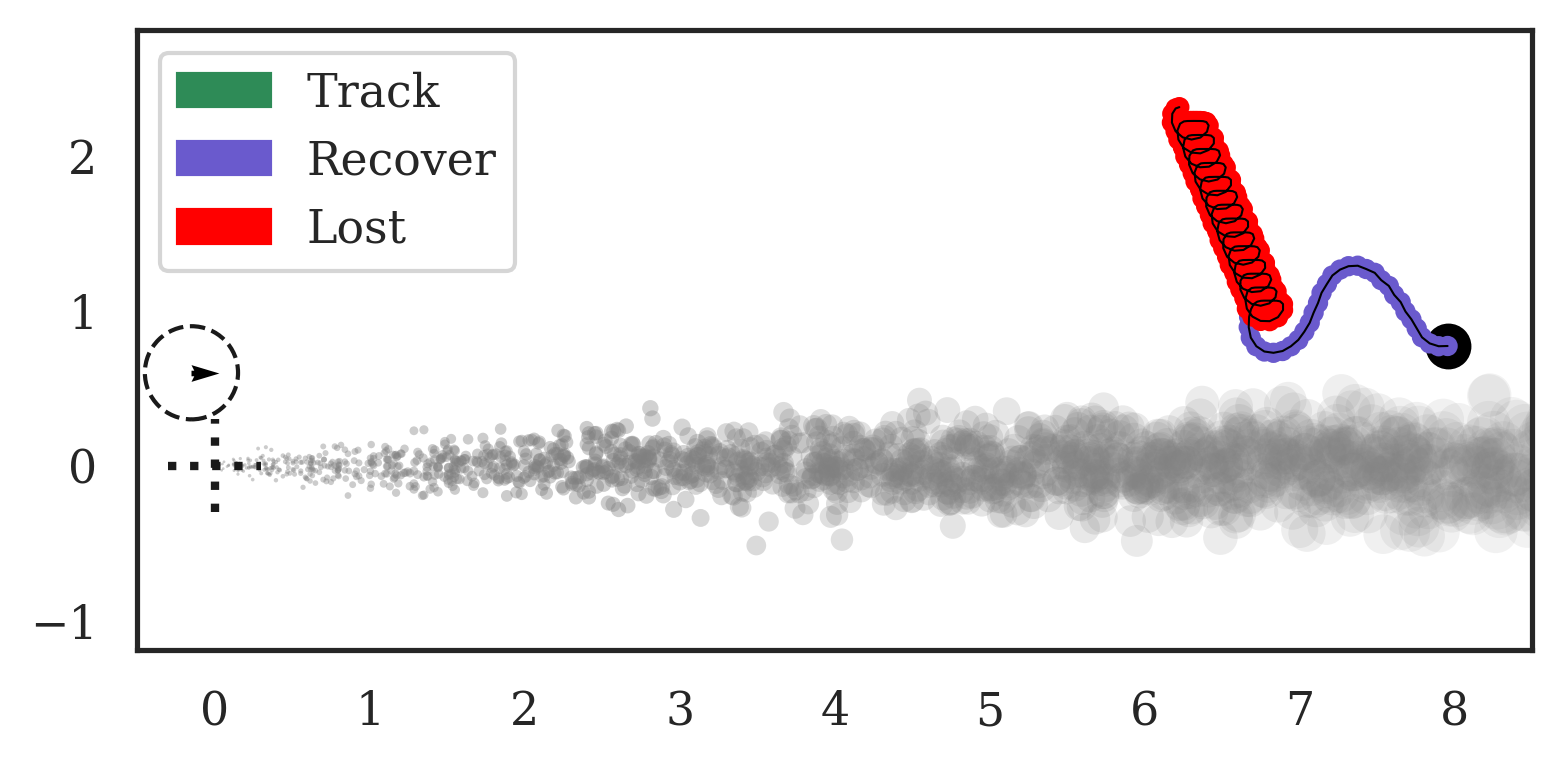}
\includegraphics[width=0.20\linewidth]{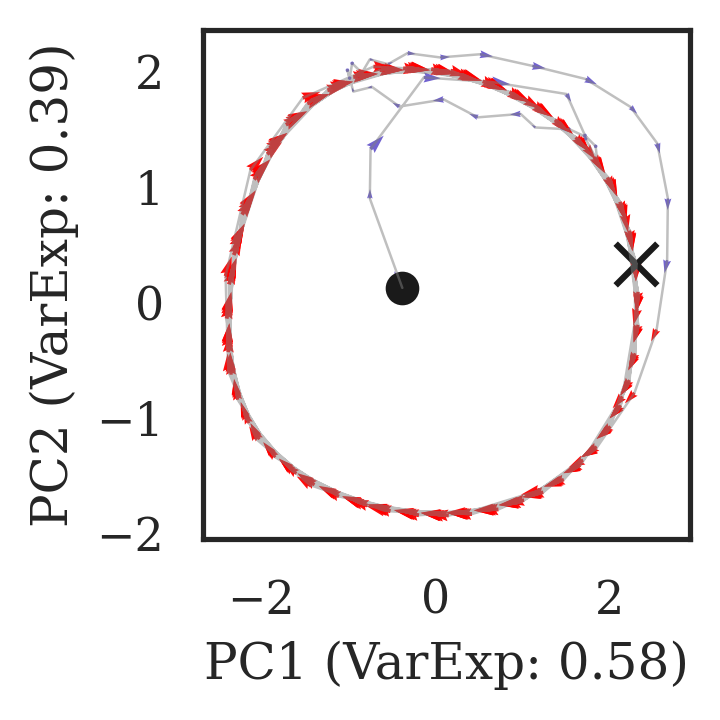} \\
\includegraphics[width=0.40\linewidth]{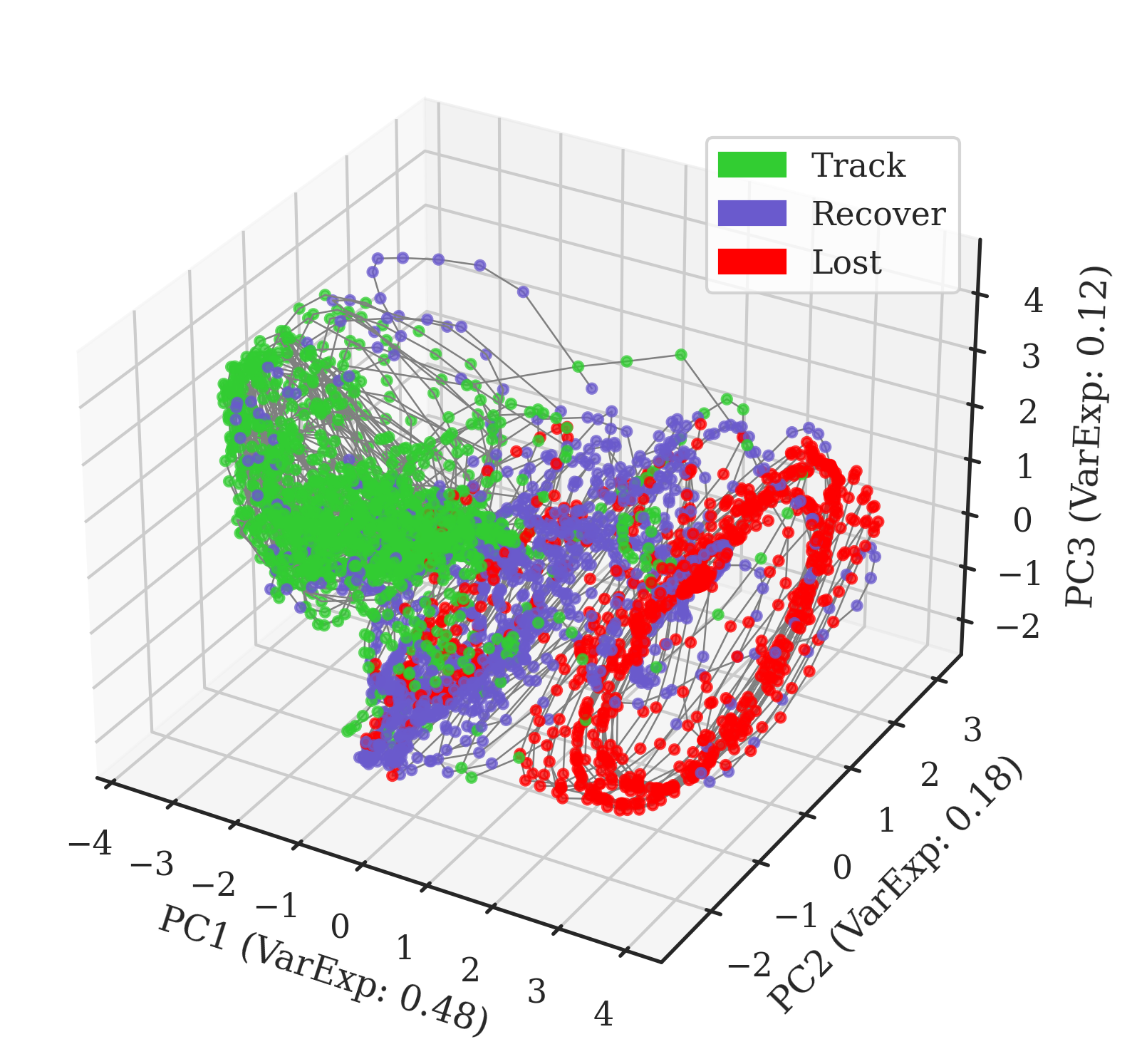}
\caption[Neural dynamics -- Agent 3]{Neural dynamics -- Agent 3 (same as Figure \ref{fig_dynamics})}
\end{figure*}

\begin{figure*}[h!]
\centering
\includegraphics[width=0.40\linewidth]{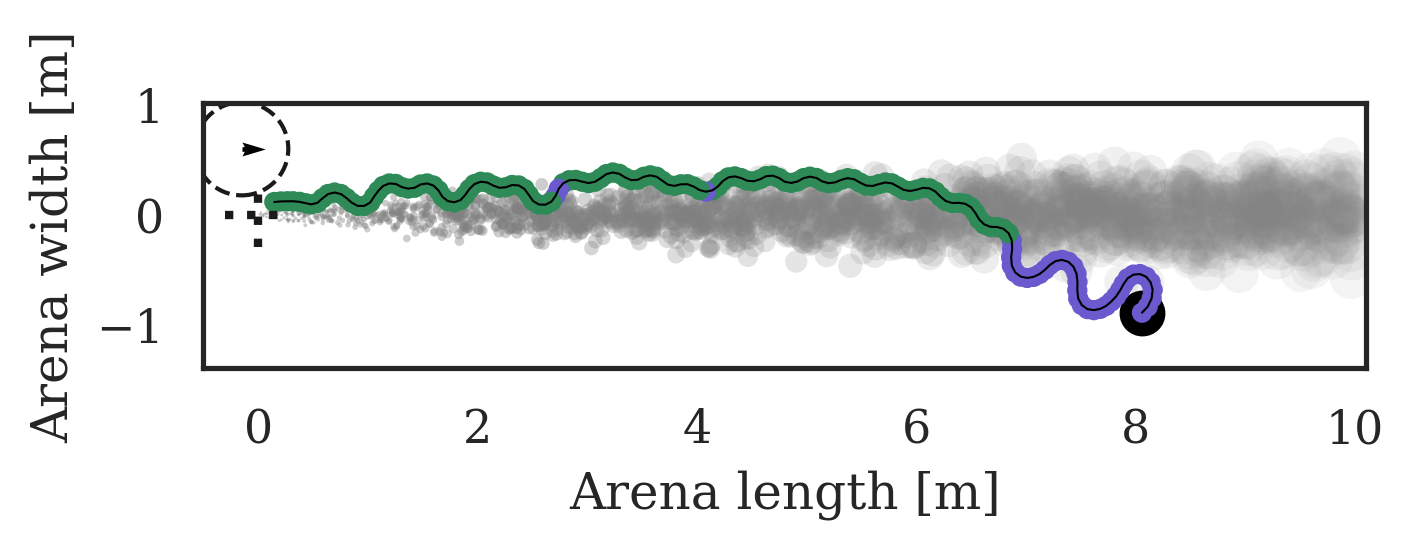}
\includegraphics[width=0.20\linewidth]{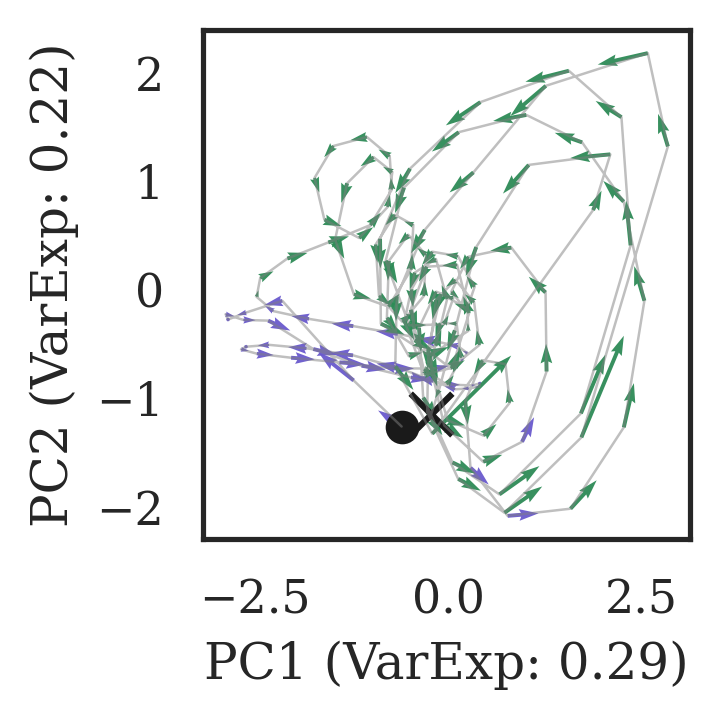} \\
\includegraphics[width=0.40\linewidth]{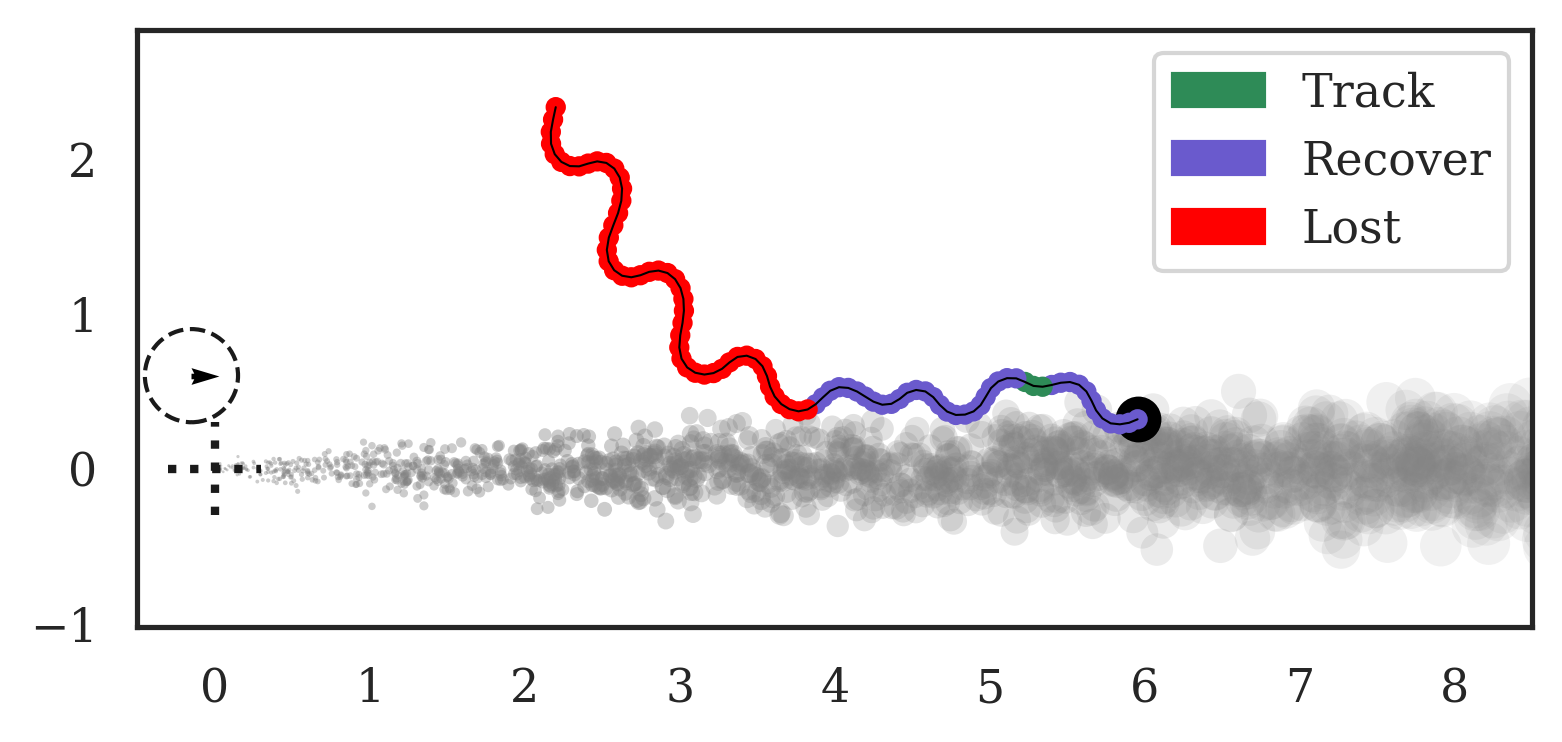}
\includegraphics[width=0.20\linewidth]{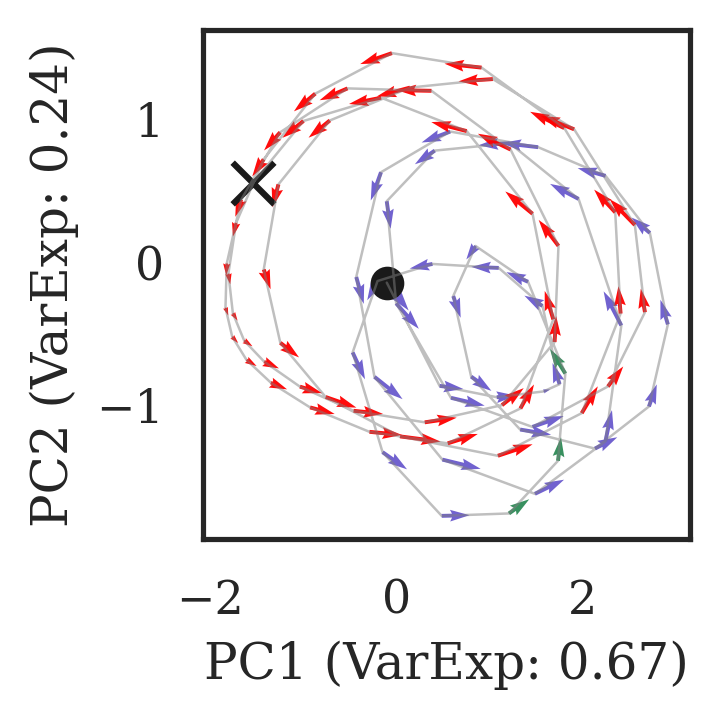} \\
\includegraphics[width=0.40\linewidth]{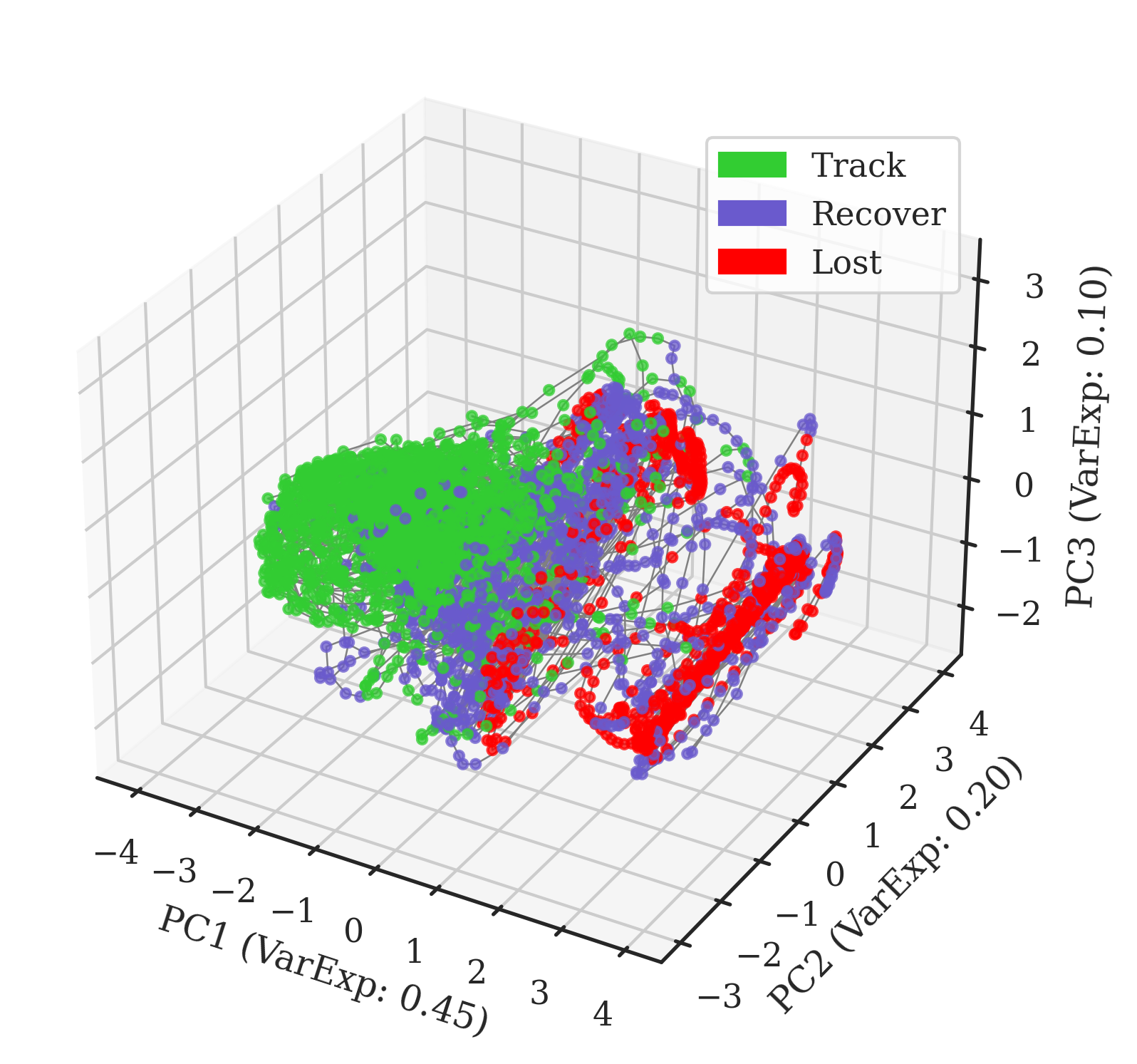}
\caption{Neural dynamics -- Agent 4}
\end{figure*}

\begin{figure*}[h!]
\centering
\includegraphics[width=0.40\linewidth]{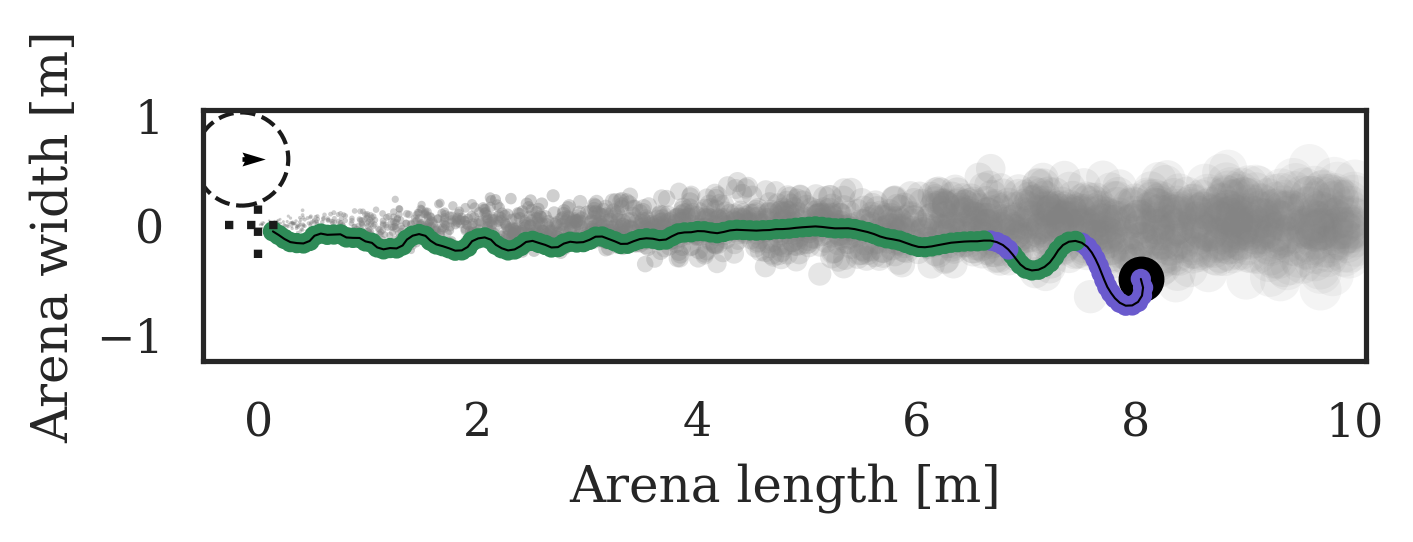}
\includegraphics[width=0.20\linewidth]{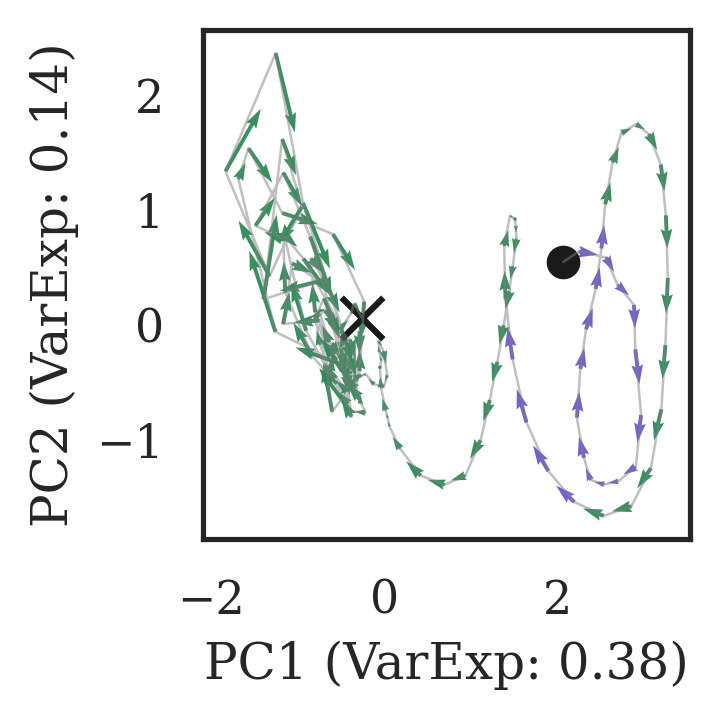} \\
\includegraphics[width=0.40\linewidth]{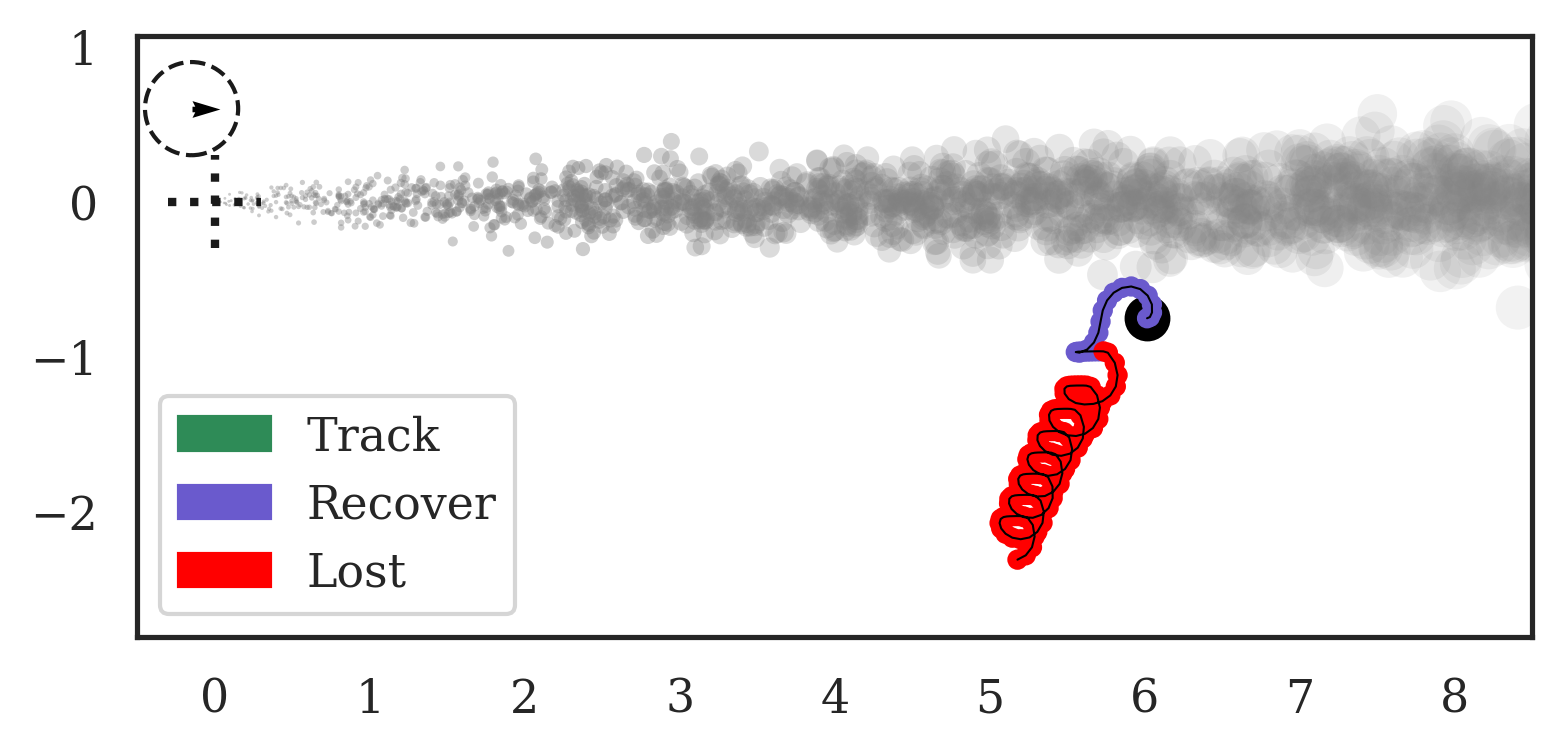}
\includegraphics[width=0.20\linewidth]{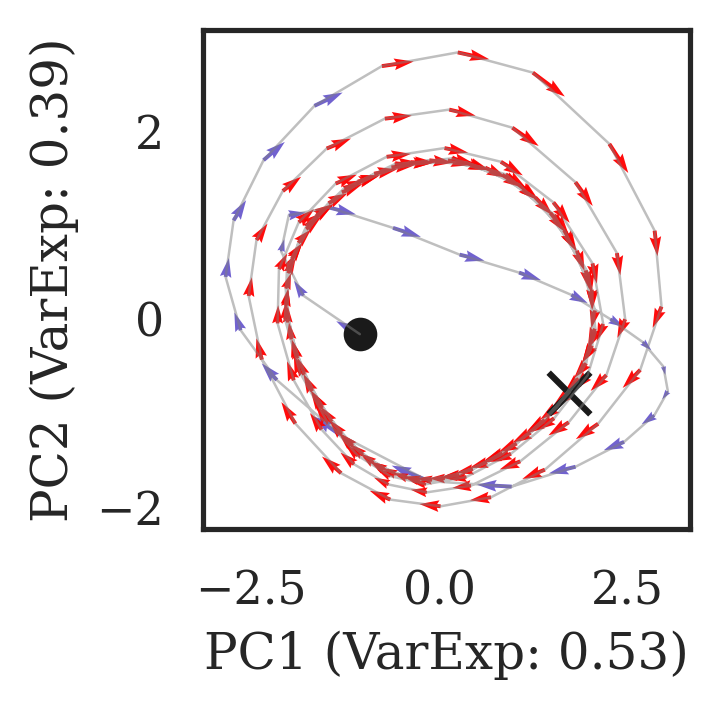} \\
\includegraphics[width=0.40\linewidth]{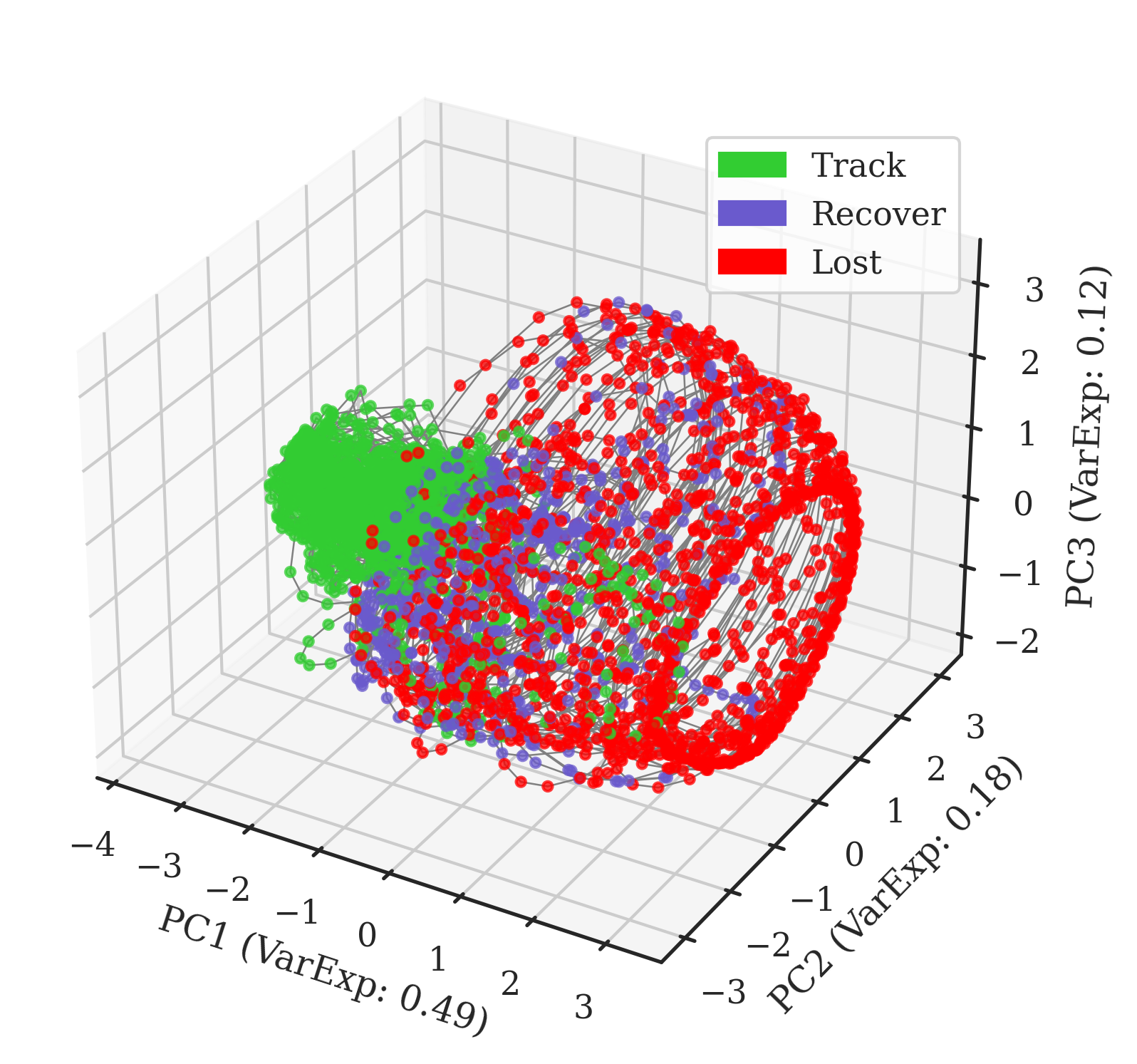}
\caption{Neural dynamics -- Agent 5}
\end{figure*}

\clearpage
\section{Transitions between neural clusters}
\label{sec_supp_ttcs}

\textbf{Transition duration calculations:}
To define entry into a neural activity regime, we first define neural activity `centroids' associated with the \textit{tracking} and \textit{lost} neural activity regimes.
These are the average of the last 1-second's neural activity from successful evaluation episodes that home in on odor source (HOME), and unsuccessful evaluation episodes where the agent straying Out Of (arena) Bounds (OOB), respectively (see Figure \ref{fig_ttcs}). 
We then define neural activity clusters associated with the HOME and OOB centroids as being comprised of all neural activity with a distance $D/2$ units from the respective centroid, where $D$ is the distance between centroids.
Finally,  for any unsuccessful tracking episode, we calculate a `time to \textit{lost}' (TTL) as the duration between the agent leaving the plume and entering the OOB cluster.
Similarly, for successfully homing episodes, we calculate a `time to \textit{track}' (TTT) as the time taken to enter the HOME cluster after entering the plume.
In calculating TTT, we exclude small excursions outside the plume where the agent is skimming the boundary of the plume and only consider excursions where the agent has entered the \textit{recovering} or \textit{lost} behavioral module.
We split TTT into two types, labeling it `time to \textit{track} not \textit{lost}' (TTT-NL) if the agent was in \textit{recovering} or `time to \textit{track} after \textit{lost}' (TTT-L) if the agent was in \textit{lost} before entering the plume. \\

\textbf{Statistical significance calculations:}
All plots use the Mann-Whitney-Wilcoxon test two-sided with Bonferroni correction, where p-value annotations indicate:  
\begin{verbatim}
ns: 5.00e-02 < p <= 1.00e+00 (not significant)
*: 1.00e-02 < p <= 5.00e-02
**: 1.00e-03 < p <= 1.00e-02
***: 1.00e-04 < p <= 1.00e-03
****: p <= 1.00e-04
\end{verbatim}

\begin{figure*}[h!]
\centering
\includegraphics[width=0.35\linewidth]{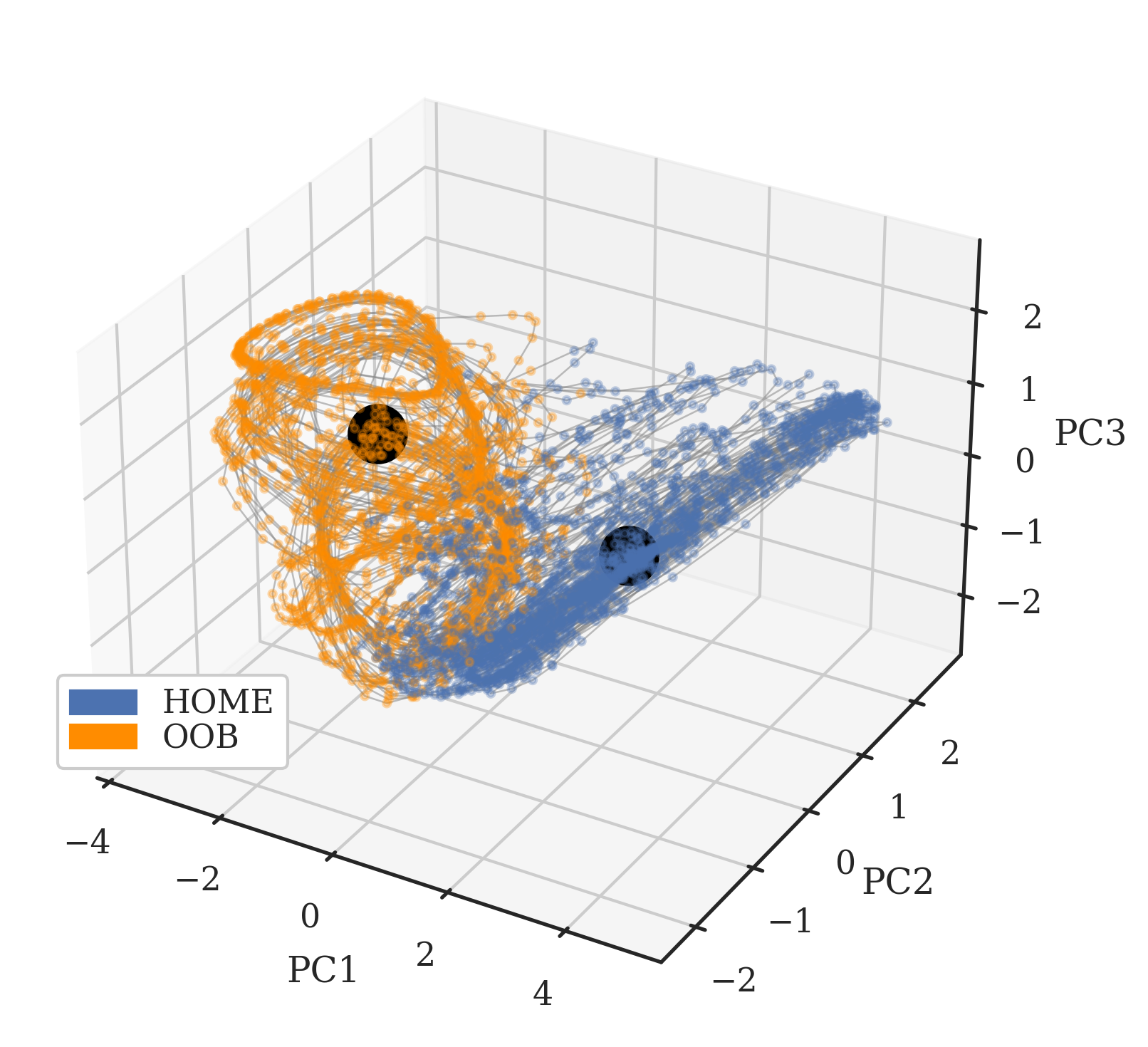} 
\includegraphics[width=0.30\linewidth]{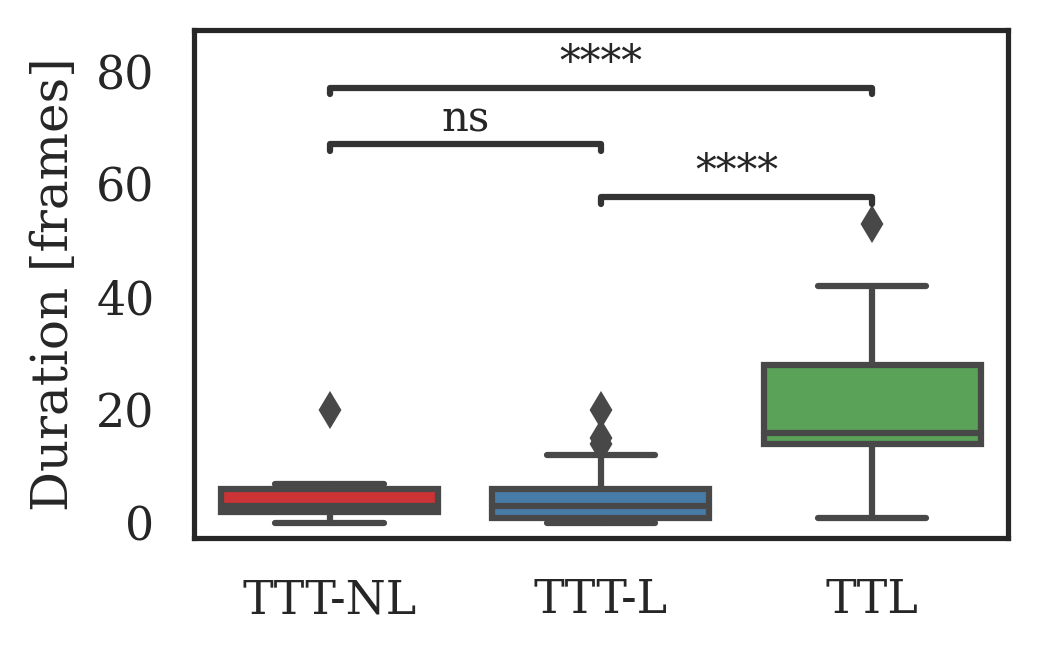}
\caption[Transitions between neural activity regimes -- Agent 1]{Transitions between neural activity regimes -- Agent 1 (compare with Agent 3 in Figure \ref{fig_ttcs})
}
\end{figure*}

\begin{figure*}[h!]
\centering
\includegraphics[width=0.35\linewidth]{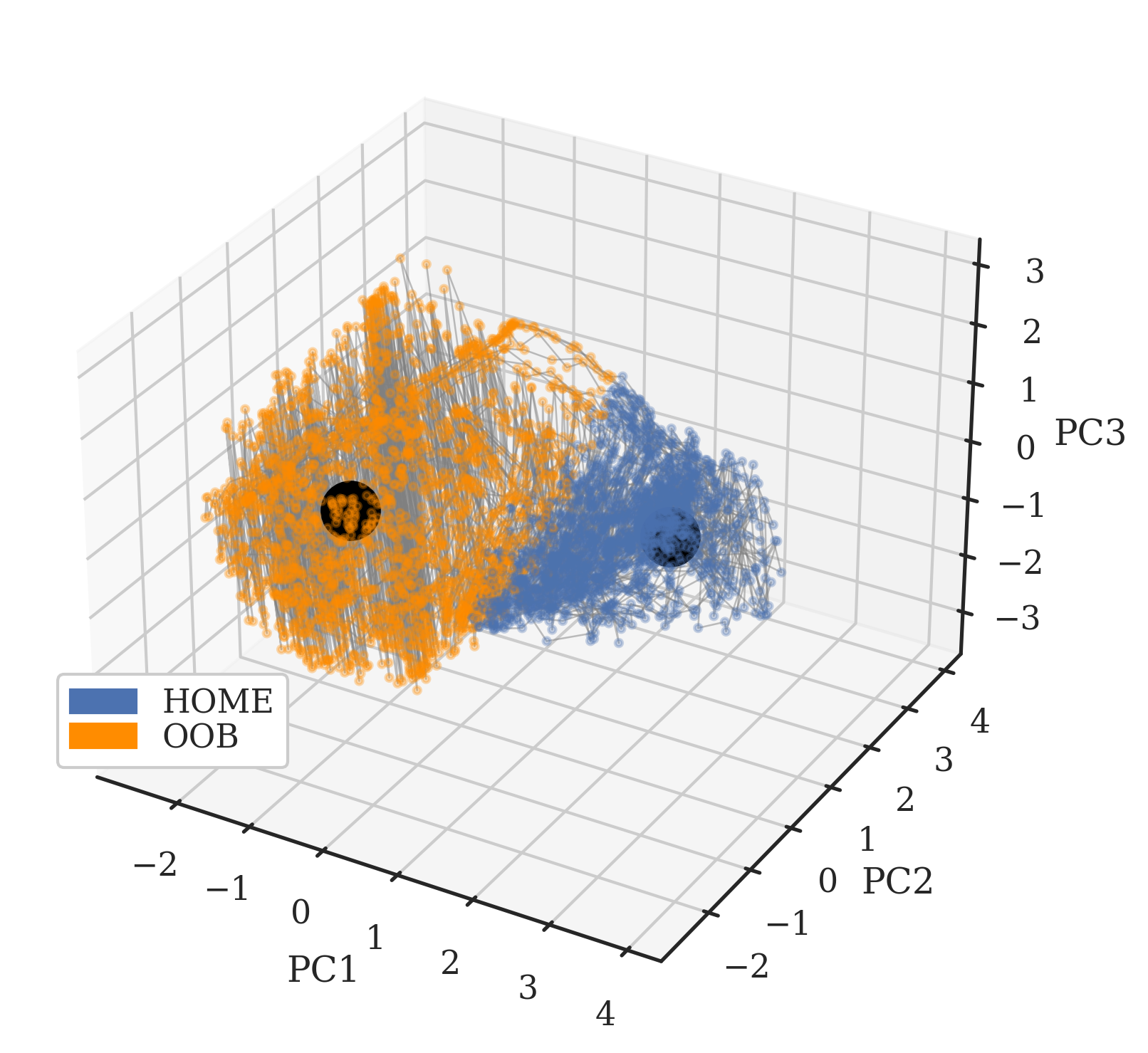} 
\includegraphics[width=0.30\linewidth]{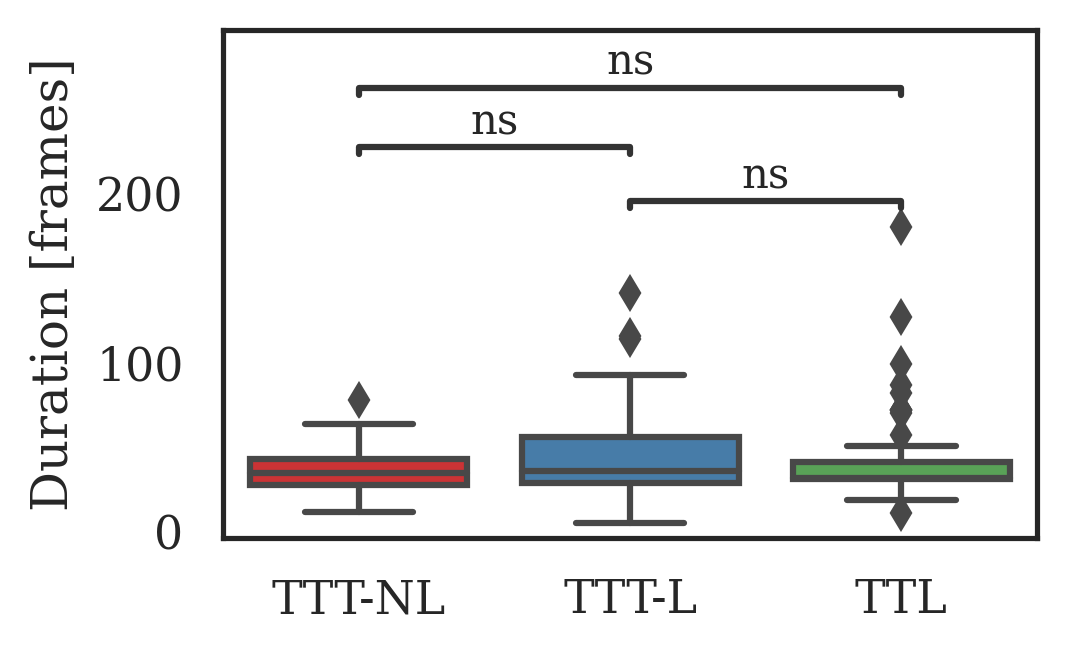}
\caption[Transitions between neural activity regimes -- Agent 2]{Transitions between neural activity regimes -- Agent 2 (NB: this agent does not follow the trend)}
\end{figure*}

\begin{figure*}[h!]
\centering
\includegraphics[width=0.35\linewidth]{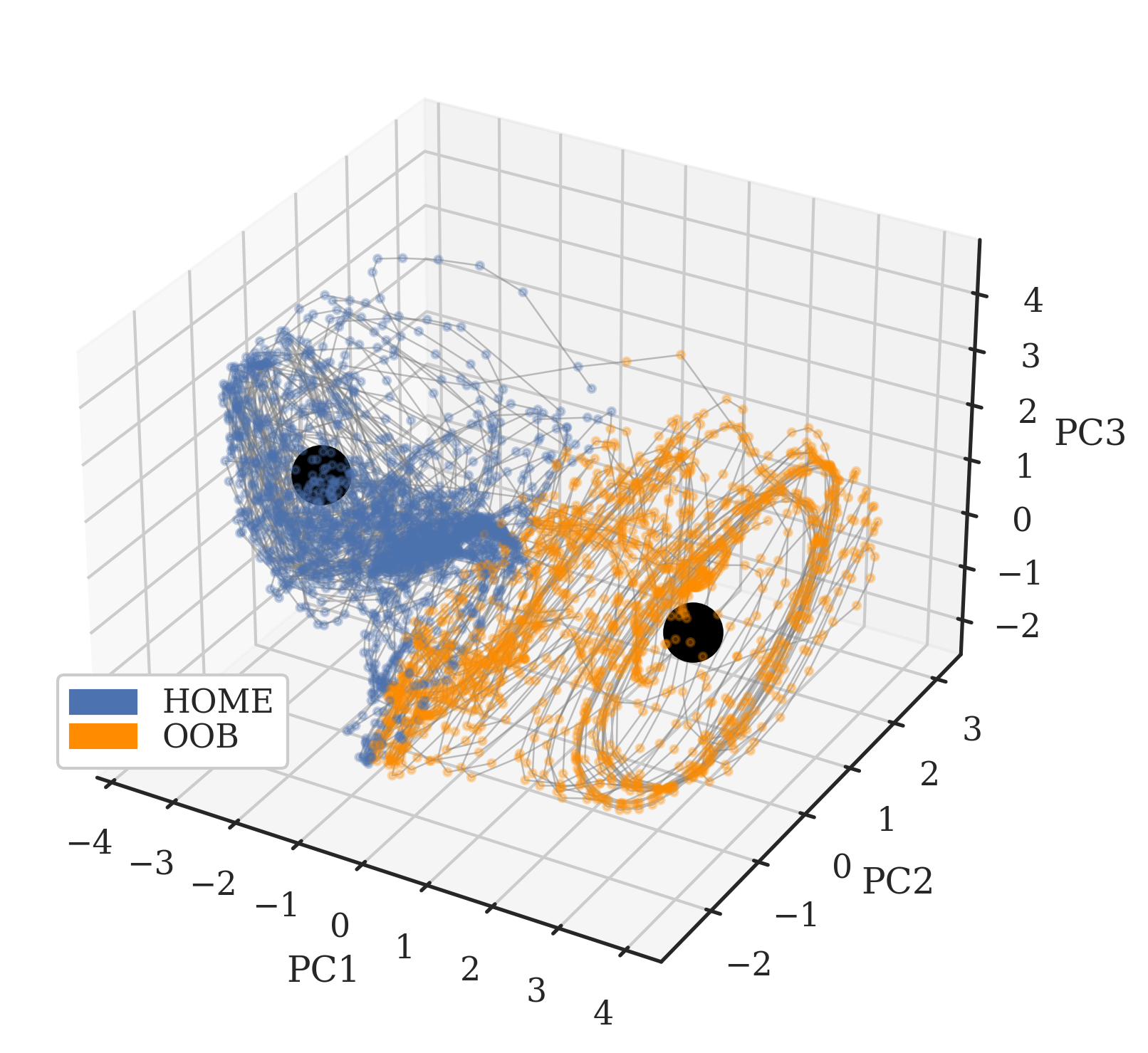} 
\includegraphics[width=0.30\linewidth]{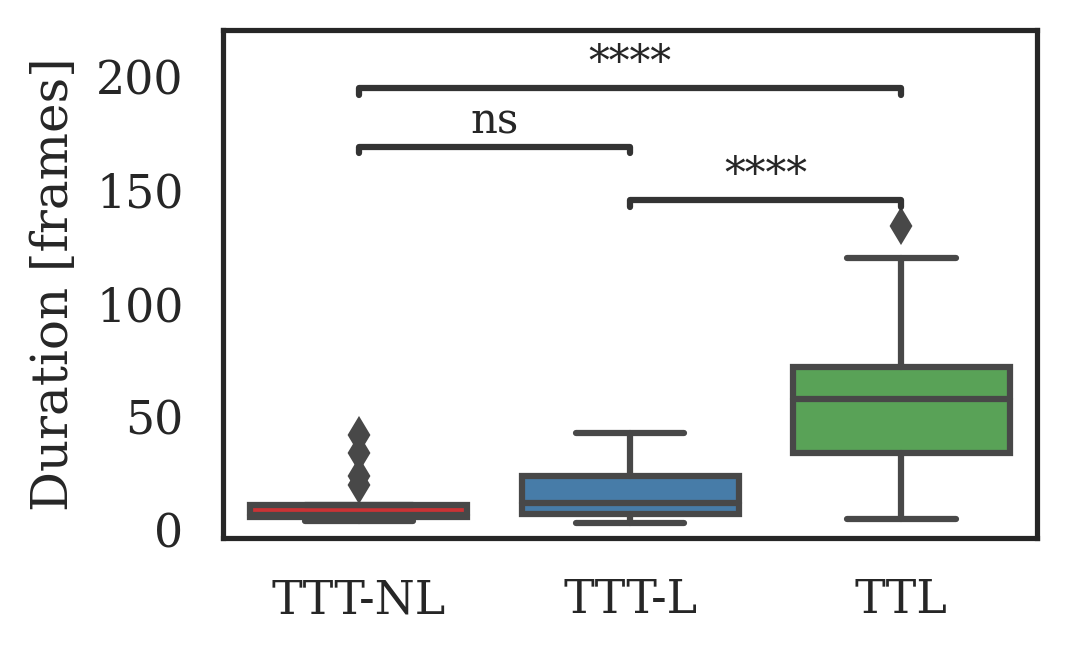}
\caption[Transitions between neural activity regimes -- Agent 3]{Transitions between neural activity regimes -- Agent 3 (same data as Figure \ref{fig_ttcs})}
\end{figure*}

\begin{figure*}[h!]
\centering
\includegraphics[width=0.35\linewidth]{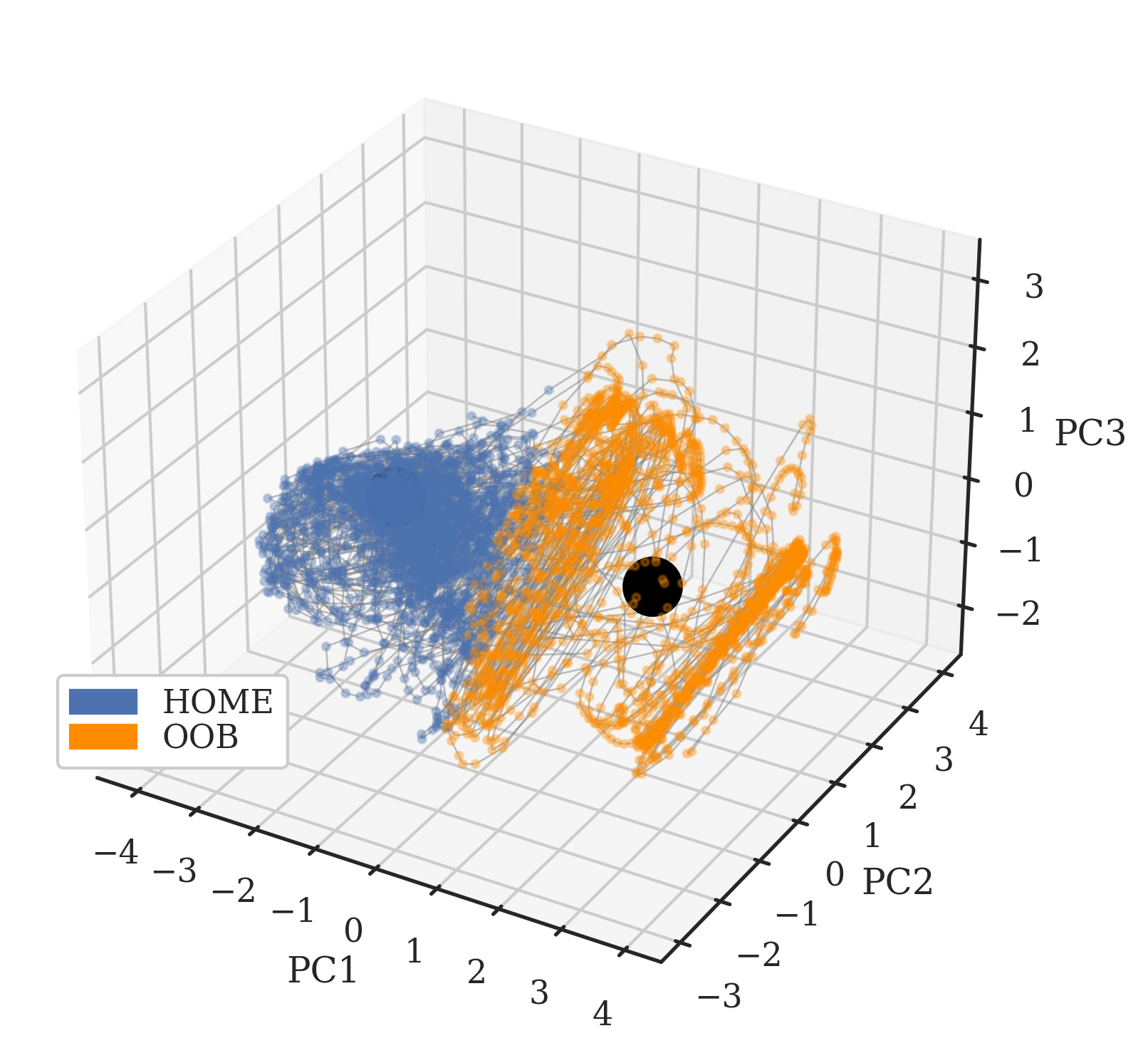} 
\includegraphics[width=0.30\linewidth]{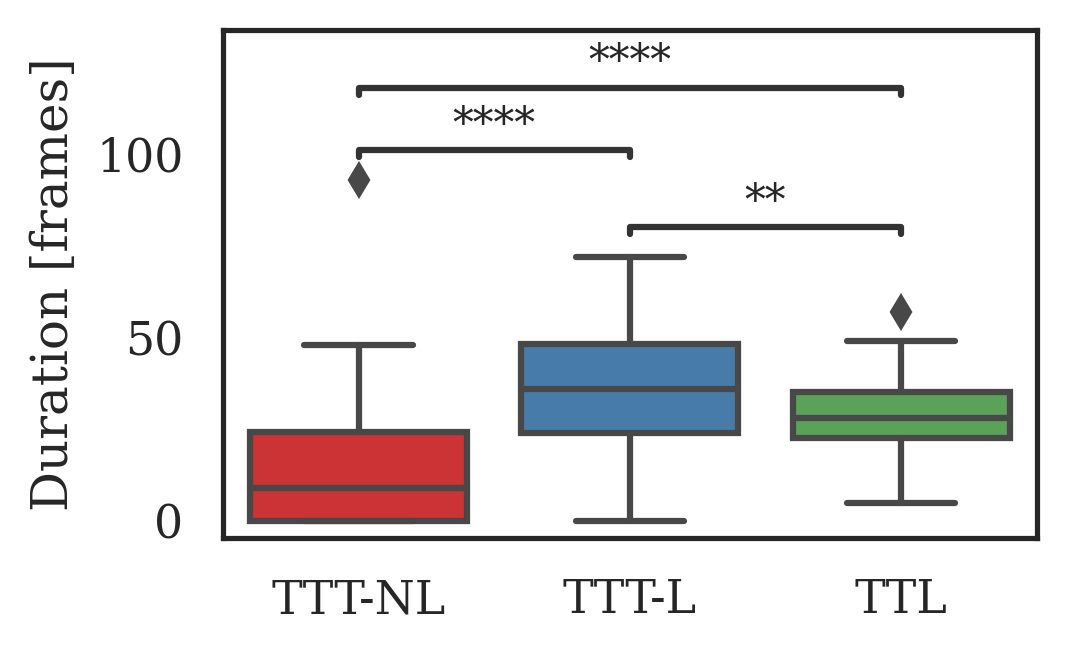}
\caption{Transitions between neural activity regimes -- Agent 4}
\end{figure*}

\begin{figure*}[h!]
\centering
\includegraphics[width=0.35\linewidth]{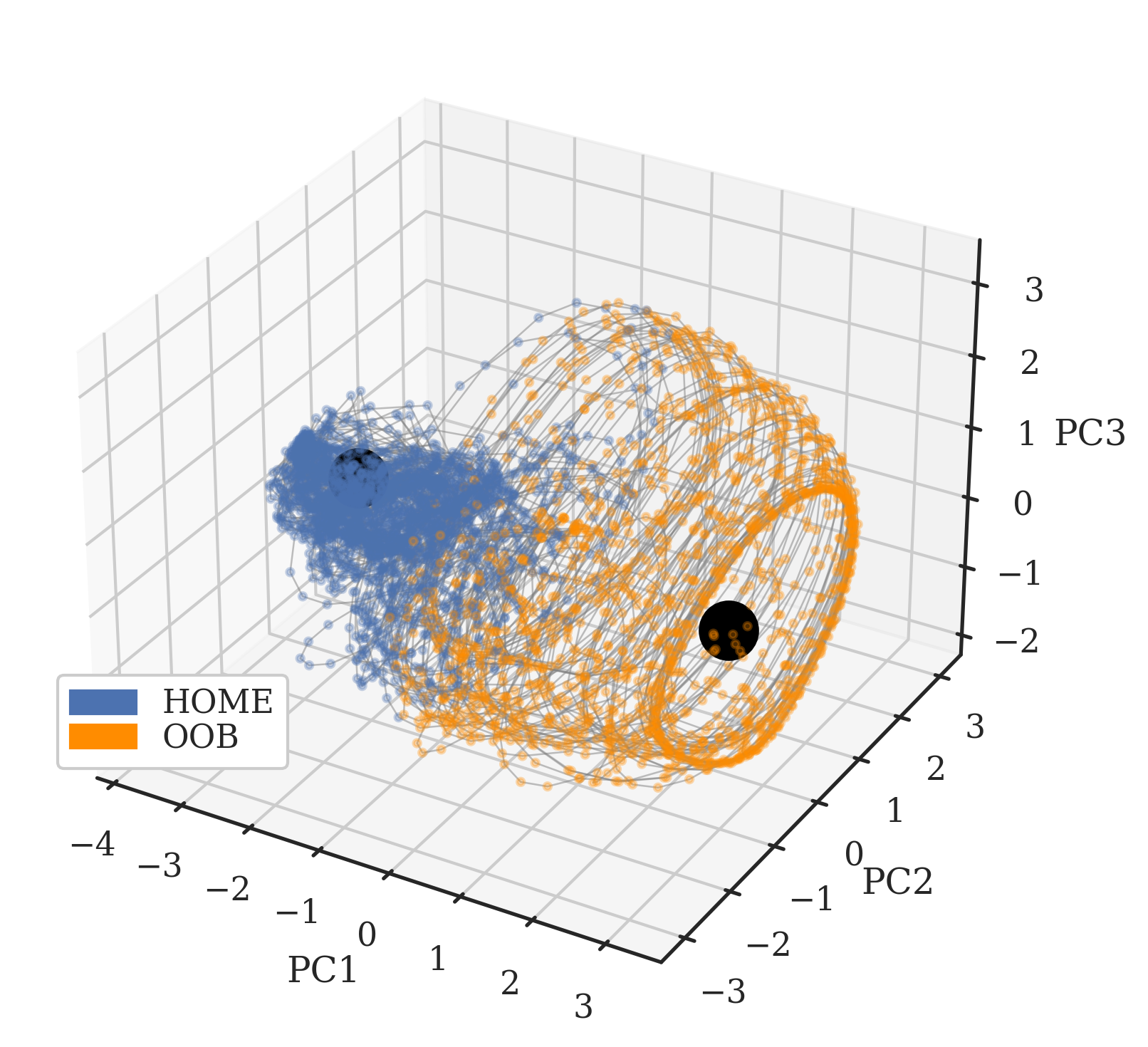} 
\includegraphics[width=0.30\linewidth]{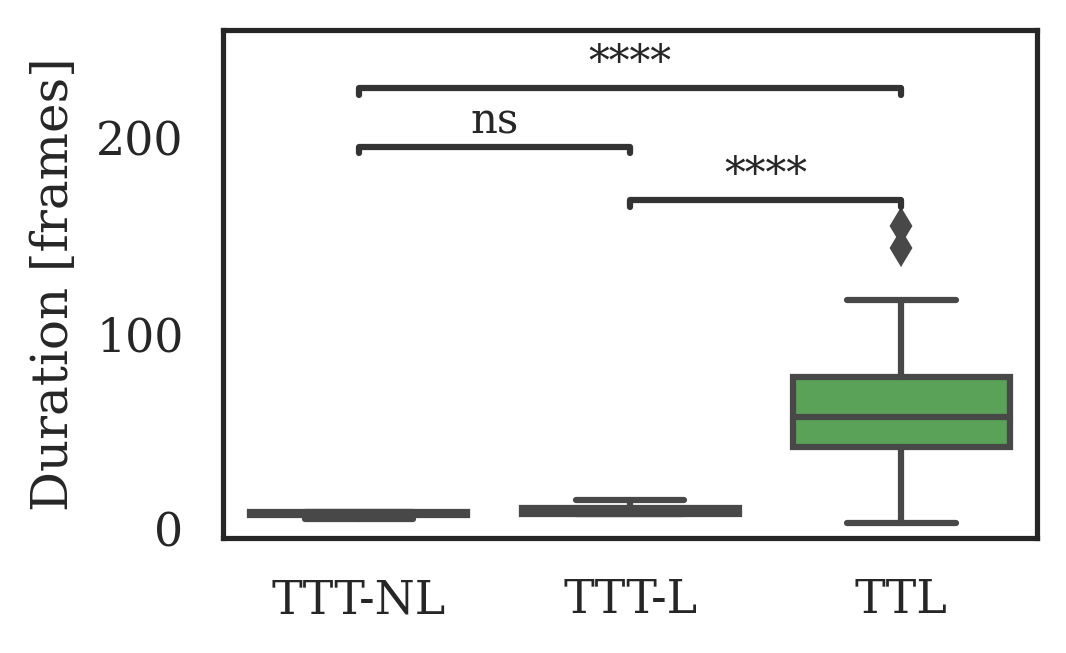}
\caption{Transitions between neural activity regimes -- Agent 5}
\end{figure*}

\clearpage
\section{RNN connectivity and stimulus integration timescales}
\label{sec_supp_eigen}

\textbf{Stimulus integration timescale $\tau_i$ calculation}:
(First see background on RNNs provided in Section \ref{sec_eigen}).
Prior literature has looked at the eigenvalues and eigenvectors of the recurrence Jacobian (and recurrence matrix) to investigate how connectivity affects the dynamics of the network \citep{rajan2006eigenvalue,maheswaranathan2019reverse}.  
Specifically \citep{maheswaranathan2019reverse} obtains the stimulus integration timescale $\tau_i$ associated with a stable eigenvalue $\lambda_i$ (i.e. $|\lambda_i| \leq 1$), by looking at the discrete-time iteration \mbox{$h_i(t)=\lambda_i^{t} h_i(0)$} that governs the integration of stimulus in the direction of eigenvector $v_i$ associated with $\lambda_i$.
They then compare this with the equivalent continuous time equation 
\mbox{$h_i(t)=h_i(0) e^{-t / \tau_i}$}, to get 
\mbox{$\tau_i = \left| ( 1/ \ln|\lambda_i| ) \right|$}.


\begin{table}[h!]
    \centering
    \begin{tabular}{ccccc}
     \hline\hline
     \textbf{Agent} & \textbf{Agent ID} & \textbf{Top 5 $\tau$s}  \\
     \hline   
        RNN 1 & 2760377 & 116.5, 81.5, 16.9, 13.5, 8.3  \\ \hline
        RNN 2 & 3199993 & 95.7, 61.7, 16.6, 12.0, 9.6  \\ \hline  
        RNN 3 & 3307e9 & 56.5, 13.0, 7.7, 6.8, 5.8  \\ \hline
        RNN 4 & 541058 & 86.4, 51.8, 15.1, 12.4, 9.7  \\ \hline
        RNN 5 & 9781ba & 86.2, 27.4, 8.6, 6.6, 5.6  \\ \hline
     \hline
    \end{tabular}
    \caption{Top 5 $\tau$s (stimulus integration timescales) for each RNN seed}
\label{table_supp_taus}
\end{table}

\begin{figure*}[h!]
\centering
\includegraphics[width=0.45\linewidth]{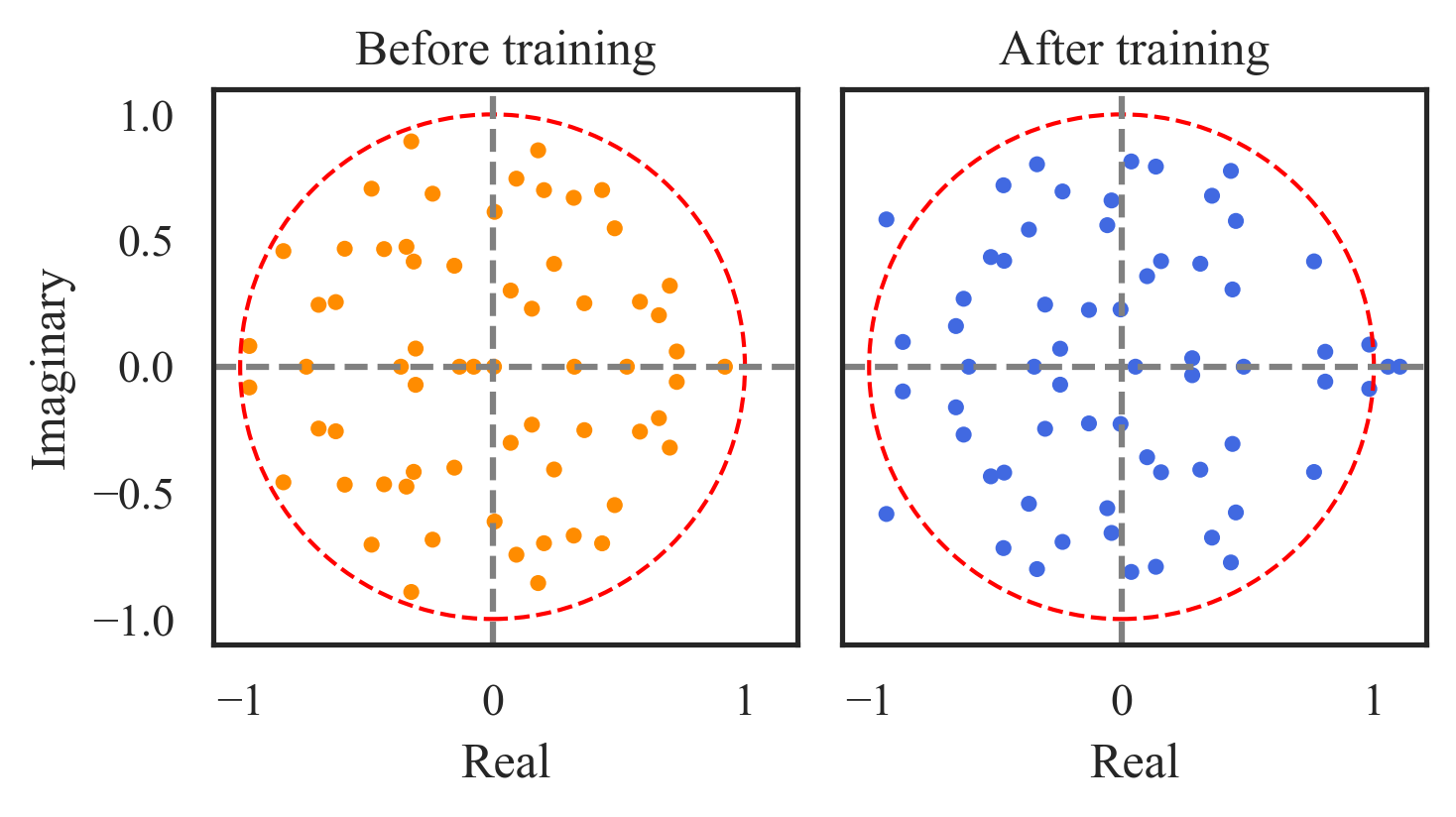}
\includegraphics[width=0.27\linewidth]{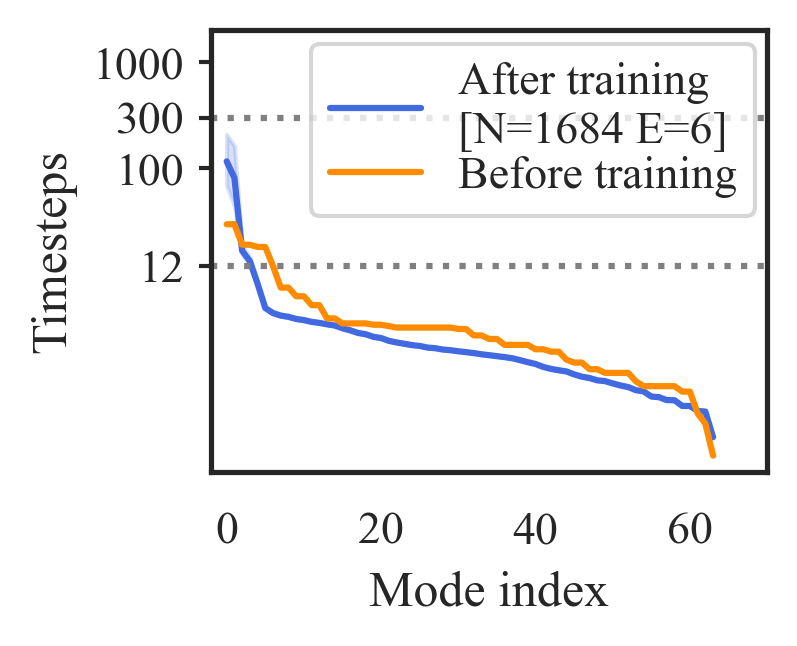}
\caption[Eigenspectra of $\bW_h$ before and after training, and stimulus integration timescales -- Agent 1]{Eigenspectra of $\bW_h$ before and after training, and stimulus integration timescales -- Agent 1 (compare with Agent 3 in Figure \ref{fig_eigen})}
\end{figure*}

\begin{figure*}[h!]
\centering
\includegraphics[width=0.45\linewidth]{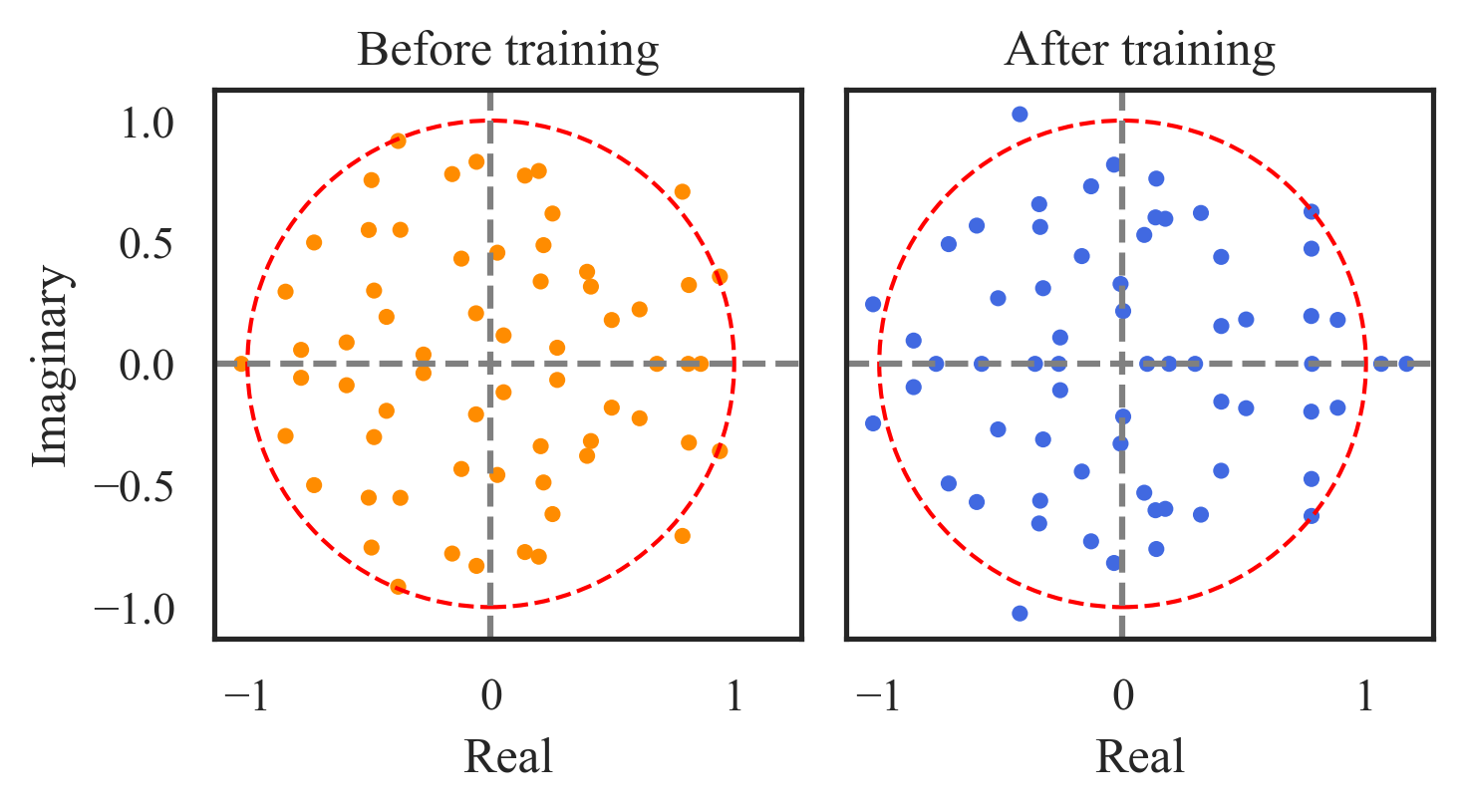}
\includegraphics[width=0.27\linewidth]{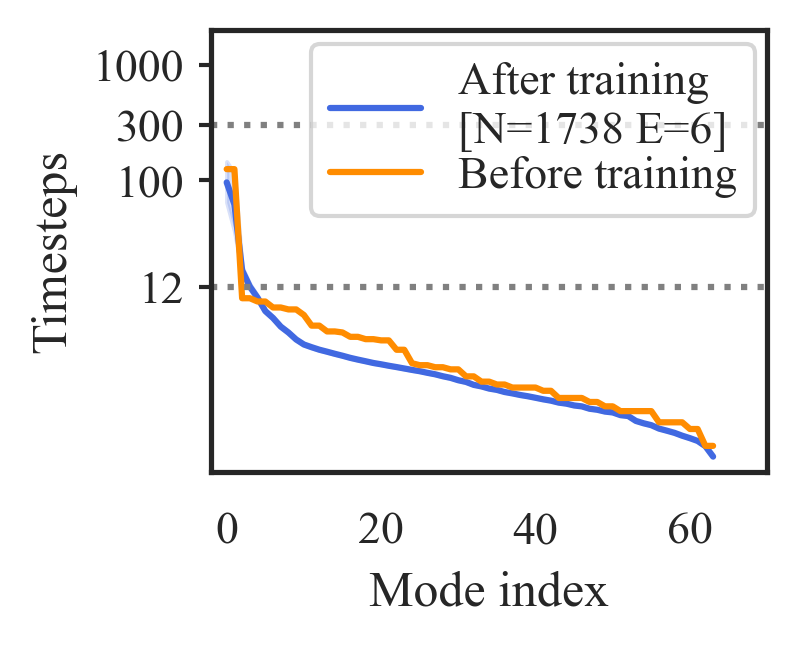}
\caption{Eigenspectra of $\bW_h$ before and after training, and stimulus integration timescales -- Agent 2}
\end{figure*}

\begin{figure*}[h!]
\centering
\includegraphics[width=0.45\linewidth]{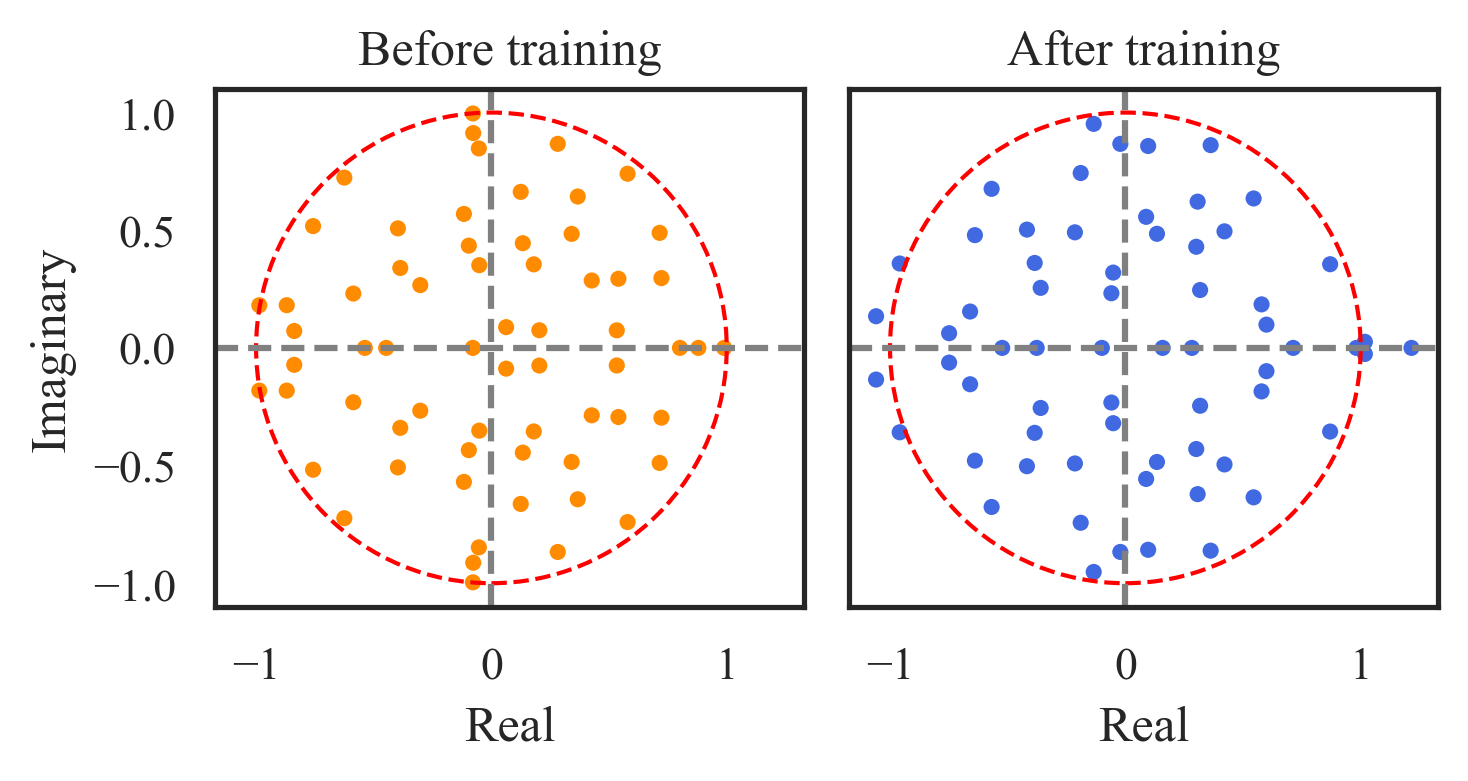}
\includegraphics[width=0.27\linewidth]{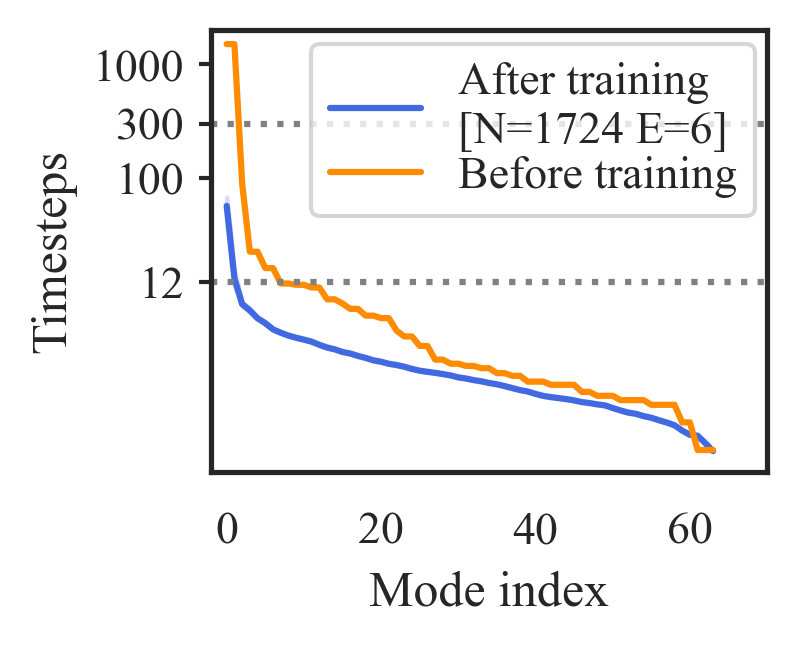}
\caption[Eigenspectra of $\bW_h$ before and after training, and stimulus integration timescales -- Agent 3]{Eigenspectra of $\bW_h$ before and after training, and stimulus integration timescales -- Agent 3 (same as Figure \ref{fig_eigen})}
\end{figure*}

\begin{figure*}[h!]
\centering
\includegraphics[width=0.45\linewidth]{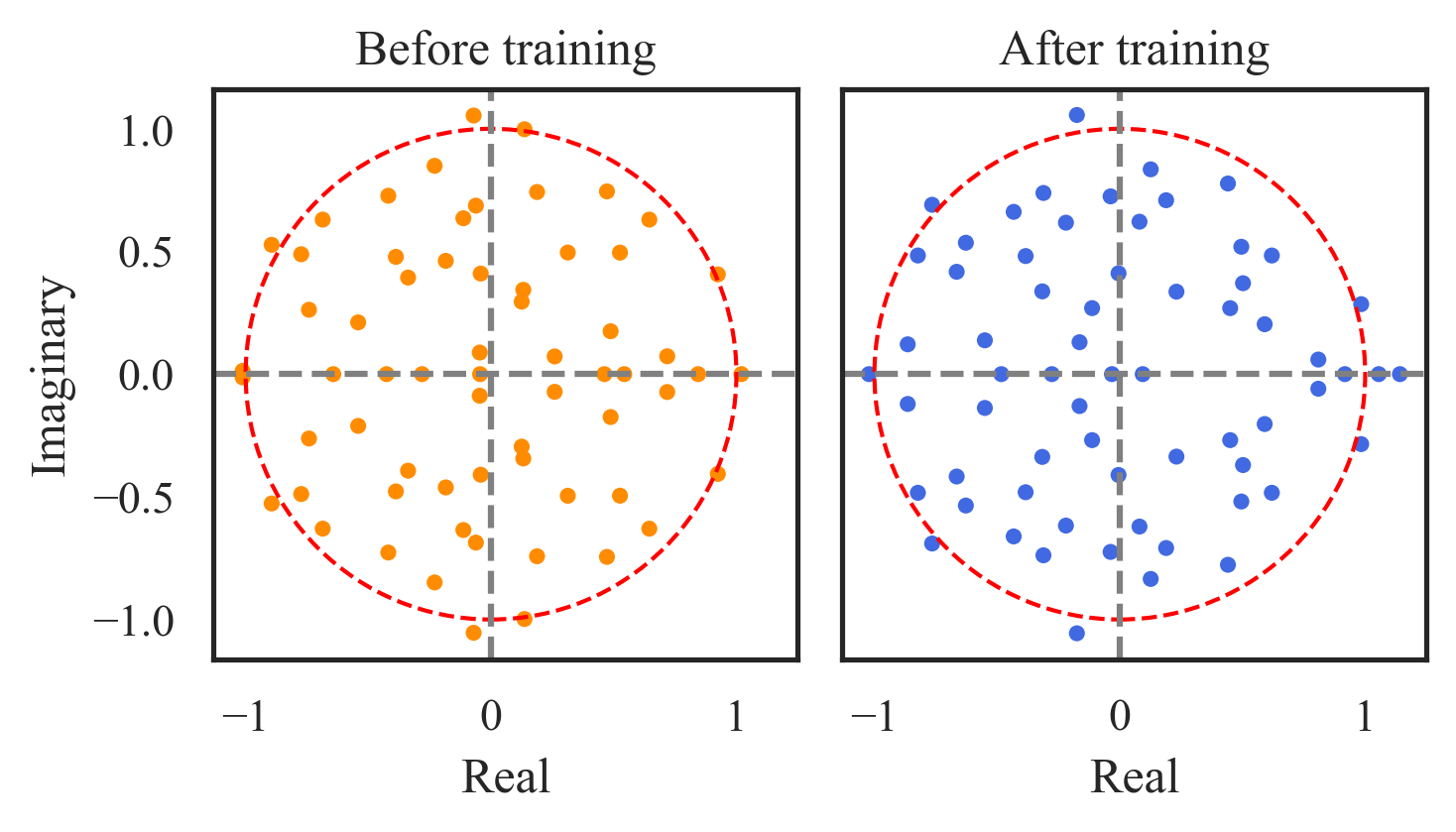}
\includegraphics[width=0.27\linewidth]{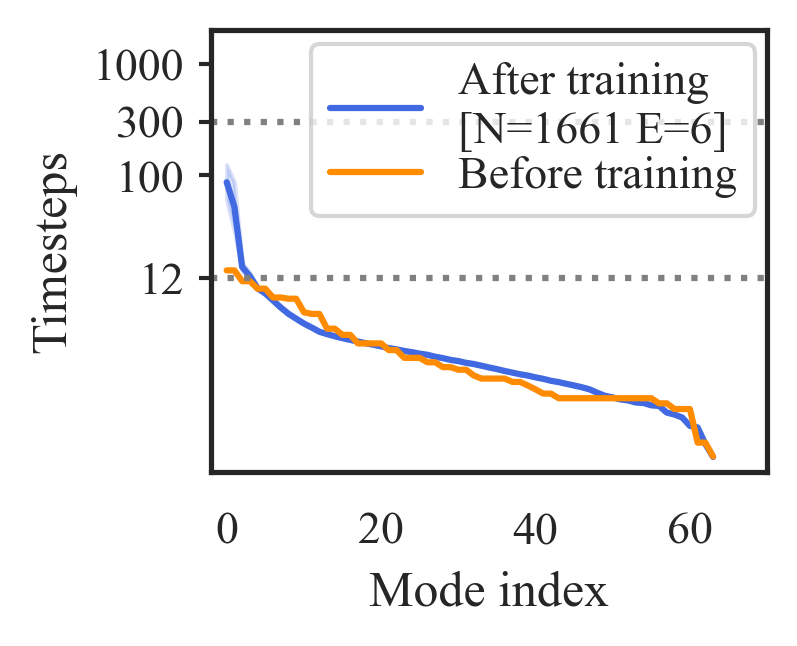}
\caption{Eigenspectra of $\bW_h$ before and after training, and stimulus integration timescales -- Agent 4}
\end{figure*}

\begin{figure*}[h!]
\centering
\includegraphics[width=0.45\linewidth]{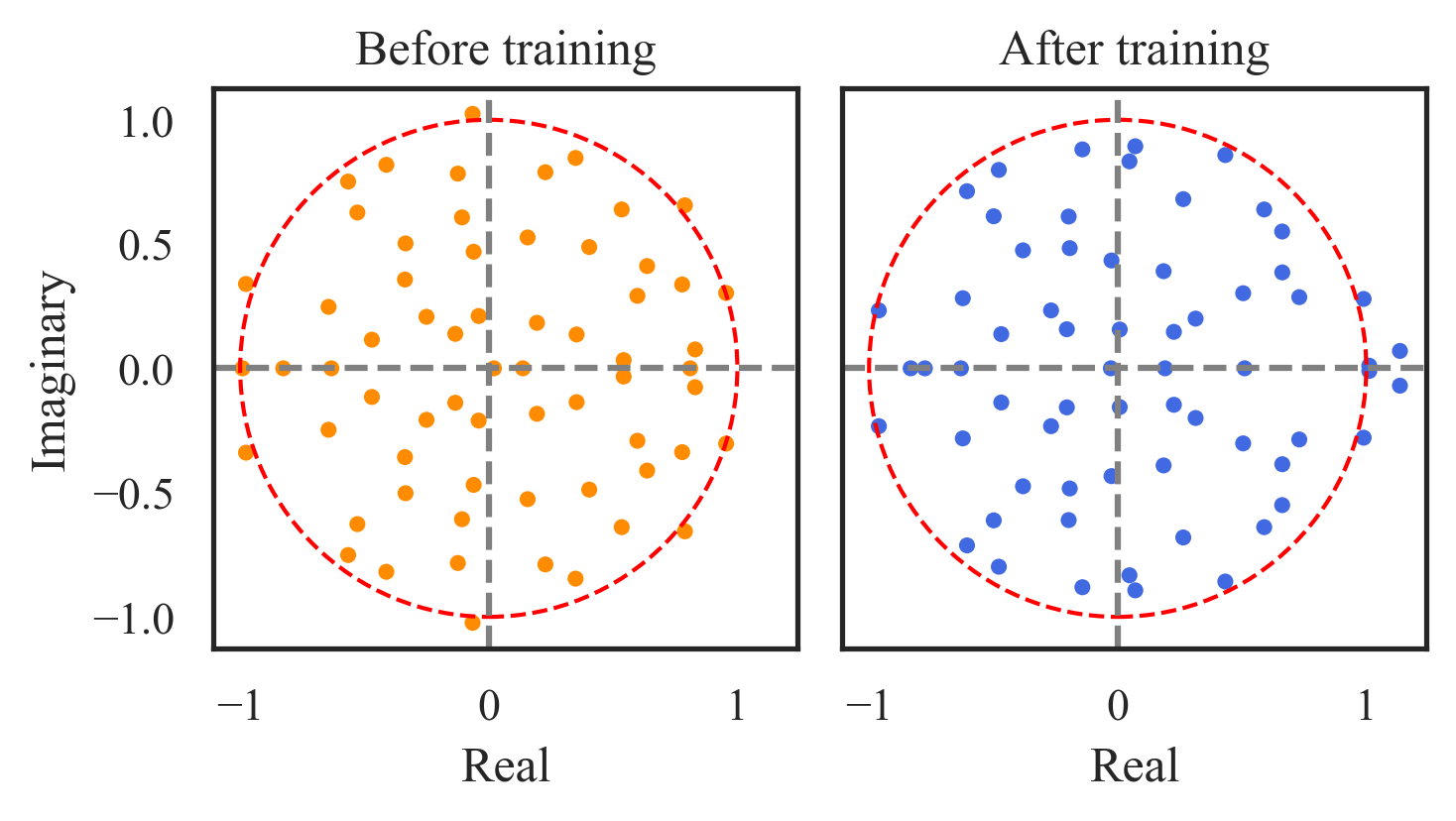}
\includegraphics[width=0.27\linewidth]{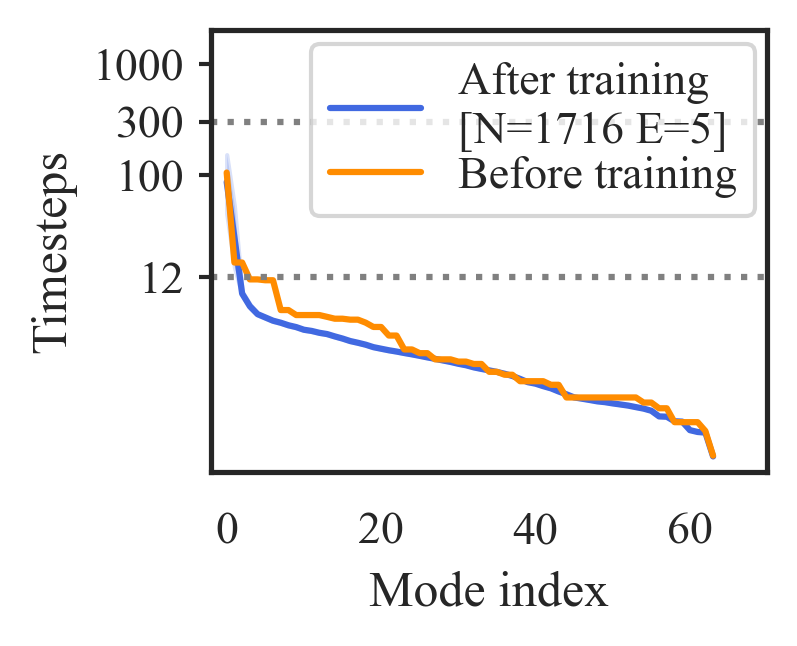}
\caption{Eigenspectra of $\bW_h$ before and after training, and stimulus integration timescales -- Agent 5}
\end{figure*}

\end{document}